\documentclass[floafix,prd,showpacs,showkeys,preprintnumbers,nofootinbib,superscriptaddress,11pt]{revtex4-1}
\usepackage[utf8]{inputenc}
\usepackage[sort&compress]{natbib}
\usepackage{ulem}
\usepackage{bm}
\usepackage{times}
\usepackage{amssymb,amsbsy,amsmath,amsfonts}
\usepackage{graphicx}
\usepackage{epstopdf}
\usepackage{float}
\usepackage{color}
\usepackage{morefloats}
\usepackage{rotating}
\usepackage{srcltx}
\usepackage{slashed}
\usepackage{subfigure}
\usepackage{multirow}
\usepackage{verbatim}
\usepackage{hyperref}
\usepackage{overpic}
\usepackage{tabularx}
\usepackage{makecell}
\usepackage{threeparttable}
\renewcommand{\arraystretch}{1.2}

\begin{document}

\title{ Three ways to decipher the nature of exotic hadrons: multiplets, three-body hadronic molecules, and correlation functions }

\author{Ming-Zhu Liu}
\affiliation{
Frontiers Science Center for Rare Isotopes, Lanzhou University,
Lanzhou 730000, China}
\affiliation{ School of Nuclear Science and Technology, Lanzhou University, Lanzhou 730000, China}

\author{Ya-Wen Pan}
\affiliation{School of Physics,  Beihang University, Beijing 102206, China}

\author{Zhi-Wei Liu}
\affiliation{School of Physics,  Beihang University, Beijing 102206, China}

\author{Tian-Wei Wu}
\affiliation{School of Science, Shenzhen Campus of Sun Yat-sen University, Shenzhen 518107, China}

\author{Jun-Xu Lu}
\affiliation{School of Physics, Beihang University, Beijing 102206, China}

\author{Li-Sheng Geng}\email{ lisheng.geng@buaa.edu.cn}
\affiliation{School of Physics, Beihang University, Beijing 102206, China}
\affiliation{Centrale Pekin, Beihang University, Beijing 100191, China}
\affiliation{Peng Huanwu Collaborative Center for Research and Education, Beihang University, Beijing 100191, China}
\affiliation{Beijing Key Laboratory of Advanced Nuclear Materials and Physics, Beihang University, Beijing 102206, China}
\affiliation{Southern Center for Nuclear-Science Theory (SCNT), Institute of Modern Physics, Chinese Academy of Sciences, Huizhou 516000, China}

\date{\today}
\begin{abstract}
In the past two decades, a plethora of hadronic states beyond the conventional quark model of $q\bar{q}$ mesons and $qqq$ baryons have been observed experimentally, which motivated extensive studies to understand their nature and the non-perturbative strong interaction.  
Since most of these exotic states are located near the mass thresholds of a pair of conventional hadrons, the prevailing picture is that they are primarily hadronic molecules.  In principle, one can verify the molecular nature of these states by thoroughly comparing their masses, decay widths, and production rates in a particular picture with experimental data. However, this is difficult or impossible. First, quantum mechanics allows for mixing configurations permitted by symmetries and quantum numbers. Second, data are relatively scarce because of their small production rates and the many difficulties in the experimental measurements. As a result, other alternatives need to be explored. This review summarizes three such approaches that can help disentangle the nature of the many exotic hadrons discovered.

In the first approach, based on the molecular interpretations for some exotic states, we study the likely existence of multiplets of hadronic molecules related by various  symmetries, such as isospin symmetry,
SU(3)-flavor symmetry, heavy quark spin/flavor symmetry, and heavy antiquark diquark symmetry, which are known to be approximately satisfied and can be employed to relate the underlying hadron-hadron interactions responsible for the formation of hadronic molecules.  The masses of these multiplets of hadronic molecules can then be obtained by solving the Lippmann-Schwinger equation. Their decay and production patterns are also related. As a result, experimental discoveries of such multiplets and confirmations of the predicted patterns will be invaluable to understanding the nature of these hadronic molecular states. 

In the second approach, starting from some hadronic molecular candidates, one can derive the underlying hadron-hadron interactions. With these interactions, one can study related three-body systems and check whether three-body bound states/resonances exist. The existence of such three-body molecules can directly verify the molecular nature of exotic hadrons of interest.  

In the third approach, one can turn to the femtoscopy technique to derive the hadron-hadron interactions, hence inaccessible. This technique provided an unprecedented opportunity to understand the interactions between unstable hadrons. Although the past focus was mainly on the light quark sector, we have seen increasing theoretical activities in the heavy quark sector in recent years. We review relevant studies and point out future directions where more effort is needed. 

Finally, to provide valuable information for present and future experiments, the decay widths and production rates of these multiplets and three-body hadronic molecules are estimated and discussed using the effective Lagrangian approaches.

\end{abstract}


\maketitle
\tableofcontents

\clearpage

\section{Introduction}

Quantum Chromodynamics (QCD), the underlying theory of the strong interaction, dictates that quarks can only exist in color singlet states, i.e., hadrons.
Theoretically, they can contain very complex structures~\cite{Bass:2004xa,Hagler:2009ni,Aidala:2012mv}. In the simplest form, hadrons are classified into mesons composed of a pair of quarks and antiquarks and baryons of three quarks.
 Hadrons with compositions different from the conventional mesons and baryons are named as exotic states including the following types: multiquark states composed of more than three quarks, glueballs composed of several gluons, and hybrids composed of both quarks and gluons.   It should be noted that there exist strong experimental signals for multiquark states such as the tetraquark states $Z_{c}(3900)$~\cite{BESIII:2013ris}, $T_{cc}(3875)$~\cite{LHCb:2021vvq}, and the pentaquark states $P_{c}(4312)$, $P_{c}(4440)$,  $P_{c}(4457)$~\cite{LHCb:2019kea}, but no unambiguous signals for either glueballs or hybrids~\cite{Morningstar:1997ff,Morningstar:1999rf,Chen:2005mg,Klempt:2007cp,Mathieu:2008me,Gregory:2012hu,Ochs:2013gi}.   Recently, the BESIII Collaboration determined the spin-parity and mass of $X(2370)$~\cite{BESIII:2023wfi},  regarded as a candidate for pseudoscalar glueballs.

The quark model has achieved great success in classifying the mesons and baryons discovered before 1960,  where the quarks are the so-called constituent quarks~\cite{Gell-Mann:1964ewy,Zweig:1964jf}. Based on the quark model, potential models,  such as the Cornell model~\cite{Eichten:1978tg} and the Godfrey-Isgur(GI) model~\cite{Godfrey:1985xj,Capstick:1986ter},  were applied to study the spectra, decays, and productions of mesons and baryons.  
The naive quark model can well describe the static properties of most ground-state hadrons,  e.g., the ground-state heavy quarknonia~\cite{Barnes:2005pb,Li:2009zu,Godfrey:2015dia}. However, for excited charmonia, the conventional quark model encounters many difficulties reproducing their masses, decay widths, and other properties.  Many of
them are generally regarded as candidates for exotic hadrons. Among them, the $X(3872)$ state discovered by the Belle Collaboration~\cite{Belle:2003nnu} may be the most well-known one, which has motivated extensive experimental and theoretical studies.  The exotic states have attracted the interest of theorists and experimentalists.   There have been many excellent reviews on studies of exotic hadrons in the past years; see, e.g., Refs.~\cite{Brambilla:2004jw,Swanson:2006st,Chen:2016qju,Hosaka:2016ypm,Lebed:2016hpi,Oset:2016lyh,Esposito:2016noz,Chen:2016spr,Richard:2016eis,Dong:2017gaw,Guo:2017jvc,Olsen:2017bmm,Ali:2017jda,Karliner:2017qhf,Guo:2019twa,Brambilla:2019esw,Liu:2019zoy,Chen:2022asf,Meng:2022ozq,Husken:2024rdk}.


To investigate the nature of these exotic states, one often calculates their mass, (partial) decay widths, and production rates in a given picture and a particular model. The potential model and quark pair creation model (often referred to as the $^3P_{0}$ model) have been successfully applied to explain the masses and strong decay widths of conventional hadrons~\cite{Eichten:1974af,Godfrey:1985xj,Capstick:1986ter,Micu:1968mk,LeYaouanc:1972vsx}. 
Most exotic states are located near the mass thresholds of hadron pairs, yet such coupled-channel effects are not considered in the conventional quark model.   In the language of the string theory, the string between the quarks will break if the distance between a pair of quarks reaches a certain length, creating a pair of light quarks from the vacuum~\cite{Bali:2000gf}.  Such quark rearrangements generate a pair of new hadrons, which can be estimated in the unquenched quark model~\cite{Zhou:2011sp,Ortega:2016mms,Luo:2019qkm,Yang:2021tvc,Deng:2023mza}.  The exotic hadrons are often explained in this picture as mixtures of hadronic molecules and conventional $q\bar{q}$ mesons or $qqq$ baryons. 
In particular,  the effects of coupled-channel hadron-hadron interactions can be considered by revising the confining potential as a screening potential in the conventional quark model~\cite{Li:2009ad,Li:2009nr,Song:2015nia}. As indicated in Ref.~\cite{Yin:2023wls},  the excitation modes of the same energy scale, e.g.,  quark radial excitations, quark pair creations, and meson creations, may compete in forming a hadron.

The compact multiquark model is another model to explain exotic hadrons~\cite{Jaffe:2003sg,Maiani:2004uc,Hogaasen:2004pm,Maiani:2004vq,Hogaasen:2005jv,Maiani:2005pe,Buccella:2006fn,Maiani:2007vr,Wu:2016vtq,Luo:2017eub,Maiani:2013nmn,Karliner:2014gca,Maiani:2014aja,Maiani:2015vwa,Maiani:2016wlq,Karliner:2016zzc,Ali:2019npk}.  
 A  simple model with only the color-magnetic interaction assigns some exotic hadrons as compact multiquark states, where the basic structure contains diquarks.    The diquarks composed of a pair of quarks are decomposed into  $3 \otimes 3=\bar{3} \oplus 6$  in terms of the SU(3)-color symmetry, where the representation $\bar{3}$ is equivalent to an anti-quark. Based on this symmetry, the parameters in the diquark model can be estimated. For instance, one can relate the masses of $\Sigma_c^{(*)}$ baryons or $\Xi_{cc}^{(*)}$ baryons with that of the doubly charmed tetraquark state $T_{cc}$~\cite{Karliner:2017qjm,Mehen:2017nrh,Cheng:2020wxa,An:2022vtg,Kim:2022mpa,Wu:2022gie}.  In this sense, the diquark model can simplify studies of multiquark systems, which otherwise are achieved by numerically solving the few-body Schr\"odinger equation~\cite{Yang:2020atz,Richard:2020zxb,Meng:2020knc}.  The MIT bag model embedding the asymptotic freedom at short distances and confinement at long distances has been applied to investigate the properties of conventional hadrons and multiquark states~\cite{DeGrand:1975cf,Jaffe:1976ig,Strottman:1979qu,Bernotas:2008bu,Bernotas:2013eia,Zhang:2021yul,Zhang:2023hmg}. 
The  QCD sum rule approach is another efficient method to handle the non-perturbative strong interaction, which has been widely applied to study the masses and decay widths of some exotic states~\cite{Lee:2007gs,Wang:2013vex,Dias:2013xfa,Qiao:2013dda,Chen:2015moa,Wang:2015epa,Chen:2017sci,Wang:2023dsm}.  Last, a model-independent first-principles method, without any ad hoc assumptions,  exists at the quark level, lattice QCD, to deal with the non-perturbative strong interaction. The QCD Lagrangian is reformulated in Euclidean space to computer correlation functions defined in the Feynman path integral formalism via numerical Monte Carlo Methods.  Lattice QCD simulations have made remarkable progress in studies of hadron spectroscopy~\cite{Dudek:2007wv,HadronSpectrum:2012gic,Dudek:2013yja,Moir:2013ub,Prelovsek:2014swa}, hadron-hadron interactions~\cite{Inoue:2010es,Mohler:2012na,Mohler:2013rwa,Ikeda:2013vwa,Prelovsek:2013cra,CLQCD:2015htz,Lang:2016hnn,Gongyo:2017fjb,Briceno:2017max,Aoki:2023qih,Alexandrou:2023cqg,Radhakrishnan:2024ihu}, electromagnetic form factors~\cite{Can:2013zpa,Can:2013tna,Bahtiyar:2015sga,Can:2015exa,Bahtiyar:2016dom,Can:2021ehb},  meson decay constants~\cite{Follana:2007uv,FermilabLattice:2011njy,Na:2012kp,Dowdall:2013tga,Boyle:2017jwu,Chen:2020qma}, hadron transition matrix elements~\cite{Na:2010uf,Bouchard:2014ypa,Meinel:2016dqj,Cooper:2020wnj,Parrott:2022rgu}, and so on.     This approach is quite promising in revealing the internal structure of some exotic states.  For instance, several lattice QCD studies~\cite{Liu:2012zya,Mohler:2013rwa,MartinezTorres:2014kpc,Bali:2017pdv,Cheung:2020mql} supported that the $DK$ interaction plays an important role in forming the exotic state $D_{s0}^{*}(2317)$, which is helpful to reveal its nature. A recent lattice QCD study~\cite{BaryonScatteringBaSc:2023zvt} provided support to the two-pole structure of the $\Lambda(1405)$ at unphysical light quark masses consistent with the theoretical expectation~\cite{Xie:2023cej,Guo:2023wes,Ren:2024frr,Zhuang:2024udv}.

A large number of theoretical studies have concluded that hadron-hadron interactions can affect the relevant conventional $q\bar{q}$ or $qqq$ bare states~\cite{Zhou:2011sp,Ortega:2016mms,Luo:2019qkm,Yamaguchi:2019vea,Yang:2021tvc}. Therefore, an outstanding issue is whether molecules mainly composed of a pair of conventional hadrons exist. One famous example is $\Lambda(1405)$, whose mass is difficult to understand in the constitute quark model as an $sud$ $\frac{1}{2}^{-}$ $p$-wave excited state~\cite{Isgur:1978xj}. However, it can be generated by the $\bar{K}N$  coupled-channel interaction in the unitary chiral approach~\cite{Kaiser:1995eg,Oset:1997it,Oller:2000fj}\footnote{Before its discovery in bubble chamber experiments  at low energy ${K}^- p$ scattering  into $\pi^- \Sigma^+$~\cite{Alston:1961zzd},  $\Lambda(1405)$  was theoretically predicted as a $\bar{K}N$ bound state~\cite{Dalitz:1959dn, Dalitz:1960du}. }.   The existence of a $\bar{K}N$ molecule has attracted long and heated discussions both experimentally and theoretically~\cite{Jido:2003cb,Lage:2009zv,Hyodo:2011ur,Guo:2012vv,Lu:2022hwm,Ahn:2003mv,HADES:2012csk,CLAS:2013rjt,J-PARCE15:2018zys,Anisovich:2020lec}, which triggered a lot of research on other likely hadronic molecules related by SU(3)-flavor symmetry~\cite{Jido:2003cb,Lage:2009zv,Hyodo:2011ur,Guo:2012vv,Lu:2022hwm}. Hadronic molecules containing heavy quarks can be traced to the proposal of $\psi(4040)$ as a $\bar{D}^{\ast}D^{\ast}$ hadronic molecule~\cite{DeRujula:1976zlg,Voloshin:1976ap}.  Although nowadays $\psi(4040)$ is widely viewed as a conventional charmonium state, the idea has inspired early theoretical studies on the possible existence of heavy hadronic molecules~\cite{Manohar:1992nd,Ericson:1993wy}. Such studies were revived after discovering $X(3872)$~\cite{Belle:2003nnu}.  The lower mass (compared to the conventional quark model prediction) and the isospin violation of the strong decays of $X(3872)$ can be easily understood, assuming that $X(3872)$ is a $D\bar{D}^{\ast}$ bound state or contains a sizable molecular component~\cite{Gamermann:2009fv,Li:2012cs,Wu:2021udi}.  

The molecular picture, where hadronic molecules are composed of conventional hadrons held together by the residual strong interaction,  is analogous to the picture where the nuclear force binds the deuteron (and atomic nuclei). The one-pion exchange mechanism was firstly applied to study deuteron-like loosely bound states by T\"ornqvist~\cite{Tornqvist:1991ks}, while the contributions of short-range interactions are necessary to bind a pair of hadrons in the charm and strangeness sectors~\cite{Tornqvist:1993ng,Swanson:2006st}. Inspired by the extensive studies of the nuclear force, the one-meson exchange theory has been applied to construct hadron-hadron potentials~\cite{Swanson:2003tb,Ding:2008gr,Liu:2008fh,Liu:2009qhy,Yang:2011wz,Liu:2019zvb},  where the $\pi$, $\sigma$, and $\rho(\omega)$ are responsible for the long-, medium-, and short-range interactions. In the One Boson Exchange (OBE) model, the $S$-wave potential is strongly related to the coupling constants of each exchanged meson and its spin-isospin quantum numbers~\cite{Rathaud:2017nkl}. However, the lattice QCD simulation of the strong interactions of the  $N\Omega$~\cite{HALQCD:2018qyu} and $\Omega_{ccc}\Omega_{ccc}$ systems~\cite{Lyu:2021qsh}  obtained bound states below their respective mass thresholds,  which is difficult to understand in the OBE mechanism. The potentials responsible for transitions between open-charm channels and hidden-charm channels, such as $\bar{D}D^* \to J/\psi \pi$~\cite{He:2017lhy} and $\bar{D}^*\Sigma_c\to J/\psi p$~\cite{Wu:2010jy,Xiao:2013yca,Roca:2015dva}   via the one-charm-meson exchange is much smaller than those via the quark potential model~\cite{Yamaguchi:2019djj},  which assumed that meson-meson scattering can occur via the quark-quark interaction through either the one-gluon exchange or the confining potential and is followed by quark rearrangements into color-singlet final states, i.e.,  the ``capture" and  ``transfer" diagrams~\cite{Barnes:1991em}. Recently, many studies argued that the vector meson exchange alone is enough to generate hadron-hadron potentials that can explain several exotic states as hadronic molecules in a unified picture~\cite{Geng:2008gx,Molina:2009ct,Lu:2017dvm,Dong:2021juy,Dong:2021bvy,Chen:2021cfl,Peng:2021hkr}.

Effective field theories (EFTs) are another widely adopted approach in exotic hadrons studies, which allow a systematic estimate of theoretical uncertainties. The downside is that it sometimes contains many unknown couplings, which need to be determined in one way or another, e.g., by fitting experimental or lattice QCD data. Chiral effective field theory(ChEFT) is one of QCD's most successful low-energy effective field theories. It is formulated in hadronic degrees of freedom and organized in chiral expansion, i.e., an expansion in powers of external momenta and light quark masses. ChEFT has been successfully applied to describe the $\pi\pi$~\cite{Gasser:1983yg,Gasser:1984gg} and $\pi N$~\cite{Gasser:1987rb} scattering, 
and then the nucleon-nucleon interaction~\cite{Weinberg:1990rz,Weinberg:1991um}, which has become the de-facto standard approach to constructing high precision nuclear forces~\cite{Epelbaum:2008ga,Machleidt:2011zz,Lu:2021gsb}. ChEFT has been extended to study exotic states in the heavy quark sector~\cite{Yao:2020bxx,Du:2021fmf,Yan:2021wdl,Meng:2022ozq} as well. Due to the many LECs that appear at higher chiral orders and the scarce experimental data for exotic states, one is often restricted to the lowest order, e.g., contact and one-pion-exchange terms~\cite{PavonValderrama:2019nbk,Du:2019pij,Du:2021zzh}. For heavy hadronic molecules, the one-pion exchange can often be treated as a perturbative correction~\cite{Lu:2017dvm}.    Nevertheless, there still exist unknown parameters, which can be estimated via relevant symmetries~\cite{Guo:2013sya,Liu:2018zzu,Yang:2020nrt} or the light meson saturation mechanism~\cite{Peng:2020xrf, Peng:2021hkr}. Both approaches can reduce the number of unknown parameters. It is worth noticing that in the chiral unitary approach (ChUA), which simplifies the scattering equation to an algebraic equation via the on-shell approximation, one can systematically study the dynamical generation of exotic hadronic molecules and the relevant invariant mass distributions~\cite{Oller:1997ti,Oset:1997it,Molina:2009ct,Molina:2010tx,Roca:2015dva,Chen:2015sxa,Oset:2016lyh,Xiao:2019aya}.

The   Breit-Wigner(BW) parameterization is generally used to analyze the invariant mass distribution of final states to extract the mass and width of a resonant state in experiments.  However, some peaks or dips in the invariant mass distributions are difficult to understand as genuine states~\cite{LHCb:2020bwg}. Instead, they can be better described as kinetic effects, such as triangle singularities or cusp effects~\cite{Rosner:2006vc,Mikhasenko:2015oxp,Aceti:2016yeb,Guo:2019twa,Dong:2020hxe,COMPASS:2020yhb}. A genuine state that interferes with coupled channels can generate a dip structure in the invariant mass distribution.  A relevant example is the dip in the $J/\psi J/\psi$ invariant mass distribution around 6.8~GeV that can be reproduced by considering coupled-channel effects~\cite{LHCb:2020bwg,Dong:2020nwy}. In Ref.~\cite{Dong:2020hxe}, Dong et al. discussed the dependence of the cusp effect on the reduced mass and the potential strength of a pair of hadrons, which helps discriminate the cusp effect from a genuine state.   On the other hand, triangle singularities always proceed via three on-shell intermediate particles in a loop diagram~\cite{Wu:2011yx, Wang:2016dtb,Sakai:2017hpg,Bayar:2016ftu}. They are located in the physical region and may produce observable effects in the invariant mass distributions~\cite{Guo:2019twa}. 
Recently, after early claims by the COMPASS Collaboration of the
``$a_1(1420)$" discovery~\cite{COMPASS:2015kdx}, it was explained as the consequence of a triangle singularity in
which the $a_1(1260)$ resonance decays into $K^*\bar{K}$, $K^*\to \pi K$ and then $K\bar{K}$ fuses to generate the
$f_0(980)$ resonance, the $\pi f_0(980)$ being the observed decay mode~\cite{Mikhasenko:2015oxp,Aceti:2016yeb,COMPASS:2020yhb}.
Therefore, it is necessary to consider kinetic effects in analyzing peaks in the experimental invariant mass distributions.  Every enhancement in the invariant mass distribution contains crucial information on the interactions, which can be employed to reveal further the inner structures or properties of related genuine states. In Ref.~\cite{Guo:2019qcn}, it is argued that the enhancement in the mass distribution of $X(3872)\gamma$ induced by the triangle diagram mechanism is helpful to determine whether $X(3872)$ is above or below the $D\bar{D}^{*0}$ mass threshold. The cusp for $\pi^0\pi^0$ scattering is expected at the $\pi^+\pi^-$ threshold~\cite{Meissner:1997gf}, which is used to extract their scattering length difference with high precision~\cite{Cabibbo:2004gq,Cabibbo:2005ez,Colangelo:2006va,Bissegger:2007yq,Batley:2009ubw,Liu:2012dv}.

Once the hadron-hadron interactions are determined, one can obtain the pole positions (mass and width) in the vicinity of the mass threshold of a pair of hadrons, and then in the molecular picture, analyze other physical observables, such as the invariant mass distributions and production rates. In Refs.~\cite{Braaten:2004fk,Braaten:2005jj,Roca:2015dva,Chen:2015sxa,Oset:2016lyh,Du:2019pij}, assuming that the weak interaction vertex can be parameterized with one single unknown parameter, the invariant mass distributions in the exclusive weak decays are used to identify the molecular nature of exotic states, where the final-state interactions dynamically generate the hadronic molecules.    
However, the internal mechanism of three-body weak decays is unknown, which is often parameterized assuming SU(3)-flavor symmetry ~\cite{Zito:2004kz,Chen:2008sw,Yang:2017nde,Geng:2018upx}, in the generalized factorization approach~\cite{Chua:2001vh,Chua:2002wn,Chua:2002yd,Geng:2005wt,Geng:2006wz,Geng:2006jt,Cheng:2012fq}, the pole model~\cite{Cheng:2001tr,Cheng:2002fp,Cheng:2005bi,Cheng:2006bn}, and the perturbative QCD approach~\cite{Wang:2016rlo,Rui:2017bgg,Zou:2020atb,Fang:2023dcy}. These three-body decays accompanied by the productions of exotic states can also be described in the triangle mechanism~\cite{Wu:2019rog,Hsiao:2019ait,Ling:2021qzl,Burns:2022uiv,Pan:2023hrk}, which usually proceed via the Cabibbo-favored two-body weak decays that can be well described through the naive factorization approach~\cite{Chau:1982da,Chau:1987tk}. Some exotic states can be produced in the inclusive process in $pp$, $e^+e^-$, $p\bar{p}$, and heavy ion collisions, which have been studied in the molecular picture with the statistical model~\cite{Andronic:2005yp,Andronic:2008gu,Andronic:2010qu,Cho:2019syk}, the coalescence model~\cite{ExHIC:2010gcb,ExHIC:2017smd,ExHIC:2017smd,Sun:2017ooe,Sun:2018mqq,Yun:2022evm}, the multiphase transport model(AMPT) model~\cite{Zhang:2020dwn}, and the PACIAE model~\cite{Xu:2021drf,Wu:2022wgn}. Moreover,  
the pentaquark states can be produced in the $J/\psi$ photoproduction off protons~\cite{Wang:2015jsa,HillerBlin:2016odx,Wang:2019krd,GlueX:2019mkq}, which can distinguish whether they are genuine states or anomalous triangle singularities.    

In the past two decades, many new hadrons have been discovered~\cite{Chen:2021ftn,Liu:2023hhl,Jia:2023upb}. Due to their proximity to the thresholds of pairs of conventional hadrons and unique properties, many can only be understood as hadronic molecules or considering final-state interactions. The rapidly evolving experimental situation has inspired a large number of theoretical works. It is impossible to cover all of them in the present review. In addition, there are already many excellent reviews in this regard.  See, e.g., Refs.~\cite{Brambilla:2004jw,Swanson:2006st,Chen:2016qju,Hosaka:2016ypm,Lebed:2016hpi,Oset:2016lyh,Esposito:2016noz,Chen:2016spr,Richard:2016eis,Dong:2017gaw,Guo:2017jvc,Olsen:2017bmm,Ali:2017jda,Karliner:2017qhf,Guo:2019twa,Brambilla:2019esw,Liu:2019zoy,Chen:2022asf,Meng:2022ozq} 
As a result, we only focus on three relatively new but less-discussed approaches, which can shed light on the
nature of exotic states, particularly whether they fit into the molecular picture and how to verify or refute such a picture. 

 The first approach relies on symmetry considerations.  Historically, the discovery of the $\Omega$ baryon helped verify the quark model, where the SU(3)-flavor symmetry plays a vital role~\cite{Gell-Mann:1964ewy}. Such an approach has been extended to study hadronic molecules. For instance,  assuming $X(3872)$ is a $J^{PC}=1^{++}$ $\bar{D}D^*$ bound state, it is natural to expect the existence of a $J^{PC}=2^{++}$ $\bar{D}^*D^*$ bound state and a  $J^{PC}=1^{++}$ $\bar{B}B^*$ bound state (denoted as $X_{b}$) according to heavy quark spin symmetry (HQSS) and heavy quark flavor symmetry (HQFS)~\cite{Molina:2009ct,Nieves:2011zz,Sun:2011uh,Nieves:2012tt,Baru:2016iwj,Liu:2019stu,Dai:2022ulk}. 
 The CMS Collaboration has searched for the $X_{b}$ in the decay channel of $\Upsilon(1S)\pi\pi$  in $pp$ collisions without success~\cite{CMS:2013ygz}, and the ATLAS Collaboration neither obtained a significant signal in the same process~\cite{ATLAS:2014mka}.      It should be noted that the isospin breaking of $X_b \to \Upsilon(1S)\pi\pi $  is not as large as $X(3872) \to J/\psi \pi \pi $ since the  $X_b$ mass is above the   $ \Upsilon(1S) \rho$ mass threshold more than $300$~MeV but the $X(3872)$ mass is quite close to the $J/\psi \rho$ mass threshold, which implies the $\Upsilon(1S)\pi\pi$ channel is not a proper channel to search for $X_b$. One should use the $\Upsilon(1S)\pi\pi \pi$ channel~\cite{Guo:2013sya,Guo:2014sca} instead. The Belle Collaboration  obtained negative results in the decay channel of $\Upsilon(1S)\omega$ in $e^+ e^-$ collisions~\cite{Belle:2014sys}. As a result, 
 based on the molecular nature of exotic states,  various symmetries,  such as
SU(3)-flavor symmetry, HQSS, HQFS, and heavy antiquark diquark symmetry(HADS) predicted the existence of their partners~\cite{Guo:2013sya,Liu:2018zzu,Yang:2020nrt}.  One can see that the existence of these partners is tied to the molecular nature of these states and the underlying hadron-hadron interactions. Thus, experimental searches for these partners can help verify or refute their molecular nature. We review the studies of several good molecular candidates and their partners related by various symmetries, including their masses, decay widths, and production mechanisms.

Along with this idea, we proposed a second approach to verify the molecular nature of exotic states. Nowadays, it is well-accepted that atomic nuclei are made of nucleons. For instance, the deuteron comprises a neutron and a proton, the triton consists of two neutrons and a proton, and the alpha particle contains two neutrons and two protons.  In practice, using the precise two-body $NN$ potential, one can reproduce most properties of the ground-state and low-lying excited states of light nuclei, where the residual three-body potential $NNN$ is also considered~\cite{Ekstrom:2015rta,Soma:2019bso,Hammer:2019poc,Hergert:2020bxy,Elhatisari:2022zrb}.~\footnote{We note that none of the existing studies can simultaneously describe medium-mass nuclei and nuclear matter in a truly ab initio way~\cite{Machleidt:2023jws,Machleidt:2024bwl}.} Such a picture has been successfully extended to studies of hypernuclei, i.e., adding one or more hyperons to the nuclear system~\cite{Hiyama:2003cu,Hiyama:2018lgs}.  We can replace the deuteron bound by the $NN$ interaction with other molecular candidates generated by certain hadron-hadron interactions. Then, we can predict the existence of three-body or even four-body hadronic molecules~\cite{Wu:2021dwy,Wu:2022ftm,Pan:2023zkl}. One example is that assuming $D_{s0}^{*}(2317)$ is a $DK$ bound state, three-body $DDK$ and four-body $DDDK$ hadronic molecules exist. These multi-hadron states rely on the underlying $DK$ interaction~\cite{Wu:2019vsy}. In other words, the  $DDK$ and $DDDK$ molecules, if discovered experimentally or on the lattice, will verify the molecular nature of $D_{s0}^{*}(2317)$.  This review discusses the studies on the masses, decay widths, and production rates of possible three-body hadronic molecules associated with some chosen two-body molecular candidates of an exotic nature.

Due to the low production rates of hadrons containing heavy quarks, direct experimental studies of hadron-hadron interactions are challenging. In recent years, it has been shown that the femtoscopic technique is a promising method for such purposes~\cite{STAR:2014dcy, STAR:2015kha, ALICE:2019hdt, ALICE:2020mfd, ALICE:2021cpv}, accessible in high energy $pp$, $pA$, and $AA$ collisions~\cite{ExHIC:2017smd, Fabbietti:2020bfg}.  Such approaches have recently been utilized in studying exotic hadronic molecular candidates~\cite{Liu:2023uly, Albaladejo:2023pzq, Khemchandani:2023xup, Ikeno:2023ojl, Torres-Rincon:2023qll, Kamiya:2022thy, Vidana:2023olz, Albaladejo:2023wmv, Liu:2023wfo, Feijoo:2023sfe,Liu:2024nac}, which show great potential in deriving the relevant hadron-hadron interactions and deciphering the nature of the many exotic hadrons discovered. We will review recent theoretical and experimental femtoscopy studies of the hadron-hadron interactions related to understanding the nature of exotic hadrons.

This review is organized as follows. In Sec.~\ref{experiment}, we briefly introduce the experimental progress on the exotic states covered in the present review.
In Sec.~\ref{multiplet molecules}, we adopt the contact-range EFT approach to assign several exotic states as hadronic molecules and predict the existence of multiplets employing various symmetries, including SU(3) flavor symmetry, HQSS,  HQFS, and HADS.   In Sec.~\ref{three-body}, based on specific good candidates of hadronic molecules and the underlying hadron-hadron interactions, we predict the corresponding three-body hadronic molecules by solving the three-body Shc\"odinger equation. In Sec.~\ref{sec:CF}, we review recent efforts in decoding the nature of specific candidates of hadronic molecules using the femtoscopic technique.
Finally, this review ends with a summary and outlook in Sec.~\ref{sum}. We relegate some of the theoretical tools, which are helpful to understanding the studies covered in this review, to the Appendices.

\section{ Experimental and theoretical progress on selected exotic hadrons}
\label{experiment}

In recent years, with ever-increasing energy and statistics and rapid developments in detection and analysis techniques, a large number of new hadrons containing heavy quarks have been discovered experimentally~\cite{Brambilla:2010cs,Brambilla:2019esw,Chen:2021ftn,Liu:2023hhl,Jia:2023upb,Johnson:2024omq}. In the following, we briefly introduce some selected exotic states that are deemed as robust hadron molecular candidates. These hadrons are mainly discovered in two modes, i.e., inclusive and exclusive processes, and the latter usually proceeds via  $b$-hadron decays. Table~\ref{productions} and Table~\ref{proexotic2} summarize the productions and decay modes of the heavy hadronic molecular candidates covered in this review. In addition, we also briefly review related theoretical studies, particularly those in the hadronic molecular picture, to put things into perspective.     

\begin{table*}[htpb]
  \caption{
 Production and decay modes  of 
$D_{s0}^*(2317)$,  $D_{s1}(2460)$, $X(3872)$, $X(4014)$, $Y(4220)$, and $Y(4360)$.    \label{productions}
}
\centering
\begin{tabular}{cccccc}
  \hline\hline State& $J^{P}$ & Inclusive process & Decay modes  & Exclusive process &  Decay modes \\ \hline
$D_{s0}^*(2317)$ & $0^{+}$ & $e^+e^-$~\cite{BaBar:2003oey,CLEO:2003ggt,Belle:2003kup,BESIII:2017vdm} & $D_s^{+}\pi^0$   & {$B \to \bar{D}^{(*)}D_{s0}^*$~\cite{Belle:2003guh,BaBar:2004yux,Belle:2015glz}}   & {$D_s^{+}\pi^0$}      
  \\ \hline  \multirow{3}{1.4cm} {$D_{s1}(2460)$}  &  \multirow{3}{0.4cm} {$1^{+}$}   &  $e^+e^-$~\cite{CLEO:2003ggt,Belle:2003kup}  & $D_s^{\ast+}\pi^0$  & \multirow{3}{3.8cm}{$B \to \bar{D}^{(*)}D_{s1}$~\cite{Belle:2003guh,BaBar:2004yux,BaBar:2006jvx}}   & \multirow{3}{1.4cm}{$D_s^{\ast+}\pi^0/D_{s}^+\gamma$} 
  \\   &   &  $e^+e^-$~\cite{Belle:2003kup}  & $D_s^{+}\gamma/D_s^{+}\pi^{+}\pi^{-}$    &  &     \\   &   &  $e^+e^-$~\cite{BaBar:2003cdx,BaBar:2006eep}  & $D_s^{+}\gamma\pi^0$    &    &  
   \\ \hline
 \multirow{11}{1.4cm} {$X(3872)$}  & \multirow{11}{0.4cm} {$1^{+}$} &$p\bar{p}$~\cite{CDF:2003cab,D0:2004zmu,CDF:2005cfq,CDF:2006ocq,CDF:2009nxk}   & $J/\psi\pi^+\pi^-$     & $B^+ \to X(3872) K^+$~\cite{Belle:2003nnu,BaBar:2004oro,BaBar:2005pcw,BaBar:2008qzi,LHCb:2013kgk,LHCb:2020fvo}  & $J/\psi\pi^+\pi^-$    \\ 
 &  &$p{p}$~\cite{LHCb:2011zzp,CMS:2013fpt}   & $J/\psi\pi^+\pi^-$     & $B^+ \to X(3872) K^+$~\cite{BaBar:2006fjg,BaBar:2008flx,Belle:2011wdj,LHCb:2014jvf,LHCb:2024tpv}  & $J/\psi \gamma/\psi(2S)\gamma$
 \\  &  &  $e^+e^-$~\cite{BESIII:2019esk}  & $\pi^0\chi_{c1}$      & $\Lambda_b^0 \to X(3872) p K^-$~\cite{LHCb:2019imv}  & $J/\psi\pi^+\pi^-$    
 \\  &  &  $e^+e^-$~\cite{BESIII:2019qvy}   & $J/\psi \omega$     & $B^+ \to X(3872) K^+$~\cite{LHCb:2015jfc}  & $J/\psi \rho^0$
 \\  &  &  $e^+e^-$~\cite{BESIII:2020nbj}   &  $\bar{D}^{*0}D^0$    & $B_s^0 \to X(3872) \phi$~\cite{LHCb:2020coc,CMS:2020eiw}  & $J/\psi\pi^+\pi^-$ 
 \\  &  &    &      & $Y(4260) \to X(3872) \gamma$~\cite{BESIII:2013fnz}  & $J/\psi\pi^+\pi^-$ 
 \\  &  &      &      & $B^+ \to X(3872) K^+$~\cite{BaBar:2006fjg}  & $J/\psi \gamma$
 \\  &  &    &   & $B^+ \to X(3872) K^+$~\cite{BaBar:2007cmo,Belle:2008fma}  & $\bar{D}^{*0}D^0$
  \\  &  &    &     & $B^+ \to X(3872) K^+$~\cite{BaBar:2010wfc}  & $J/\psi \omega$
  \\  &  &   &      & $B^+ \to X(3872) K^+$~\cite{Belle:2006olv}  & $D^0 \bar{D}^0 \pi^0$
    \\  &  &   &      & $B^+ \to X(3872) \pi^+\pi^-$~\cite{LHCb:2023reb}  & $J/\psi\pi^+\pi^-$
  \\ $X(4014)$  &$?^{?}$  &$ \gamma\gamma$~\cite{Belle:2021nuv}    & $\psi(2S)\gamma$\\ \hline
 \multirow{5}{1.4cm} {$Y(4220)$}  & \multirow{5}{0.4cm} {$1^{-}$} &$e^+e^-$  & $J/\psi\pi^+\pi^-$~\cite{BESIII:2016bnd},~$h_c \pi^+\pi^-$~\cite{BESIII:2016adj}   &    &   \\ 
 &  &$e^+e^-$   & ~$\psi(2S)\pi^+\pi^-$~\cite{BESIII:2017tqk,BESIII:2021njb}    &     \\ 
 &  &$e^+e^-$   &$\omega \chi_{c0}$~\cite{BESIII:2014rja,BESIII:2019gjc}, ~$J/\psi K^+ K^-$~\cite{BESIII:2022joj}       &      
   \\ 
 &  &$e^+e^-$   &$\pi^+D^0D^{\ast-}$~\cite{BESIII:2018iea}, ~$\pi^{+}D^{*0}D^{*-}$~\cite{BESIII:2023cmv}    &    
  \\ 
 &  &$e^+e^-$   &$J/\psi K_{s}^0K_{s}^0$~\cite{ BESIII:2022kcv},~$J/\psi \eta$~\cite{BESIII:2020bgb,BESIII:2023tll}    &      \\ \hline
  \multirow{5}{1.4cm} {$Y(4360)$}  & \multirow{5}{0.4cm} {$1^{-}$} &$e^+e^-$  & $\psi(2S)\pi^+\pi^-$~\cite{BaBar:2006ait,Belle:2007umv,BaBar:2012hpr,Belle:2014wyt,BESIII:2017tqk}  &    &   \\ 
 &  &$e^+e^-$   &   $J/\psi \pi^+\pi^-$~\cite{BESIII:2016bnd},~$h_c \pi^+\pi^-$~\cite{BESIII:2016adj}     &     \\ 
 &  &$e^+e^-$   &$J/\psi \eta$~\cite{BESIII:2020bgb,BESIII:2023tll}     &      
   \\ 
 &  &$e^+e^-$   &  $\pi^+\pi^- D^+D^-$~\cite{BESIII:2022quc}   &    
  \\ 
 &  &$e^+e^-$   &  $\gamma \chi_{c2}$~\cite{BESIII:2021yal}  &    
\\ \hline \hline
\end{tabular}
\end{table*}

\subsection{$D_{s0}^*(2317)$ and $D_{s1}(2460)$}

In 2003, the BaBar Collaboration discovered a narrow state near 2.32 GeV in the $D_{s}^{+}\pi^{0}$ invariant mass distribution~\cite{BaBar:2003oey}, named $D_{s0}^{*}(2317)^+$. It was subsequently confirmed by the CLEO~\cite{CLEO:2003ggt} and Belle Collaboration~\cite{Belle:2003guh}. In addition, a new state near  2.46 GeV, named $D_{s1}(2460)$, was also discovered~\cite{CLEO:2003ggt,Belle:2003guh}. The BESIII Collaboration  observed $D_{s0}^{*}(2317)$ in the process of $e^+ e^- \to D_{s}^{\ast+}D_{s0}^{*}(2317)^-$ with a statistical significance  more than $5\sigma$~\cite{BESIII:2017vdm}.  Assuming  $D_{s0}^*(2317)$ and $D_{s1}(2460)$  as the $c\bar{s}$ charmed strange mesons of  $J^P=0^+$ and $J^P=1^+$,  their masses are lower by around  160 and 70 MeV than the prediction of the GI  model~\cite{Godfrey:1985xj}\footnote{According to the GI model, there exist two $P$-wave $1^+$ $c\bar{s}$ states, $^1P_1$  and $^3P_1$, with masses of $2.53$ and $2.57$ GeV.   } .  Therefore, $D_{s0}^{*}(2317)$ and $D_{s1}(2460)$  are often considered exotic states.  
In addition to its dominant decay mode  $D_{s}^{+}\pi^{0}$~\cite{BaBar:2004yux,BaBar:2006eep}, the upper limit of the branching fraction of  the $D_{s0}^{\ast}(2317)^+$ decaying into $D_{s}^{+}\gamma$, $D_{s}^{\ast+}\gamma$, and $D_{s}^{\ast+}\pi^0$
were obtained by the Belle~\cite{Belle:2003kup,Belle:2003guh} and BaBar~\cite{BaBar:2004gtd}  Collaborations. The decays of $D_{s1}(2460)^+$ into $D_s^+\gamma$, $D_s^+\pi^0\gamma$\footnote{The significant decay mode of $D_s^+\pi^0\gamma$ is from $D_s^{*+}\pi^0$~\cite{BaBar:2003cdx} }, and $D_s^+ \pi^+\pi^-$ were observed by the BaBar Collaboration~\cite{BaBar:2004gtd,BaBar:2006eep,BaBar:2006jvx}. The upper limit of the branching fraction of the decay $D_{s1}(2460)^+ \to D_{s}^+\pi^0$ was obtained by the Belle Collaboration~\cite{Belle:2003kup}.  Moreover, the Belle Collaboration observed $D_{s0}^*(2317)$ in the  decay process $\bar{B}^0 \to [D_{s0}^{*}(2317)^+ \to D_s^+ \pi^0 ]K^{-}$~\cite{Belle:2004dpt}. In Table~\ref{productions}, we summarise the  production and decay modes of   $D_{s0}^*(2317)$ and $D_{s1}(2460)$. 
We note that the $D_{s0}^{\ast}(2317)$ and $D_{s1}(2460)$ are only discovered in $e^+e^-$ collisions, while they have not been observed in $pp$, $pA$, and $AA$ collisions.

Because of their unique properties, the $D_{s0}^*(2317)$ and $D_{s1}(2460)$ have attracted intensive discussions on their nature. In Ref.~\cite{Song:2015nia}, the authors found that the masses of  $D_{s0}^*(2317)$ and  $D_{s1}(2460)$ do not agree with the experimental data, even adding the screening potential to the conventional quark model, where the Coulomb potential and linear potential are modified because in the energy region near the mass threshold of a pair of hadrons, the linear potential is screened by the vacuum polarization effects of dynamical fermions~\cite{Li:2009ad,Li:2009zu}. The effect of vacuum polarization can also be described by hadron loops, which have been considered on top of the conventional quark model~\cite{Kalashnikova:2005ui,Suzuki:2005ha,Barnes:2007xu,Ortega:2009hj,Danilkin:2010cc,Ferretti:2013faa,Ferretti:2013vua}. Therefore,  if the $DK$ and $D^*K$ components were embodied into the conventional quark model, the mass puzzle of  $D_{s0}^*(2317)$ and  $D_{s1}(2460)$ can be resolved~\cite{Albaladejo:2018mhb,Yang:2021tvc,Luo:2021dvj,Hao:2022vwt,Yang:2023tvc,Ni:2023lvx}, which indicates that the $D^{(*)}K$ components play an important role in forming $D_{s0}^*(2317)$ and  $D_{s1}(2460)$.    
The molecular picture of  $D_{s0}^*(2317)$ and  $D_{s1}(2460)$ can not only explain their mass puzzle but also reconcile their mass splitting~\cite{Barnes:2003dj, Gamermann:2006nm, Gamermann:2007fi, Guo:2006fu,Xie:2010zza,Cleven:2010aw,Guo:2015dha,Wu:2019vsy}.
 A bound state below the $DK$ mass threshold has been identified by lattice QCD simulations~\cite{Liu:2012zya,Mohler:2013rwa,Lang:2014yfa,Bali:2017pdv,Alexandrou:2019tmk}. The couplings between a single channel and the relevant bare state are included in the potential of single-channel scattering~\cite{Hyodo:2011qc,Yamaguchi:2019seo,Yamaguchi:2019vea,Kinugawa:2023fbf,Terashima:2023tun}.   
 Consequently, with the $D^{(*)}K$ potentials supplemented by the  $c\bar{s}$ core couplings to the $D^{(*)}K$ components, $D_{s0}^*(2317)$ and $D_{s1}(2460)$ can be dynamically generated~\cite{Weinberg:1965zz,MartinezTorres:2014kpc,Albaladejo:2016hae,Song:2022yvz}. With the scattering length and effective range  of the $DK$ scattering obtained  by   several  lattice QCD  groups~\cite{Liu:2012zya,Bali:2017pdv,Cheung:2020mql},   it is found that the $DK$ molecular component  accounts for more than $70\%$ of the physical $D_{s0}^*(2317)$ wave function~\cite{MartinezTorres:2014kpc,Albaladejo:2018mhb,Yang:2021tvc,Guo:2023wkv,Gil-Dominguez:2023huq}.

In addition to the studies on their masses,  the decay and production rates of $D_{s0}^*(2317)$ and $D_{s1}(2460)$ have been extensively discussed.        
The dominant decays of $D_{s0}^*(2317)^{+}$ and  $D_{s1}(2460)^+$  to $D_{s}^+\pi^0$ and $D_{s}^{*+}\pi^0$   imply that  $D_{s0}^*(2317)^{+}$ and $D_{s1}(2460)^+$  are  quite  narrow states  since the decays  of $D_{s0}^*(2317)^{+} \to D_{s}^+\pi^0$ and $D_{s1}(2460)^+\to D_{s}^{*+}\pi^0$   break isospin.   The partial decay widths of $D_{s0}^*(2317) \to D_{s}\pi $ and   $D_{s1}(2460)  \to D_{s}^*\pi$ are estimated  to be  tens of keV assuming  $D_{s0}^*(2317)$ and  $D_{s1}(2460)$ as  $c\bar{s}$ excited states~\cite{Fayyazuddin:2003aa,Ishida:2003gu,Wei:2005ag,Nielsen:2005zr,Song:2015nia}, while hundreds of keV assuming them as hadronic molecules~\cite{Faessler:2007gv,Faessler:2007us,Fu:2021wde}, which  can be used to discriminate the nature of  $D_{s0}^*(2317) $ and   $D_{s1}(2460)$~\cite{Guo:2023wkv}.   
    In Ref.~\cite{Faessler:2007cu}, Faessler et al. calculated the decays $B\to D_{s0}^*(2317)\bar{D}^{(*)}$ and $B\to D_{s1}(2460)\bar{D}^{(*)}$  assuming   $D_{s0}^*(2317)$ and  $D_{s1}(2460)$ as $D^{(*)}K$ molecules~\cite{Faessler:2007cu}. The results are consistent but a bit smaller than the experimental data. Assuming  $D_{s0}^*(2317)$ and  $D_{s1}(2460)$  as  $c\bar{s}$  excited states, the decays $B\to D_{s0}^*(2317)\bar{D}^{(*)}$ and $B\to D_{s1}(2460)\bar{D}^{(*)}$ were investigated as well. However, the results suffer from large uncertainties~\cite{Colangelo:1991ug,Veseli:1996yg,Cheng:2003sm,Colangelo:2005hv,Cheng:2006dm,Segovia:2012yh}.  Moreover, the branching fractions  of the decays $\Lambda_b \to \Lambda_c D_{s0}^*(2317)[D_{s1}(2460)] $  are  larger than those of   $B \to \bar{D}^{(*)}D_{s0}^*(2317)[D_{s1}(2460)]   $~\cite{Datta:2003yk,Liu:2024xbw}, which indicate that their production rates in $\Lambda_b$ decays are  larger than those in  $B$ decays.

 
 The production yield of a hadron in electron-positron and heavy ion collisions reflects its internal structure\footnote{The production of a hadron can be factorized into short-range and long-range contributions, while the precision of theoretical estimations (Monte Carlo simulations) for short-range contribution is at an order-of-magnitude level.  Therefore, to improve the reliability of predictions, one needs to calculate other physical observables, such as the ratio between the production rates, which could cancel some uncertainties~\cite{Guo:2014sca,Qin:2020zlg,Wu:2022wgn}.       }. It is an important observable for discriminating the different interpretations of exotic states. The productions of $D_{s0}^*(2317)$ and $D_{s1}(2460)$ in $e^+e^-$ and $AA$ collisions have been investigated. Using the coalescence model~\cite{ExHIC:2010gcb}, the yields of $D_{s0}^*(2317) $ as an excited  $c\bar{s}$ state, a compact multiquark state $c\bar{s}q\bar{q}$, and a $DK$  hadronic molecule in heavy ion collisions are estimated. Ref.~\cite{Albaladejo:2016hae} argued that the enhancement in the  $DK$ invariant mass distribution of the weak decay $B \to DDK$ implies the 
existence of $D_{s0}^*(2317)$ as a $DK$ bound state and claimed that using the mass distribution can probe the molecular nature of an exotic state.   Identifying  $D_{s0}^*(2317) $ and   $D_{s1}(2460)$ as  $D^{(*)}K$ hadronic molecules and  $c\bar{s}$ excited states, Wu et al. estimated their production rates in $e^+ e^-$ collisions~\cite{Wu:2022wgn}. The hadronic molecular picture is more consistent with the experimental data~\cite{BaBar:2006eep}. 

\subsection{$X(3872)$ and selected $X$ states}

In 2003, the Belle Collaboration discovered a charmonium-like state named $X(3872)$ in the $J/\psi \pi^+ \pi^-$ mass distribution of the decay  $B\to K J/\psi \pi^+ \pi^-$~\cite{Belle:2003nnu}, later confirmed by many Collaborations, e.g., BaBar~\cite{BaBar:2004iez}, CDF~\cite{CDF:2003cab}, D0~\cite{D0:2004zmu}, CMS~\cite{CMS:2013fpt}, LHCb~\cite{LHCb:2011zzp}, and BESIII~\cite{BESIII:2013fnz}. Moreover, it was measured in many other decay modes, including   $\bar{D}^0{D}^0\pi^0$~\cite{Belle:2006olv}, $\bar{D}^{\ast0}{D}^0$~\cite{BaBar:2007cmo,Belle:2008fma,BESIII:2020nbj}, $J/\psi\gamma$~\cite{Belle:2005lfc,BaBar:2006fjg,BaBar:2008flx}, $J/\psi \omega$~\cite{BESIII:2019qvy}, $J/\psi \rho^0$~\cite{LHCb:2015jfc}, $\chi_{c1}\pi^0$~\cite{BESIII:2019esk,Belle:2019jeu}, and $\psi(2S)\gamma$~\cite{LHCb:2024tpv}. The BaBar Collaboration reported the absolute branching fraction $\mathcal{B}(X(3872)\to J/\psi \pi^+ \pi^-)=4.1\pm 1.3 \%$~\cite{BaBar:2019hzd}.   It is worth noting that the $X(3872)$ is observed not only in $e^{+}e^{-}$ collisions, but also in $pp$ collisions~\cite{LHCb:2011zzp}, and $p\bar{p}$ collisions~\cite{CDF:2003cab}\footnote{The signal for $X(3872)$  in heavy ion collision(Pb-Pb) is $4.2~\sigma$~\cite{CMS:2021znk}.  }.     Interestingly, $X(3872)$ is also produced in both the $\Lambda_b$ decay~\cite{LHCb:2019imv} and $B_s$ decay~\cite{CMS:2020eiw,LHCb:2023reb}. The average mass and width of $X(3872)$  are $m=3871.65\pm0.06$~MeV and $\Gamma=1.19\pm0.21$~MeV~\cite{ParticleDataGroup:2020ssz}. The mass is lower than the GI model prediction for the first radial excited $P$-wave $\chi_{c1}$ charmonium~\cite{Godfrey:1985xj}.   

The branching ratios of the $X(3872)$ decay play an important role  in 
revealing its nature. The branching ratio of $Br[X(3872)\to J/\psi  \pi \pi \pi (\omega)  ]/Br[X(3872)\to J/\psi\pi\pi(\rho)]$  is measured to be around 1~\cite{Belle:2005lfc,BaBar:2010wfc,BESIII:2019qvy,LHCb:2022jez}, which points to isospin violation in the wave function of the $X(3872)$.  Thus, it is difficult to understand regarding the $X(3872)$ as a pure $c\bar{c}$ charmonium state. It is, however, readily explained in the molecular picture composed of the neutral and charged $\bar{D}^*D$ channels. The isospin violation is enhanced due to kinematical reasons, as the effective phase space available in the case of the $\rho$ is much larger than that in the case of the $\omega$~\cite{Suzuki:2005ha,Gamermann:2009fv,Gamermann:2009uq,Li:2012cs}. The branching fraction ratios $\mathcal{B}(B^+ \to K^+ X(3872))/\mathcal{B}(B^0 \to K^0 X(3872))$~\cite{Belle:2008oko,Belle:2011vlx} and  $\mathcal{B}(X(3872) \to \chi_{c2} \gamma )/\mathcal{B}(X(3872) \to \chi_{c1} \gamma )$~\cite{Belle:2013ewt} play an important role in clarifying the nature of $X(3872)$ because different models can yield distinct predictions. Moreover, the ratios of the production of $X(3872)$ to the production of $\psi(2S)$ measured experimentally, i.e.,  $\mathcal{B}(X(3872) \to \psi(2S) \gamma )/\mathcal{B}(X(3872) \to \psi \gamma )$~\cite{BaBar:2008flx,Belle:2011wdj,LHCb:2014jvf,LHCb:2024tpv},  $\mathcal{B}(B_s^0 \to X(3872)\pi^+\pi^-)\times \mathcal{B}(X(3872)\to J/\psi \pi^+\pi^-)$/$\mathcal{B}(B_s^0 \to \psi(2S)\pi^+\pi^-)\times \mathcal{B}(\psi(2S)\to J/\psi \pi^+\pi^-)=(6.8\pm1.1\pm0.2)\times 10^{-2}$~\cite{LHCb:2023reb}, $\mathcal{B}(B^+ \to X(3872) K^+)\times \mathcal{B}(X(3872)\to J/\psi \pi^+\pi^-)$/$\mathcal{B}(B^+ \to \psi(2S) K^+)\times \mathcal{B}(\psi(2S)\to J/\psi \pi^+\pi^-)=(3.69\pm0.07\pm0.06)\times 10^{-2}$~\cite{LHCb:2020fvo}, and $\mathcal{B}(\Lambda_b^0  \to X(3872) p K^-)\times \mathcal{B}(X(3872)\to J/\psi \pi^+\pi^-)$/$\mathcal{B}(\Lambda_b^0  \to \psi(2S) p K^-)\times \mathcal{B}(\psi(2S)\to J/\psi \pi^+\pi^-)=(5.4\pm 1.1 \pm0.2)\times 10^{-2}$~\cite{LHCb:2019imv}, are likely to tell us more on the nature of $X(3872)$.   The lineshape of $X(3872)$ is precisely analyzed by the  BESIII~\cite{BESIII:2023hml}, LHCb~\cite{LHCb:2020xds}, and Belle~\cite{Belle:2023zxm} Collaborations, which are very helpful in constraining various theoretical interpretations of its nature. 

In 2004, the Belle Collaboration observed a state around 3940 MeV in the $J/\psi \omega$ invariant mass distribution of the  $B\to J/\psi \omega K$ decay~\cite{Belle:2004lle}, which was later confirmed by the BaBar Collaboration in the same process but with the mass determined to be  3915 MeV~\cite{BaBar:2007vxr}. In 2009, the Belle Collaboration observed a state near 3915 MeV in the  $\gamma\gamma\to J/\psi\omega$ reaction~\cite{Belle:2009and}. Later, the BaBar Collaboration determined its quantum numbers to be  $J^{PC}=0^{++}$~\cite{BaBar:2012nxg}. In 2020, the LHCb Collaboration observed a similar state $\chi_{c0}(3930)$  in the $D^{+}D^{-}$ mass distribution of the $B^{+}\to D^{+}D^{-}K^{+} $ decay~\cite{LHCb:2020pxc,LHCb:2020bls}.  In the Review of Particle Physics (RPP)~\cite{ParticleDataGroup:2020ssz}, all these states are referred to as  $\chi_{c0}(3915)$ and  viewed as a candidate for the $\chi_{c0}(2P)$ charmonium~\cite{Duan:2020tsx,Duan:2021bna}\footnote{As proposed by  Guo and Meissner~\cite{Guo:2012tv}, 
if $X(3915)$ is assigned  as the $\chi_{c0}(2P)$ charmonium, there exist several issues:   1)  The partial decay width of $X(3915) \to J/\psi \omega$ is too lager for the Okubo-Zweig-Iizuka
(OZI) suppressed decay; 2) There exists no significant signal for the decay $X(3915) \to \bar{D}D$; 3) The mass splitting between $\chi_{c2}(2P)$ and $\chi_{c0}(2P)$  is too small.     }. 
The LHCb Collaboration recently reported a charmonium state named $X(3960)$ with $J^{PC}=0^{++}$ in the $D_{s}^{+}D_{s}^{-}$ mass distribution of the $B^{+}\to D_{s}^{+}D_{s}^{-}K^{+} $ decay. Its mass and width are determined to be $m=3955\pm6\pm11$ MeV and $\Gamma=48\pm17\pm10$ MeV~\cite{LHCb:2022aki,LHCb:2022dvn}. In the same energy region, a charmonium $\chi_{c2}(3930)$ with  $J^{PC}=2^{++}$ was discovered by the Belle~\cite{Belle:2005rte},  BarBar~\cite{BaBar:2010jfn}, and LHCb~\cite{LHCb:2019lnr,LHCb:2020pxc} Collaborations, which indicate that the mass splitting between $\chi_{c2}(2P)$ and $\chi_{c0}(1P)$ is only around $10$ MeV, inconsistent with the predictions of the GI model~\cite{Godfrey:1985xj}.  

In 2008, the CDF Collaboration observed a resonance $X(4140)$ in the $J/\psi\phi$ mass distribution in the decay of $B\to J/\psi \phi K$ with the mass and width being $m=4143.0\pm2.9\pm1.2$ MeV and $\Gamma=11.7^{+8.3}_{-5.0}\pm3.7$ MeV~\cite{CDF:2009jgo}. Later, the CDF Collaboration updated their analysis of $B\to J/\psi \phi K$ and confirmed the existence of $X(4140)$~\cite{CDF:2011pep},  which was also  confirmed by the CMS~\cite{CMS:2013jru} and  D$0$~\cite{D0:2013jvp,D0:2015nxw} Collaborations. Moreover, another new state with a  mass of $4274.4^{+8.4}_{-6.7}\pm1.9$~MeV and a width of $32.3^{+21.9}_{-15.3}\pm7.6$~MeV is observed by the CDF Collaboration~\cite{CDF:2011pep}. All of them were later confirmed, and a new state $X(4685)$ was discovered by the LHCb Collaboration~\cite{ LHCb:2016axx,LHCb:2021uow}.  

The discovery of $X(3872)$ motivated intensive discussions on its nature and other $X$ states.  According to the recent studies~\cite{Ortega:2009hj,Yamaguchi:2019vea,Duan:2021alw,Kinugawa:2023fbf,Deng:2023mza,Song:2023pdq},  $X(3872)$ contains two components, e.g., a $\bar{D}^*D$ molecular component and  a $\chi_{c1}(2P)$ charmonium component. The former one accounts for more than $80\%$ of the $X(3872)$ wave function according to the recent analysis of the BESIII Collaboration~\cite{BESIII:2023hml}, the theoretical analysis of the LHCb lineshape~\cite{Esposito:2021vhu,Baru:2021ldu}, and the recent lattice QCD simulations~\cite{Li:2024pfg}, which indicates that $X(3872)$ couples strongly to the $\bar{D}^*D$ channel.  Another recent work concluded that the non-molecular component of $X(3872)$ is not dominant judging from the scattering length $a$ and effective range $r$~\cite{Song:2023pdq}. Therefore, by identifying $X(3872)$ as a state with predominant $\bar{D}^*D$  hadronic molecular component, several puzzles associated with $X(3872)$ can be resolved. The mass deviation from the conventional quark state can be easily understood in the  $\bar{D}^*D$ molecular picture~\cite{Swanson:2003tb,Voloshin:2003nt,AlFiky:2005jd,Liu:2008fh,Gamermann:2009fv,Sun:2011uh,Nieves:2012tt,Guo:2013sya,Wang:2013daa,Karliner:2015ina,Liu:2019stu,Wang:2020dgr}.  Studying the $\bar{D}^*D$ interaction, lattice QCD simulations found a bound state below the $\bar{D}^*D$ mass threshold, which can be identified as $X(3872)$~\cite{Prelovsek:2013cra}. One should note that whether the $X(3872)$ is located above or below the $\bar{D}^{*0}D^0$ mass threshold has not been determined experimentally, which is key to clarifying its nature.

\begin{table}[ttt]
\centering
\caption{Absolute branching fractions of the $X(3872)$ decays obtained in Ref.~\cite{Li:2019kpj}. 
\\ Source: Data taken from Ref.~\cite{Li:2019kpj}.  
\label{resultsbsto3872}
}
\begin{tabular}{c c c | c c c c c c c}
  \hline \hline
   decay modes    &~~~~ Branching fractions($10^{-2}$)    
         \\ \hline 
      $X(3872) \to J/\psi \pi^+\pi^-$    & ~~~~ $4.1^{+1.9}_{-1.1}$    
        \\  
        $X(3872) \to D^{*0}\bar{D}^0+c.c$    & ~~~~ $52.4^{+25.3}_{-14.3}$    
        \\      $X(3872) \to J/\psi \gamma$    & ~~~~ $1.1^{+0.6}_{-0.3}$    
        \\
        $X(3872) \to \psi(2S) \gamma$    & ~~~~ $2.4^{+1.3}_{-0.8}$    
        \\      $X(3872) \to \chi_{c1} \pi^0$    & ~~~~ $3.6^{+2.2}_{-1.6}$    
        \\    $X(3872) \to J/\psi \omega$    & ~~~~ $4.4^{+2.3}_{-1.3}$    
        \\
        $X(3872) \to$Unknown    & ~~~~ $31.9^{+18.1}_{-31.5}$  \\ 
  \hline \hline
\end{tabular}
\end{table}

Many decay modes of the $X(3872)$ have been measured. In addition to its total decay width, $\Gamma=1.19\pm 0.21$~MeV\footnote{This value is extracted by the BW parameterization, while it becomes much smaller using the  Flatt$\acute{\text{e}}$ parameterization~\cite{LHCb:2020xds,BESIII:2023hml}.   }, obtained from a global analysis of the Belle, BaBar, BESII, and LHCb measurements~\cite{Li:2019kpj}, the absolute branching fractions of the $X(3872)$ decays are given in Table~\ref{resultsbsto3872}.  The remaining branching fractions of $X(3872)$ is  $31.9\%$, which implies that other partial decays of $X(3872)$ have not been detected. The ratios of the branching fractions of the $X(3872)$ decays are of great value in revealing its nature because, in the ratios, both experimental and theoretical uncertainties are reduced.      
Assuming $X(3872)$ as a pure excited charmonium state, the ratio of $\Gamma[X(3872)\to J/\psi \rho] / \Gamma[X(3872)\to J/\psi \omega ]$ is heavily suppressed in the isospin limit, quite different from the experimental data~\cite{Belle:2005lfc,BaBar:2010wfc,BESIII:2019qvy}, which can, however, be naturally described in the molecular picture~\cite{Braaten:2005ai,Gamermann:2009fv,Ortega:2009hj,Hanhart:2011tn,Li:2012cs,Takeuchi:2014rsa,Zhou:2017txt,Wu:2021udi}. In Ref.~\cite{Wu:2021udi}, with the molecular interpretation of $X(3872)$  the ratio of $\Gamma[X(3872) \to  \pi^0 \chi_{cJ}] / \Gamma[X \to \pi^+\pi^- J/\psi]$ with $J=0,1,2$ measured by the BESIII Collaboration~\cite{BESIII:2019esk}  is well described. For the radiative decays, Dong et al. argued that including the $c\bar{c}$ component into the $\bar{D}^*D$ hadronic molecule is necessary to explain the ratio $\Gamma[X(3872) \to  \gamma  \psi] / \Gamma[X \to  \gamma \psi(2S)]$~\cite{Dong:2009uf}, which agrees with the expectation for a pure charmonium~\cite{Lahde:2002wj,Barnes:2005pb,Barnes:2003vb,Li:2009zu,Badalian:2012jz} but inconsistent with that for a pure molecule~\cite{Swanson:2004pp}. From the theoretical analyses of these ratios,  one can obtain conclusions on the structure of $X(3872)$ similar to those obtained by analyzing its mass.

The $X(3872)$  has been observed in many modes. First, we examine the productions of $X(3872)$ in $b$-flavored hadron decays. Within the molecular picture, the absolute branching fractions of the decays $B^+ \to X(3872) K^+$  and $B^0  \to X(3872) K^0$ and their ratio can be explained~\cite{Wu:2023rrp}. At the same time, they are difficult to understand in the case of a pure charmonium state~\cite{Meng:2005er}. In particular, the branching fractions of the decays $B \to X(3872) K^*$ and   $B_s \to X(3872) \phi$ are explained in the molecular picture of $X(3872)$ in a unified manner~\cite{Wu:2023rrp,Liang:2023jxh}. The productions of $X(3872)$ in $Y$ states can also help probe its nature.
In Ref.~\cite{Guo:2013zbw}, Guo et al. proposed to produce   $X(3872)$ in the most promising channel $Y(4260)\to X(3872)\gamma$ in  $e^+e^-$ collisions via  the triangle diagram mechanism, where $Y(4260)$ and $X(3872)$ are $\bar{D}D_{1}$ and $\bar{D}D^*$ molecules, which has been verified by the BESIII Collaboration~\cite{BESIII:2013fnz}. Recently,  $X(3872)$ is observed in the process of $e^+e^-\to \omega X(3872)$ by the BESIII Collaboration~\cite{BESIII:2022bse}, which is proposed to proceed via an intermediate highly excited vector charmonium~\cite{Qian:2023taw}. The measurements of the differential cross sections for the productions of $J/\psi$ and $X(3872)$ states in the decay channels $J/\psi \pi^+\pi^-$  are performed by the CMS, LHCb, and ATLAS Collaborations, which can be well described assuming $X(3872)$ as a mixture of a bare  $c\bar{c}$ state and a $\bar{D}^*D$  molecule in the non-relativistic QCD approach~\cite{Cisek:2022uqx}.      

From the studies of the mass, width, and productions of $X(3872)$,  one can conclude that it contains a large $\bar{D}D^*$ component. Similarly, other $X$ states are also expected to couple strongly to a pair of charmed mesons. In Refs.~\cite{Ortega:2017qmg,Duan:2020tsx,Deng:2023mza}, it was argued that the couplings of the excited charmonia  $\chi_{c0}(2P)$ and $\chi_{c2}(2P)$  to a pair of charmed mesons lead to two physical states around $3.9$ GeV, i.e., $X(3915)$ and $Z(3930)$. From the conventional quark model,   $X(3915)$ and $Z(3940)$ are assigned as the  charmonia $\chi_{c0}(2P)$~\cite{Liu:2009fe} and $\chi_{c2}(2P)$~\cite{Wang:2013lpa}. 
  Up to now,  the first excited $\chi_{c2}$ state  seems unambiguously established, but 
there exist several issues in identifying  $X(3915)$ as the $\chi_{c0}(2P)$ charmonium~\cite{Guo:2012tv}.  Because of these issues,  $X(3915)$  was proposed as a compact tetraquark or a hadronic molecule~\cite{Olsen:2014maa,Baru:2017fgv}, as well as identifying  $X(3915)$ and  $Z(3930)$  as the same  $\chi_{c2}(2P)$ charmonium~\cite{Zhou:2015uva}. On the other hand, studying the coupled-channel potentials of $D\bar{D}$ and $D_{s}\bar{D}_s$, lattice QCD simulations found two bound states near these two thresholds~\cite{Prelovsek:2020eiw}. Recently, considering all possible channels,   another lattice QCD group found the existence of two resonant states below the $\bar{D}^*D^*$ mass threshold with $J^{PC}=0^{++}$ and $J^{PC}=2^{++}$, but no pole in the vicinity of the $D\bar{D}$ and $D_{s}\bar{D}_s$ mass thresholds~\cite{Wilson:2023anv,Wilson:2023hzu}.  
The newly discovered charmonium state $X(3960)$ is  a candidate neither for $\chi_{c0}(2P)$ nor  for $\chi_{c0}(3P)$, but is explained as an enhancement in the $D_s\bar{D}_s$ invariant mass distribution due to the existence of a  $D_s\bar{D}_s$ molecular state~\cite{Bayar:2022dqa,Xin:2022bzt,Mutuk:2022ckn}, a mixture of $c\bar{c}$ and $D_{s}^+D_s^{-}$~\cite{Chen:2023eix} or a compact tetraquark state~\cite{Agaev:2022pis}. Mixing hadronic molecules and excited charmonium states makes the energy region $3.9$ GeV complex. Recently, assuming $X(3915)$ and  $\chi_{c2}(2P)$  as the same state and $X(3960)$ as the $D_s^+D_s^-$ molecule, Ji et al. showed that the pionless ETF can well describe the $\bar{D}D$ and $D_s^+D_s^-$ invariant mass distributions in the $\gamma\gamma$ and $B$ decay processes, and predicted four $J^{PC}=0^{++}$ hadronic molecules accounting for the structures in the relevant energy region~\cite{Ji:2022uie}. 
As for the $\chi_{cJ}(3P)$ states with $J=0,1,2$, the predicted  masses are around $4.2$~GeV in the conventional quark model~\cite{Li:2009zu}. In Refs.~\cite{Chen:2016iua,Hao:2019fjg,Duan:2021alw}, $X(4140)$ is regarded as a candidate for $\chi_{c1}(3P)$, while in Refs.~\cite{Lu:2016cwr,Ferretti:2020civ,Deng:2023mza}, $X(4274)$ is regarded as a candidate for $\chi_{c1}(3P)$. However, there are no good candidates for $\chi_{c0}(3P)$ and $\chi_{c2}(3P)$ experimentally.

\subsection{$Y(4230)$, $Y(4360)$, and selected $Y$  states}

Since the observation of $J/\psi$ in 1974~\cite{E598:1974sol,SLAC-SP-017:1974ind}, more and more vector charmonium-like states were discovered experimentally, while some of them can not be easily identified as $J^{PC}=1^{--}$ $c\bar{c}$ charmonia predicted by the GI model, denoted by $Y$ states. The most famous vector charmonium-like state $Y(4260)$ is firstly discovered by the BaBar Collaboration in the $J/\psi \pi \pi$ mass distribution of the $e^+ e^- \to \gamma_{ISR} \pi^+ \pi^- J/\psi$ process ~\cite{BaBar:2005hhc}, which was later confirmed by the CLEO~\cite{CLEO:2006ike} and Belle~\cite{Belle:2007dxy} Collaborations. On the other hand, the Belle Collaboration found another state near $4.008$ GeV, in addition to $Y(4260)$~\cite{Belle:2007dxy}. With more precise measurements of the $e^+ e^- \to  \pi^+ \pi^- J/\psi$ process, the   BESIII Collaboration found that the original $Y(4260)$ state splits into two states $Y(4220)$ and $Y(4320)$, and the former is consistent with the  $Y(4260)$ reported by the previous experiment~\cite{BESIII:2016bnd}. In the process of $e^+ e^- \to  \omega \chi_{c0}$, the BESIII Collaboration observed a resonant  state near $4.22$ GeV~\cite{BESIII:2014rja,BESIII:2019gjc}, in agreement with the $Y(4220)$, which was also confirmed in the processes $e^+ e^- \to  D^0 D^{\ast-}\pi^{+}$~\cite{BESIII:2018iea},  $e^+ e^- \to h_{c} \pi^{-}\pi^{+}$~\cite{BESIII:2016adj}, $ e^+ e^- \to \pi^{+}D^{*0}D^{*-}$~\cite{BESIII:2023cmv}, $e^+ e^- \to J/\psi K^+ K^-$~\cite{BESIII:2022joj}, $e^+ e^- \to \psi(2S) \pi^{-}\pi^{+}$ \cite{BESIII:2017tqk}, and $e^+e^- \to J/\psi \eta$~\cite{BESIII:2020bgb,BESIII:2023tll} by the BESIII Collaboration as shown in Table~\ref{productions}. Moreover, the BaBar and Belle Collaborations tried to search for $Y(4220)$ in the  decay of $B^+\to K^+ \pi^+ \pi^- J/\psi $, but without success~\cite{BaBar:2005xmz,Belle:2019pfg}. In the measured  $R$ value scan by the BES Collaboration, the peaks around the energies of  $\psi(3770)$, $\psi(4040)$, $\psi(4160)$,  and $\psi(4415)$ are  pronounced~\cite{BES:2007zwq,BES:2009ejh}, while no evidence for $Y(4230)$ and $Y(4360)$, neither in the exclusive $e^+e^- \to $ open charm processes~\cite{Belle:2006hvs,Belle:2007qxm,Belle:2009dus,BESIII:2024ths}\footnote{The $Y(4320)$ was  observed in the process $e^+e^- \to D^{*0}D^{*-}\pi^+ $ by the BESIII Collaboration~\cite{BESIII:2023cmv}.   }, which implies  the exotic nature of    $Y(4230)$ and $Y(4360)$~\cite{Chen:2016qju}.

In addition to $Y(4220)$ and $Y(4260)$, other excited vector charmonium states were also needed to explain the experimental data. The BaBar Collaboration observed a resonant state $Y(4320)$ in the process $e^+ e^- \to \pi^+ \pi^- \psi(2S)$~\cite{BaBar:2006ait}. This process has been reanalyzed by the Belle Collaboration later, finding that there exist two structures at $4.36$ and $4.66$ GeV, denoted as $Y(4360)$ and $Y(4660)$~\cite{Belle:2007umv}, which were later confirmed by the BaBar Collaboration~\cite{BaBar:2012hpr}. The  BESIII Collaboration observed the  $Y(4360)$ in  the $e^+ e^- \to  \pi^+ \pi^- J/\psi(2S)$ process at center-of-mass
energies from 4.008 to 4.600 GeV~\cite{BESIII:2017tqk}, and  observed   both  $Y(4360)$ and $Y(4660)$  in the same process  at larger center-of-mass
energies from 4.0076 to 4.6984 GeV~\cite{BESIII:2021njb}. In particular, the Belle Collaboration observed a state at $4.63$~GeV, named $Y(4630)$, in the $\Lambda_{c}^{+}\bar{\Lambda}_{c}^{-}$ mass distribution~\cite{Belle:2008xmh}, consistent with the $Y(4660)$ observed in the process $e^+ e^- \to \psi(2S) \pi^{-}\pi^{+}$. Unlike the Belle Collaboration, the BESIII Collaboration did not observe the enhancement structure $Y(4630)$  by  
measuring the cross sections of the $e^+ e^- \to \Lambda_c \bar{\Lambda}_c$ process~\cite{BESIII:2023rwv}. Recently, some highly excited charmonium states, such as $Y(4500)$~\cite{BESIII:2022joj,BESIII:2023cmv,BESIII:2023wqy} and $Y(4700)$~\cite{BESIII:2022kcv,BESIII:2023wqy},  were discovered by the BESIII Collaboration. The studies show that the $Y$ states have strong couplings to hidden-charm final states, while the $Y$ states decaying into open-charm mesons have also been observed recently~\cite{BESIII:2018iea,BESIII:2023cmv}.

For vector charmonium states, the orbital angular momentum between the charmed quark and antiquark can be $L=0$ or $L=2$, mixing $S$-wave and $D$-wave. The $S-D$ mixing  mechanism was applied to study the $\rho-\pi$ puzzle~\cite{Rosner:2001nm},  the di-electron widths of highly  excited vector charmonium-like states~\cite{Badalian:2008dv}, and the production rates of vector charmonia in $B$ decays~\cite{Gao:2006yu,Yang:2018mkn}. The $S-D$ mixing mechanism was also applied to study the mass spectra of vector charmonium-like states.      Ref.~\cite{Wang:2019mhs} claimed that the $4S-3D$ mixing results in two vector charmonium states, $Y(4220)$ and $Y(4380)$. The latter one is a prediction. In the same scheme,  the $5S-4D$ mixing results in two states, $Y(4415)$ and $Y(4500)$. 
In Ref.~\cite{Chen:2015bft}, Chen et al. assigned the $Y(4260)$ and $Y(4360)$ as  fake states via the  Fano-like interference phenomena.  Regarding the quark constituents, the $Y$ states strongly couple to a pair of $S$-wave charmed meson and $P$-wave charmed meson. Therefore, $Y(4260)$ is assigned as a $\bar{D}D_1$ bound state~\cite{Ding:2008gr,Liu:2013vfa,Cleven:2013mka,Wang:2013cya,Dong:2014zka,Qin:2016spb,Dong:2019ofp}.~\footnote{We note that, however, Ref.~\cite{MartinezTorres:2009xb} explained $Y(4260)$ as a $J/\psi K\bar{K}$ three-body molecule, similar to the picture where the $X(2175)$ state is a $\phi K\bar{K}$ three-body molecule~\cite{MartinezTorres:2008gy}. } The $Y(4260)$  splits into $Y(4220)$ and $Y(4360)$, which can be nicely arranged into a doublet of bound states of $\bar{D}D_1$ and  $\bar{D}^*D_1$~\cite{Ji:2022blw,Peng:2022nrj,Wang:2023ivd}. These two states can not be accommodated in the $c\bar{c}$ spectrum~\cite{Deng:2023mza}. Moreover, the unquenched quark model shows that  $\psi(3770)$, $\psi(4040)$, $\psi(4160)$,  and $\psi(4415)$ dominantly couple to $\bar{D}D$, $\bar{D}^*D^*$, $\bar{D}^*D^*$, and $\bar{D}^*D_1$~\cite{Deng:2023mza}. As the masses of excited vector charmonia reach $4.6$ GeV, other configurations, such as charmed baryon and antibaryon pairs or three-body charmed mesons, should be considered on top of the $c\bar{c}$ configurations.

\subsection{$Z_{c}(3900)$, $Z_{c}(4020)$, and $Z_{cs}(3985)$}

Next, we turn to the $Z_{c}$ states. 
In 2013, the BESIII Collaboration reported a hidden-charm tetraquark state $Z_{c}(3900)$ in the $\psi\pi^{\pm}$ mass distribution of $e^{+}e^{-}\to J/\psi \pi^{+}\pi^{-}$~\cite{BESIII:2013ris}, which was reported by the Belle Collaboration several days later~\cite{Belle:2013yex}.  The $Z_{c}(3900)$  was also observed in the  $(D\bar{D}^{\ast})^{\pm}$ invariant mass distribution~\cite{BESIII:2013qmu}. Its neutral partner $Z_{c}(3900)^{0}$ was also observed~\cite{Xiao:2013iha}.
We note that the $Z_{c}(3900)$ is an isovector state and located close to the $D\bar{D}^*$ mass threshold.  Another state $Z_{c}(4020)$ near the mass threshold of  $D^*\bar{D}^*$ was discovered in the $\pi^{\pm}h_{c}$ mass distribution of $e^{+}e^{-}\to h_{c} \pi^{+}\pi^{-}$ ~\cite{BESIII:2013ouc} and the $(D^*\bar{D}^{\ast})^{\pm}$ mass distribution of $e^{+}e^{-}\to (D^*\bar{D}^{\ast})^{\pm}\pi^{\pm}$ ~\cite{BESIII:2013mhi} by the BESIII Collaboration. The Belle Collaboration reported a charged charmonium-like structure at $M_{\pi^\pm \psi(2S)}=4.05$ GeV in the $Y(4360)$ decays with $3.5$~$\sigma$~\cite{Belle:2014wyt}. We note that the $Z_c(3900)$ has not been seen in $b$ hadron decays such as $\bar{B}^0 \to (J/\psi \pi^+)K^-$ by the Belle Collaboration~\cite{Belle:2014nuw} and $B^0 \to (J/\psi \pi^+)\pi^- $ by the LHCb Collaboration~\cite{LHCb:2014vbo}.  The Belle Collaboration has reported the upper limit of $Z_{c}(3900)^{0}$ and $Z_{c}(4020)^{0}$ in $B$ decays, i.e., $\mathcal{B}(B^{\pm}\to K^{\pm}Z_c(3900)^{0})\mathcal{B}(Z_c(3900)^{0}\to \eta_c \pi^+\pi^-)<4.7\times 10^{-5}$ and $\mathcal{B}(B^{\pm}\to K^{\pm}Z_c(4020)^0)\mathcal{B}(Z_c(4020)^0\to \eta_c \pi^+\pi^-)<1.6\times 10^{-5}$~\cite{Belle:2015yoa}. 
In the semi-inclusive weak decays of $b$-favored hadrons, 
the D0 Collaboration first observed $Z_{c}(3900)$  in  the $J/\psi \pi^{\pm}$ mass distribution in  $p\bar{p}$ collisions~\cite{D0:2018wyb}, which is correlated to the $J/\psi \pi^+ \pi^-$ system in the invariant mass spectrum including $Y(4230)$ and $Y(4360)$.
Recently, the BESIII Collaboration observed  a state $Z_{cs}(3985)$ near the $D_{s}^{-}{D}^{\ast0}/D_{s}^{\ast-}{D}^{0}$ mass threshold  in the $K^+$ recoil mass distribution of the $e^{+}e^{-}\to D_{s}^{-}{D}^{\ast0}K^{+}/D_{s}^{\ast-}{D}^{0}K^{+}$ process~\cite{BESIII:2020qkh}, which is often viewed as the SU(3)-flavor partner of $Z_{c}(3900)$. Very recently, the $Z_{cs}(4000)$ is observed in the $J/\psi K^+$ mass distribution in the decay of $B^+ \to J/\psi \phi K^+$ by the LHCb Collaboration~\cite{LHCb:2021uow}. The masses of $Z_{cs}(4000)$ and $Z_{cs}(3985)$ are consistent with each other, but the width of the former is much larger than that of the latter. There are heated discussion on whether they are the same states~\cite{Yang:2020nrt,Meng:2021rdg,Ortega:2021enc,Han:2022fup,Wang:2023vtx,Yu:2023nmb,Wua:2023ntn}.

 Another charged hidden-charm tetraquark state is the $Z_{c}(4430)$ discovered by the Belle Collaboration in the $\pi\psi(2S)$ mass distribution of the decay $B \to \pi^{\pm} K\psi(2S)$~\cite{Belle:2007hrb}. Although this state was not confirmed by the BaBar Collaboration~\cite{BaBar:2008bxw}, 
it was confirmed in the same process by the LHCb Collaboration seven years later~\cite{LHCb:2014zfx}.  
One should note that much progress has been made in searching for the $Z_{c}$ states in exclusive $b$ decays.  The Belle Collaboration observed two states $Z_{c1}(4050)$ and $Z_{c2}(4250)$ in the $\pi^-\chi_{c1}$ mass distribution of the $B^{0}\to K^{+}\pi^-\chi_{c1}$ decay~\cite{Belle:2008qeq}. These two states are not confirmed by the BaBar Collaboration~\cite{BaBar:2011hrz}. The $Z_{c}(4200)$ was observed by the Belle Collaboration in the $J/\psi\pi^+$ mass distribution in the decay  $\bar{B}^0 \to J/\psi \pi^+ K^-$~\cite{Belle:2014nuw}, and the LHCb Collaboration later observed it in the same channel~\cite{LHCb:2019maw}, which was recently confirmed by the LHCb Collaboration in the   $\psi(2S)\pi^+$ mass distribution in the decay  $ B^+ \to \psi(2S)K^+ \pi^+ \pi^- $~\cite{LHCb:2024cwp}. 
In the decay of $B^0 \to \eta_{c}(1S)\pi^- K^+$, the LHCb Collaboration observed a state $Z_{c}(4100)$ in the $\eta_{c}(1S)\pi^-$ mass distribution~\cite{LHCb:2018oeg}.   Due to the low significance of $Z_{c}(4100)$ in the LHCb data, more data are required to confirm this state.      
 The Belle Collaboration observed two charged bottomonium-like resonant states,  $Z_{b}(10610)$ and  $Z_{b}(10650)$, in five different decay channels, $\pi^{\pm}\Upsilon(nS)$~$(n=1,2,3)$ and $h_{b}(mP)\pi^{\pm}$~$(m=1,2)$~\cite{Belle:2011aa}, which are close to the $\bar{B}^*B$ and $\bar{B}^*B^*$ mass thresholds. Like  $Z_{c}(3900)$ and $Z_{c}(4020)$,  $Z_{b}(10610)$ and  $Z_{b}(10650)$  were also observed by the Belle Collaboration in the open-bottom channels of $\bar{B}^*B$ and $\bar{B}^*B^*$, and the branching fractions of  $Z_b$ open-bottom decays are larger than those of the $Z_b$ hidden-bottom decays ~\cite{Belle:2015upu}.  

Given that $Z_c(3900)$ and $Z_c(4020)$ are located in the vicinity of  the $\bar{D}^*D$ and $\bar{D}^*D^*$ mass thresholds, they are  expected to be the  $\bar{D}^*D$ and $\bar{D}^*D^*$ resonant states~\cite{Zhang:2013aoa,Cui:2013yva,Albaladejo:2015lob,Chen:2023def}.  The isovector $\bar{D}^*D^{(*)}$ potentials are less attractive than the isoscalar  $\bar{D}^*D^{(*)}$ potentials, and therefore there are no bound states in the isovector sector~\cite{Sun:2011uh,Chen:2014afa}.  However, if the next-to-leading order potentials are taken into account, one can obtain the molecules above the $\bar{D}^*D$ and $\bar{D}^*D^*$ mass thresholds~\cite{Wang:2020dko,Meng:2020ihj}. Considering SU(3)-flavor symmetry, the molecules above the $D_s^*\bar{D}$ and  $D_s^*\bar{D}^*$ mass thresholds are expected~\cite{Meng:2020ihj,Wang:2020htx,Meng:2021rdg,Du:2022jjv}, the former of which corresponds to $Z_{cs}(3985)$ discovered by the BESIII Collaboration. Similarly, with heavy quark flavor symmetry,  two hadronic molecules above the $\bar{B}^*B$ and $\bar{B}^*B^*$  mass thresholds may correspond to $Z_{b}(10610)$ and  $Z_{b}(10650)$~\cite{Cleven:2013sq,Dong:2012hc,Wang:2014gwa,Chen:2015ata,Wang:2020dko}. In the QCD sum rule approach, $Z_c(3900)$, $Z_{c}(4020)$, and $Z_{cs}(3985)$ are assigned as compact tetraquark states~\cite{Wua:2023ntn}, in agreement with Refs.~\cite{Dias:2013xfa,Wang:2013exa,Wang:2013llv,Wang:2019hnw,Deng:2014gqa,Anwar:2018sol,Wang:2020iqt,Ferretti:2021zis}.   Since the isovector $\bar{D}^*D^{(*)}$ potentials are not strong enough to form bound states, they may manifest themselves as cusp structures in the vicinity of the $\bar{D}^*D^{(*)}$ mass thresholds~\cite{Wang:2013cya,Chen:2013coa,Liu:2013vfa,Aceti:2014uea,HALQCD:2016ofq,Ge:2021sdq,Ikeno:2020mra}.  It should be noted that a pure kinematic
two-body threshold cusp, without a near-threshold pole, is unlikely to produce a pronounced and narrow peak in
the mass distribution of the elastic channel of the pair of open heavy-flavor mesons, which has the
threshold in the vicinity of the observed peak~\cite{Bugg:2008wu,Guo:2014iya}. As indicated in Ref.~\cite{Dong:2020hxe}, a pronounced peak would imply a large scattering length or a large reduced mass, implying the existence of a near-threshold pole.  However, the situation becomes more complicated when triangle singularities arise at the same near-threshold region. The triangle singularity refers to the  branch point   in a three-point loop function~\cite{Liu:2015taa,Szczepaniak:2015eza}, and its effect is sensitive to the kinematic variables~\cite{Wang:2013hga}.

 \begin{table*}[ttt]
  \caption{
 Production and decay modes  of  $Z_c(3900)$, $Z_c(4020)$, $P_c(4312)$, $P_c(4440)$, $P_c(4457)$, $P_c(4380)$, $P_{cs}(4338)$, $P_{cs}(4459)$, and $T_{cc}(3875)^+$.
    \label{proexotic2}
}
\centering
\begin{tabular}{cccccc}
  \hline\hline State& $J^{P}$ & Inclusive process & Decay modes  & Exclusive process &  Decay modes 
\\ \hline  \multirow{5}{1.4cm} {$Z_c(3900)$}  & \multirow{5}{0.4cm} {$1^{+}$} & $e^+e^- \to J/\psi \pi^0 \pi^0$  & $J/\psi \pi^0$~\cite{BESIII:2015cld}  &  $Y(4260) \to J/\psi \pi^+ \pi^-$   & $J/\psi\pi^\pm$~\cite{BESIII:2013ris,Belle:2013yex}  \\ 
 &  & $e^+e^- \to \pi^{\pm}(\bar{D}D^*)^{\mp}$   &  $(\bar{D}D^*)^{\mp}$~\cite{BESIII:2013qmu,BESIII:2015pqw}  &   $Y(4160) \to J/\psi \pi^+ \pi^-$   & $J/\psi\pi^\pm$~\cite{Xiao:2013iha}    \\ 
 &  &$e^+e^- \to \pi^{0}(\bar{D}D^*)^{0}$    &$(\bar{D}D^*)^{0}$~\cite{BESIII:2015ntl} & $Y(4220) \to J/\psi \pi^0 \pi^0$    &    $J/\psi \pi^0$~\cite{BESIII:2020oph}  
   \\ 
 &  &$e^+e^-$   &  $\pi^+\pi^- D^+D^-$~\cite{BESIII:2022quc}   &    
  \\ 
 &  &$e^+e^-$   &  $\gamma \chi_{c2}$~\cite{BESIII:2021yal}  &     \\
  &  &$e^+e^- \to \pi^{\pm}Z_c^{\mp}(3900)$   &  $\rho^{\mp}\eta_{c}$~\cite{BESIII:2019rek}  &     \\  \hline
\multirow{4}{1.4cm} {$Z_c(4020)$}  & \multirow{4}{0.4cm} {$1^{+}$} & $e^+e^- \to h_c \pi^+ \pi^-$  & $h_c \pi^\pm$~\cite{BESIII:2013ouc}  &     &   \\ 
 &  & $e^+e^- \to h_c\pi^0\pi^0$   &  $h_c\pi^0$~\cite{BESIII:2014gnk}  &    &    \\ 
 &  &$e^+e^- \to \pi^{0}(\bar{D}^*D^*)^{0}$    &$(\bar{D}^*D^*)^{0}$~\cite{BESIII:2015tix} &     &   
   \\  &  &$e^+e^- \to \pi^{\pm}(\bar{D}^*D^*)^{\mp}$    &$(\bar{D}^*D^*)^{\mp}$~\cite{BESIII:2013mhi} &     &   
  \\ \multirow{2}{1.4cm}{$Z_{cs}(3985)$}  & \multirow{2}{0.4cm}{$1^{+}$}  &$e^+e^-\to K^{+}(D_s^-D^{*0}+D_s^{*-}D^0) $    & $RM(K^+)$~\cite{BESIII:2020qkh}     \\ 
   &  &$e^+e^-\to K_S^{0}(D_s^+D^{*-}+D_s^{*+}D^-) $    & $RM(K_S^0)$~\cite{BESIII:2022qzr}
   \\ \hline
$P_c(4312)$ & $1/2^{+}$ & &  &  $\Lambda_b  \to J/\psi p \bar{K}  $    & $J/\psi p$~\cite{LHCb:2019kea} 
  \\ $P_c(4380)$ & $3/2^{+}$    & &     & $\Lambda_b  \to J/\psi p \bar{K}  $    & $J/\psi p$~\cite{LHCb:2015yax}   \\
  $P_c(4440)$ & $?^{?}$ & &  & $\Lambda_b  \to J/\psi p \bar{K}  $     & $J/\psi p$~\cite{LHCb:2019kea}  
  \\ $P_c(4457)$  &$?^{?}$  & & &  $\Lambda_b  \to J/\psi p \bar{K}  $    & $J/\psi p$~\cite{LHCb:2019kea}  
   \\ \hline
$P_{cs}(4338)$ & $1/2^{+}$ &  &  &  $B  \to J/\psi \Lambda \bar{p} $   & $J/\psi \Lambda$~\cite{LHCb:2022ogu} 
  \\ $P_{cs}(4459)$ & $1/2^{+}$ &  &   &  $\Xi_b  \to J/\psi \Lambda \bar{K}  $   & $J/\psi \Lambda$~\cite{LHCb:2020jpq}
   \\ \hline
$T_{cc}(3875)^+$ & $1^{+}$ &$pp $   & $D^0D^0\pi^+$~\cite{LHCb:2021vvq}  
\\ \hline \hline
\end{tabular}
\end{table*}

\subsection{$P_{c}(4380)$, $P_{c}(4312)$, $P_{c}(4440)$, and $P_{c}(4457)$}

In 2015, the LHCb Collaboration observed two hidden-charm pentaquark states, $P_{c}(4380)$ and $P_{c}(4450)$, in the $J/\psi p$ mass distribution of the $\Lambda_{b}\to J/\psi p K$~\cite{LHCb:2015yax} decay, which are also analyzed in the full amplitude fit to the decay of $\Lambda_{b}\to J/\psi p \pi$ with a low significance~\cite{LHCb:2016lve}. In the updated analysis of a large data sample in 2019, the  $P_{c}(4450)$ splits into two states $P_{c}(4440)$ and $P_{c}(4457)$, and in addition, a new state $P_{c}(4312)$ was observed in the same process~\cite{LHCb:2019kea}. The broad $P_c(4380)$ observed in 2015 was not confirmed in the 2019 analysis, while a theoretical analysis of the LHCb data indicated the existence of a narrow  $P_c(4380)$ with $1.7\sigma$ significance~\cite{Du:2019pij}.   The quest for the $P_c(4312)$  in the $\eta_c p$ mass distribution of the $\Lambda_{b}\to \eta_c p K$ decays is performed by the LHCb Collaboration, while no evidence is found~\cite{LHCb:2020kkc}.  In 2020, the LHCb Collaboration observed a hidden-charm pentaquark state with strangeness $P_{cs}(4459)$ in the $J/\psi \Lambda$ mass distribution in the decay of $\Xi_{b}\to J/\psi \Lambda K$~\cite{LHCb:2020jpq}. Such a state can also be viewed as a mixture of two pentaquark states near the mass of  $P_{cs}(4459)$. 
 The hidden-charm pentaquark state $P_{c}(4337)$ was observed in the $J/\psi p$ and $J/\psi \bar{p}$ mass distribution of the decay  $B_{s}\to J/\psi p \bar{p}$~\cite{LHCb:2021chn}. Very recently, the LHCb Collaboration reported a hidden-charm pentaquark state with strangeness $P_{cs}(4338)$ in the $J/\psi\Lambda$ mass distribution of the decay $B\to J/\psi \Lambda \bar{p}$~\cite{LHCb:2022ogu}. 

The recent studies~\cite{Xiao:2019aya,Du:2019pij,Lin:2019qiv,Sakai:2019qph,Pan:2023hrk} show that the $\bar{D}^{(*)}\Sigma_{c}^{(*)}$, $\bar{D}^{(*)}\Lambda_c$, $J/\psi p$, and $\eta_c p$ channels all contribute to the formation of the hidden-charm pentaquark molecules, where the $\bar{D}^{(*)}\Sigma_{c}^{(*)}$  channels play the dominant role.
From the perspective of the one-meson exchange theory, the potentials of $\bar{D}^{(*)}\Sigma_{c}^{(*)} \to \bar{D}^{(*)}\Sigma_{c}^{(*)}$,  
$\bar{D}^{(*)}\Sigma_{c}^{(*)} \to \bar{D}^{(*)}\Lambda_c$,  and $\bar{D}^{(*)}\Lambda_c \to \bar{D}^{(*)}\Lambda_c$ are induced by light meson exchanges,  while those of  $\bar{D}^{(*)}\Sigma_{c}^{(*)} \to J/\psi p$
and $\bar{D}^{(*)}\Sigma_{c}^{(*)} \to \eta_c p$ are from the exchanges of charmed mesons, which are difficult to accommodate in a unified model since the charmed meson exchange may not work at short distances~\cite{Yamaguchi:2019djj}.  Due to the uncertainties in the off-diagonal potentials, the partial decays of hidden-harm pentaquark molecules suffer significant uncertainties. Assuming that  $P_c(4440)$ and $P_c(4457)$  are predominantly generated by the  $\bar{D}^*\Sigma_c$ interactions,  there are two scenarios for their spins:  the $P_c(4440)$ and $P_c(4457)$ as $\tfrac{1}{2}^-$ and $\tfrac{3}{2}^-$ $\bar{D}^* \Sigma_c$ molecules (scenario A) or $\tfrac{3}{2}^-$ and $\tfrac{1}{2}^-$ $\bar{D}^* \Sigma_c$ molecules (scenario B). As shown in Ref.~\cite{Pan:2023hrk},  once the coupled-channel potentials are considered, Scenario A is favored over Scenario B, which differs from the single-channel case~\cite{Liu:2019tjn}.  In the heavy quark limit, the contact potentials of the $\bar{D}^{(*)}\Sigma_{c}^{(*)}$ system can be parameterized by two parameters $C_a$ and $C_b$, corresponding to the spin-spin independent and spin-spin dependent terms. The ratio of $C_b$ to $C_a$ in the single-channel case is determined as $-0.176$ in scenario A and $0.158$ in scenario B~\cite{Liu:2019zvb}, consistent with that in the couple-channel case~\cite{Pan:2023hrk}. Ref.~\cite{Burns:2022uiv} argued that the visible structure of $P_{c}(4457)$ is due to the effect of triangle singularity, while the ratio $C_b/C_a$ is around $0.5$, where the spin-spin term is a bit large\footnote{ For the     $\bar{D}^{(*)}D^{(*)}$ and  $\bar{D}^{(*)}\Sigma_c^{(*)}$ systems, we obtain the ratio of $C_b/C_a$ in the range of $0.1-0.3$ according to the light meson saturation approach~\cite{Liu:2019zvb,Liu:2020tqy}. }.   The two-body decays of pentaquark molecules are investigated in the contact EFT approach, which can also be studied via the triangle mechanism~\cite{Xiao:2019mvs,Lin:2019qiv}. The three-body decays of pentaquark molecules have also been estimated, and the decays of pentaquark molecules with $\Sigma_c^*$ baryons account for a large proportion of their total widths~\cite{Xie:2022hhv}.          
In addition to the molecular interpretations,  there exist other explanations for these pentaquark states, e.g.,  hadro-charmonia~\cite{Eides:2019tgv}, compact pentaquark states~\cite{Ali:2019npk,Wang:2019got,Cheng:2019obk,Weng:2019ynv,Zhu:2019iwm,Pimikov:2019dyr,Ruangyoo:2021aoi}, 
virtual states~\cite{Fernandez-Ramirez:2019koa}, triangle singularities~\cite{Nakamura:2021qvy}\footnote{ Before 2019,  triangle singularities were applied to explain the lineshape of hidden-charm pentaquark states~\cite{Guo:2015umn,Liu:2015fea,Bayar:2016ftu}. Very recently, some interesting triangle and box singularities are predicted in the energy region close to the $\bar{D}^{(*)}\Lambda_c$ threshold in the processes of $P_c$ decays or $\bar{D}^{(*)}\Sigma_c^{(*)}$ rescattering into  $J/\psi p$ and $J/\psi p \pi$~\cite{Duan:2023dky}.}, and cusp effects~\cite{Burns:2022uiv}. With SU(3)-flavor symmetry, the  $\bar{D}^{(*)}\Sigma_{c}^{(*)}$  molecules indicate the existence of the  $\bar{D}^{(*)}\Xi_{c}^{\prime(*)}$ molecules~\cite{Xiao:2019gjd,Liu:2020hcv,Xiao:2021rgp,Wang:2022mxy}, but no bound states near the $\bar{D}^{(*)}\Xi_c$ mass thresholds~\cite{Liu:2020hcv}. However,  $P_{cs}(4338)$ and  $P_{cs}(4459)$ discovered by the LHCb Collaboration are in the vicinity of mass thresholds of $\bar{D}\Xi_c$ and $\bar{D}^*\Xi_c$. There may exist some mechanism to increase the strength of the $\bar{D}^{(*)}\Xi_c$ potentials, such as coupled-channel effects and  SU(3)-flavor symmetry breaking effects~\cite{Feijoo:2022rxf}.

\subsection{$T_{cc}(3875)$} 

 In 2017, the LHCb Collaboration observed a doubly charmed baryon $\Xi_{cc}^{++}$ in the $\Lambda_{c}^{+}K^{-}\pi^{+}\pi^{+}$ mass distribution in $pp$ collisions~\cite{LHCb:2017iph}. In 2020, the LHCb Collaboration observed a narrow doubly charmed tetraquark state $T_{cc}$ in the $DD\pi$ mass distribution in $pp$ collisions~\cite{LHCb:2021vvq}. The difference between its mass and the threshold $D^{*+}D^0$  and its width are of the order of hundreds of keV~\cite{LHCb:2021auc}, which is the most precisely measured exotic state.  Such states containing a pair of charmed quarks are explicitly exotic and have attracted much attention.

Since the doubly charmed tetraquark state lies close to the $DD^*$ mass threshold, the discussions on the $DD^*$  potential motivated several lattice QCD investigations~\cite{Padmanath:2022cvl,Chen:2022vpo,Lyu:2023xro}, indicating that there exists a virtual state at the   $DD^*$  mass threshold. According to Refs.~\cite{Liu:2019stu,Peng:2023lfw}, the isoscalar  $DD^*$ potential is less attractive than the isoscalar  $\bar{D}D^*$ potential, which is likely to generate a weakly bound state below the $DD^*$ mass threshold.  Identifying the doubly charmed tetraquark state $T_{cc}$ as a  $DD^*$  molecule, the decay widths of $T_{cc}$ can be reasonably explained as well~\cite{Ling:2021bir,Meng:2021jnw,Feijoo:2021ppq,Chen:2021vhg,Qiu:2023uno,Zhang:2024dth,Sun:2024wxz}. It is worth noting that the precise mass of $T_{cc}$ inspired studies of high order contributions to the $DD^*$ potentials in EFTs~\cite{Zhai:2023ejo,Dai:2023mxm,Wang:2022jop}, as well as the impact of the left-hand cut in the $DD^*$ scattering~\cite{Du:2023hlu,Wang:2023iaz,Meng:2023bmz,Hansen:2024ffk}. The compositeness analysis indicates that the newly discovered $T_{cc}$ is dominantly a $DD^*$ molecule~\cite{Albaladejo:2021vln,Du:2021zzh,Kinugawa:2023fbf,Wang:2023ovj,Dai:2023kwv,Dai:2023cyo}.   In Refs.~\cite{Weng:2021hje,Wu:2022gie,Song:2023izj,Li:2023wug,Ma:2023int,Mutuk:2023oyz}, the compact tetraquark picture  for the $T_{cc}$ was investigated. In Ref.~\cite{Chen:2021vhg},   considering the isospin breaking in the $D^{*+}D^0$ and $D^{+}D^{*0}$ channels, Chen et al. employed the OBE model to reproduce the $T_{cc}$ mass, and then predicted a resonant state $T_{cc}^{\prime}$ with a larger mass and width, where $T_{cc}$ and $T_{cc}^{\prime}$ dominantly couple to  $D^{*+}D^0$ and $D^{+}D^{*0}$, respectively. In the isospin limit, Wang adopted the QCD sum rules to calculate the masses of isoscalar and isovector $DD^*$ molecules, finding that the mass of the isovector molecule is larger than that of the isoscalar one~\cite{Xin:2021wcr}.

\section{Multiplets of hadronic molecules}
\label{multiplet molecules}

\subsection{Symmetries}
This review only focuses on the hadronic molecular interpretation of the exotic states introduced in the previous section. If these states are dominantly hadronic molecules, one can deduce the underlying hadron-hadron interactions from their masses and related properties. Using symmetry arguments, these interactions can be extended to those between pairs of other hadrons. 

It is well known that symmetries play a crucial role in particle and nuclear physics. The  Eightfold Way discovered by the Gell-Mann and Ne'eman paved the way to understanding the baryons and mesons~\cite{Neeman:1961jhl,Gell-Mann:1962yej}. The breaking of the symmetry underlying the  Eightfold Way resulted in the Gell-Mann-Okubo formula~\cite{Gell-Mann:1962yej,Okubo:1961jc}, predicting the mass of the $\Omega$ baryon, which was later verified experimentally~\cite{Barnes:1964pd}. In 1964, Gell-Mann~\cite{Gell-Mann:1964ewy} and Zweig~\cite{Zweig:1964jf} successfully classified the ground-state light mesons and baryons using the SU(3)-flavor symmetry. Symmetries can also help relate hadronic molecules; thus, they play a vital role in understanding them. The SU(3)-flavor symmetry, HQSS, HQFS, and HADS have been widely applied to study hadronic molecules. Symmetries also manifest themselves in the potentials between pairs of hadrons. This can be best demonstrated in the contact-range EFT. As a result, in this review, we take the contact-range potentials to discuss hadronic molecules and the symmetry implications for the existence of multiplets of hadronic molecules. 

EFTs are low-energy theories that apply to specific systems within a specified energy range, where the underlying high-energy theory for these systems is more difficult to solve.  The interactions of a given system are expressed by the Lagrangian, which contains all the terms satisfying the relevant symmetry requirements. The terms are ordered by a small expansion parameter $M_{s}/M_{h}$, where $M_{s}$ and $M_{h}$ represent the soft scale and hard scale.   
The advantage of EFTs is that they can systematically improve the results according to an appropriate power counting rule, and one can estimate the uncertainties of any given order.  In the following, we briefly introduce the kinds of EFTs relevant to the present review. For details on these EFTs, one can refer to the comprehensive reviews~\cite{Meissner:1993ah,Epelbaum:2008ga,Machleidt:2011zz,Guo:2017jvc,Hammer:2019poc}.

QCD is part of the Standard Model of particle physics that deals with quarks,  gluons, and their strong interactions.  The strong interaction is weak at short distances or high momentum transfer, referred to as the asymptotic freedom. At the same time, it is strong at long distances or low energies, leading to the confinement of quarks into colorless objects, i.e., hadrons, which dictates that the strong interaction is non-perturbative at low energies. ChEFT is one of the most successful theories for the non-perturbative strong interaction.     
It is based on chiral symmetry and its breaking pattern, which strongly constrains the interactions between hadrons consisting of light quarks ($u$, $d$, and to a lesser extent $s$). In recent years, it has been extended to the heavy quark sector. A recent review on this topic is Ref.~\cite{Meng:2022ozq}. Chiral symmetry is explicitly broken due to the non-vanishing quark masses. In addition,  a spontaneously broken global symmetry implies the existence of (massless) Nambu-Goldstone bosons with the quantum numbers of the broken generators.  The pion masses are not exactly zero because the up and down quark masses are not exactly zero, i.e., explicit chiral symmetry breaking.    ChEFT has traditionally been applied to understand hadron-hadron interactions in the light-quark sector, where the relevant degrees of freedom are hadrons and pions. Because the interactions with the Nambu-Goldstone bosons must vanish in the chiral limit, the low-energy expansion of the chiral Lagrangian is arranged in powers of derivatives and the pion mass. The hard scale of ChEFT is the chiral symmetry breaking scale, $\Lambda_{\chi}\approx 1$ GeV, and the soft scale $Q$ is a small external momentum or the pion mass.  As the nucleon is embodied into  ChEFT, the above power counting rule is destroyed due to the non-vanishing nucleon mass in the chiral limit, which can be solved by treating the nucleon as a heavy static source (i.e., Heavy Baryon ChPT(HBChPT))~\cite{Jenkins:1990jv,Bernard:1992qa} or adopting the so-called extended on-mass-shell scheme~\cite{Fuchs:2003qc,Geng:2013xn}.

The effective field theory dealing with heavy hadrons is the heavy quark effective field theory (HQET)~\cite{Isgur:1989vq,Isgur:1990yhj},  where the hard scale and soft scale are the masses of heavy quarks $m_{Q}$ ($Q=c,b$) and QCD scale $\Lambda_{QCD}$, leading to a new expansion parameter $\Lambda_{QCD}/m_{Q}$.  Since $m_{Q}$  is larger than $\Lambda_{QCD}$, i.e., $\Lambda_{QCD}/m_{Q} \ll 1$, the heavy quark acts like a static color source, similar to the HBChPT. A heavy quark inside a hadron shares the velocity $v$ as the hadron.  Thus, its momentum can be written as $p=m_Qv+k$, where $k$ is a small residual momentum of the order of $\Lambda_{QCD}$. In this picture, the heavy quark field $\Psi$ is decomposed into  
\begin{eqnarray}
\Psi(x)=e^{-i m_Qv\cdot x}(h_v(x)+H_v(x)),  
\end{eqnarray}
where $h_v(x)$ and $H_v(x)$ are defined as 
\begin{eqnarray}
h_v(x)=e^{-i m_Qv\cdot x}\frac{1+v \!\!\!\slash}{2}\Psi(x),  \\ \nonumber
H_v(x)=e^{-i m_Qv\cdot x}\frac{1-v \!\!\!\slash}{2}\Psi(x).
\end{eqnarray}
In the rest frame of the heavy quark, $h_v$ and $H_v$ correspond to the upper and lower components of $\Psi$, also called the large and small components.  The heavy quark part of the QCD Lagrangian $\mathcal{L}_{QCD}=\bar{\Psi}i D \!\!\!\!\slash \Psi -m_Q\bar{\Psi}\Psi $ with $D_{\mu} =\partial_{\mu}-i g T^a A^a_{\mu}  $ is simplified as 
\begin{eqnarray}
\mathcal{L}_{eff}=\bar{h}_v i v \cdot D h_v + \bar{h}_v i D_{\perp} \!\!\!\!\!\!\!\!\slash ~~~~ \frac{1}{i v\cdot D +2 m_Q - i\varepsilon  }i D_{\perp} \!\!\!\!\!\!\!\!\slash ~~~~ h_v.
\end{eqnarray}
The lowest order of the above  effective Lagrangian  in $1/m_{Q}$ is written as 
\begin{eqnarray}
\mathcal{L}_{eff}^{0}=\bar{h}_v (i v^{\mu} \partial_{\mu}+  g v^{\mu} T^a A^a_{\mu} ) h_v,
\end{eqnarray}
which exhibits the crucial features of HQET: the quark-gluon coupling is independent of the quark's spin, and the Lagrangian is independent of the heavy quark flavor, showing the heavy quark spin-flavor symmetry, i.e.,  HQSS and HQFS, which are broken by the $1/m_{Q}$-contributions. The heavy quark symmetry dictates that the strong interaction is independent of the heavy quark spin, which provides a natural explanation for the mass difference of $(D, D^*)$ and $(B,B^*)$ and those of their baryon counterparts. It is important to note that the effective Lagrangian describing the interactions between heavy mesons and Nambu-Goldstone bosons should satisfy heavy quark symmetry, Lorentz invariance, and chiral symmetry~\cite{Yan:1992gz}. The fields of a pair of $(D, D^*)$ mesons are written as a superfield, i.e., ${H}=\frac{1+ v\!\!\!\slash }{2}(D^{*\mu}\gamma_{\mu}- D\gamma_5)$ and  $\bar{H}=(D^{*\mu\dag}\gamma_{\mu}+ D^{\dag}\gamma_5)\frac{1+ v\!\!\!\slash }{2}$~\cite{Wise:1992hn}. Thus, the effective Lagrangian is 
\begin{eqnarray}
 \mathcal{L}&=&-i Tr  \bar{H}_a v^{\mu} \partial_{\mu}  H_a     +  \frac{i}{2}Tr \bar{H}_a v^{\mu} (\xi^{\dag}\partial_{\mu}\xi+ \xi \partial_{\mu} \xi^{\dag} )_{ba} H_b  \\ \nonumber  &+&  g \frac{i}{2}  Tr \bar{H}_a  H_b \gamma_{\mu}\gamma_5 (\xi^{\dag}\partial_{\mu}\xi- \xi \partial_{\mu} \xi^{\dag} )_{ba} + \cdots,
\end{eqnarray}
where $\xi$ is the pseudoscalar octet meson field, and the ellipsis denote  terms with more derivatives.    To include the effect of HQSS breaking, the color magnetic moment term is included 
\begin{eqnarray}
 \mathcal{\delta L}=\frac{\lambda_2}{m_Q}Tr \bar{H}_a \sigma^{\mu\nu}\sigma_{\mu\nu} H_b. 
\end{eqnarray}
The heavy quark symmetry has been applied to analyze the exclusive semileptonic decays of  $B\to D l v_{l}$ and $B\to D^* l v_{l}$.  
The transitions of $B$ mesons into $D$ mesons  are parameterised as~\cite{Casalbuoni:1996pg,Caprini:1997mu} 
\begin{eqnarray}
 \langle D(v^\prime) | \bar{c} \gamma^{\mu} b | B(v) \rangle &=& \sqrt{m_{D}m_{B}} \xi(v\cdot v^{\prime})(v + v^{\prime})_{\mu} \\ \nonumber   \langle D^*(v^\prime,\varepsilon) | \bar{c} \gamma^{\mu} b | B(v) \rangle &=& \sqrt{m_{D}m_{B}} i\xi(v\cdot v^{\prime})\varepsilon_{\mu\nu\alpha\beta}\varepsilon^{*\nu} v^{\prime \alpha} v^{\beta}
 \\ \nonumber   \langle D^*(v^\prime,\varepsilon) | \bar{c} \gamma^{\mu} b | B(v) \rangle &=& \sqrt{m_{D}m_{B}} \xi(v\cdot v^{\prime})[(1+ v\cdot v^{\prime})\varepsilon_{\mu}^*- (\varepsilon^* \cdot v) v_{\mu}^{\prime}],
\end{eqnarray}
where $\xi(v\cdot v^{\prime})$ is the Isgur-Wise function, which indicates that the heavy quark symmetry constrains the form factors~\cite{Casalbuoni:1992dx,Cheng:1996if}.

HQET only separates the scales $\Lambda_{QCD}$ and $m_{Q}$. In the heavy quark limit,  the relative momentum scale of two heavy quarks in a doubly heavy baryon is characterized by  $m_{Q}v$, where $m_Q$ is the mass of two heavy quarks and  $v$ their velocity~\cite{Brambilla:2005yk,Hu:2005gf}.  On the other hand, the energy scale that governs the interaction between the heavy diquark and light one is $\Lambda_{QCD}$. For $m_{Q}v \gg \Lambda_{QCD}$, a pair of heavy quarks behaves like a point-like particle referred to as a diquark in the antitriplet or sextet color configuration. The interaction of the antitriplet field with the light quark is similar to that of the heavy antiquark with a light quark in the $D$ and $B$ mesons.  Consequently,  the behavior of a pair of heavy quarks is similar to that of a heavy anti-quark regarding the color degree of freedom in the heavy quark limit.    This symmetry is often called HADS, first proposed in Ref.~\cite{Savage:1990di}.  Using the HQET one can derive the relationship of mass splittings between single charmed mesons and doubly charmed baryons, e.g., $m_{D^*}-m_{D}=\frac{3}{4}(m_{\Xi_{cc}^*}-m_{\Xi_{cc}})$~\cite{Lewis:2001iz,Hu:2005gf,Brambilla:2005yk}, which has been confirmed by a series of lattice QCD studies~\cite{Padmanath:2015jea,Chen:2017kxr,Alexandrou:2017xwd,Mathur:2018rwu}.  The coupling of a doubly heavy baryon to a pion can be
related to that of a charmed meson to a pion via HADS, e.g.,  $g_{\Xi_{QQ}\Xi_{QQ}\pi}=-\frac{1}{3}g_{H_{Q}H_{Q}\pi}$~\cite{Liu:2018euh}. Using the quark model, one can also obtain such a relation, e.g.,  $g_{\Xi_{QQ}\Xi_{QQ}\pi}=-\frac{5}{12}g_{H_{Q}H_{Q}\pi}$~\cite{Liu:2018bkx}, consistent with the relation obtained from HADS.   We note that the HADS has not been explicitly verified experimentally. 

At last, we discuss the SU(3)-flavor symmetry. In general, the uncertainty of SU(3)-flavor symmetry is estimated to be $m_{s}/\Lambda_{QCD}\sim 30\%$.  In terms of the recent lattice QCD results,  the ratio of the decay constant of the $\pi$ meson to that of the $K$ meson is   $f_{K}/f_{\pi}\approx 1.19$~\cite{FlavourLatticeAveragingGroup:2019iem}, which indicates that the breaking of SU(3)-flavor symmetry is about  $19\%$. Therefore, the breaking of SU(3)-flavor symmetry is around  $20\%\sim 30\%$.  
SU(3)-flavor symmetry has also been adopted to study the branching fractions of the nonleptonic decays of charmed baryons and shown to be approximately satisfied~\cite{Lu:2016ogy,Geng:2017mxn,He:2018joe,Geng:2018plk,Geng:2018upx,Jia:2019zxi,Geng:2019awr,Geng:2023pkr}. Motivated by the exotic states discovered in $b$-flavored hadron decays,  SU(3)-flavor symmetry is employed to discuss their relationships~\cite{Huang:2022zsy,Qin:2022nof,Li:2023kcl,Han:2023teq}.   The SU(3)-flavor symmetry is also used in studying hadron-hadron interactions, such as hyperon-nucleon interactions~\cite{Aoki:2011ep,Li:2016paq,Ren:2016jna,Song:2018qqm} and hyperon-hyperon interactions~\cite{Inoue:2010es,Yamaguchi:2016kxa,Richard:2016eis,Li:2018tbt,Kamiya:2021hdb,Green:2021qol}. Recently, inspired by the discoveries of the tetraquark state $Z_{cs}(3985)$ as well as the pentaquark states  $P_{cs}(4459)$ and $P_{cs}(4338)$,  SU(3)-flavor symmetry is employed to predict their symmetry partners~\cite{Chen:2016ryt,Xiao:2019gjd,Wang:2019nvm,Peng:2020hql,Liu:2020hcv,Meng:2020ihj,Yang:2020nrt,Wang:2020htx,Chen:2021cfl,Zhu:2021lhd,Yan:2021tcp,Du:2021bgb}.

Based on the above symmetries, one can find the relationship between hadron-hadron potentials, which can help reduce the number of unknown parameters of the potentials and reduce the uncertainties of theoretical studies. In this review, we mainly focus on the contact potentials of EFTs (see the Appendix for details). Once  the hadron-hadron potentials are obtained, one can analyze the physical observables of relevant scattering  processes by solving the Schr\"odinger equation or 
Lippmann-Schwinger(LS) equation. Due to the scarcity of scattering data, one usually employs the masses and widths of exotic states near the mass thresholds of a pair of hadrons as inputs. In recent years,  lattice QCD simulations have become reliable for extracting hadron-hadron scattering information.

\subsection{Mass spectra}

In this section, we discuss the likely existence of multiplets of hadronic molecules in the $D^{(*)}K$,  $\bar{D}^{(\ast)}\Sigma_{c}^{(\ast)}$, $\bar{D}^{(\ast)}{D}^{(\ast)}$, and $\bar{\Sigma}_c^{(\ast)}\Sigma_{c}^{(\ast)}$ systems, where HQSS correlates the states belonging to the same multiplet. Moreover, we discuss the multiplets of hadronic molecules involving the excited charmed mesons or baryons related by HQSS. Based on the established  HQSS multiplets of hadronic molecules, we predict other hadronic molecules using the SU(3)-flavor symmetry, HADS, and HQFS.  

\subsubsection{$D^{(*)}K$ molecules}

\begin{table}[!h]
\centering
\caption{Binding energies ($B$ in units of  MeV) and mass spectra ($M$ in units of MeV) of isoscalar  $DK$ and $D^*K$ molecules.    
  }
\label{mass2317}
\begin{tabular}{cccccc}
\hline\hline
molecule & $I$ & $J^{P}$ 
& $B$ (MeV)
& $M$ (MeV )    \\
  \hline
  $D K$ & $0$ & $0^{+}$  & $45$ & Input  \\
  $D^*K$ & $0$ & $1^{+}$  & 48 & $2456$  \\
  \hline \hline
\end{tabular}
\end{table}

Identifying $D_{s0}^*(2317)$ and $D_{s1}(2460)$ as $DK$ and $D^*K$ molecules, their mass splitting can be naturally explained~\cite{Chen:2004dy}. Since the $D$ meson and the $D^*$ meson are related to each other by HQSS, $D_{s0}^*(2317)$ and $D_{s1}(2460)$  are viewed as a doublet of hadronic molecules. In terms of the meson exchange theory~\cite{Liu:2020nil} and chiral perturbation theory~\cite{Altenbuchinger:2013vwa}, the $DK$ and $D^*K$ potentials are the same, which implies that the contact-range potentials of  $DK$ and $D^*K$  are the same, denoted by $C_a$~\cite{Hu:2020mxp}.  With  the mass of $D_{s0}^*(2317)$, the unknown parameter $C_a$ is determined to be $62.61$~GeV$^{-2}$ for a cutoff $\Lambda=1$~GeV, and then the mass of the $D^*K$ molecule is predicted as shown in Table~\ref{mass2317},  which satisfies HQSS. In the molecular picture~\cite{Kolomeitsev:2003ac,Guo:2006rp,Altenbuchinger:2013vwa,Kong:2021ohg,Liu:2023cwk}, $D_{s1}(2460)$ is considered to be the counterpart of $D_{s0}^*(2317)$.  Recent studies showed that the compositeness of $D_{s0}^*(2317)$ and $D_{s1}(2460)$ is at least $70\%$ and $50\%$~\cite{MartinezTorres:2014kpc,Albaladejo:2018mhb,Yang:2021tvc,Song:2022yvz,Guo:2023wkv,Gil-Dominguez:2023huq}, where the impact of other components on the $D^{(*)}K$ scattering can come from the CDD pole or the form factors. Replacing the $K$ meson by the $K^*$ meson, one would 
 naturally  expect multiplets of hadronic molecules of $DK^*$~\cite{Gamermann:2007fi} and $D^*K^*$~\cite{Molina:2010tx}.

 \begin{figure}[!h]
    \centering
    \includegraphics[width=11cm]{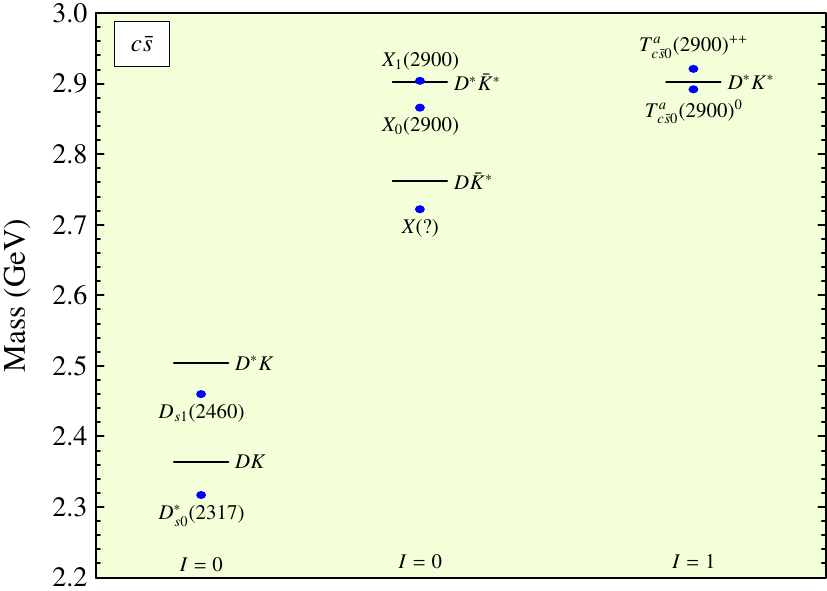}\\
    \caption{Locations of charmed-strange hadronic molecular candidates with respect to the $D^{(*)}K$,  $D^{(*)}\bar{K}^*$, and $D^*K^*$ mass thresholds.  }\label{dsspectrum}
\end{figure}

 With the $G$-parity transformation, one can relate the $D^{*}\bar{K}^*$ system with the $D^{*}K^*$ system. Several enhancements have been discovered near the $D^{*}\bar{K}^*$ mass thresholds~\cite{LHCb:2020bls,LHCb:2022sfr}.     
 The open-charm tetraquark state $X_{0}(2866)$ is assigned as the isoscalar  $D^{*}\bar{K}^*$ bound state with $J^P=0^+$~\cite{Liu:2020nil,Chen:2020aos,Huang:2020ptc,Hu:2020mxp,Xiao:2020ltm,Kong:2021ohg}. However, it is still controversial for the existence of the two molecules of $J^P=1^+$ $D^{*}\bar{K}^*$ and $J^P=2^+$  $D^{*}\bar{K}^*$.  The OBE model indicates that the  $D^{*}\bar{K}^*$ potential with $J^P=0^+$  is more attractive than that of  $D^{*}\bar{K}^*$ with $J^P=2^+$~\cite{Liu:2020nil}. However, it is opposite to the results of the local hidden gauge approach~\cite{Molina:2020hde}. The difference can be traced to the different prescriptions of the spin-spin interactions of the two hadrons. Very recently, assuming that the hadron-hadron interaction at the quark level is only generated by the $u$ and $d$ quarks while the $s$ quark is treated as a heavy spectator, Wang et al. assigned the newly discovered  $T_{cs0}(2900)$ and  $T_{c\bar{s}0}^a(2900)$ as the SU(2) partners of  $T_{cc}$ and $Z_c(3900)$ in the molecular picture~\cite{Wang:2023hpp}. We show the locations of the above exotic states with respect to the mass thresholds of $D^{(*)}K$,  $D^{(*)}\bar{K}^*$, and $D^*K^*$ in Fig.~\ref{dsspectrum}.     

One can relate a charmed meson to a doubly charmed baryon through HADS in the heavy quark limit.   Therefore, it can relate the $\bar{\Xi}_{cc}^{(*)} K$ system to the above $D^{(*)} K$ system via HADS. The existence of $\bar{\Xi}_{cc}^{(*)} K$ molecules was firstly predicted by Guo et al.~\cite{Guo:2011dd} based on the SELEX measurement of $\Xi_{cc}$~\cite{SELEX:2002wqn}. Later, the $\bar{\Xi}_{cc}^{(*)} K$ potentials and corresponding hadronic molecules were investigated based on the proposal of $D_{s0}^*(2317)$ and $D_{s1}(2460)$ as $DK$ and $D^*K$ molecules ~\cite{Guo:2017vcf,Yan:2018zdt,Meng:2018zbl,Wu:2020rdg,Wang:2022aga,Liang:2023scp}.  In Ref.~\cite{Yan:2018zdt}, Yan et al. pointed out that the $\bar{\Xi}_{cc}^{(*)} K$ molecules can mix with excited $\Omega_{cc}$ baryons.  Moreover, in terms of HQFS, one would expect the existence of  $BK$ and $B^*K$ molecules~\cite{Kolomeitsev:2003ac,Guo:2006fu,Guo:2006rp,Faessler:2008vc,Cleven:2010aw,Guo:2011dd,Colangelo:2012xi,Altenbuchinger:2013vwa,Lang:2015hza,Fu:2021wde}, the bottom partners of  $D_{s0}^*(2317)$ and $D_{s1}(2460)$, i.e., $B_{s0}^*$ and $B_{s1}$. It is worth noting that the lattice QCD predictions on the $B_{s0}^*$ and $B_{s1}$ masses~\cite{Lang:2015hza} are consistent with ChPT~\cite{Guo:2006fu,Guo:2006rp,Guo:2011dd,Colangelo:2012xi,Altenbuchinger:2013vwa}.   Currently, there is no experimental sign for the existence of these bottom mesons.  Their future discovery will surely help verify the molecular nature of $D_{s0}^*(2317)$ and $D_{s1}(2460)$.  With HQSS and HQFS,  Guo et al. predicted the existence of a series of hadronic molecules composed of kaon and $P$-wave charmed/bottom mesons~\cite{Guo:2011dd}.

\subsubsection{$\bar{D}^{(\ast)}\Sigma_{c}^{(\ast)}$ molecules}

\begin{table*}[htpb]
  \caption{
    Binding energies (in units of MeV) for the pentaquarks composed of
   a charmed baryon and a charmed antimeson, obtained in the contact-range EFT with the two constants fixed by reproducing  $P_{c}(4440)$ and $P_{c}(4457)$ as $1/2^-$ and $3/2^-$ molecules (scenario A) and $3/2^-$ and $1/2^-$ (scenario B), for cutoffs of 0.5 and 1 GeV. The $15\%$ uncertainty is estimated by the heavy quark spin symmetry breaking for the contact-range potentials. 
    \label{tab:predictions-Pc}
}
\centering
\begin{tabular}{cccccc}
  \hline\hline State& $J^{P}$ &Threshold & $\Lambda$(GeV) & B.E (Scenario A)&B.E (Scenario B)\\ \hline
$P_{c}(4312)$($\bar{D}\Sigma_{c}$) & $\tfrac{1}{2}^{-}$ &4320.8&  0.5(1)
  &  $8.8^{+5.0}_{-4.2}$ $(7.6^{+9.2}_{-6.0})$ &  $14.4^{+6.5}_{-5.8}$ $(12.8^{+11.8}_{-8.6})$
  \\
  \hline $P_{c}(?)$($\bar{D}\Sigma_{c}^{\ast}$) &$\tfrac{3}{2}^{-}$ &4385.3& 0.5(1)
  &$9.1^{+5.0}_{-4.3}$ $(8.1^{+9.4}_{-6.3})$ & $14.7^{+6.6}_{-5.9}$ $(13.5^{+12.0}_{-8.9})$  \\ \hline
  $P_{c}(4440)$($\bar{D}^{\ast}\Sigma_{c}$) &? &4462.2& 0.5(1)  &
  Input($\tfrac12^{-}$)     &Input($\tfrac32^{-}$)
  \\
  \hline $P_{c}(4457)$($\bar{D}^{\ast}\Sigma_{c}$)& ?
  &4462.2& 0.5(1) & Input($\tfrac32^{-}$) & Input($\tfrac12^{-}$)
  \\
  \hline $P_{c}(?)$($\bar{D}^{\ast}\Sigma_{c}^{\ast}$) &$\tfrac12^{-}$ &4526.7& 0.5(1)
  & $25.6^{+9.0}_{-8.4}$ $(26.3^{+16.6}_{-13.7})$ & $3.0^{+2.8}_{-2.0}$ $(3.4^{+6.3}_{-3.2})$
  \\
  \hline $P_{c}(?)$($\bar{D}^{\ast}\Sigma_{c}^{\ast}$) &$\tfrac32^{-}$ &4526.7& 0.5(1)
  & $15.8^{+6.7}_{-6.1}$ $(15.9^{+12.8}_{-9.8})$ & $10.0^{+5.2}_{-4.5}$ $(10.0^{+10.2}_{-7.2})$
  \\
  \hline $P_{c}(?)$($\bar{D}^{\ast}\Sigma_{c}^{\ast}$) &$\tfrac52^{-}$&4526.7& 0.5 (1)
  & $3.0^{+2.8}_{-2.0}$ $(3.4^{+6.3}_{-3.2})$ & $25.6^{+9.0}_{-8.4}$ $(26.3^{+16.6}_{-13.7})$
\\ \hline \hline
\end{tabular}
\end{table*}

In the heavy quark limit, the  $\bar{D}^{(\ast)}\Sigma_{c}^{(\ast)}$ system develops seven states. Its contact-range potentials are parameterized by two unknown constants, i.e., $C_a$ and $C_b$.   
Assuming the three hidden-charm pentaquark states $P_{c}(4312)$, $P_{c}(4440)$, and $P_{c}(4457)$ as  $\bar{D}^{(\ast)}\Sigma_{c}$ bound states, one can fully determine $C_a$ and $C_b$, and then predict the masses of the other four states. 
 In Ref.~\cite{Liu:2019tjn}  two scenarios are proposed:  the $P_c(4440)$ and $P_c(4457)$ as $\tfrac{1}{2}^-$ and $\tfrac{3}{2}^-$ $\bar{D}^* \Sigma_c$ molecules (scenario A) or $\tfrac{3}{2}^-$ and $\tfrac{1}{2}^-$ $\bar{D}^* \Sigma_c$ molecules (scenario B). The $P_{c}(4312)$ determines a favorable scenario. To show the impact of the cutoff in the form factor on the results,  the cutoff is varied from $0.5$ GeV to $1$ GeV. In addition,  a $15\%$ uncertainty originating from HQSS breaking was considered, which is estimated by the expansion  $\mathcal{O}(\Lambda_{QCD}/m_{c})$ with $\Lambda_{QCD}\sim 200-300$ MeV and $m_{c}=1.5$ GeV~\cite{Pan:2019skd}.

    In Table~\ref{tab:predictions-Pc},  the mass spectra of the $\bar{D}^{(\ast)}\Sigma_{c}^{(\ast)}$ molecules in both scenarios are presented, where the values inside and outside the parentheses correspond to the cutoff for $\Lambda=1$~GeV and $\Lambda=0.5$~GeV. One finds that the results only weakly depend on the cutoff, and one can not discriminate the two scenarios using the $P_c(4312)$ mass as the only criterion.  Nevertheless, in addition to $P_{c}(4312)$, $P_{c}(4440)$, and $P_{c}(4457)$, four more molecules are predicted, leading to a complete HQSS multiplet of hadronic molecules~\cite{Liu:2019tjn}. Such a multiplet is later confirmed by many other theoretical studies~\cite{Xiao:2019aya,Sakai:2019qph,Yamaguchi:2019seo,Valderrama:2019chc,Meng:2019ilv,Du:2019pij,Yalikun:2021bfm,Xie:2022hhv}.
    Given the existence of such a multiplet, searches for the other four molecules are crucial to verify the molecular nature of $P_{c}(4312)$, $P_{c}(4440)$, and $P_{c}(4457)$, which offers a model-independent approach to check the molecular picture of the pentaquark states.    
  In addition to the  $\bar{D}^{(\ast)}\Sigma_c$ channels, the $\bar{D}\Lambda_{c1}(2595)$ channel was claimed to play a nonnegligible role in generating the molecular pentaquark states~\cite{Burns:2019iih}. In Ref.~\cite{Peng:2020gwk}, Peng et al. used the contact-range approach to study the impact of the $\bar{D}\Lambda_{c1}(2595)$ channel on the  $\bar{D}^{(\ast)}\Sigma_c$ molecules, suggesting that there 
   may exist two peak structures around $P_{c}(4457)$ with different parity. However, In the OBE model, the  $\bar{D} \Lambda_{c1}(2595)$ channel is less relevant in forming the molecules~\cite{Yalikun:2021bfm}.

    It is evident that in the molecular picture, the spins of $P_c(4440)$ and $P_{c}(4457)$  can not be resolved in the contact-range EFT. As a result, we turn to the phenomenological model, i.e., the OBE model. In our study, after removing the delta term of the spin-spin interaction,  we used a unified cutoff to reproduce the masses of $P_{c}(4312)$, $P_{c}(4440)$, and $P_{c}(4457)$ in the $\bar{D}^{(*)}\Sigma_c$ molecular picture, and predict their HQSS partners, in agreement with the results of scenario B of the contact-range EFT approach, which favors the spins of $P_{c}(4440)$ and $P_{c}(4457)$ as  $3/2$ and $1/2$~\cite{Liu:2019zvb}, Such a preference is in agreement with the analysis of the LHCb invariant mass distributions by the pionful EFT \cite{Du:2019pij,Du:2021fmf}. We note that whether to remove the delta term of the spin-spin interaction remains open. In Ref.~\cite{Yalikun:2021bfm}, the authors did not completely remove the delta term rather than introduced a parameter that controls the delta contribution to calculate the masses of the $\bar{D}^{(\ast)}\Sigma_{c}^{(\ast)}$ molecules in both scenarios. Up to now, the spins of  $P_{c}(4440)$ and $P_{c}(4457)$ and the nature of the pentaquark states remain undetermined.    Therefore, it is crucial to study these puzzles using other approaches.  In addition to searching for the HQSS partners mentioned above, the existence of their SU(3)-flavor partners can also help resolve this issue.

\begin{table}[ttt]
\centering
\caption{Binding energies of ${D}^{(\ast)}\Sigma_{c}^{(\ast)}$ molecules with $I=1/2$ }
\label{dk5}
\begin{tabular}{cccccc}
\hline\hline
Molecule  & $I$ & $J^{P}$  & B.E (MeV) &  Mass (MeV) \\
\hline
${D}\Sigma_c$ & $\tfrac{1}{2}$ & $\tfrac{1}{2}^-$ &
$31.7^{+16.6}_{-13.9}$  &4289.3 \\
\hline
${D}\Sigma_c^*$ & $\tfrac{1}{2}$ & $\tfrac{3}{2}^-$ &
$32.5^{+16.8}_{-14.1}$ & $4352.5$ \\
\hline
${D}^*\Sigma_c$ & $\tfrac{1}{2}$ & $\tfrac{1}{2}^-$
&  $18.4^{+11.9}_{-9.3}$  & $4444.6$  \\
${D}^*\Sigma_c$ & $\tfrac{1}{2}$ & $\tfrac{3}{2}^-$
&  $57.4^{+24.8}_{-21.9}$ & $4405.6$  \\
\hline
${D}^*\Sigma_c^*$ & $\tfrac{1}{2}$ & $\tfrac{1}{2}^-$
&  $19.2^{+12.7}_{-9.8}$ & $4507.8$ \\
${D}^*\Sigma_c^*$ & $\tfrac{1}{2}$ & $\tfrac{3}{2}^-$
&  $32.1^{+17.0}_{-14.2}$& $4494.9$  \\
${D}^*\Sigma_c^*$ & $\tfrac{1}{2}$ & $\tfrac{5}{2}^-$
& $61.4^{+25.9}_{-23.7}$ &$4465.6$\\
  \hline\hline
\end{tabular}
\end{table}
    
With the $G$-transformation, one can obtain the ${D}^{(\ast)}\Sigma_{c}^{(\ast)}$ potentials from the  $\bar{D}^{(\ast)}\Sigma_{c}^{(\ast)}$ ones. The only difference in the OBE  potentials is the sign of the $\pi$ and $\omega$ exchange potentials. One can straightforwardly obtain the mass spectrum of the ${D}^{(\ast)}\Sigma_{c}^{(\ast)}$ system in terms of the  obtained  $\bar{D}^{(\ast)}\Sigma_{c}^{(\ast)}$ mass spectrum as shown in Table~\ref{dk5}~\cite{Liu:2020nil}, which has been investigated after the doubly charmed tetraquark state $T_{cc}$ was discovered by the LHCb Collaboration~\cite{Chen:2021kad,Shen:2022zvd,Liu:2023clr}.     
One can see that the  ${D}^{(\ast)}\Sigma_{c}^{(\ast)}$ states bind more than the $\bar{D}^{(\ast)}\Sigma_{c}^{(\ast)}$ ones. Such a feature has been observed in other systems. For instance,  the $\bar{D}^{(*)}D^{(*)}$ states bind more than the  ${D}^{(*)}D^{(*)}$ ones  and the $\bar{\Sigma}_c^{(*)}{\Sigma}_c^{(*)}$  states more than the   
${\Sigma}_c^{(*)}{\Sigma}_c^{(*)}$ ones~\cite{Peng:2020xrf}. Refs.~\cite{Dong:2021juy,Dong:2021bvy,Chen:2021spf,Wang:2023hpp,Wang:2023eng} have investigated the hadronic molecules near the thresholds of the hidden-charm and open-charm hadron-hadron systems.

     \begin{table}[ttt]
\caption{Bound states of a singly charmed baryon and a singly charmed antimeson, obtained in the contact-range EFT  with the two constants fixed by reproducing  $P_{c}(4440)$ and $P_{c}(4457)$ as $1/2^-$ and $3/2^-$ molecules (scenario A) and $3/2^-$ and $1/2^-$ (scenario B),
with cutoffs of 0.5 and 1 GeV,  and the $25\%$ uncertainty caused by SU(3)-flavor symmetry and heavy quark spin symmetry breaking for the contact-range potentials.  \label{resutlspcs} }
\centering
\begin{tabular}{cccccccc}
\hline\hline State& $J^{P}$  & $\Lambda$(GeV)   & B. E(A) & Mass(A) &  B. E(B) & Mass(B)   \\
\hline $\bar{D}\Xi_{c}^{\prime}$ &$1/2^{-}$ &
1(0.5) & $8.5^{+17.4}_{-8.4}(9.3_{-6.7}^{+8.7})$  &4437(4436)  & $14.0^{+21.7}_{-12.8}(14.9_{-9.3}^{+11.4})$    & 4431(4430)  
\\
  \hline
$\bar{D}\Xi_{c}^{\ast}$ &$3/2^{-}$ & 1(0.5)&
$9.0^{+17.7}_{-8.8}(9.5_{-6.7}^{+7.8})$ &4504(4504)   & $14.7^{+21.9}_{-13.3}$($15.2_{-9.4}^{+11.4}$)   &   4499(4498)    
\\ \hline
$\bar{D}^{\ast}\Xi_{c}^{\prime}$ &$1/2^{-}$ &1(0.5)  & $23.4^{+27.0}_{-18.9}(22.5_{-12.3}^{+14.2})$     &4563(4564)  &   $5.6^{+14.3}_{\dag}$$(5.2_{-4.9}^{+6.4})$  &4581(4581) 
\\
$\bar{D}^{\ast}\Xi_{c}^{\prime}$& $3/2^{-}$
 &1(0.5) & $5.6^{+14.3}_{\dag}(5.2_{-4.3}^{+6.4})$ & 4581(4581)  & $23.4^{+27.0}_{-18.8}$($22.5_{-12.3}^{+14.2}$)  &  4563(4564)
 
\\
\hline $\bar{D}^{\ast}\Xi_{c}^{\ast}$ &$1/2^{-}$
&1(0.5)& $28.0^{+29.4}_{-21.4}(26.3_{-13.7}^{+15.5})$ & 4627(4628)  & $4.0^{+12.5}_{\dag}$ $(3.3_{-3.0}^{+5.1})$  & 4651(4651)  
\\
 $\bar{D}^{\ast}\Xi_{c}^{\ast}$ &$3/2^{-}$
& 1(0.5)& $17.2^{+23.2}_{-14.9}(16.4_{-9.8}^{+11.6})$ &4637(4638)  & $11.1^{+18.9}_{-10.5}$$(10.5_{-7.2}^{+9.1})$   &  4643(4644)  
\\
 $\bar{D}^{\ast}\Xi_{c}^{\ast}$ &$5/2^{-}$
& 1(0.5)& $4.0^{+12.5}_{\dag}(3.3_{-3.0}^{+5.1})$ & 4651(4651)  & $28.0^{+29.4}_{-21.4}$$(26.3_{-13.7}^{+15.5})$  & 4627(4628)   
\\
\hline\hline
\end{tabular}
\end{table}

With  SU(3)-flavor symmetry, one can relate the $\bar{D}^{(\ast)}\Sigma_{c}^{(\ast)}$ system to the $\bar{D}^{(\ast)}\Xi_{c}^{\prime(\ast)}$ system, and the latter also generates seven states dictated by HQSS.   Before the discovery of the $P_{cs}(4459)$ by the LHCb Collaboration,  the  mass spectra  of the $\bar{D}^{(\ast)}\Xi_{c}^{\prime(\ast)}$ system  were predicted in several theoretical works~~\cite{Xiao:2019gjd,Peng:2019wys,Wang:2019nvm}. Since the contact-range potentials of the $\bar{D}^{(\ast)}\Xi_{c}^{\prime(\ast)}$ system are the same as those of the $\bar{D}^{(\ast)}\Sigma_{c}^{(\ast)}$ system, one can easily obtain the mass spectrum of the $\bar{D}^{(\ast)}\Xi_{c}^{\prime(\ast)}$ system~\cite{Liu:2020hcv}. In Table~\ref{resutlspcs}, we present the masses of the $\bar{D}^{(\ast)}\Xi_{c}^{\prime(\ast)}$ molecules  for scenarios A and B, consistent with Refs.~\cite{Xiao:2019gjd,Wang:2019nvm}. Utilizing the OBE model,  Chen et al. arrived at similar conclusions~\cite{Chen:2022onm}. The experimental searches for $\bar{D}^{(\ast)}\Xi_{c}^{\prime(\ast)}$ molecules are crucial to understand the molecular  nature of $P_{c}(4312)$, $P_{c}(4440)$, and $P_{c}(4457)$.

   \begin{figure}[!h]
\begin{center}
\begin{overpic}[scale=.60]{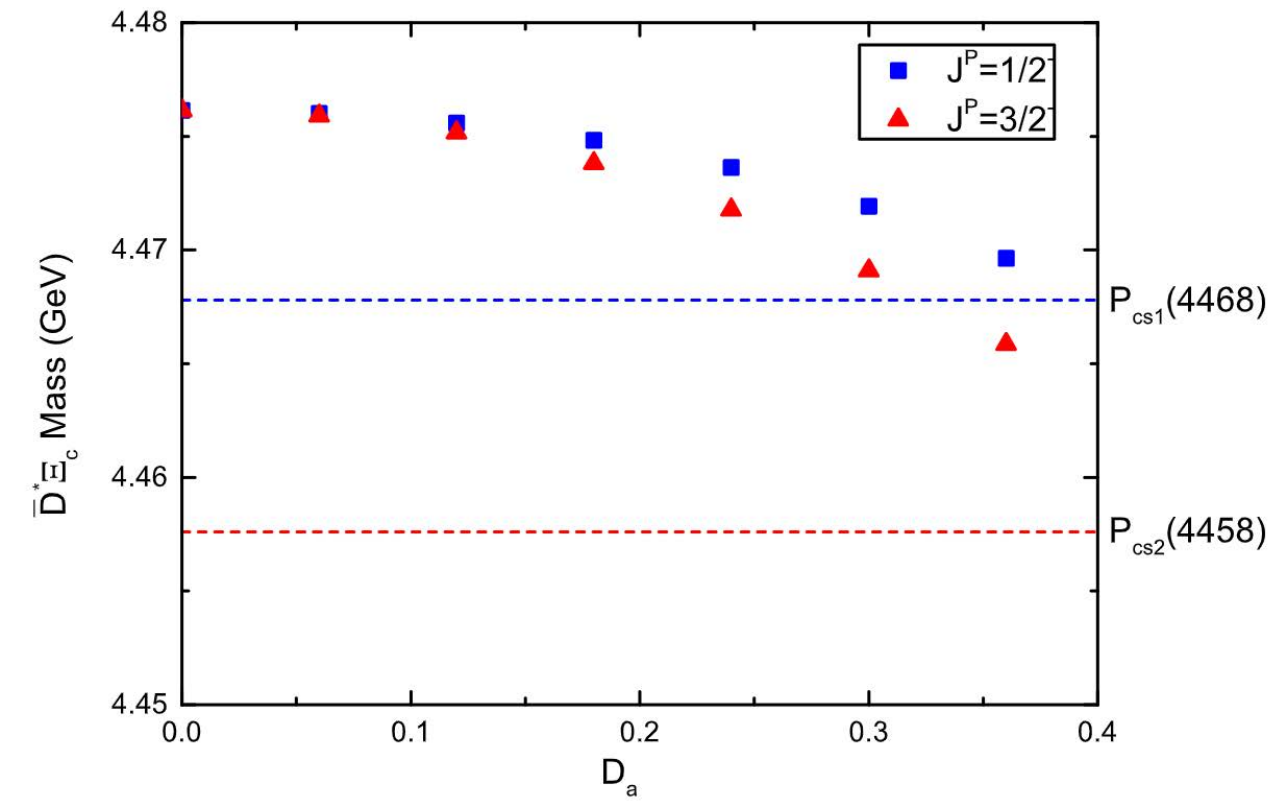}
\end{overpic}
\caption{$J^{P}=1/2^{-}$ $\bar{D}^{\ast}\Xi_{c}$  and $J^{P}=3/2^{-}$ $\bar{D}^{\ast}\Xi_{c}$ masses in scenario A  as a function of the parameter $D_a$, which describes the strength of the elastic potentials: $V=D_a\times F_{1/2}^{\prime}$       }
\label{SA}
\end{center}
\end{figure}

\begin{figure}[!h]
\begin{center}
\begin{overpic}[scale=.60]{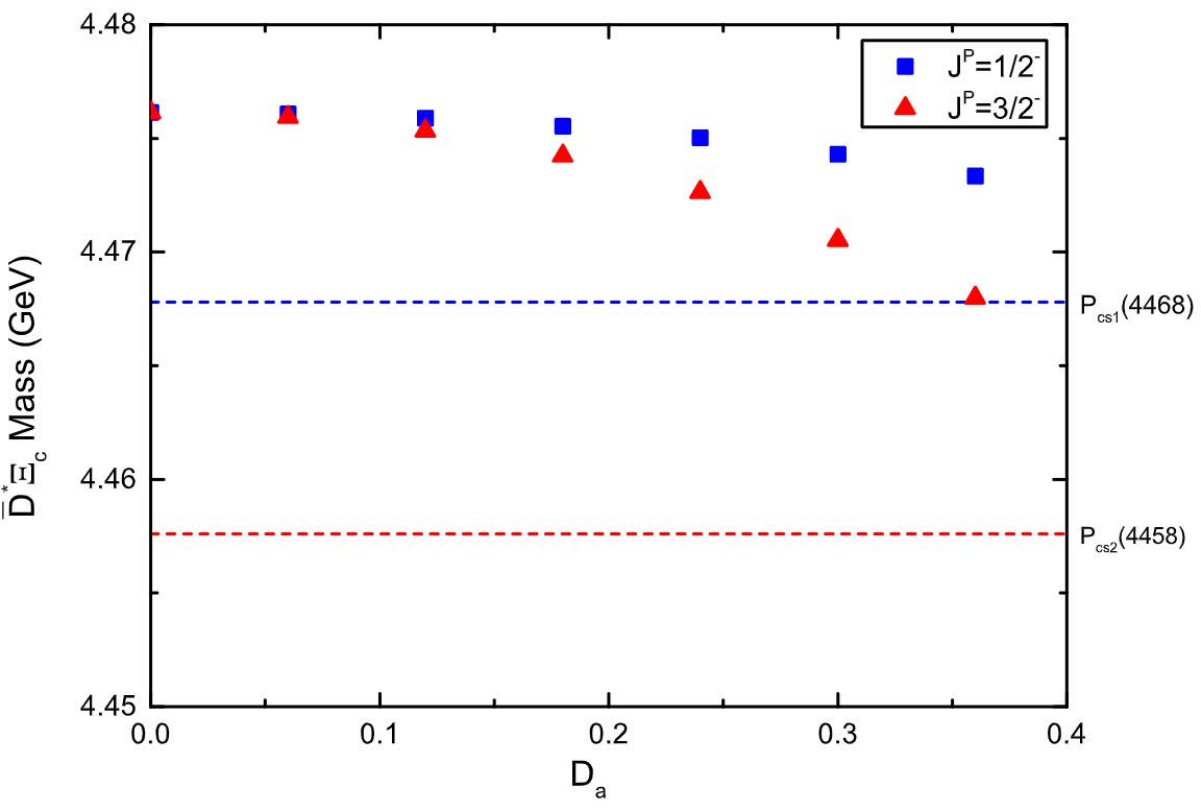}
\end{overpic}
\caption{Masses of $J^{P}=1/2^{-}$ and $J^{P}=3/2^{-}$ $\bar{D}^{\ast}\Xi_{c}$ molecules in scenario B  as a function of the parameter $D_a$, which describes the strength of the elastic potentials: $V=D_a\times F_{1/2}^{\prime}$       }
\label{SB}
\end{center}
\end{figure}

   \begin{figure}[!h]
    \centering
    \includegraphics[width=11cm]{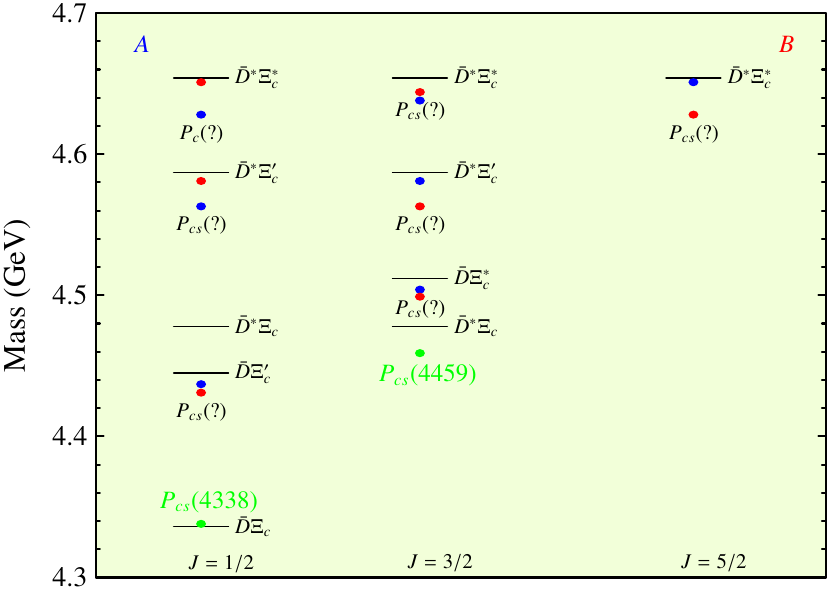}\\
    \caption{ Spectrum of  $\bar{D}^{(*)}\Xi_{c}^{\prime(*)}$  hadronic molecules in Scenario A~(shown in blue) and Scenario B~(shown in red), and the 
 locations of hidden-charm strangeness pentaquark hadronic molecular candidates with respect to the  $\bar{D}^{(\ast)}\Xi_c$ mass thresholds. }\label{pcsspectrum}
\end{figure}

   We note that the hidden-charm strange pentaquark states $P_{cs}(4459)$ and $P_{cs}(4338)$ discovered by the LHCb Collaboration are near the $\bar{D}^*\Xi_c$ and $\bar{D}\Xi_c$  mass thresholds as shown in Fig.~\ref{pcsspectrum}. As a result, it is necessary to study the $\bar{D}^{(\ast)}\Xi_c$  system. 
 In the heavy quark limit, the $\bar{D}^{(\ast)}\Xi_c$ system develops three states, and the contact-range potentials for all three states are the same. Assuming the $P_{cs}(4459)$ as a $\bar{D}^*\Xi_c$ bound state, it is natural to expect one more $\bar{D}^*\Xi_c$ bound state and a  $\bar{D}\Xi_c$ bound state with the same binding energy, where the $\bar{D}\Xi_c$ molecule corresponds to $P_{cs}(4338)$. In this proposal, the binding energy of the $\bar{D}^*\Xi_c$ bound state is around $20$~MeV, which indicates that the binding energy of the $\bar{D}\Xi_c$ bound state should also be around $20$ MeV, quite different from the experimental mass of $P_{cs}(4338)$.  Such a naive expectation of a $\bar{D}^{(\ast)}\Xi_c$  multiplet in the single-channel approximation is not consistent with the current experimental data.  One needs to consider coupled-channel effects to understand these pentaquark states better. The meson exchange theory tells that the coupled channels are $\bar{D}^*\Xi_{c}-\bar{D}^*\Xi_{c}^{\prime}$ and $\bar{D}^*\Xi_{c}-\bar{D}\Xi_{c}^{\ast}$ for $J=1/2$ and $J=3/2$, but there is no coupling to $\bar{D}\Xi_c$.  As depicted in Fig.~\ref{pcsspectrum}, the $\bar{D}^*\Xi_{c}$ mass threshold is close to the $\bar{D}^*\Xi_{c}^{\prime}$ mass threshold, and the $\bar{D}^*\Xi_{c}$ mass threshold is close to the $\bar{D}\Xi_{c}^{\ast}$ mass threshold. There are likely strong couplings between these channels.    Therefore, we proposed $P_{cs}(4338)$ as a $\bar{D}\Xi_c$ bound state and fixed the contact-range potentials of the $\bar{D}^{(\ast)}\Xi_c$ system.

  \begin{figure}[ttt]
    \centering
    \includegraphics[width=17cm]{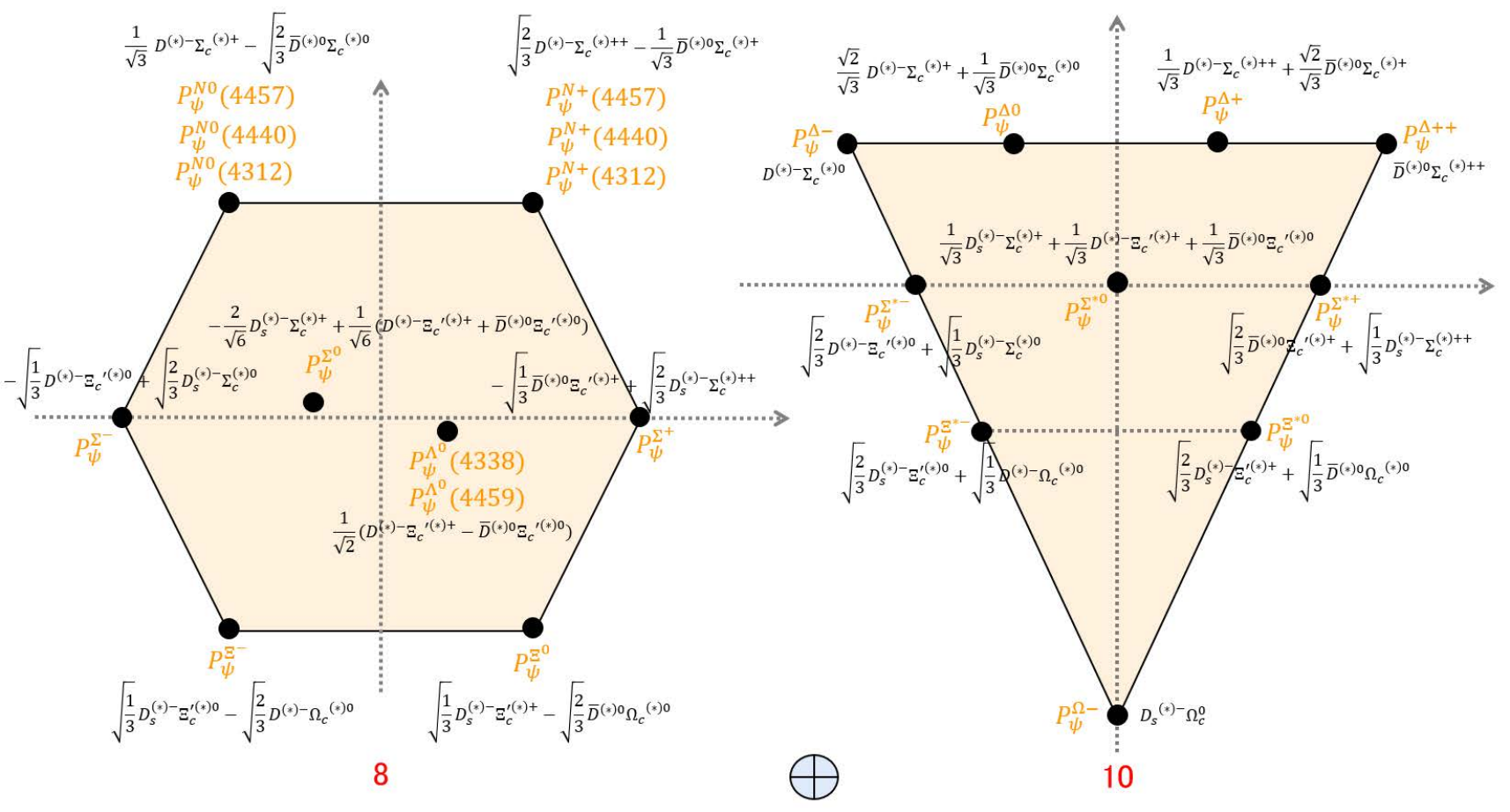}\\
    \caption{  $P_\psi^N$ and  $P_{\psi}^{\Lambda}$ as part of SU(3)-flavor multiplet of hidden-charm pentaquark hadronic molecules, and $P_{\psi}^{N}(4312)$,  $P_{\psi}^{N}(4440)$ and  $P_{\psi}^{N}(4457)$ as part of HQSS multiplet of hadronic molecules  $P_\psi^N$  as well as $P_{\psi}^{\Lambda}(4338)$ and $P_{\psi}^{\Lambda}(4459)$ as part of HQSS multiplet of hadronic molecules $P_{\psi}^{\Lambda}$.     }\label{dbarsigmacmultiplet}
\end{figure}

 The contact-range potentials of the   $\bar{D}^*\Xi_{c}^{\prime}$ and $\bar{D}\Xi_{c}^*$  system are determined by the $\bar{D}^{(\ast)}\Sigma_c^{(\ast)}$ system  for scenarios A and B. The off-diagonal contact potentials parameterized by $D_a$ are difficult to estimate.  Facing this situation, we require that the off-diagonal potentials be in a reasonable region. That is to say, the higher pole should always exist, and the imaginary part of the higher pole\footnote{The  $\bar{D}^*\Xi_{c}^{\prime}-\bar{D}^*\Xi_{c}$ coupled channel would generate two poles: a higher pole and a lower pole, dominantly coupling to the $\bar{D}^*\Xi_{c}^{\prime}$ channel and the  $\bar{D}^*\Xi_{c}$ channel, respectively. From  SU(3)-flavor symmetry, the higher pole can be seen as the counterpart of  $P_{c}(4440)$ and $P_{c}(4457)$.} is smaller than those of $P_{c}(4440)$ and $P_{c}(4457)$. One can see that the degeneracy of the two $\bar{D}^*\Xi_c$ states gradually lifts as $D_a$  increases in scenarios A and B as shown in Fig.~\ref{SA} and Fig.~\ref{SB}. The two states may correspond to the structures around $P_{cs}(4459)$. In both scenarios, we find that the $J^P=3/2^-$ $\bar{D}^*\Xi_c$ molecule is more bound than the $J^P=1/2^-$ $\bar{D}^*\Xi_c$ molecule. As a result, we obtain another HQSS multiplet of hadronic molecules composed of $\bar{D}^{(\ast)}\Xi_c$ and $\bar{D}^{(\ast)}\Xi_{c}^{\prime(\ast)}$ as shown in Fig.~\ref{pcsspectrum}. Refs.~\cite{Karliner:2022erb,Wang:2022mxy} argued for the existence of $\bar{D}^{(*)}\Xi_c$  molecules, while the mechanism for the breaking of two degenerate $\bar{D}^*\Xi_c$ states was not discussed. In Ref.~\cite{Giachino:2022pws},  Giachino et al. not only considered the effect of coupled channels but also $S-D$ wave mixing to assign the $P_{cs}(4338)$ and two structures around $P_{cs}(4459)$ as  $\bar{D}^{(\ast)}\Xi_c$ molecules in the meson exchange model, which to some extent explains the binding mechanism of $\bar{D}^{(*)}\Xi_c$  molecules.  The $D_{s}^{(\ast)}\Lambda_c$ contribution to the  $\bar{D}^{(*)}\Xi_c$  molecules are also investigated in Refs.~\cite{Chen:2022wkh,Giachino:2022pws,Feijoo:2022rxf}. One can see that the  $\bar{D}^{(\ast)}\Xi_c$ channels, $\bar{D}^{(\ast)}\Xi_{c}^{\prime(\ast)}$ channels, and $D_{s}^{(*)}\Lambda_c$ channels all play a role in forming the hidden-charm strange molecules, while the role of each component is ambiguous due to the large uncertainties in the potentials.

\begin{table}[ttt]
 \caption{
  Heavy- and light-flavor symmetry partners of
  the LHCb pentaquark trio, $P_c(4312)$, $P_c(4440)$, and
  $P_c(4457)$ (or $P_{c1}$, $P_{c2}$, $P_{c3}$ for short).
  These include the five-flavor pentaquarks with quark contents
  ${\bar b} c s d u$ and $b {\bar c} s d u$. The masses of the $P_{c}$s partners are assigned with a $20\%$ uncertainty originating from the SU(3) flavor symmetry breaking.
}
\label{tab:partners1}
\begin{tabular}{cccccc|cccccc}
\hline \hline
Molecule & $I$ & $S$ & $B_P$ & $M_P$ & Partner
&
Molecule & $I$ & $S$ & $B_P$ & $M_P$ & Partner 
\\
  \hline
  $\bar{D} \Sigma_c$ & $\tfrac{1}{2}$ & \phantom{+}$0$ & Input & Input
  & $P_{c1}$
  &
  ${B} \Sigma_c$ & $\tfrac{1}{2}$ & \phantom{+}$0$ &
  $27.5^{+9.5}_{-8.0}$ & $7710.5^{+8.0}_{-9.5}$ & $P_{c1}$
  \\
  $\bar{D}^* \Sigma_c$ & $\tfrac{1}{2}$ & \phantom{+}$0$ & Input & Input
  & $P_{c2}$
  &
  ${B}^* \Sigma_c$ & $\tfrac{1}{2}$ & \phantom{+}$0$ &
  $43.6^{+10.6}_{-9.3}$ & $7734.6^{+9.3}_{-10.6}$
  & $P_{c2}$
  \\
  $\bar{D}^* \Sigma_c$ & $\tfrac{1}{2}$ & \phantom{+}$0$ & Input & Input
  & $P_{c3}$
  &
  ${B}^* \Sigma_c$ & $\tfrac{1}{2}$ & \phantom{+}$0$ &
  $18.6^{+7.6}_{-6.0}$ & $7759.7^{+6.0}_{-7.6}$
  & $P_{c3}$
  \\
  \hline
  $\bar{D} \Xi_c'$ & $0$ & $-1$ &
  $9.6^{+10.4}_{-7.3}$ & $4436.3^{+7.3}_{-10.4}$ & $P_{c1}$
  &
  ${B} \Xi_c'$ & $0$ & $-1$ &
  $29.0^{+18.4}_{-16.1}$ & $7829.0^{+16.1}_{-18.4}$ & $P_{c1}$
  \\
  $\bar{D}^* \Xi_c'$ & $0$ & $-1$ &
  $22.9^{+15.9}_{-13.1}$ & $4564.7^{+13.1}_{-15.9}$ & $P_{c2}$
  &
  ${B}^* \Xi_c'$ & $0$ & $-1$ &
  $45.3^{+23.1}_{-21.0}$ & $7857.8^{+21.0}_{-23.1}$ & $P_{c2}$ 
  \\
  $\bar{D}^* \Xi_c'$ & $0$ & $-1$ &
  $5.4^{+7.7}_{-4.7}$ & $4581.8^{+4.7}_{-7.7}$ & $P_{c3}$
  &
  ${B}^* \Xi_c'$ & $0$ & $-1$ &
  $19.8^{+14.5}_{-12.3}$ & $7883.3^{+12.3}_{-14.5}$ & $P_{c3}$
  \\
  \hline
  $\bar{D} \Sigma_b$ & $\tfrac{1}{2}$ & \phantom{+}$0$
  & $20.2^{+5.3}_{-4.7}$ & $7660.1^{+4.7}_{-5.3}$ & $P_{c1}$
  &
  ${B} \Sigma_b$ & $\tfrac{1}{2}$ & \phantom{+}$0$
  & $48.4^{+22.7}_{-18.3}$ & $11044.2^{+18.3}_{-22.7}$ & $P_{c1}$
  \\
  $\bar{D}^* \Sigma_b$ & $\tfrac{1}{2}$ & \phantom{+}$0$
  & $37.5^{+7.3}_{-6.5}$ & $7784.2^{+6.5}_{-7.3}$ & $P_{c2}$
  &
  ${B}^* \Sigma_b$ & $\tfrac{1}{2}$ & \phantom{+}$0$
  & $67.5^{+24.9}_{-28.2}$ & $11070.3^{+28.2}_{-24.9}$ & $P_{c2}$
  \\
  $\bar{D}^* \Sigma_b$ & $\tfrac{1}{2}$ & \phantom{+}$0$
  & $14.3^{+4.7}_{-4.0}$ & $7807.4^{+4.0}_{-4.7}$ & $P_{c3}$
  &
  ${B}^* \Sigma_b$ & $\tfrac{1}{2}$ & \phantom{+}$0$
  & $37.1^{+19.2}_{-15.2}$ & $11100.7^{+15.2}_{-19.2}$ & $P_{c3}$
  \\
  \hline
  $\bar{D} \Xi_b'$ & $0$ & $-1$
  & $20.4^{+14.6}_{-12.3}$ & $7782.6^{+12.3}_{-14.6}$ & $P_{c1}$
  &
  ${B} \Xi_b'$ & $0$ & $-1$
  & $48.8^{+29.0}_{-25.1}$ & $11165.7^{+25.1}_{-29.0}$ & $P_{c1}$
  \\
  $\bar{D}^* \Xi_b'$ & $0$ & $-1$
  & $37.8^{+20.3}_{-18.2}$ & $7907.2^{+18.2}_{-20.3}$ & $P_{c2}$
  &
  ${B}^* \Xi_b'$ & $0$ & $-1$
  & $68.0^{+33.7}_{-30.1}$ & $11191.8^{+33.7}_{-30.1}$ & $P_{c2}$
  \\
  $\bar{D}^* \Xi_b'$ & $0$ & $-1$
  & $14.5^{+11.9}_{-9.7}$ & $7929.6^{+9.7}_{-11.9}$ & $P_{c3}$
  &
  ${B}^* \Xi_b'$ & $0$ & $-1$
  & $37.5^{+24.5}_{-20.8}$ & $11222.2^{+20.8}_{-24.5}$ & $P_{c3}$
  \\
  \hline \hline
\end{tabular}
\end{table}

 \begin{table}[ttt]
 \caption{
  Heavy- and light-flavor symmetry partners of
  the LHCb pentaquark trio, $P_c(4312)$, $P_c(4440)$ and
  $P_c(4457)$ (or $P_{c1}$, $P_{c2}$, $P_{c3}$ for short).
  These include the five-flavor pentaquarks with quark contents
  ${\bar b} c s d u$ and $b {\bar c} s d u$. The masses of the $P_{c}$s partners are assigned a $20\%$ uncertainty originating from the SU(3) flavor symmetry breaking.}\label{tab:partners2}
\begin{tabular}{cccccc|cccccc}

\hline \hline
Molecule & $I$ & $S$ & $B_P$ & $M_P$ & Partner
&
Molecule & $I$ & $S$ & $B_P$ & $M_P$ & Partner 
\\
  \hline
  $\bar{D} \Sigma_b^*$ & $\tfrac{1}{2}$ & \phantom{+}$0$
  & $19.1^{+4.3}_{-3.7}$ & $7680.7^{+3.7}_{-4.3}$ & $\frac{3}{2}$ &    $B \Sigma_c^*$ & $\tfrac{1}{2}$ & \phantom{+}$0$
  & $27.0^{+8.6}_{-7.3}$ & $7770.6^{+7.3}_{-8.6}$ & $\frac{3}{2}$ 
  \\
  $\bar{D}^* \Sigma_b^*$ & $\tfrac{1}{2}$ & \phantom{+}$0$
  & $41.9^{+7.8}_{-7.2}$ & $7799.2^{+7.2}_{-7.8}$ & $\frac{1}{2}$ &   $B^* \Sigma_c^*$ & $\tfrac{1}{2}$ & \phantom{+}$0$
  & $49.1^{+11.7}_{-10.4}$ & $7793.7^{+10.4}_{-11.7}$ & $\frac{1}{2}$ 
  \\
  $\bar{D}^* \Sigma_b^*$ & $\tfrac{1}{2}$ & \phantom{+}$0$
  & $29.2^{+6.3}_{-5.6}$ & $7811.9^{+5.6}_{-6.3}$ & $\frac{3}{2}$ &   $B^* \Sigma_c^*$ & $\tfrac{1}{2}$ & \phantom{+}$0$
  & $35.6^{+9.8}_{-8.5}$ & $7807.3^{+8.5}_{-9.8}$ & $\frac{3}{2}$ 
  \\
  $\bar{D}^* \Sigma_b^*$ & $\tfrac{1}{2}$ & \phantom{+}$0$
  & $11.1^{+4.3}_{-3.6}$ & $7830.0^{+3.6}_{-4.3}$ & $\frac{5}{2}$ &  $B^* \Sigma_c^*$ & $\tfrac{1}{2}$ & \phantom{+}$0$
  & $15.6^{+7.0}_{-5.8}$ & $7827.3^{+5.8}_{-7.0}$ & $\frac{5}{2}$ 
  \\
  \hline
  $\bar{D} \Xi_b^*$ & $0$ & $-1$ &
  $21.6^{+17.7}_{-9.7}$ & $7799.5^{+9.7}_{-17.7}$ & $\frac{3}{2}$ &  $B \Sigma_b^*$ & $\tfrac{1}{2}$ & \phantom{+}$0$
  & $46.9^{+21.0}_{-16.9}$ & $11065.2^{+16.9}_{-21.0}$ & $\frac{3}{2}$
  \\
  $\bar{D}^* \Xi_b^*$ & $0$ & $-1$
  & $42.2^{+21.7}_{-19.6}$ & $7920.2^{+19.6}_{-21.7}$ & $\frac{1}{2}$ &  $B^* \Sigma_b^*$ & $\tfrac{1}{2}$ & \phantom{+}$0$
  & $72.9^{+26.0}_{-21.8}$ & $11084.3^{+21.8}_{-26.0}$ & $\frac{1}{2}$
  \\
  $\bar{D}^* \Xi_b^*$ & $0$ & $-1$ &
  $29.4^{+17.4}_{-15.3}$ & $7932.9^{+15.3}_{-17.4}$ & $\frac{3}{2}$ &  $B^* \Sigma_b^*$ & $\tfrac{1}{2}$ & \phantom{+}$0$
  & $57.2^{+23.0}_{-18.9}$ & $11100.0^{+18.9}_{-23.0}$ & $\frac{3}{2}$
  \\
  $\bar{D}^* \Xi_b^*$ & $0$ & $-1$ &
  $11.3^{+10.5}_{-8.3}$ & $7951.1^{+8.3}_{-10.5}$ & $\frac{5}{2}$ &   $B^* \Sigma_b^*$ & $\tfrac{1}{2}$ & \phantom{+}$0$
  & $32.4^{+18.3}_{-14.3}$ & $11124.8^{+14.3}_{-18.3}$ & $\frac{5}{2}$    
    \\
  \hline
  $B \Xi_b^*$ & $0$ & $-1$
  & $47.2^{+27.5}_{-23.8}$ & $11186.1^{+23.8}_{-27.5}$ & $\frac{3}{2}$ &   $B \Xi_c^*$ & $0$ & $-1$
  & $28.4^{+17.6}_{-15.4}$ & $7896.7^{+15.4}_{-17.6}$ & $\frac{3}{2}$ 
   \\
  $B^* \Xi_b^*$ & $0$ & $-1$
  & $73.3^{+35.3}_{-31.7}$ & $11205.2^{+31.7}_{-35.3}$ & $\frac{1}{2}$ &  $B^* \Xi_c^*$ & $0$ & $-1$
  & $50.9^{+24.9}_{-22.7}$ & $7919.4^{+22.7}_{-24.9}$ & $\frac{1}{2}$ 
    \\
  $B^* \Xi_b^*$ & $0$ & $-1$
  & $57.6^{+30.6}_{-27.0}$ & $11220.9^{+27.0}_{-30.6}$ & $\frac{3}{2}$  &  $B^* \Xi_c^*$ & $0$ & $-1$
  & $37.1^{+20.5}_{-18.4}$ & $7933.2^{+18.4}_{-20.5}$ & $\frac{3}{2}$ 
   \\
  $B^* \Xi_b^*$ & $0$ & $-1$
  & $32.7^{+23.0}_{-19.3}$ & $11245.8^{+19.3}_{-23.0}$ & $\frac{5}{2}$   &  $B^* \Xi_c^*$ & $0$ & $-1$
  & $16.7^{+13.4}_{-11.2}$ & $7953.6^{+11.2}_{-13.4}$ & $\frac{5}{2}$ 
  \\  \hline \hline
\end{tabular}
\end{table}

According to the SU(3)-flavor symmetry, two-body antiheavy-meson-heavy-baryon states can be decomposed
into $3 \otimes 6 = 8 \oplus 10$, i.e., into the octet
and decuplet representations as shown in Fig.~\ref{dbarsigmacmultiplet}. It is obvious that the $\bar{D}^{(*)}\Sigma_{c}^{(*)}$ system with $I=1/2$ and $\bar{D}^{(\ast)}\Xi_{c}^{\prime(\ast)}$ system with $I=0$ belong to the octet representation and do not mix with the systems belonging to the decuplet representation. The $\bar{D}^{(\ast)}\Xi_{c}^{\prime(\ast)}$ system with $I=1$ mixes with the $\bar{D}_s^{(\ast)}\Sigma_c^{(\ast)}$ system,  where the proportion of each component is given by the SU(3)
Clebsch-Gordan coefficients~\cite{Kaeding:1995vq}.  Due to the absence of information on the molecules in the decuplet representation, we only focus on the molecules in the octet representation. 
Employing HQFS and SU(3)-flavor symmetry,  a larger family of pentaquark molecules is predicted in Table~\ref{tab:partners1}, where we only show the results for scenario A of Ref.~\cite{Peng:2019wys}. Moreover, in Table~\ref{tab:partners2},  we predict the HQSS partners of Table~\ref{tab:partners1}.   In Ref.~\cite{Peng:2019wys},  we assumed that the octet representation is the low energy configuration to predict a larger set of pentaquark molecules. Among these pentaquark molecules, the 
pentaquark states with five different flavors~\cite{Shen:2022rpn} and hidden-charm double strangeness~\cite{Marse-Valera:2022khy,Wang:2020bjt} were investigated. In particular, 
unified models have been proposed to describe hidden-charm and hidden-charm strange pentaquark states as meson-baryon molecules~\cite{Chen:2021cfl, Chen:2022onm,Yan:2021nio,Yan:2022wuz,Yang:2022ezl}.

\begin{figure}[!h]
\begin{center}
\includegraphics[width=4.0in]{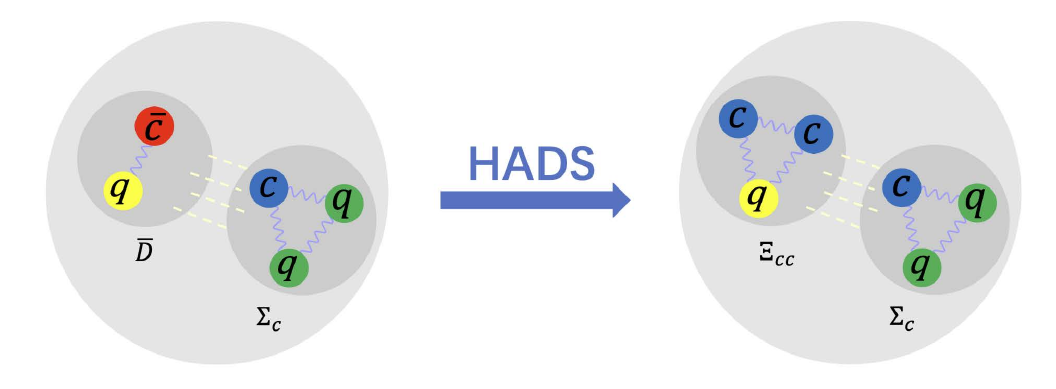}
\caption{HADS relates the $\bar{D}^{(\ast)}\Sigma_{c}^{(\ast)}$ system with the $\Xi_{cc}^{(\ast)}\Sigma_{c}^{(\ast)}$ system.   }
\label{xiccsigmac}
\end{center}
\end{figure}

In the following, we discuss the applications of HADS in the $\bar{D}^{(\ast)}\Sigma_{c}^{(\ast)}$ system. 
 Via HADS, one can relate the $\bar{D}^{(\ast)}\Sigma_{c}^{(\ast)}$ system with the $\Xi_{cc}^{(\ast)}\Sigma_{c}^{(\ast)}$ system  as shown in Fig.~\ref{xiccsigmac}. In the heavy quark limit,  the contact-range potentials of the $\Xi_{cc}^{(\ast)}\Sigma_{c}^{(\ast)}$ system share the same low-energy constants $C_{a}$ and $C_{b}$ as those of the $\bar{D}^{(\ast)}\Sigma_{c}^{(\ast)}$ system. Therefore, with the same inputs  as the  $\bar{D}^{(\ast)}\Sigma_{c}^{(\ast)}$ system,  we can predict the mass spectrum of the $\Xi_{cc}^{(\ast)}\Sigma_{c}^{(\ast)}$ system.  The $\Xi_{cc}^{\ast}$ baryon has not yet been discovered; therefore, its mass is unknown. As a result, we turn to lattice QCD for help. Lattice QCD 
simulations support the HADS relation $m_{\Xi_{cc}^*}-m_{\Xi_{cc}}=\frac{3}{4}(m_{D^*}-m_{D})$~\cite{Padmanath:2015jea,Chen:2017kxr,Alexandrou:2017xwd,Mathur:2018rwu}, which helps determine the mass of  $\Xi_{cc}^*$ as $m_{\Xi_{cc}^*}=3727$ MeV. The violation of HADS is expected to be $\Lambda_{QCD}/m_{Q}v$, instead of $\Lambda_{QCD}/m_{Q}$ in HQSS. Therefore, the uncertainty of HADS is $30\%\sim40\%$ in the charm sector and   $15\%\sim20\%$ in the bottom sector~\cite{Guo:2013xga}.  In Table~\ref{tab:predictionssigmacc}, we present the mass spectrum of the $\Xi_{cc}^{(\ast)}\Sigma_{c}^{(\ast)}$ system with a $25\%$ uncertainty in scenario A and scenario B. One can see a complete HQSS multiplet of hadronic molecules composed of triply charmed dibaryons, consistent with the results of the OBE model~\cite{Pan:2020xek}.

\begin{table}[!h]
  \caption{Mass spectrum (in units of MeV) of triply charmed  dibaryons composed of a doubly charmed
    baryon ($\Xi_{cc}$, $\Xi_{cc}^*$) and a singly charmed baryon
    ($\Sigma_c$, $\Sigma_c^*$) for cutoffs $\Lambda=0.5$ GeV and $\Lambda=1$ GeV.  The $25\%$ uncertainty induced by HADS is considered in the contact-range potentials of the $\Xi_{cc}^{(\ast)}\Xi_{c}^{\prime(\ast)}$ system. 
   \label{tab:predictionssigmacc}
  }
\centering
\begin{tabular}{ccccccccccccc}
  \hline\hline State& $J^{P}$ &Threshold & $\Lambda$(GeV) &B.E(Scenario A)&B.E(Scenario B)\\
\hline $\Xi_{cc}\Sigma_{c}$ &$0^{+}$ &6074.9& 0.5(1)
& $10.0^{+7.9}_{-6.4}$ $(17.9^{+20.8}_{-14.4})$&  $30.4^{+15.5}_{-14.2}$ $(43.2^{+33.1}_{-27.2})$
\\
$\Xi_{cc}\Sigma_{c}$ &$1^{+}$ &6074.9&  0.5(1)
&  $18.4^{+11.3}_{-9.9}$ $(28.3^{+26.3}_{-20.1})$&  $20.7^{+12.1}_{-10.8}$ $(31.2^{+27.7}_{-21.6})$
\\
\hline $\Xi_{cc}\Sigma_{c}^{\ast}$ &$1^{+}$ &6139.5& 0.5(1)
&  $11.3^{+8.4}_{-7.0}$ $(20.0^{+25.4}_{-17.2})$&  $29.6^{+15.1}_{-13.8}$ $(42.8^{+32.7}_{-26.9})$
\\
  $\Xi_{cc}\Sigma_{c}^{\ast}$ &$2^{+}$&6139.5&0.5(1) &
$20.0^{+11.8}_{-10.4}$ $(30.7^{+27.2}_{-21.2})$ &$20.0^{+11.8}_{-10.4}$ $(30.8^{+27.3}_{-21.3})$
\\
\hline   $\Xi_{cc}^{\ast}\Sigma_{c}$ &$1^{+}$ &6180.9& 0.5(1)&
$28.2^{+14.7}_{-13.4}$ $(41.0^{+32.0}_{-26.0})$ & $12.2^{+8.8}_{-7.4}$ $(21.0^{+22.4}_{-16.2})$
\\
 $\Xi_{cc}^{\ast}\Sigma_{c}$ &$2^{+}$ &6180.9& 0.5(1)&
$10.2^{+8.0}_{-6.5}$ $(18.5^{+21.0}_{-14.8})$ &$30.7^{+15.5}_{-14.2}$ $(44.1^{+33.3}_{-27.5})$
\\
\hline $\Xi_{cc}^{\ast}\Sigma_{c}^{\ast}$ &$0^{+}$ &6245.5& 0.5(1)&
$35.0^{+16.8}_{-15.6}$ $(50.2^{+35.6}_{-29.9})$ & $7.6^{+6.7}_{-5.3}$ $(15.8^{+19.2}_{-13.0})$
\\
 $\Xi_{cc}^{\ast}\Sigma_{c}^{\ast}$ &$1^{+}$ &6245.5& 0.5(1)&
$29.9^{+15.2}_{-13.9}$ $(43.7^{+32.9}_{-27.2})$ & $11.5^{+8.5}_{-7.0}$ $(20.7^{+22.0}_{-15.9})$
\\
 $\Xi_{cc}^{\ast}\Sigma_{c}^{\ast}$ &$2^{+}$ &6245.5&0.5(1)&
$20.3^{+11.8}_{-10.5}$ $(31.6^{+27.5}_{-21.6})$ & $20.3^{+11.8}_{-10.5}$ $(31.6^{+27.5}_{-21.6})$
\\
 $\Xi_{cc}^{\ast}\Sigma_{c}^{\ast}$ &$3^{+}$ &6245.5& 0.5(1)&
$7.6^{+6.7}_{-5.3}$ $(15.8^{+19.2}_{-13.0})$ & $35.0^{+16.8}_{-15.6}$ $(50.2^{+35.6}_{-30.0})$
\\
\hline\hline
\end{tabular}
\end{table}
 
One can see that the triply charmed dibaryon molecules are tied to the molecular nature of the pentaquark states. In other words, if one can obtain information on the $\Xi_{cc}^{(\ast)}\Sigma_{c}^{(\ast)}$ molecules, one can probe the nature of the pentaquark states. We do not expect to see experimental discoveries soon because of the low production rates of the triply charmed dibaryon molecules. Luckily, there exist lattice QCD simulations of the $\Xi_{cc}^{(\ast)}\Sigma_{c}^{(\ast)}$ system~\cite{Junnarkar:2019equ}. However, the lattice QCD results suffer from large uncertainties. If, in the future, lattice QCD simulations can yield the mass splitting of the $\Xi_{cc}\Sigma_{c}$ doublet, the uncertainties can be largely reduced. The relative sign for the mass splitting of the $\Xi_{cc}\Sigma_{c}$ doublet is the most crucial information because the sign of the mass splitting of the $\Xi_{cc}\Sigma_{c}$ doublet is
opposite in scenarios A and B. As a result, one can obtain the spin configuration of $P_{c}(4440)$ and $P_{c}(4457)$ from the mass splitting of the $\Xi_{cc}\Sigma_{c}$ doublet~\cite{Pan:2019skd}, which is a model-independent approach to determine the spins of $P_{c}(4440)$ and $P_{c}(4457)$ in the molecular picture. Ref.~\cite{Wang:2019gal} found that the mass of  $J^{P}=0^+ ~\Xi_{cc}\Sigma_c$ is larger than that of $J^{P}=1^+ ~\Xi_{cc}\Sigma_c$  in the  QCD sum rule approach, which agrees with the OBE model~\cite{Chen:2018pzd} but differs from the OBE model with the delta potential removed~\cite{Pan:2020xek}. Similar results are found for the   $\bar{D}^{(*)}\Sigma_{c}^{(*)}$ OBE potentials~\cite{Liu:2019zvb}. These studies motivated discussions on whether the delta potential should be removed~\cite{Yalikun:2021bfm}.   
Similarly, in terms of HADS, one can relate the $\bar{D}^{(\ast)}\Xi_{c}^{\prime(\ast)}$ system with the $\Xi_{cc}^{(\ast)}\Xi_{c}^{\prime(\ast)}$ system, which is the same as relating the $\Xi_{cc}^{(\ast)}\Sigma_{c}^{(\ast)}$ system with the $\Xi_{cc}^{(\ast)}\Xi_{c}^{\prime(\ast)}$ system via the SU(3)-flavor symmetry. Some states of the $\Xi_{cc}^{(\ast)}\Xi_{c}^{\prime(\ast)}$ system  were  investigated in the OBE model~\cite{Chen:2018pzd}.  Considering HQFS to 
the potential and using the same cutoff \footnote {It should be cautious about making predictions with HQFS. 
As discussed in Ref.~\cite{Baru:2018qkb}, the application of  HQFS beyond the $c\bar{c}$ sector has a limitation regarding model dependence within the contact-range EFT framework. The impact of the cutoff variation in the form factor on the predictions is larger as the system's reduced mass increases, implying that it is impossible to make model-independent predictions with HQFS. To reduce the uncertainty of HQFS in EFTs, obtaining the potential dependence on the reduced mass is necessary~\cite{AlFiky:2005jd,Chen:2018pzd,Peng:2019wys}.     }, we obtain the mass spectrum of  the $\Xi_{bb}^{(\ast)}\Sigma_{b}^{(\ast)}$ system, which are shown in Table~III of Ref.~\cite{Pan:2020xek}.

 \begin{figure}[!h]
\begin{center}
\includegraphics[width=3.7in]{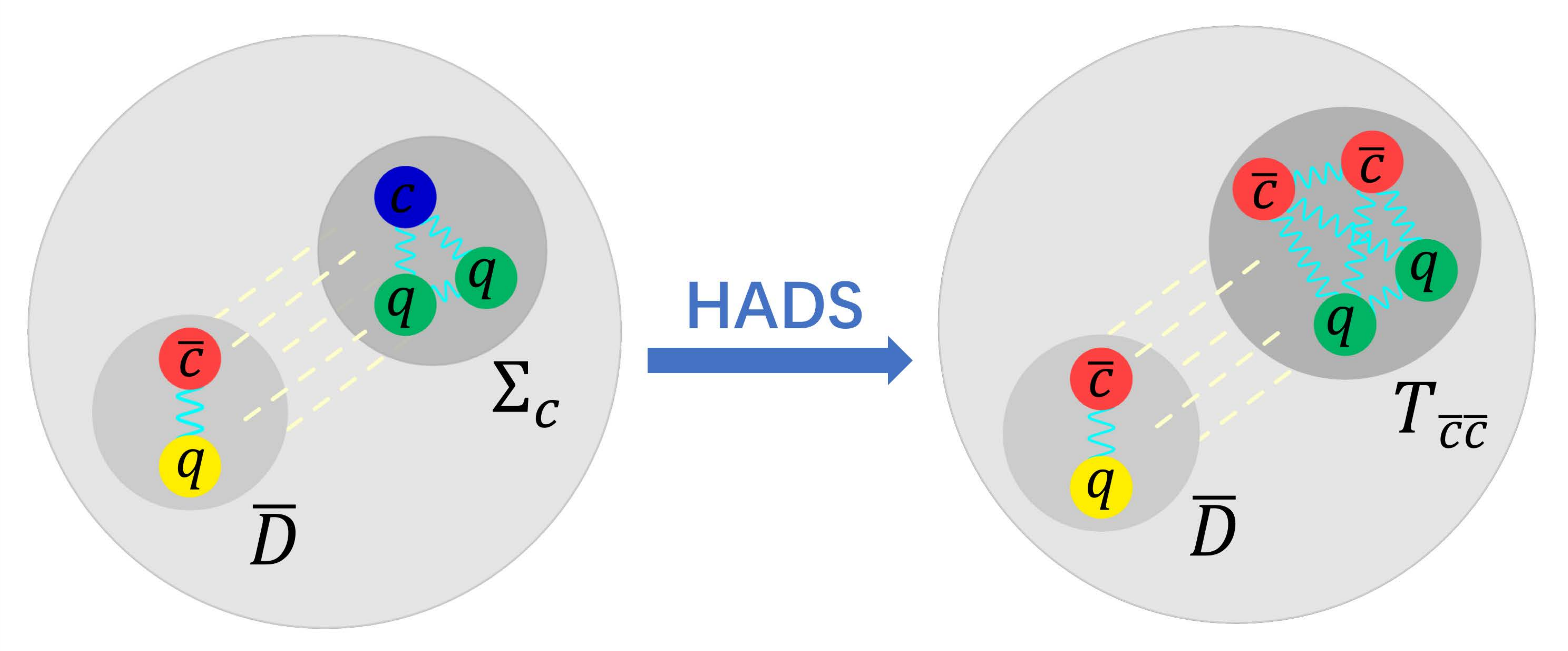}
\caption{HADS relates the $\bar{D}^{(\ast)}\Sigma_{c}^{(\ast)}$ system with the $\bar{D}^{(\ast)}T_{\bar{c}\bar{c}}^{(\ast)}$ system.   }
\label{DTcc}
\end{center}
\end{figure}

 \begin{table}[ttt]
\centering
\caption{Masses (in units of MeV) of  $T_{\bar{c}\bar{c}}^{1}$, $T_{\bar{c}\bar{c}}^{2}$, and $T_{\bar{c}\bar{c}}^{3}$ in several models and their averages. \label{masses}
}
\label{results}
\begin{tabular}{c c c c c c c c}
  \hline \hline
    Tetraquark    &~~~~\cite{Cheng:2020wxa}  &~~~\cite{Kim:2022mpa}   &~~~\cite{Zhang:2021yul}  &~~~\cite{Weng:2021hje}   &~~~ A.V
         \\ \hline 
       $T_{\bar{c}\bar{c}}^{0}$   &~~~~ 3999.8&~~~4132 &~~~4032  &~~~3969.2  &~~~4033.3
         \\   $T_{\bar{c}\bar{c}}^{1}$   &~~~~ 4124.0&~~~4151 &~~~~4117 &~~~4053.2 &~~~4111.3
         \\  $T_{\bar{c}\bar{c}}^{2}$ &~~~~ 4194.9&~~~4185 &~~~~4179&~~~4123.8&~~~4170.7 
         \\   
  \hline \hline
\end{tabular}
\end{table}

Since the distance between the anticharmed and charmed quarks in the $\bar{D}^{(\ast)}\Sigma_{c}^{(\ast)}$ system are far away, these charmed quarks and anticharmed quarks can be safely related to a pair of charmed quarks and anticharmed quarks via HADS, which respectively correspond to the $\Xi_{cc}^{(\ast)}\Sigma_{c}^{(\ast)}$  and $\bar{D}^{(\ast)}T_{\bar{c}\bar{c}}^{(\ast)}$ systems.  
HADS relates the $\Sigma_{c}^{(\ast)}$ baryons with the doubly charmed tetraquark states  $T_{\bar{c}\bar{c}}$. Refs.~\cite{Mehen:2017nrh,Cheng:2020wxa,Wu:2022gie}  investigated  the mass spectrum  of the $T_{\bar{c}\bar{c}}$ states via HADS.  
We note that the $T_{\bar{c}\bar{c}}$ states predicted via HADS are different from the $T_{cc}(3875)$ discovered by the LHCb Collaboration.  In terms of HADS, the isospin of  $T_{\bar{c}\bar{c}}$  equals 1, while isospin 0 is favored for the $T_{cc}(3875)$ experimentally. Using HADS, we relate the  $\bar{D}^{(\ast)}\Sigma_{c}^{(\ast)}$ system with the $\bar{D}^{(\ast)}T_{\bar{c}\bar{c}}^{(\ast)}$ system as shown in Fig.~\ref{DTcc}, where the masses of $T_{\bar{c}\bar{c}}$ are taken to be the average of several theoretical predictions as shown in Table~\ref{masses}~\cite{Cheng:2020wxa,Kim:2022mpa,Zhang:2021yul,Weng:2021hje}. In the heavy quark limit, the $\bar{D}^{(\ast)}\Sigma_{c}^{(\ast)}$ and  $\bar{D}^{(\ast)}T_{\bar{c}\bar{c}}^{(\ast)}$ systems share the same low-energy constants. In Table~\ref{resultsDTCC},  we present the  mass spectrum  of  the $\bar{D}^{(\ast)}T_{\bar{c}\bar{c}}^{(\ast)}$ system in scenarios A and B.  The $\bar{D}^{(\ast)}T_{\bar{c}\bar{c}}^{(\ast)}$ molecules belong to the new kind of hadronic molecules composed of a conventional hadron and a compact tetraquark state~\cite{Pan:2022whr}. The $\bar{D}^{(\ast)}T_{\bar{c}\bar{c}}^{(\ast)}$  molecules contain the same quark contents as the $\bar{D}\bar{D}\bar{D}^{\ast}$ molecules and $\bar{\Omega}_{ccc}p$ hadronic atom. However, these three kinds of molecules are bound by different mechanisms. Experimental searches for $\bar{D}^{(\ast)}T_{\bar{c}\bar{c}}^{(\ast)}$ molecules would help verify the molecular nature of the pentaquark states and the existence of compact tetraquark states.

\begin{table}[!h]
\centering
\caption{Binding energies (in units of MeV) of the $\bar{D}^{(\ast)}T_{\bar{c}\bar{c}}^{(\ast)}$ molecules in scenarios A and B obtained with single-channel potentials. The numbers inside and outside the brackets correspond to $\Lambda=0.75$ GeV and $\Lambda=1.5$ GeV. The superscripts and subscripts are obtained by allowing for a $25\%$ breaking of HADS.   
}
\label{resultsDTCC}
\begin{tabular}{c c c c c c c c}
  \hline \hline
     Molecule   &~~~~ $J^P$&  Threshold   &~~~~B.E.(Scenario A)   &~~~~B.E.(Scenario B) 
         \\ \hline   $\bar{D}T_{\bar{c}\bar{c}}^{0}$   &~~~~ $0^-$ & 5900.3 &~~~~   $24.1_{-22.5}^{+39.4}(16.5_{-12.1}^{+16.1})$  &~~~  $ 32.1_{-28.3}^{+44.9}(23.3_{-15.4}^{+19.3}) $
         \\  $\bar{D}T_{\bar{c}\bar{c}}^{1}$  &~~~~ {$1^-$}&  5978.3 &~~~~$24.7_{-23.0}^{+39.7}(16.8_{-12.2}^{+16.2})$   &~~~   $32.8_{-28.8}^{+45.3}(23.6_{-15.5}^{+19.4})$
         \\  $\bar{D}T_{\bar{c}\bar{c}}^{2}$   &~~~~ $2^-$&  6037.7 &~~~~ $25.2_{-23.3}^{+40.0}(17.0_{-12.3}^{+16.2})$  &~~~   $33.4_{-29.1}^{+45.5}(23.8_{-15.6}^{+19.4}) $ 
         \\   $\bar{D}^{\ast}T_{\bar{c}\bar{c}}^{0}$   &~~~~ $1^-$& 6042.3  &~~~~$29.5_{-26.2}^{+42.1}(18.6_{-13.0}^{+16.7})$   &~~~  $ 38.2_{-32.0}^{+47.6}(25.6_{-16.3}^{+19.9})$
        \\   $\bar{D}^{\ast}T_{\bar{c}\bar{c}}^{1}$  &~~~~ {$0^-$}& 6120.3   &~~~~$43.7_{-35.3}^{+50.7}(29.7_{-18.0}^{+21.6})$ &~~~   $26.1_{-23.8}^{+39.6}(15.6_{-11.4}^{+15.2}) $
         \\   $\bar{D}^{\ast}T_{\bar{c}\bar{c}}^{1}$  &~~~~ {$1^-$}& 6120.3  &~~~~$36.8_{-31.0}^{+46.6}(24.1_{-15.6}^{+19.2})$  &~~~ $ 32.4_{-28.1}^{+43.8}(20.6_{-13.9}^{+17.6})$
         \\$\bar{D}^{\ast}T_{\bar{c}\bar{c}}^{1}$  &~~~~ {$2^-$}&   6120.3  &~~~~$24.2_{-22.3}^{+38.2}(14.0_{-10.6}^{+14.4})$  &~~~   $46.1_{-36.7}^{+52.1}(31.6_{-18.8}^{+22.4}) $
         \\
         $\bar{D}^{\ast}T_{\bar{c}\bar{c}}^{2}$  &~~~~ {$1^-$}&  6179.7  &~~~~ $51.7_{-39.8}^{+55.0}(35.8_{-20.5}^{+24.0})$   &~~~  $20.9_{-19.8}^{+35.7}(11.2_{-9.0}^{+12.8})$
         \\   $\bar{D}^{\ast}T_{\bar{c}\bar{c}}^{2}$  &~~~~ {$2^-$}&  6179.7  &~~~~$37.4_{-31.3}^{+46.8}(24.3_{-15.6}^{+19.3}) $  &~~~ $32.9_{-28.5}^{+44.0}(20.8_{-14.0}^{+17.7})$ 
         \\$\bar{D}^{\ast}T_{\bar{c}\bar{c}}^{2}$  &~~~~ {$3^-$}&    6179.7     &~~~~$19.0_{-18.3}^{+34.3}(9.8_{-8.1}^{+12.0})$   &~~~  $54.3_{-41.3}^{+56.4}(37.8_{-21.3}^{+24.8}) $
         \\
  \hline \hline
\end{tabular}
\end{table}

In summary, identifying the $P_c(4312)$, $P_c(4440)$, and $P_c(4457)$ as the $\bar{D}^{(*)}\Sigma_c$ bound states in two spin assignments, a complete HQSS multiplet of $\bar{D}^{(*)}\Sigma_c^{(\ast)}$ hadronic molecules are predicted by a series of works~\cite{Liu:2019tjn,Xiao:2019aya,Du:2019pij,Liu:2019zvb,Yamaguchi:2019seo,Sakai:2019qph,Du:2021fmf}, which implies the existence of fine structures in hadronic molecules. As indicated in Refs.~\cite{Peng:2020gwk,Yalikun:2021bfm},  the  $P$-wave  $\bar{D}\Lambda_{c1}$ potential may play a role in forming $P_{c}(4457)$, but not break the multiplet structure. However, identifying $P_{c}(4457)$ as the $S$-wave bound state of $\bar{D}\Lambda_{c1}$~\cite{Burns:2019iih}, the multiplet structure would be broken. Such a scenario can be discriminated by the partial decay width of $P_{c}(4457) \to P_{c}(4312) \pi$. The width of the $\bar{D}^*\Sigma_c$ molecule is only $100$~keV~\cite{Ling:2021lmq} but that of the $\bar{D}\Lambda_{c1}$ molecule is $11$~MeV~\cite{Wu:2024bvl}.   The spin order of  $P_c(4440)$ and $P_c(4457)$ has not been determined so far. Many theoretical works calculated relevant physical observables of pentaquark states to determine the favorable scenario~\cite{Chen:2019asm,Xiao:2019aya,Du:2019pij,Liu:2019zvb,Zhang:2023czx}. A model-independent approach is proposed to determine the spins of  $P_c(4440)$ and $P_c(4457)$ from the mass splitting of the $\Xi_{cc}\Sigma_c$ doublet~\cite{Pan:2019skd}.       
Based on the  $\bar{D}^{(*)}\Sigma_c^{(\ast)}$  multiplet  picture, several molecular multiplets are predicted:  the ${D}^{(*)}\Sigma_c^{(\ast)}$ multiplet via the $G$-parity transformation~\cite{Liu:2020nil,Chen:2021kad,Shen:2022zvd,Liu:2023clr},    the   $\bar{D}^{(\ast)}\Xi_{c}^{\prime(\ast)}$ multiplet via the SU(3)-flavor symmetry~\cite{Xiao:2019gjd,Peng:2019wys,Wang:2019nvm,Liu:2020hcv,Chen:2021cfl,Chen:2022onm,Wang:2023eng}, $B^{(*)}\Sigma_c^{(\ast)}$, $\bar{D}^{(*)}\Sigma_b^{(\ast)}$, $B^{(*)}\Sigma_b^{(\ast)}$ multiplet via HQFS~\cite{Peng:2019wys,Wang:2019ato}, and $B^{(\ast)}\Xi_{c}^{\prime(\ast)}$, $\bar{D}^{(\ast)}\Xi_{b}^{\prime(\ast)}$, and $B^{(\ast)}\Xi_{b}^{\prime(\ast)}$ multiplets via the SU(3)-flavor symmetry and HQFS~\cite{Peng:2019wys}. In addition,  with HADS  the $\Xi_{cc}^{(\ast)}\Sigma_{c}^{(\ast)}$~\cite{Pan:2019skd} and $\Xi_{cc}^{(\ast)}\Xi_{c}^{\prime(\ast)}$ multiplets composed of the triply charmed dibaryons, as well as a new kind of $\bar{D}^{(\ast)}T_{\bar{c}\bar{c}}^{(\ast)}$~\cite{Pan:2022whr}  hadronic molecules composed of compact doubly charmed tetraquark states and charmed mesons are predicted. Experimental searches for or lattice QCD simulations of these multiplets of hadronic molecules would be helpful in verifying the molecular nature of these pentaquark states.

\subsubsection{ $\bar{D}^{(\ast)}D^{(\ast)}$ molecules }

 A pair of ${D}^{(\ast)}$ and  $\bar{D}^{(\ast)}$ mesons  can couple to  six $S$-wave meson pairs with given $J^{PC}$.  In the heavy quark limit, the potentials of the $\bar{D}^{(\ast)}D^{(\ast)}$ system are only dependent on the spin of the pair of light quarks, i.e., $j_l=0$ and $j_l=1$, which combine the above six states into two multiplets with $j_l=0$ and $j_l=1$~\cite{Guo:2017jvc}\footnote{Strictly speaking,  the two multiplets are not related to each other via HQSS. As indicated in Refs.~\cite{Voloshin:2016cgm,Hu:2024hex}, these two multiplets are only related to each other considering the Light Quark Spin Symmetry, which means that the interaction between the heavy mesons does not depend
on the total angular momentum of the light degrees of freedom. }.  With HQSS,  the contact-range potentials of the $\bar{D}^{(\ast)}D^{(\ast)}$ system are parameterized by two parameters $C_a$ and $C_b$~\cite{AlFiky:2005jd,Nieves:2012tt,Guo:2013sya,Mutuk:2018zxs}.  In the isoscalar sector,  the $X(3872)$ is assumed as a $\bar{D}D^{\ast}$ bound state with the quantum numbers $J^{PC}=1^{++}$~\cite{Tornqvist:2004qy,Braaten:2004fk,Eichten:2005ga,AlFiky:2005jd,Nieves:2012tt,Guo:2013sya,Artoisenet:2009wk,Liu:2009qhy,Gamermann:2009uq,Lee:2009hy,Prelovsek:2013cra,Karliner:2015ina,Liu:2019stu,Yamaguchi:2019vea}, which can determine the sum of $C_a$ and $C_b$.  The contact-range potential of the $J^{PC}=1^{++}$$\bar{D}D^{\ast}$ system
is the same as  that of the $J^{PC}=2^{++}$$\bar{D}^{\ast}D^{\ast}$ system. As a result, HQSS predicts a  $J^{PC}=2^{++}$$\bar{D}^{\ast}D^{\ast}$ bound state with a binding energy of $5$ MeV, consistent with the OBE model~\cite{Liu:2019stu}. In Ref.~\cite{Baru:2016iwj}, Baru et al. demonstrated that when all the partial waves and coupled-channels are included,  the HQSS predictions are robust even taking into account the OPE interaction.  However, one can not fully determine the mass spectrum of the $\bar{D}^{(\ast)}D^{(\ast)}$ system because there exists only one input. Given this situation, one can use the light meson saturation approach for help. For the $\bar{D}^{(*)}\Sigma_c^{(\ast)}$ system, with the light meson saturation approach, one can obtain the ratio of $C_b$ to $C_a$ as
\begin{eqnarray}
  \frac{C_b^{{\rm sat} (V)}}{C_a^{{\rm sat} (V+\sigma)}} \simeq 0.123 \, ,
\end{eqnarray}
which is consistent with the reference value determined in the contact-range EFT approach for scenario B, i.e., $  \frac{C_b}{C_a}=0.158 $. One can see that the light meson saturation approach works well in estimating the ratio of  $C_b$ to $C_a$.

\begin{table}[!h]
\centering
\caption{ Scattering lengths ($a$ in units of fm), binding energies ($B$ in units of  MeV if bound states exist) and mass spectra ($M$ in units of MeV) of prospective isoscalar heavy antimeson-meson molecules.  The $\dag$ symbol indicates that there is no bound state. The subscript $V$ denotes the existence of a virtual state. The uncertainties originate from the HQSS breaking of the order $15\%$.
  }
\label{tab:comparison-EFT1}
\begin{tabular}{cccccc}
\hline\hline
Molecule & $I$ & $J^{PC}$  & $a$ (fm)
& $B$ (MeV)
& $M$ (MeV )    \\
  \hline
  $D\bar{D}$ & $0$ & $0^{++}$ &{$-20.5^{+(27.5,\infty)}_{+(-\infty,17.2)}$} & $0.1^V$ & $3733.9^{V}$  \\
  \hline
  $D \bar{D}^{\ast}+ D^{\ast}\bar{D}$ & $0$ & $1^{++}$
  & $2.7_{-0.7}^{+2.8}$  & Input & Input  \\
  $D \bar{D}^{\ast}- D^{\ast}\bar{D}$ & $0$ & $1^{+-}$ & $-1.3_{-0.8}^{+0.4}$  & $7.9^{V}$ & $3868.1^{V}$    \\
  \hline
  $D^{\ast}\bar{D}^*$ & $0$ & $0^{++}$  & $-0.3^{-0.1}_{+0.1}$   & $\dag$ & $\dag$ \\
  $D^{\ast}\bar{D}^*$ & $0$ & $1^{+-}$    &$-1.5_{-1.0}^{+0.5}$  &$6.3^{V}$ & $4011.7^{V}$  \\
   $D^{\ast}\bar{D}^*$ & $0$ & $2^{++}$  & $2.5_{-0.6}^{+1.9}$  &$4.9^{+5.3}_{-3.6}$ & $4013.1$   \\
  \hline \hline
\end{tabular}
\end{table}

Using the light meson saturation approach, one can estimate the ratio of $C_{b}$ to $C_{a}$ of  $\bar{D}^{(\ast)}D^{(\ast)}$, i.e., $C_{b}/C_{a}=0.35$~\cite{Liu:2020tqy}. This shows that the spin-spin term of  the $\bar{D}^{(\ast)}D^{(\ast)}$ interaction plays a more important role than those of the $\bar{D}^{(*)}\Sigma_c^{(\ast)}$ system. With the sum of $C_a$ and $C_b$, one can fix  $C_a$ and $C_b$.  In Table~\ref{tab:comparison-EFT1}, we present the mass spectrum of  the $\bar{D}^{(\ast)}D^{(\ast)}$ system and their scattering lengths. For the convention of the scattering length, we refer to Eq.(\ref{ERE Formula}). For some channels with no bound states,  we search for virtual poles in the unphysical sheet close to the real axis. The potentials for the bound states are stronger than those for the virtual states.  It is worth noting that the ratio estimated in the light meson saturation mechanism suffers from unquantified uncertainties. Therefore, the results for the $\bar{D}^{(\ast)}D^{(\ast)}$ system need further investigations.  Ref.~\cite{Ji:2022uie} systematically analysed the $\bar{D}^{(\ast)}D^{(\ast)}$ molecules. The results depend on the cutoff in the form factors and the interpretation for the newly discovered exotic state $X(3960)$.  In the isoscalar sector, in addition to the $J^{PC}=1^{++}$$\bar{D}D^{\ast}$ 
and  $J^{PC}=2^{++}$$\bar{D}^{\ast}D^{\ast}$ molecules, the  $D\bar{D}$  channel is close to forming a bound state, but it is bound in other approaches~\cite{Ji:2022uie,Dong:2021juy,Prelovsek:2020eiw}. A recent lattice QCD calculations found no bound or resonant state of $D\bar{D} $ in the vicinity of the $D\bar{D} $ threshold~\cite{Wilson:2023hzu}.  We note that in the local hidden gauge approach~\cite{Molina:2009ct}, these states bind more than their counterparts in other approaches.

In the heavy quark limit, the contact-range potentials of the $\bar{B}^{(*)}B^{(*)}$ and ${B}^{(*)}D^{(*)}$ systems are the same as those of the $\bar{D}^{(*)}D^{(*)}$ system.  Therefore, 
assuming $X(3872)$ as a $\bar{D}^*D$ bound state, we predict the mass spectrum and scattering lengths of the $\bar{B}^{(*)}B^{(*)}$ and ${B}^{(*)}D^{(*)}$ systems in Table~\ref{tab:comparison-EFT2}.~\footnote{We note that similar results are obtained in Ref.~\cite{Ozpineci:2013zas} using heavy quark symmetry and the local hidden gauge approach
.} These systems are more bound because the reduced masses are larger. Among these molecules, the $X_b$ as a $\bar{B}^*B$ bound state is viewed as the HQFS partner of $X(3872)$. The existence of $X_b$ is crucial to 
establishing the molecular nature of $X(3872)$~\cite{Guo:2013sya,Sun:2011uh,Liu:2019stu}.  The same can be said about $X_{2}(4013)$.   Experimental searches for  $X_b$ and $X_2$ are ongoing, but no conclusive results exist yet~\cite{Belle:2014sys,Belle:2021nuv}.    

\begin{table}[!h]
\centering
\caption{ Same as Table~\ref{tab:comparison-EFT1} but for the $\bar{B}^{(*)}B^{(*)}$ and  ${B}^{(*)}B^{(*)}$ systems.
  }
\label{tab:comparison-EFT2}
\begin{tabular}{cccccc}
\hline\hline
Molecule & $I$ & $J^{PC}$  & $a$ (fm)
& $B$ (MeV)
& $M$ (MeV )    \\
  \hline
  $B\bar{B}$ & $0$ & $0^{++}$ & $ 1.1 $ & $ 20.5  $ & $ 10538.4  $  \\
  \hline
  $B \bar{B}^{\ast}+ B^{\ast}\bar{B}$ & $0$ & $1^{++}$
  &   $1.0 $  & 38.3  & 10565.8   \\
  $B \bar{B}^{\ast}- B^{\ast}\bar{B}$ & $0$ & $1^{+-}$ & $1.6  $  & $ 6.0  $ & $10598.2 $    \\
  \hline
  $B^{\ast}\bar{B}^*$ & $0$ & $0^{++}$  & $ -3.5 $   & $ 0.5^{V} $ & $ 10648.9^{V} $ \\
  $B^{\ast}\bar{B}^*$ & $0$ & $1^{+-}$    &$ 1.6$  &$ 6.0 $ & $ 10643.4 $  \\
   $B^{\ast}\bar{B}^*$ & $0$ & $2^{++}$  & $1.0 $  &$ 38.5 $ & $ 10610.9 $   \\
  \hline \hline
\end{tabular}
\end{table}

  \begin{figure}[!h]
    \centering
    \includegraphics[width=16cm]{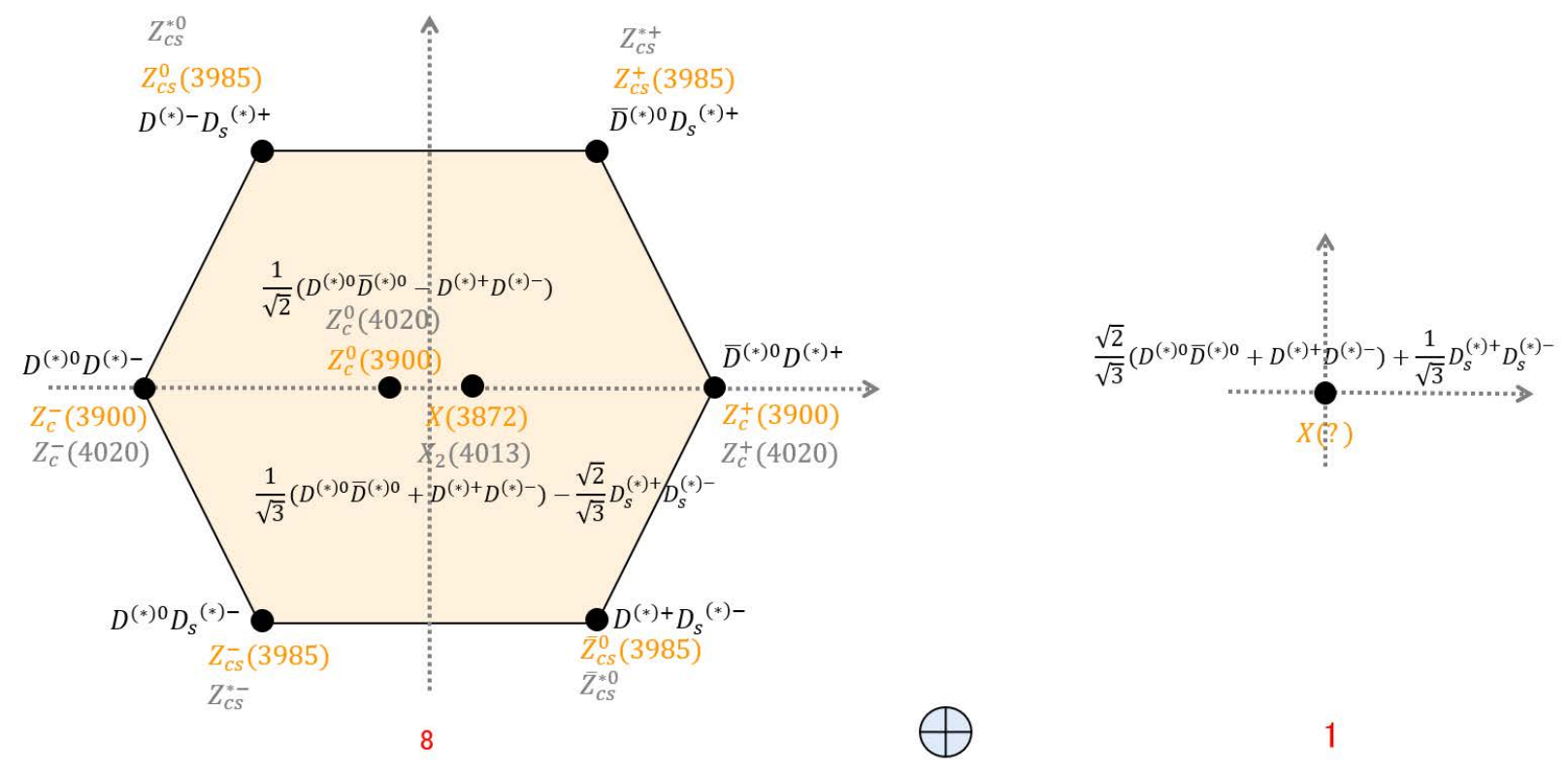}\\
    \caption{  $Z_c(3900)$, $Z_{cs}(3985)$, $X(3872)$ as parts of the SU(3)-flavor  $\bar{D}_{(s)}^{(*)}{D}_{(s)}^{(*)}$ multiplet of hadronic molecules, and the representations in grey referring to their corresponding HQSS partners.     }\label{ddbarmultpic}
\end{figure}

\begin{figure}[ttt]
    \centering
    \includegraphics[width=11cm]{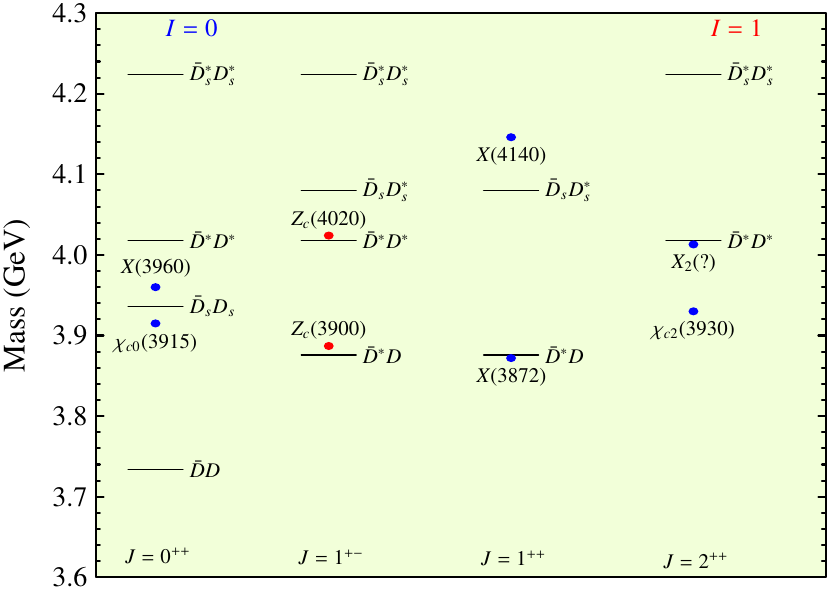}\\
    \caption{ Locations  of  isoscalar~(blue) and isovector~(red)   hadronic molecular candidates  with respect to the   $\bar{D}^{(*)}D^{(*)}$  and   $\bar{D}_s^{(*)}{D}_s^{(*)}$  mass thresholds.   }\label{ddbarspectrum}
\end{figure}

In the SU(3)-flavor symmetry limit, the $\bar{P}^{(\ast)}P^{(\ast)}$ system contains both a singlet and an octet irreducible representation as shown in Fig.~\ref{ddbarmultpic}. In both representations, they contain $\bar{D}^{(\ast)}D^{(\ast)}$ and $\bar{D}_s^{(\ast)}D_s^{(\ast)}$ components, which indicates that the $\bar{D}_s^{(\ast)}D_s^{(\ast)}$ system affects the behaviors of  isoscalar hadronic molecules.      In Refs.~\cite{Hidalgo-Duque:2012rqv,Meng:2020cbk,Ji:2022uie,Ji:2022vdj}, the authors employ SU(3)-flavor symmetry and HQSS to study the molecules near the  $\bar{D}^{(\ast)}D^{(\ast)}$   and $\bar{D}_s^{(\ast)}D_s^{(\ast)}$ mass thresholds, where the results depend sensitively on the inputs.  
In general, with ideal mixing, the octet and singlet configurations specify the $\bar{D}_s^{(\ast)}D_s^{(\ast)}$ and  $\bar{D}^{(\ast)}D^{(\ast)}$ components~\cite{Liu:2009qhy}.  In the  meson exchange theory, the strength of the $\bar{D}_s^{(\ast)}D_s^{(\ast)}$ potentials  is  weaker than that of the isoscalar $\bar{D}^{(\ast)}D^{(\ast)}$ potentials~\cite{Liu:2009qhy,Liu:2017mrh}, leading to the nonexistence of $\bar{D}_s^{(\ast)}D_s^{(\ast)}$ bound states~\cite{Dong:2021juy}. Recently, lattice QCD simulations obtained a $\bar{D}_sD_s$ bound state~\cite{Prelovsek:2020eiw}, which is also found in the local hidden gauge approach~\cite{Bayar:2022dqa}. It can be seen that the precise $\bar{D}^{(\ast)}D^{(\ast)}$   and $\bar{D}_s^{(\ast)}D_s^{(\ast)}$ interactions, and particularly the SU(3) flavor symmetry breaking, are important to probe the nature of exotic states near their mass thresholds. In particular, the impact of compact configurations such as excited charmonium components and compact tetraquark components on the   $\bar{D}^{(*)}D^{(*)}$  and   $\bar{D}_s^{(*)}{D}_s^{(*)}$  mass thresholds, often expressed as effective potentials, were not included in our studies.   
In particular, the $\chi_{c0}(3915)$  and the newly discovered $X(3960)$  are located in the vicinity of the  $\bar{D}_sD_s$ mass threshold as shown in Fig.~\ref{ddbarspectrum}. This 
indicates that the couplings between the $\bar{D}_sD_s$ channel and the bare $\chi_{c0}(2P)$ component may affect the physical state around this energy region.

\begin{figure}[!h]
\begin{center}
\includegraphics[width=4.0in]{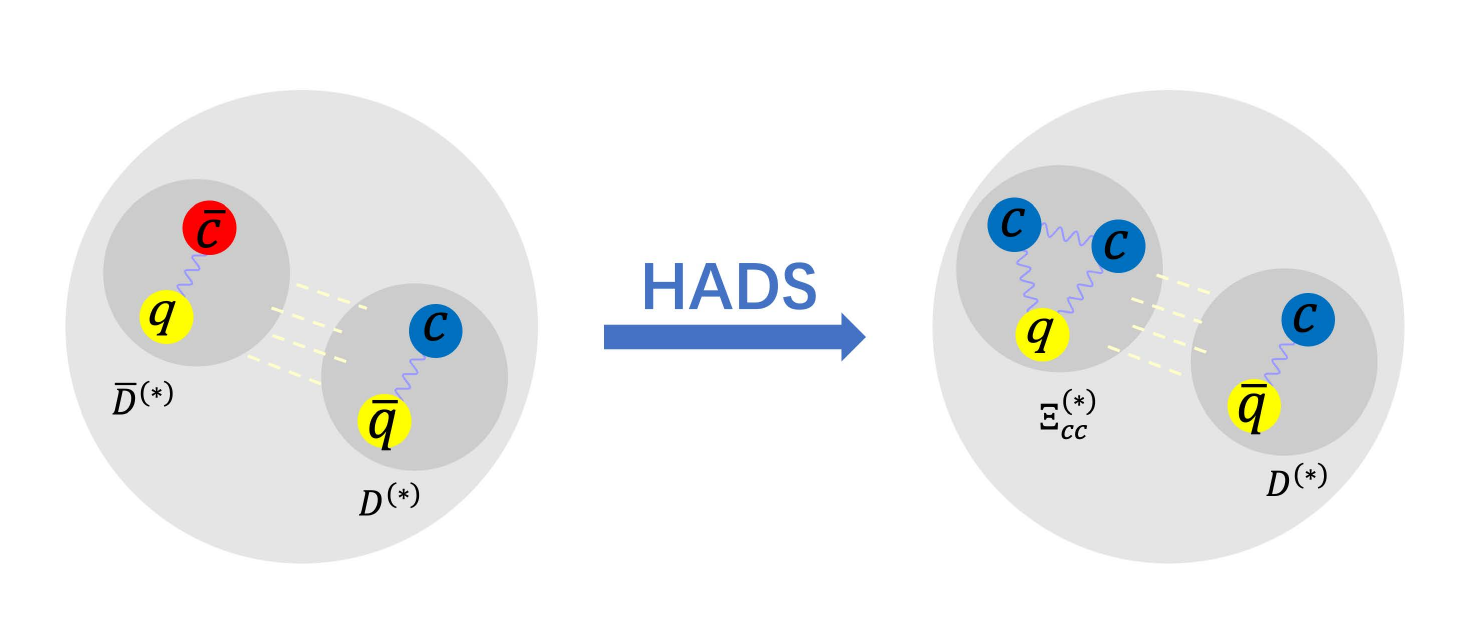}
\caption{HADS relates the $\bar{D}^{(\ast)}\Sigma_{c}^{(\ast)}$ system with the $\bar{D}^{(\ast)}T_{\bar{c}\bar{c}}^{(\ast)}$ system.   }
\label{DDtoDXi}
\end{center}
\end{figure}

Via HADS, one can relate the $D^{(\ast)}\bar{D}^{(\ast)}$ system with the $D^{(\ast)}\Xi_{cc}^{(\ast)}$ system, as shown in Fig.~\ref{DDtoDXi}.  In the heavy quark limit, the  $D^{(\ast)}\bar{D}^{(\ast)}$ and  $D^{(\ast)}\Xi_{cc}^{(\ast)}$ systems share the same couplings $C_{a}$ and $C_{b}$. With the couplings obtained,  one can calculate the binding energies and scattering lengths of the $D^{(\ast)}\Xi_{cc}^{(\ast)}$ system, which are shown in Table~\ref{tab:comparisonDXicc}.  We predict two bound states $J^{P}=1/2^{-}$ $ {D}^*\Xi_{cc}$ and $J^{P}=5/2^{-}$ $ {D}^*\Xi_{cc}^*$, consistent with the results of  Refs.~\cite{Guo:2013xga, Chen:2017jjn,Liu:2018bkx}. The minimum quark content of the $D^{(\ast)}\Xi_{cc}^{(\ast)}$ molecules are $ccc$, which  could mix with excited fully charmed  baryons.  One can see that the $D^{(\ast)}\Xi_{cc}^{(\ast)}$ states are more bound than  the $D^{(\ast)}\bar{D}^{(\ast)}$ states  due to the larger reduced masses of the $D^{(\ast)}\Xi_{cc}^{(\ast)}$ system. In terms of HQFS, one can expect a more bound system composed of  ${B}^{(\ast)}\Xi_{bb}^{(\ast)}$.      

\begin{table}[!h]
\centering
\caption{ Scattering lengths ($a$ in units of fm), binding energies ($B$ in units of MeV, if bound states exist) and mass spectra ($M$ in units of MeV) of prospective isoscalar heavy meson-baryon molecules.  The superscript $V$ denotes the existence of virtual states, the $\dag$ symbol denotes that the system is unbound, and the $?$ symbol indicates that the system is bound with the central value of the potential and can become unbound if the potential is made less attractive within the range specified in the text.  The uncertainties originate from the HADS breaking of the order $30\%$.
  }
\label{tab:comparisonDXicc}
\begin{tabular}{cccccc}
\hline\hline
Molecule & $I$ & $J^{PC}$  & $a$ (fm)
& $B$ (MeV)
& $M$ (MeV )   \\
  \hline
  $D\Xi_{cc}$ & $0$ & $1/2^{-}$ & $3.3^{-9.1}_{-1.5}$ & $2.0^{+7.3}_{\dag}$ & ?    \\
  $D\Xi_{cc}^{\ast}$ & $0$ & $3/2^{-}$   &  $3.2^{-9.5}_{-1.4}$
  & $2.1^{+7.5}_{\dag}$ & ?                \\  
  $D^{\ast}\Xi_{cc}$ & $0$ & $1/2^{-}$ & $1.8_{-0.5}^{+3.5}$   & $8.7^{+13.3}_{-8.1}$ & 5621.3  \\
   $D^{\ast}\Xi_{cc}$ & $0$ & $3/2^{-}$  &$4.7^{-8.0}_{-2.7}$   &  $0.9^{+5.3}_{\dag}$ & $?$  \\
  $D^{\ast}\Xi_{cc}^{\ast}$ & $0$ & $1/2^{-}$ &$-0.9_{-0.9}^{+0.4}$   & $11.2^{V}$ & $5724.6^{V}$   \\
  $D^{\ast}\Xi_{cc}^{\ast}$ & $0$ & $3/2^{-}$ & $22.8_{-22.0}^{-24.6}$   &$0^{+2.8}_{\dag}$ & $?$  \\
 $D^{\ast}\Xi_{cc}^{\ast}$ & $0$ & $5/2^{-}$ &$1.6_{-0.4}^{+1.7}$   &$12.9^{+15.9}_{-10.9}$ & $5723.2$    \\
  \hline \hline
\end{tabular}
\end{table}

Except the isospin-isospin factors,    the expression of the leading-order $\bar{D}^{(\ast)}D^{(\ast)}$  contact potentials for the isovector sector are the same as those for the isoscalar sector~\cite{Nieves:2012tt,Ji:2022uie}.    With the light meson saturation mechanism, it is found that the isovector $\bar{D}^{(\ast)}D^{(\ast)}$ potentials are less attractive than their isoscalar counterparts.   In the local hidden gauge approach~\cite{Aceti:2014uea}, Aceti et al. argued that due to the OZI rule, 
 the light meson exchange for the $\bar{D}^*D$ system with $I=1$  is suppressed. The two-pion exchange contributions are much smaller than the heavy meson exchange contributions, which shows that the isovector $\bar{D}^*D$ potential is weak.  In addition, the OBE model showed that the isovector $\bar{D}^*D$ system can not form a bound state~\cite{Liu:2019stu}.   In Ref.~\cite{He:2017lhy}, He et al. employed the meson exchange model to obtain the potentials of the $J/\psi\pi-\bar{D}^*D$ system and assigned the $Z_c(3900)$ as a virtual state. They found that the inelastic potential  $J/\psi\pi-\bar{D}^*D$  plays a minor role.   In Ref.~\cite{Wang:2020dko},    Wang et al. employed the chiral EFT approach to obtain the  $\bar{D}^*D$ and  $\bar{D}^*D^*$ potentials up to the next-to-leading order. Then, they fitted the experimental invariant mass distributions. Later, they introduced the  $J/\psi\pi$ channel, and  reproduced the masses and widths of $Z_c(3900)$ and $Z_c(4020)$ with the contact potentials~\cite{Meng:2020ihj}. In Ref.~\cite{Du:2022jjv}, Du et al. argued that  $Z_c(3900)$ and $Z_c(4020)$ could be either virtual states or resonant states with contact potentials by fitting the experimental data. Very recently, a more comprehensive analysis of the $\pi\pi$, $J/\psi\pi$, and $\bar{D}^*D$ mass distributions of the BESIII data favors the resonant interpretation for $Z_{c}(3900)$~\cite{Chen:2023def}, where they took into account the triangle singularities in the open charm triangle diagram, $\bar{D}^*D-J/\psi \pi$ final-state interaction, and the $\pi\pi-K\bar{K}$ final-state interaction.  Therefore,  virtual states or resonant states are likely to exist in the isovector  $\bar{D}^{(\ast)}D^{(\ast)}$ system. Using the compositeness rule in Ref.~\cite{Matuschek:2020gqe}, the compositeness component of $Z_{c}(3900)$ was estimated to be around $40\% \sim 50\%$~\cite{Yan:2023bwt,Chen:2023def}, indicating that the components other than the molecular ones may play an important role in forming  $Z_{c}(3900)$.

 \begin{table}[ttt]
\centering
\caption{ Pole positions of $\bar{D}D^*$,  $\bar{D}^*D^*$,  $\bar{D}D_{s}^*$, and $\bar{D}^*D_s^*$ with $J^{PC}=1^{+-}$ for a cutoff of $\Lambda=1$ GeV. The subscript V denotes the existence of virtual states. 
  }
\label{poleDD}
\begin{tabular}{cccccc}
\hline\hline
Molecule & $I$ & $J^{PC}$  & $m-i\frac{\Gamma}{2}$ (MeV)
& Experimental data (MeV)
    \\
  \hline
  $D \bar{D}^{\ast}$ & $1$ & $1^{+-}$
  & Input  & 3888.4-14.2$i$   \\
  $D^{\ast} \bar{D}^{\ast}$ & $1$ & $1^{+-}$ & 4029.2-12.9$i$  & 4024.1-6.5$i$    \\
  $D_{s}\bar{D}^*$ & $\frac{1}{2}$ & $1^{+-}$  & 3988.5-13.2$i$   & 3982.5-6.4$i$ \\
  $D_{s}^*\bar{D}^*$ & $\frac{1}{2}$ & $1^{+-}$    &4131.5-12.0$i$  &  -  
   \\
  $B \bar{B}^{\ast}$ & $1$ & $1^{+-}$
  & $10602^{V}$  & 10608.4-7.8$i$   \\
  $B^{\ast} \bar{B}^{\ast}$ & $1$ & $1^{+-}$ & $10648^{V}$  & 10653.2-7.2$i$     \\
  \hline \hline
\end{tabular}
\end{table}

 In principle, with only  $C_a$ and $C_{b}$,  the contact potentials can not generate resonant states in the single-channel case. If a $q^2$ term is included, i.e.,  $C_{a}+ D_{a}\cdot q^2$~\cite{Albaladejo:2015lob,Yang:2020nrt,Meng:2020ihj,Xie:2022lyw}, resonances can be dynamically  generated.  Experimentally, the two tetraquark states $Z_{c}(3900)$ and $Z_{c}(4020)$ are located above the $\bar{D}^*D$ and $\bar{D}^*D^*$ mass thresholds,  which can be naturally interpreted as $\bar{D}^*D^{(*)}$ resonant states~\cite{Wang:2020dko,Meng:2020ihj}.  Assuming $Z_{c}(3900)$ as a $J^{PC}=1^{+-}$$\bar{D}D^*$ resonance, one can determine $C_{a}=-7$ GeV$^{-2}$  and $D_{a}=-220$ GeV$^{-4}$ for a cutoff $\Lambda=1$ GeV.  Via HQSS and SU(3)-flavor symmetry, one can relate the $\bar{D}D^*$ channel with the $\bar{D}^*D^*$,  $\bar{D}D_{s}^*$, and $\bar{D}^*D_s^*$ channels. Therefore, with the parameters obtained,  the molecular states in the $\bar{D}^*D^*$,  $\bar{D}D_{s}^*$, and $\bar{D}^*D_s^*$ systems in Table~\ref{poleDD}, where the $\bar{D}^*D^*$ and  $\bar{D}D_{s}^*$ resonant states likely correspond to the exotic states of  $Z_c(4020)$~\cite{BESIII:2013ouc} and $Z_{cs}(3985)$ discovered by the BESIII Collaboration~\cite{BESIII:2020qkh}. In contrast, the $\bar{D}^*D_s^*$ resonant state has not yet been discovered. Here we only discuss the resonant states with the quantum numbers $J^{PC}=1^{+-}$.  
Moreover, with HQFS, one can relate the  $\bar{D}D^*$ and $\bar{D}^*D^*$ systems with the $\bar{B}B^*$ and $\bar{B}^*B^*$ systems. With the same inputs, the poles near the $\bar{B}B^*$ and $\bar{B}^*B^*$ mass thresholds are predicted,  transiting into virtual states, consistent with Ref.~\cite{Wang:2020dko}.  Since  the $\bar{B}^*B^{(*)}$ reduced mass is larger, the isovector $\bar{B}^*B^{(*)}$ molecules are more bound, resulting in virtual $\bar{B}B^*$ and $\bar{B}^*B^*$ states.    
We note that other channels composed of a charmonium state and a light meson may also play a role in generating these molecules.

\begin{table}[ttt]
\centering
\caption{ Masses of $ {D}D^*$,  $ {D}^*D^*$,  $ {D}D_{s}^*$, and $ {D}^*D_s^*$ with $J^{P}=1^{+}$ for a cutoff of $\Lambda=1$ GeV. 
  }
\label{poleTcc}
\begin{tabular}{cccccc}
\hline\hline
Molecule & $I$ & $J^{PC}$  & $m$ (MeV)
& Experimental data (MeV)
    \\
  \hline
  $D {D}^{\ast}$ & $0$ & $1^{+}$
  & Input  & 3874.73   \\
  $D^{\ast}{D}^{\ast}$ & $0$ & $1^{+}$ & 4015.3  & -    \\
  \hline
  $D_{s}{D}^*$ & $\frac{1}{2}$ & $1^{+}$  & 3975.2   & - \\
  $D_{s}^*{D}^*$ & $\frac{1}{2}$ & $1^{+}$    & 4118.3  &-  
   \\
  \hline
  $B {B}^{\ast}$ & $0$ & $1^{+}$
  & 10551.2  & -  \\
  $B^{\ast} {B}^{\ast}$ & $0$ & $1^{+}$ & 10596.1  & -    \\
  \hline \hline
\end{tabular}
\end{table}

Finally, we discuss the open-charm $D^{(*)}D^{(*)}$  molecules.  Because of the Bose-Einstein statistics, the quantum numbers of the $D^{(*)}D^{(*)}$ system satisfy $(-1)^{I+S+L+1}=1$. In the isoscalar sector, there exist only two states in contrast with the  $\bar{D}^{(*)}D^{(*)}$ system, i.e., $J^{P}=1^{+}$ $DD^*$ and $J^{P}=1^{+}$ $D^*D^*$.   
In the heavy quark limit, the $DD^*$ contact interaction is the same as that of $D^*D^*$. Unlike its hidden-charm counterpart, the isospin-breaking effect is expected to be small due to the smaller mass splitting of the two components of the $D^*D$ channel.   
 Identifying the
$T_{cc}(3875)$ discovered by the LHCb Collaboration  as a $DD^*$ bound state, one can determine the coupling $C_{DD}=-18.215$ GeV$^{-2}$ for a cutoff $\Lambda=1$ GeV. With the coupling obtained,  one predicts a  $D^*D^*$ bound state with a binding energy of $1.9$ MeV, in agreement with Refs.~\cite{Albaladejo:2021vln,Dai:2021vgf,Du:2021zzh}.  In Ref.~\cite{Du:2021zzh}, the $DD^*$ potentials are determined by fitting the LHCb data, where the effect of three-body cuts are considered. Because the $T_{cc}$ mass is quite close to the $DD\pi$ mass threshold, the impact of  three-body cuts on the pole position has been widely discussed~\cite{Du:2023hlu,Meng:2023bmz,Wang:2023iaz,Zhang:2024dth,Hansen:2024ffk,Du:2024snq,Abolnikov:2024key,Dawid:2024dgy}.

With SU(3)-flavor symmetry, we obtain two  $D_sD^*$ and $D_s^*D^*$ bound states with binding energies of 1.6 MeV and 2.6 MeV as shown in Table~\ref{poleTcc}, consistent with the results of the chiral unitary approach~\cite{Dai:2021vgf}. The $D_sD^*$ and $D_s^*D^*$ systems can bind via the one-kaon exchange.  On the other hand, there is essentially no light-meson exchange allowed for the $D_s^*\bar{D}$ system. Therefore, the strength of the  $D_s^*\bar{D}$ interaction is very weak~\cite{Chen:2020yvq,Ikeno:2020mra}, which indicates there are no bound states around the $D_s^*\bar{D}$ mass threshold.      
Along the same line, with HQFS,  two $B^*B$  and  $B^*B^*$  bound states are predicted  with binding energies of
 $53.0$ and $53.3$~MeV.   Regarding quark rearrangements, the doubly heavy tetraquark molecules are also likely to be compact doubly heavy tetraquark states. The quark model predictions for the doubly charmed tetraquark state are above the $D^*D$ mass threshold~\cite{Karliner:2017qjm,Mehen:2017nrh,Eichten:2017ffp}, which supports the molecular interpretation for the $T_{cc}(3875)$. The lattice QCD simulations of the $D^*D$ interaction resulted in a virtual state~\cite{Padmanath:2022cvl,Chen:2022vpo,Lyu:2023xro,Whyte:2024ihh}, which indicates that the  $D^*D$ interaction is attractive but not very strong. It should be noted that there would be a bound state below the $D^*D$ threshold if the $m_{\pi}$ mass approaches its physical mass~\cite{Lyu:2023xro}, consistent with the EFT analysis~\cite{Abolnikov:2024key}.  However, the quark model predictions for the doubly bottomed tetraquark state are below the $BB$ mass threshold~\cite{Weng:2021hje,Meng:2021yjr}. The light meson saturation mechanism shows that the hidden-charm dimeson molecules are more bound than the open-charm dimeson molecules. However, the hidden-charm pentaquark states are less bound than their open-charm partners~\cite{Liu:2020nil,Shen:2022zvd}. Compared with the hidden-charm hadronic molecules, the widths of open-charm hadronic molecules are narrower due to fewer decay modes. Meanwhile, the production mechanism of open-charm hadronic molecules is less than that of hidden-charm hadronic molecules since the formation of the  $cc$ diquark is more difficult than those of the $c \bar{c}$.       
The above studies provide two universal rules regarding the strengths of the potentials of a given system.

In the heavy quark limit, the interactions between a pair of heavy flavor hadrons generated by the dynamic of a pair of $qq$ or $q\bar{q}$ ($q$ referring to up or down quarks) in the heavy flavor hadrons, can be generally classified into a spin-spin dependent term $C_a$ and a spin-spin independent term $C_b$, which can be  saturated by the one-boson exchange potential, i.e.,                  
\begin{eqnarray}
\label{67890}
C_{a}&\propto& -\frac{g_{\sigma1}^2}{m_{\sigma}^2}-\frac{g_{v1}^2}{m_{v}^2}(\eta-\vec{\tau}_{1}\cdot\vec{\tau}_{2}),    \\ \nonumber
C_{b}&\propto& -\frac{f_{v1}^2}{6 M^2}(\eta-\vec{\tau}_{1}\cdot\vec{\tau}_{2}),
\end{eqnarray}
where  $\eta=1$ for the $q\bar{q}$ system and  $\eta=-1$ for the $q{q}$ system. One can see that the potentials for the systems with $q\bar{q}$ are more attractive than those with $q{q}$ as they contain the same isospin and angular momentum, such as $V_{DK}^{I=0} > V_{D\bar{K}}^{I=0} $ and  $V_{D\bar{D}}^{I=1} > V_{DD}^{I=1} $, consistent with Refs.~\cite{Guo:2015dha,Karliner:2015ina,Sakai:2017avl,Ding:2020dio,Wang:2023hpp}  Thus, one can generalize that  systems with a pair of a light quark and a light antiquark experience stronger attraction than those with a pair of light quarks or light antiquarks. In addition, the isospin factor in Eq.~(\ref{67890}) increases as the isospin increases, such as  $\vec{\tau}_{1}\cdot\vec{\tau}_{2}=-3$ for $I=0$ and $\vec{\tau}_{1}\cdot\vec{\tau}_{2}=1$ for $I=1$ as the isospin of each hadron equals to $1/2$,  which indicates that the hadron-hadron potentials with lower isospin are stronger,  such as $V_{DK}^{I=0} > V_{D{K}}^{I=1} $ and  $V_{D\bar{D}}^{I=0} > V_{D\bar{D}}^{I=1} $, consistent with Refs.~\cite{Guo:2015dha,Karliner:2015ina,Sakai:2017avl,Dong:2021bvy,Ding:2020dio,Wang:2023hpp}.    Thus, one can conclude that for a given system, its potential becomes stronger as its total isospin decreases.

\subsubsection{$\bar{D}^{(\ast)}D_{1,2}$ and $B_c\bar{B}_c$ molecules}

In the previous section, we discussed the final-state contributions of a pair of ground-state charmed mesons to the $X$ and $Z$  charmonium-like states. Due to the angular momentum conservation, we note that their contributions to the $Y$ charmonium-like states are minor. On the other hand, the $Y$ charmonium-like states with the quantum numbers  $J^{PC}=1^{--}$ strongly couple to either a pair of ground-state and excited charmed mesons, such as $\bar{D}^{(\ast)}D_{1,2}(D_{0,1})$,  or a pair of charmed baryon and anti-baryon (denoted by $B_c\bar{B}_c$), such as $\Lambda_c\bar{\Lambda}_c$ and $\Sigma_c\bar{\Sigma}_c$. In the heavy quark limit, the $P$-wave charmed mesons are classified as two doublets of  $(D_0, D_1(2430))$ and $(D_1(2420), D_2)$, while the widths of the former one are so broad that they are not good candidates for the constituents of hadronic molecules. In the following, we discuss only two such kinds of molecules: $\bar{D}^{(*)}D_{1,2}$ and $\Sigma_{c}^{(*)}\bar{\Sigma}_{c}^{(\ast)}$.

\begin{table}[!h]
\centering
\caption{ Mass spectrum of the $\bar{D}^{(*)}D_{1,2}$ system. The fourth column assumes $Y(4230)$ as the  $J^{PC}=1^{--}$ $\bar{D}D_{1}$, and fifth column assumes $Y(4360)$ as the $J^{PC}=1^{--}$ $\bar{D}^*D_{1}$ bound state. The sixth column considers both assumptions in fourth column and fifth column.         The superscript $V$ indicates a virtual state solution.   
}
\label{resultsDD12}
\begin{tabular}{c c c c c c c c c}
  \hline \hline
     Molecule   & $J^{PC}$~~~~  &  Wave function  &    Mass(A)~(MeV)   &   ~ Mass(B)~(MeV)   &    ~~ Mass(\cite{Peng:2022nrj})    &    ~~ Mass(\cite{Liu:2024ziu})
         \\ \hline   $\bar{D}D_{1}$   &$1^{-+}$ ~~~~  & $\frac{1}{\sqrt{2}}(\bar{D}D_{1}+{D}\bar{D}_{1})$  &~~~~ $-$   &~~~~ $-$     &~~~~ $-$     &~~~~ $4287.1$
         \\    $\bar{D}D_{1}$   &$1^{--}$ ~~~~  & $\frac{1}{\sqrt{2}}(\bar{D}D_{1}-{D}\bar{D}_{1})$  &~~~~  Input   &~~~~ $4021.2$  &~~~~  Input   &~~~~  Input
                  \\    $\bar{D}D_{2}$   &$2^{-+}$ ~~~~  & $\frac{1}{\sqrt{2}}(\bar{D}D_{2}+{D}\bar{D}_{2})$  &~~~~ $-$   &~~~~ $4311.7$     &~~~~ $-$  &~~~~ $4275.2$ 
                  \\    $\bar{D}D_{2}$   &$2^{--}$ ~~~~  & $\frac{1}{\sqrt{2}}(\bar{D}D_{2}-{D}\bar{D}_{2})$  &~~~~  $4317.5 $  &~~~~  $4223.3 $    &~~~~  $4323.2$ &~~~~ $4275.2$
                           \\    $\bar{D}^*D_{1}$   &$0^{-+}$ ~~~~  & $\frac{1}{\sqrt{2}}(\bar{D}^*D_{1}+{D}^*\bar{D}_{1})$  &~~~~   $-$  &~~~~   $-$  &~~~~   $-$  &~~~~ $4365.4$
       \\    $\bar{D}^*D_{1}$   &$0^{--}$ ~~~~  & $\frac{1}{\sqrt{2}}(\bar{D}^*D_{1}-{D}^*\bar{D}_{1})$  &~~~~  $-$  &~~~~   $-$    &~~~~  $4404.6$  &~~~~ $4381.6$
              \\    $\bar{D}^*D_{1}$   &$1^{-+}$ ~~~~  & $\frac{1}{\sqrt{2}}(\bar{D}^*D_{1}-{D}^*\bar{D}_{1})$  &~~~~ $-$   &~~~~   $-$  &~~~~   $-$  &~~~~ $4377.7$
              \\    $\bar{D}^*D_{1}$   &$1^{--}$ ~~~~  & $\frac{1}{\sqrt{2}}(\bar{D}^*D_{1}+{D}^*\bar{D}_{1})$  &~~~~   $4428.5 $     &~~~~  Input     &~~~~  Input   &~~~~  Input
                           \\    $\bar{D}^*D_{1}$   &$2^{-+}$ ~~~~  & $\frac{1}{\sqrt{2}}(\bar{D}^*D_{1}+{D}^*\bar{D}_{1})$  &~~~~   $4418.5 $  &~~~~   $4320.9 $    &~~~~ $-$    &~~~~ $4377.7$ 
                               \\    $\bar{D}^*D_{1}$   &$2^{--}$ ~~~~  & $\frac{1}{\sqrt{2}}(\bar{D}^*D_{1}-{D}^*\bar{D}_{1})$  &~~~~ $4369.0 $   &~~~~   $4180.3 $   &~~~~   $4430.0$   &~~~~ $4369.5$  
  \\    $\bar{D}^*D_{2}$   &$1^{-+}$ ~~~~  & $\frac{1}{\sqrt{2}}(\bar{D}^*D_{2}+{D}^*\bar{D}_{2})$  &~~~~  $-$    &~~~~  $-$    &~~~~  $4428^{V}$     &~~~~ $4416.9$ 
  \\    $\bar{D}^*D_{2}$   &$1^{--}$ ~~~~  & $\frac{1}{\sqrt{2}}(\bar{D}^*D_{2}-{D}^*\bar{D}_{2})$  &~~~~   $-$     &~~~~  $-$   &~~~~  $-$  &~~~~ $4408.7$    \\    $\bar{D}^*D_{2}$   &$2^{--}$ ~~~~  & $\frac{1}{\sqrt{2}}(\bar{D}^*D_{2}+{D}^*\bar{D}_{2})$ &~~~~ $4466.0 $ &~~~~ $4396.2 $    &~~~~  $4456^{V}$    &~~~~ $4400.3$ 
  \\    $\bar{D}^*D_{2}$   &$2^{-+}$ ~~~~  & $\frac{1}{\sqrt{2}}(\bar{D}^*D_{2}-{D}^*\bar{D}_{2})$  &~~~~ $- $ &~~~~ $-$   &~~~~ $-$    &~~~~ $4424.8$ 
   \\    $\bar{D}^*D_{2}$   &$3^{-+}$ ~~~~  & $\frac{1}{\sqrt{2}}(\bar{D}^*D_{2}+{D}^*\bar{D}_{2})$  &~~~~ $-$    &~~~~ $-$   &~~~~ $-$    &~~~~ $4435.9$ 
  \\    $\bar{D}^*D_{2}$   &$3^{--}$ ~~~~  & $\frac{1}{\sqrt{2}}(\bar{D}^*D_{2}-{D}^*\bar{D}_{2})$  &~~~~ $4251.3$  &~~~~ $3862.8$   &~~~~ $4270$   &~~~~ $4387.1$ 
      \\
  \hline \hline
\end{tabular}
\end{table}

 In the heavy quark limit, the contact-range potentials of the $\bar{D}^{(*)}D_{1,2}$ system are characterized by four parameters~\cite{Guo:2017jvc,Peng:2022nrj}.  The contact potentials of $\bar{D}^{(*)}D_{1,2}$ system are shown in Eq.~(\ref{DbarD1}).     In terms of the products of spin operators,  they can be written as 
\begin{eqnarray}
V=D_{a} + S_{1}\cdot S_{2} D_{b} + S_{12} \cdot S_{12} C_{a} + Q_{ij} Q_{ij} C_b,      
\end{eqnarray}
where $D_{a}$ and $D_b$ characterise the potentials of the direct scattering  process, and $C_a$ and $C_b$ the cross scattering  process. Then the contact potentials of $\bar{D}^{(*)}D_{1,2}$ system are rewritten  in Eq.~(\ref{DbarD1}).

 In principle, one needs experimental data to fix these parameters. However, in the $R$-value scan, only the states $Y(4220)$ and $Y(4360)$ are difficult to fit into the conventional charmonium states~\cite{BES:2001ckj}, which are around the $\bar{D}D_{1}$  and $\bar{D}^{\ast}D_{1}$ mass thresholds. For a cutoff of $\Lambda=1$ GeV, the parameters of the $\bar{D}D_{1}$ and  $\bar{D}^*D_{1}$ contact-range potentials are determined to be $-34.3$~GeV$^{-2}$ and $-32.0$~GeV$^{-2}$.  The $\bar{D}D_{1}$ and  $\bar{D}^*D_{1}$ contact-range potentials are similar, which might be accidental because they belong to different light quark multiplets~\cite{Guo:2017jvc} and do not satisfy HQSS.  $Y(4220)$ and $Y(4360)$ are expected to be the $\bar{D}D_{1}$  and $\bar{D}^{\ast}D_{1}$ molecules~\cite{Peng:2022nrj,Ji:2022blw,Wang:2023ivd}.     
With the molecular assumptions for $Y(4220)$ and $Y(4360)$,  one can not fix all four parameters of the $\bar{D}^{(*)}D_{1,2}$ system. We further resort to the light meson saturation approach, which determines the ratios of these parameters.  
With the couplings in Ref.~\cite{Peng:2021hkr},  the ratios are estimated to be 
\begin{eqnarray}
\label{couplingratios}
\frac{D_{b}^{sat}}{D_{a}^{sat}}=0.74,~~~ \frac{C_{a}^{sat}}{D_{a}^{sat}}=-1.07,~~~\frac{C_{b}^{sat}}{D_{a}^{sat}}=-0.28. 
\end{eqnarray}
One can see that $C_{a}$ and $D_{b}$ are more than $50\%$ of $D_{a}$, which induces large mass splittings between the multiplet of hadronic molecules. Assuming $Y(4220)$ as the $J^{PC}=1^{--}$ $\bar{D}D_{1}$ bound state, the flavor wave function is $\frac{1}{\sqrt{2}} (\bar{D}D_{1} - {D}\bar{D}_{1} ) $ and the contact-range potential is $D_a-C_a$. Together with the ratio obtained in the light meson saturation approach, one further obtains $D_{a}=-16.57$~GeV$^{-2}$ and $C_{a}=17.73$~GeV$^{-2}$.  At last, one obtains 
$D_{b}=-12.26$~GeV$^{-2}$ and $C_{b}=4.64$~GeV$^{-2}$.   Identifying $Y(4360)$ as a $\bar{D}^*D_1$ bound state,  one can obtain the sum of $D_{a}-\frac{5}{6}D_{b}-\frac{1}{4}C_{a}-\frac{5}{4}C_{b}=-32.0$~GeV$^{-2}$.  With the ratios obtained in the light meson saturation approach, one can determine  $D_{a}=-31.97$~GeV$^{-2}$, $D_{b}=-23.66$~GeV$^{-2}$, $C_{a}=34.21$~GeV$^{-2}$, and $C_{b}=8.95$~GeV$^{-2}$.  We show the mass spectrum of the $\bar{D}^*D_{1,2}$ system for the two cases in Table~\ref{resultsDD12}.  Case A refers to the former fitted parameters, and Case B to the latter. The molecules in Case B are more bound such that the $Y(4220)$ mass shifts to $4021$~MeV. Assuming that $Y(4220)$ and $Y(4360)$ are $\bar{D}D_{1}$ and $\bar{D}^*D_{1}$ bound states, the difference of Case A and Case B can be attributed to the unknown couplings between excited charmed mesons and light mesons. The ratios in Eq.~(\ref{couplingratios}) determined by the light meson saturation approach strongly relate to the couplings between excited charmed mesons and light mesons. As indicated in Ref.~\cite{Peng:2021hkr}, the couplings $f_{v}^{\prime}$ and $h_v$ are in the range of    $(2.6\sim 3.6)\cdot2.9$  and    $(7.4\sim 15.4)\cdot2.9$, respectively, where their  central values  are taken to predict the ratios in Eq.~(\ref{couplingratios}).  It is worth noting that these states $J^{PC}=2^{--}$ $\bar{D}D_{2}$, $J^{PC}=2^{--}$ $\bar{D}^*D_{1}$, $J^{PC}=2^{-+}$ $\bar{D}^*D_{1}$, $J^{PC}=2^{--}$ $\bar{D}^*D_{2}$,  and  $J^{PC}=3^{--}$ $\bar{D}^*D_{2}$ are  bound in both cases.

In Ref.~\cite{Peng:2022nrj}, Peng et al. argued that the terms $C_a$ and $C_b$ are strongly correlated, while the others are negligible. Then assuming $Y(4220)$ and $Y(4360)$ as the $J^{PC}=1^{--}$ $\bar{D}D_{1}$ and  $J^{PC}=1^{--}$ $\bar{D}^*D_{1}$ bound states, respectively,  they predicted the mass spectra of the $\bar{D}^{(*)}D_{1,2}$ system as shown in Table~\ref{resultsDD12}, close to the results of Case A. In Ref.~\cite{Liu:2024ziu}, assuming the terms $D_b$ and $C_b$ are negligible and  $Y(4220)$ and $Y(4360)$ as the $J^{PC}=1^{--}$ $\bar{D}D_{1}$ and  $J^{PC}=1^{--}$ $\bar{D}^*D_{1}$ bound states, the mass spectra of  the $\bar{D}^{(*)}D_{1,2}$ system  were predicted in Table~\ref{resultsDD12}.  In this case, a complete multiplet of HQSS hadronic molecules is obtained, consistent with Ref.~\cite{Dong:2021juy}.  It should be noted that the predicted $J^{PC}=1^{--}$ $\bar{D}^*D_{2}$ bound state in Refs.~\cite{Dong:2021juy,Liu:2024ziu} is likely to correspond to the $\psi(4415)$, as suggested in Refs.~\cite{Ji:2022blw,Wang:2023ivd}.                            In Ref.~\cite{Dong:2019ofp}, Dong et al. adopted the meson exchange model to assign $Y(4260)$ as a $J^{PC}=1^{--}$ $\bar{D}D_1$ bound state and predicted its $C$-parity partner, a $J^{PC}=1^{-+}$ $\bar{D}D_1$ bound state with a large binding energy. 
With SU(3)-flavor symmetry, the vector charmonium state  $Y(4630)$ is explained as a $\bar{D}_s^*D_{s1}(2536)$ molecule~\cite{He:2019csk,Yang:2021sue}, where the $\eta$ and $\phi$ exchanges are responsible for the $\bar{D}_s^*D_{s1}(2536)$  interactions. In Refs.~\cite{Dong:2021juy,Peng:2022nrj}, a complete multiplet of $\bar{D}_s^{(*)}D_{s1,2}$ were predicted.

In addition, the molecules composed of the ground-state charmed meson $D$ and the exotic charmed mesons  $D_{s0}^*(2317)$/$D_{s1}(2460)$, which are assigned as the $DK$/$D^*K$ molecules in the aforementioned discussions,    are likely to exist.  
 In Ref.~\cite{SanchezSanchez:2017xtl},  the potentials of $DD_{s0}^{*}(2317)$ and  $DD_{s1}(2460)$  are generated via the one kaon exchange,  and their binding energies are calculated.  Three-body hadrons could also form these two molecules, $DDK$ and $DD^*K$, regarding quark contents, which will be discussed in the next section.

Several  $Y$ states around $4.6$ GeV were  discovered experimentally,  including $Y(4660)$~\cite{Belle:2007umv,BaBar:2012hpr,Belle:2014wyt,BESIII:2021njb,BESIII:2023cmv}. In principle, the vector charmonium states located in this energy region can decay into  $\bar{\Lambda}_c\Lambda_c$. The Belle Collaboration observed a state named as  $Y(4630)$ in the $\bar{\Lambda}_c\Lambda_c$ mass distribution~\cite{Belle:2008xmh}, which was not confirmed by the BESIII Collaboration~\cite{BESIII:2023rwv}. It is necessary to study the hadronic molecules composed of a pair of charmed and anticharmed baryons.  In Ref.~\cite{Qiao:2005av}, Qiao assigned the $Y(4260)$ as a $\bar{\Lambda}_c\Lambda_c$ baryonium to explain the decay properties of $Y(4260)$. In Refs.~\cite{Wan:2019ake,Wang:2021qmn}, utilizing QCD sum rules the authors found that the mass of the $\bar{\Lambda}_c\Lambda_c$ baryonium is above its threshold. In Refs.~\cite{Chen:2011cta,Chen:2013sba}, Chen et al. employed the  Heavy
Baryon ChPT to investigate likely bound states below the $\bar{\Lambda}_c\Lambda_c$ threshold. In Ref.~\cite{Deng:2013aca}, Deng et al. used the color flux-tube model to assign $Y(4360)$ as a $\bar{\Lambda}_c\Lambda_c$ bound state. In Refs.~\cite{Chen:2017vai,Song:2022yfr,Cao:2024hnn}, the authors found the existence of a bound state below the $\bar{\Lambda}_c\Lambda_c$ threshold using the OBE model, consistent with the contact-range EFT~\cite{Lu:2017dvm,Dong:2021juy}. In Ref.~\cite{Cao:2019wwt}, by fitting the Belle and BESIII data, the authors found the existence of a virtual state below the $\bar{\Lambda}_c\Lambda_c$  threshold.  The theoretical analysis shows that the  $\bar{\Lambda}_c\Lambda_c$ potential is quite uncertain. More precise experimental data or lattice QCD calculations would help reduce such uncertainties and understand nearby vector charmonium states.

In this section, we investigate the mass spectrum of the  $\Sigma_c^{(\ast)}\bar{\Sigma}_c^{(\ast)}$ system. The contact-range potentials of the $\Sigma_{c}^{(*)}\bar{\Sigma}_{c}^{(\ast)}$ system are parameterized with three parameters: $E_a$, $E_b$, and $E_c$, 
 \begin{eqnarray}
V=E_a+ S_{1}\cdot S_{2} E_b +  Q_{ij} Q_{ji} E_c,
\end{eqnarray}
 where  $E_a$, $E_b$, and $E_c$  represent the electric charge, magnetic dipole, and electric quadrupole terms, which are constructed by the products of irreducible tensors built
from the light-spin operators.

\begin{table}[!h]
\centering
\caption{ Likely bound states in the $\Sigma_{c}^{(*)}\bar{\Sigma}_{c}^{(\ast)}$ system.  
}
\label{resultsSigmacSigmac}
\begin{tabular}{c c c c c c c c}
  \hline \hline
     Molecule   & $J^{PC}$~~~~  &  Potential   &    Attractive?~~~
         \\ \hline   $\Sigma_{c}\bar{\Sigma}_{c}$   &$0^{-+}$ ~~~~  & $E_{a}-\frac{4}{3}E_b$&~~~~   ?
         \\    $\Sigma_{c}\bar{\Sigma}_{c}$   &$1^{--}$~~~~  & $E_{a}+\frac{4}{9}E_b$&~~~~   likely
                \\    $\bar{\Sigma}_{c}^*{\Sigma}_{c}$   & $1^{--}$~~~ & $E_{a}-\frac{11}{9}E_b+5E_c$&~~~~   ?
                \\    $\bar{\Sigma}_{c}^*{\Sigma}_{c}$   & $1^{-+}$~~~ & $E_{a}-E_b-5E_c$&~~~~   ?
                 \\    $\bar{\Sigma}_{c}^*{\Sigma}_{c}$   & $2^{-+}$~~~ & $E_{a}+\frac{1}{3}E_b-E_c$&~~~~   likely
                \\    $\bar{\Sigma}_{c}^*{\Sigma}_{c}$   & $2^{--}$~~~ & $E_{a}+E_b+E_c$&~~~~   likely
          \\    $\bar{\Sigma}_{c}^*{\Sigma}_{c}^{\ast}$   & $0^{-+}$~~~ & $E_{a}-\frac{5}{3}E_b+5E_c$&~~~~   ?
            \\    $\bar{\Sigma}_{c}^*{\Sigma}_{c}^{\ast}$   & $1^{--}$~~~ & $E_{a}-\frac{11}{9}E_b+E_c$&~~~~   ? 
                        \\    $\bar{\Sigma}_{c}^*{\Sigma}_{c}^{\ast}$   & $2^{-+}$~~~ & $E_{a}-\frac{1}{3}E_b-3E_c$&~~~~   ? 
             \\    $\bar{\Sigma}_{c}^*{\Sigma}_{c}^{\ast}$   & $3^{--}$~~~ & $E_{a}+E_b+E_c$&~~~~   likely
      \\
  \hline \hline
\end{tabular}
\end{table}

 Since no experimental data is available, one can not determine $E_{a}$ and $E_{b}$. Often, one resorts to phenomenological models such as the OBE model to predict the mass spectrum of the $\Sigma_c^{(\ast)}{\Sigma}_c^{(\ast)}$ system~\cite{Ling:2021asz}. On the other hand, relying on the light meson saturation mechanism, one can also understand the $\Sigma_c^{(\ast)}\bar{\Sigma}_c^{(\ast)}$  system qualitatively. 
 Following Ref.~\cite{Peng:2020xrf}, the couplings of $E_{a}$, $E_{b}$, and $E_{c}$ are saturated in the following way
\begin{eqnarray}
E_{a}^{sat(\sigma)}(\Lambda\sim m_{\sigma})&\propto& -\frac{g_{\sigma2}^2}{m_{\sigma}^2},    \\ \nonumber
E_{a}^{sat (V)}(\Lambda\sim m_{V})&\propto& \frac{g_{v2}^2}{m_{v}^2}(\eta+\vec{T}_{1}\cdot\vec{T}_{2}),  \\ \nonumber
E_{b}^{sat (V)}(\Lambda\sim m_{V})&\propto& \frac{f_{v2}^2}{6 M^2}(\eta+\vec{T}_{1}\cdot\vec{T}_{2}), \\ \nonumber
E_{c}^{sat (V)}(\Lambda\sim m_{V})&\propto& \frac{h_{v}^2}{36 M^4}m_{V}^2(\eta+\vec{T}_{1}\cdot\vec{T}_{2}), 
\label{SigmacSigmac}
\end{eqnarray}
where $\eta=+1$ is for the $\Sigma_{c}^{(*)}{\Sigma}_{c}^{(\ast)}$ system and $\eta=-1$ is for the $\Sigma_{c}^{(*)}\bar{\Sigma}_{c}^{(\ast)}$ system.  
With the relevant values  of Table~\ref{tab:couplingsDS} and the coupling $h_{v}=\eta g_{v}$ with $\eta=0$~\cite{Liu:2019zvb}, we obtain the  ratios of the $\Sigma_{c}^{(*)}\bar{\Sigma}_{c}^{(\ast)}$ system:
\begin{eqnarray}
\frac{E_{b}^{sat}}{E_{a}^{sat}}=0.19,~~~ \frac{E_{c}^{sat}}{E_{a}^{sat}}=0,
\end{eqnarray}
which indicates that the $E_{c}$ term can be safely neglected.
Moreover, we obtain the ratios of the $\Sigma_{c}^{(*)}{\Sigma}_{c}^{(\ast)}$ system: 
\begin{eqnarray}
\frac{E_{b}^{sat}}{E_{a}^{sat}}=0.10,~~~ \frac{E_{c}^{sat}}{E_{a}^{sat}}=0.
\end{eqnarray}
 The $E_a$ term provides attraction. According to the light meson saturation mechanism, the $E_{b}$ term also provides attraction~\cite{Peng:2020xrf}. As shown in Table~\ref{resultsSigmacSigmac}, we can anticipate that the $J^{PC}=1^{--}~\bar{\Sigma}_{c}\Sigma_{c}$,  $J^{PC}=2^{--}~\bar{\Sigma}_{c}\Sigma_{c}^{*}$, $J^{PC}=2^{-+}~\bar{\Sigma}_{c}\Sigma_{c}^{*}$, and  $J^{PC}=3^{--}~\bar{\Sigma}_{c}^{\ast}\Sigma_{c}^{*}$ systems bind, where the state $J^{PC}=3^{--}~\bar{\Sigma}_{c}^{\ast}\Sigma_{c}^{*}$ is the most bound.  The $\Sigma_c^{(\ast)}\bar{\Sigma}_c^{(\ast)}$ system can be decomposed into isospin eigenstates with isospin $0$, $1$, or $2$. Referring to the light meson saturation approach, we find that the $\Sigma_c^{(\ast)}\bar{\Sigma}_c^{(\ast)}$ system is more bound as the isospin decreases.    In this picture, one can estimate the potential strength of the $\Sigma_c^{(\ast)}\bar{\Sigma}_c^{(\ast)}$  system, which is helpful to understand the relevant hadron-hadron interactions and explore $Y$ states near the $\Sigma_c^{(\ast)}\bar{\Sigma}_c^{(\ast)}$  mass threshold. In Ref.~\cite{Dong:2021juy}, Dong et al. calculated the complete spectra of the $\Sigma_c^{(\ast)}\bar{\Sigma}_c^{(\ast)}$ system, showing that the binding energies of 10 states are almost degenerate.   In Ref.~\cite{Wang:2024riu}, with the $\Sigma_c^{(\ast)}\bar{\Sigma}_c^{(\ast)}$ potential derived from quark-level potentials, they obtained four bound states, corresponding to four `` likely " states in Table~\ref{resultsSigmacSigmac}. In Refs.~\cite{Lee:2011rka,Yang:2018amd,Song:2022svi},  the $\Sigma_c^{(\ast)}\bar{\Sigma}_c^{(\ast)}$ potentials are investigated in the OBE model and found to be consistent with the analysis of the contact-range EFT. In terms of the light meson saturation mechanism, we find that the ratio of $E_b$/$E_a$ for the $\Sigma_c^{(\ast)}\bar{\Sigma}_c^{(\ast)}$  system is larger than that for the $\Sigma_c^{(\ast)}{\Sigma}_c^{(\ast)}$ system, which hints at a large mass splitting for the $\Sigma_c^{(\ast)}\bar{\Sigma}_c^{(\ast)}$ system than for the $\Sigma_c^{(\ast)}{\Sigma}_c^{(\ast)}$ system. Moreover, the $\Sigma_c^{(\ast)}\bar{\Sigma}_c^{(\ast)}$  molecules are more bound than the $\Sigma_c^{(\ast)}{\Sigma}_c^{(\ast)}$ molecules.

\subsection{Strong and radiative decays}

The decay patterns of hadronic molecules are essential in revealing their nature.  The likely decay modes of the predicted multiplets can help plan future experiments. In this review, we only focus on strong decays and radiative decays.~\footnote{We note that weak decays of hadronic molecules have recently been explored since the decay of the $D\bar{B}$ molecular candidate can only proceed via the weak interaction~\cite{Liu:2024lmv}.}  The decay modes include two-body decays and three-body decays. The three-body decays of hadronic molecules can proceed via one of their unstable constituents at tree level,  and two-body decays of hadronic molecules are illustrated via the triangle diagram. The effective lagrangian approach can calculate the tree and triangle diagrams.     In addition,  the width of a hadronic molecule can be obtained from the unitary amplitude.    

\subsubsection{Three-body decays }
 
 \begin{figure}[!h]
\begin{center}
\includegraphics[width=6.5in]{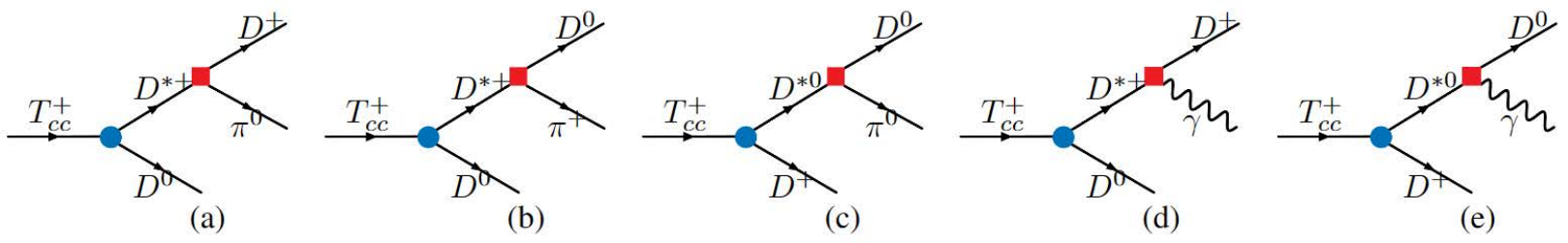}
\caption{  Tree-level diagrams for the strong decays of  $T_{cc}^{+}\to D^{+}\pi^{0}(D^{\ast+})D^{0}$~(a), $T_{cc}^{+}\to D^{0}\pi^{+}(D^{\ast+})D^{0}$~(b),  and  $T_{cc}^{+}\to D^{0}\pi^{0}(D^{\ast0})D^{+}$~(c) as well as  radiative decays of  $T_{cc}^{+}\to D^{+}\gamma(D^{\ast+})D^{0}$~(d) and $T_{cc}^{+}\to D^{0}\gamma(D^{\ast0})D^{+}$~(e).   }
\label{tree12}
\end{center}
\end{figure}

The decay of the doubly charmed tetraquark state $T_{cc}(3875)$ is a typical example of three-body decays. As a $D^*D$ bound state,  it can decay into $DD\pi(\gamma)$ via an off-shell $D^*$ decaying into $D\pi(\gamma)$ as shown in Fig.~\ref{tree12}. One should note that this is the dominant decay mode of $T_{cc}$ since two-body decays are not allowed for $T_{cc}$ as a $D^*D$ bound state due to parity and momentum conservations.   We employ the Lagrangian approach to calculate the decay widths of $T_{cc}\to DD\pi(\gamma) $.    For the  $D^*D$ system, the difference  between the $D^{*0}D^+$ and $D^{*+}D^0$  mass thresholds is about $1$~MeV, which indicates that the isospin breaking in the $T_{cc}$ couplings to $D^{*0}D^+$ and $D^{*+}D^0$ is small. Therefore, we estimate the $T_{cc}$ couplings to its constituents $D^*D$ by the compositeness condition in the isospin limit~\cite{Ling:2021bir}, consistent with Refs~\cite{Feijoo:2021ppq,Albaladejo:2021vln,Du:2021zzh}. Then, we estimated the sum of its radiative and strong decay widths to be $63$~keV,  consistent with the experimental data. In other words, our results support the molecular interpretation for $T_{cc}(3875)$.  Considering final-state interactions, the decay modes of Fig.~\ref{tree12} can result in triangle diagrams, which are of the next-to-leading order in the Language of EFTs.  As a result, the contribution of such triangle diagrams can be treated perturbatively. In Ref.~\cite{Yan:2021wdl}, the contribution of the triangle diagram is estimated to be $20\%$ of the tree diagram.

The impact of three-body threshold $DD\pi$ on the $T_{cc}$ pole position and the line shape of the $DD\pi$ invariant mass distribution are investigated in Ref.~\cite{Du:2021zzh},   where the three-body unitarity is satisfied when simultaneously considering the one-pion exchange between the $D$ and $D^*$ mesons and the finite width of the $D^*$ meson. In Ref.~\cite{Dai:2023mxm}, the one-pion exchange and final-state interactions were taken into account in the EFT framework with three-body contribution treated as a perturbative correction, the calculated leading-order width of    $T_{cc}$  was consistent with the experimental measurement, larger than Ref.~\cite{Du:2021zzh}, which indicates that neglecting three-body unitarity can overestimate the $T_{cc}$ width. In Ref.~\cite{Albaladejo:2021vln}, without the three-body effect, the obtained width of $T_{cc}$ is larger than Ref.~\cite{Du:2021zzh}, but the invariant mass distributions are well described, indicating the small impact of three-body dynamics on the lineshape.

Considering HQSS, it is natural to expect a $D^*D^*$ bound state. Because the 
$D^*D^*$ molecules are located below the mass threshold by about $2$ MeV,  the phase space for its decay into $D^*D\pi$ and $D\pi D\pi$  is small. Therefore, the $D^*D^*$ molecule dominantly decays into  $D^*D $.  In Ref.~\cite{Dai:2021vgf}, Dai et al. estimated the partial decay width to be around $2$ MeV, much larger than the $T_{cc}$ width. The widths of three-body decays and four-body decays of the $D^*D^*$ molecules are estimated to be around $65$~keV, much smaller than that of the two-body decay~\cite{Jia:2022qwr,Jia:2023hvc}. The decay modes of the $D^*D^*$ molecule are more abundant than those of the $D^*D$ molecule. According to SU(3)-flavor symmetry, the decay modes of the   $D^*D_s $ and   $D_s^*D^*$  molecules are similar to those of the  $D^*D$ and  $D^*D^*$ molecules.

It should be  noted  that the $X(3872)$ is located almost exactly in the $\bar{D}^{*0}D^{0}$ mass threshold. Whether the $X(3872)$ is located above or below the $\bar{D}^{*0}D^{0}$ mass threshold remains undetermined, which would heavily affect its decay behavior.  Regarding 
the  $X(3872)$ as a $\bar{D}^{*0}D^{*0}$-$D^{*-}D^{+}$ molecule, the widths of the three-body decay of  $ X(3872) \to \bar{D}D\pi(\gamma)$ are estimated to be tens of keV, a bit smaller than that of $T_{cc}$ as a $D^*D$ molecule~\cite{Guo:2014hqa,Schmidt:2018vvl,Dai:2019hrf,Meng:2021kmi}, where the three-body partial widths are only from the neutral component of $X(3872)$.  Taking into account the $\bar{D}D$ final-state interaction in Ref.~\cite{Guo:2014hqa},  the three-body partial decay widths can be quite different if     
there is a pole around the $\bar{D}D$ threshold. In addition, Baru et al. found that including the three-body $\bar{D}D\pi$ effects, the partial width of $X(3872)\to \bar{D}D\pi $ is reduced by about a factor of $2$~\cite{Baru:2011rs}.

 \begin{figure}[ttt]
\begin{center}
\includegraphics[width=6.5in]{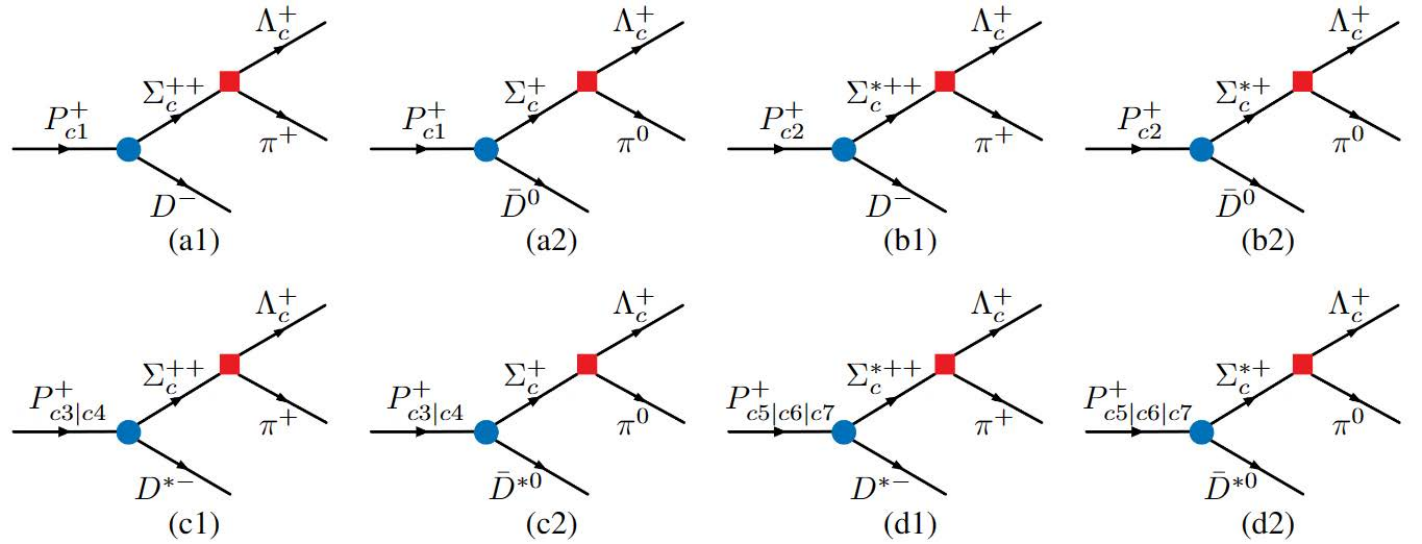}
\caption{  Tree-level diagrams for the  decays of $P_{c1}\to \bar{D}\Sigma_c \to \bar{D}\Lambda_{c}\pi$ (a), $P_{c2}\to \bar{D}\Sigma_c^{\ast} \to \bar{D}\Lambda_{c}\pi$ (b), $P_{c3|c4}\to \bar{D}^{\ast}\Sigma_c \to \bar{D}^{\ast}\Lambda_{c}\pi$ (c), and $P_{c5|c6|c7}\to \bar{D}^{\ast}\Sigma_c^{\ast} \to \bar{D}^{\ast}\Lambda_{c}\pi$ (d).  }
\label{tree1}
\end{center}
\end{figure}

\begin{table}[ttt]
\centering
\caption{Partial decay widths (in units of MeV) of $P_c^+ \to {D^{(\ast)-}} \Lambda_c^+ \pi^+$ and $P_c^+ \to \bar{D}^{(\ast)0} \Lambda_c^+ \pi^0$ in Fig.~\ref{tree1} in scenario A and scenario B for  a cutoff $\Lambda=1.5$ GeV.
}
\label{tab:width1}
\begin{tabular}{c c c c c}
  \hline \hline
      Scenario  &~~~~   A   &~~~~    A    &~~~~   B   &~~~~    B
         \\   Mode   &~~~~  $D^{(\ast)-} \Lambda_c^+ \pi^+$ &~~~~ $\bar{D}^{(\ast)0} \Lambda_c^+ \pi^0$ &~~~~  ${D^{(\ast)-}} \Lambda_c^+ \pi^+$ &~~~~ $\bar{D}^{(\ast)0} \Lambda_c^+ \pi^0$
         \\ \hline $P_{c1}$    &~~~~  0.051 &~~~~ 0.201  &~~~~  0.030 &~~~~ 0.129
         \\  $P_{c2}$    &~~~~  2.670 &~~~~ 2.859  &~~~~ 2.164 &~~~~ 2.251 
         \\  $P_{c3}$    &~~~~  0.002 &~~~~ 0.038  &~~~~  0.566 &~~~~ 1.964
         \\  $P_{c4}$    &~~~~  0.189 &~~~~ 0.655  &~~~~  0.001 &~~~~ 0.013
         \\  $P_{c5}$    &~~~~  2.751 &~~~~ 2.435  &~~~~  7.087 &~~~~ 6.490
         \\  $P_{c6}$    &~~~~  3.567 &~~~~ 3.188   &~~~~  4.539 &~~~~ 4.156
         \\  $P_{c7}$    &~~~~  2.991 &~~~~ 2.739   &~~~~  1.352 &~~~~ 1.195 \\
  \hline \hline
\end{tabular}
\end{table}

The three-body decays of the $\bar{D}^{(\ast)}\Sigma_{c}^{(\ast)}$ molecules can proceed as shown in Fig.~\ref{tree1}. The $\bar{D}^{(\ast)}\Sigma_{c}^{(\ast)}$ molecules are denoted by $P_{c1}$, $P_{c2}$, $\cdots$, $P_{c7}$, following the order of Scenario A in Table~\ref{tab:predictions-Pc}. The binding energies of the $\bar{D}^{\ast}\Sigma_{c}^{(\ast)}$ molecules relative to their mass thresholds are around 4-20 MeV~\cite{Liu:2019tjn,ParticleDataGroup:2020ssz}, while the mass thresholds of the $\bar{D}\pi\Sigma_{c}^{(\ast)}$ are 5 MeV less than those of the $\bar{D}^{\ast}\Sigma_{c}^{(\ast)}$, which implies that the decays of $\bar{D}^{\ast}\Sigma_{c}^{(\ast)}$ molecules into  $\bar{D}\pi\Sigma_{c}^{(\ast)}$ via off-shell $\bar{D}^{\ast}$ mesons are heavily suppressed. The phase space for the $\Sigma_{c}^{(\ast)}\to \Lambda_{c}\pi$ is more than 30 MeV so that the tree-level decays of $\bar{D}^{(\ast)}\Sigma_{c}^{(\ast)} \to \bar{D}^{(\ast)}\left(\Sigma_{c}^{(\ast)}\to \Lambda_{c}\pi\right) \to \bar{D}^{(\ast)}\Lambda_{c}\pi$ are allowed,  as shown in Fig.~\ref{tree1}. Using the masses obtained in Ref.~\cite{Liu:2019tjn}, we predict the decays of pentaquark molecules into $\bar{D}^{(\ast)}\Lambda_{c}\pi$ as shown in Table~\ref{tab:width1}.  The decay width of $P_{c}(4312)$  into $\bar{D}\Lambda_{c}\pi$ is about hundreds of keV, which accounts for $2\%-3\%$ of its total width.   The decay width of $P_{c}(4457)$  into $\bar{D}\Lambda_{c}\pi$ is up to several MeV, while for $P_{c}(4440)$  it is only tens of keV.  The partial decay width of $P_{c}(4457)$   accounts for tens of percent of its total width, while for $P_{c}(4440)$, it accounts for less than one percent. In addition, we predict the partial decay widths of the other four molecules into $\bar{D}^{(\ast)}\Lambda_{c}\pi$, which are about several MeV, some of which are even up to tens of MeV, in agreement with the results of Ref.~\cite{Lin:2019qiv}. In Ref.~\cite{Du:2021fmf}, the effect of three-body channels $\bar{D}^{(*)}\Lambda_c\pi$ was embodied into the full unitary amplitude, which implies that the three-body partial decay widths  are included in their total widths. The three-body dynamics for the $\bar{D}^{(\ast)}\Sigma_{c}^{(\ast)}$ molecules is investigated in a phenomenological model~\cite{Lin:2023ihj}. Taking into account the final-state rescattering of Fig.~\ref{tree1}, the tree diagrams will convert into the triangle diagrams as shown in Fig.~2  of Ref.~\cite{Xie:2022hhv}, which only contribute to the widths of $P_c$ states by several keV at most. 
Similarly, the $\bar{D}^{(\ast)}\Xi_{c}^{\prime(*)}$ molecules can  
decay into $\bar{D}^{(\ast)}\Xi_c\pi$, and the partial decay widths should be similar to the $P_c$ states in terms of SU(3) flavor symmetry.  However, the LHCb Collaboration found no significant signal of $P_{c}(4312)$, $P_{c}(4440)$, and $P_{c}(4457)$ in the $\bar{D}^{(*)}\Lambda_c \pi$ mass distributions~\cite{LHCb:2024pnt}.

\begin{figure}[ttt]
\begin{center}
\includegraphics[width=4.0in]{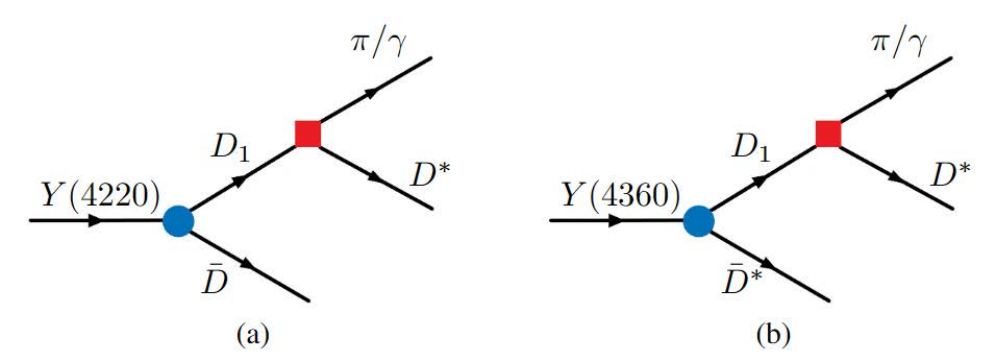}
\caption{  Tree diagrams of  $Y(4220) \to  D\bar{D}^* \pi (\gamma) $ and  $Y(4360) \to  D^*\bar{D}^* \pi (\gamma) $.  }
\label{tree4220}
\end{center}
\end{figure}

Assuming $Y(4220)$ and $Y(4360)$ as the  $\bar{D}D_1$ and $\bar{D}^*D_1$ bound states, they can decay into $\bar{D}D^* \pi(\gamma)$ and $\bar{D}^*D^* \pi(\gamma)$ via an off-shell $D_1$ meson as shown in Fig.~\ref{tree4220}, where we assume that  $D_1$ meson decay into $D^* \pi$ via the $D$-wave.     The decay widths  $Y(4220) \to  D\bar{D}^* \pi (\gamma) $ and  $Y(4360) \to  D^*\bar{D}^* \pi (\gamma) $  are shown in Table~\ref{widths4220and4360}, indicating that the radiative decays of $Y(4220)$ and $Y(4360)$ are  smaller than those of pionic decays by one order of magnitude.    In Ref.~\cite{Dong:2019ofp}, Dong et al. calculated the width of the partial decay  $Y(4260) \to \pi D^*\bar{D}$ to be $1.9$~MeV, consistent with ours.   In Ref.~\cite{Qin:2016spb}, Qin et al. assumed that the $Y(4260)$ is composed of a mixture of a  $\bar{D}D_1$ bound state and a compact $c\bar{c}$ core and simulated the $D^*\bar{D}$ invariant mass distribution of the process $e^+e^- \to Y(4260) \to \pi D^*\bar{D}$. They found that the partial widths of the three-body decay   $Y(4260) \to \pi D^*\bar{D}$ are $56.5$ MeV for $S$-wave and $8.3$~MeV for $D$-wave.       In Ref.~\cite{Wang:2023dsm}, Wang estimated the partial decay width of $Y(4220) \to \bar{D}D^* \pi $ up to be $0.12$ MeV assuming $Y(4220)$ as the compact tetraquark state,   smaller than that obtained assuming $Y(4220)$  as the hadronic molecule, which indicate that there exists prominent difference for the width of the decay  $Y(4220) \to \bar{D}D^* \pi $ in two scenarios of  $Y(4220)$. The  $Y(4220)$ and $Y(4360)$ can decay into the final states with hidden-charm number  $J/\psi \pi\pi$, $J/\psi K K$, and $h_c \pi\pi$ as well~\cite{Wang:2023dsm}, which generally proceed via the triangle mechanism, e.g., the final states of  $\bar{D}^{(*)}$ and $D^*$ in Fig.~\ref{tree4220} rescattering into $J/\psi \pi$ and $h_c\pi$.

\begin{table}[!h]
    \centering
    \caption{Widths (in units of MeV) of the partial decays of  $Y \to \bar{D}^{(*)}D^{*}\pi(\gamma)$ and $Y \to  {D}^{(*)}\bar{D}^{*}\pi(\gamma)$.   \label{widths4220and4360}}
    \begin{tabular}{c|cccc}
    \hline\hline    
          Molecules   &~~ $\bar{D}^{(*)}D^* \pi$   &~~~~${D}^{(*)}\bar{D}^* \pi$ &~~ $\bar{D}^{(*)}D^* \gamma$   &~~~~${D}^{(*)}\bar{D}^* \gamma$\\  \hline   
        $Y(4220) $   &~~ $1.84$   &~~~~$1.84$ &~~ $0.12$  &~~~~$0.12$ \\
        $Y(4360)$    &~~ $3.21$   &~~~~$3.21$ &~~ $0.21$  &~~~~$0.21$         
         \\  
    \hline \hline
    \end{tabular}
\end{table}

\subsubsection{Two-body decays}

 Two methods can be employed to calculate the two-body decay widths.  The first is to construct coupled-channel potentials constrained by symmetries, such as the chiral Lagrangian with SU(3)-flavor symmetry~\cite{Oller:1997ti,Oset:1997it,Jido:2003cb} and SU(4) symmetry~\cite{Gamermann:2006nm}, the local hidden gauge Lagrangian with SU(2)-flavor symmetry~\cite{Molina:2008jw},  SU(3)-flavor symmetry~\cite{Geng:2008gx}  or  SU(4)-flavor symmetry~\cite{Wu:2010jy,Molina:2009ct},  the contact-range potentials with HQSS~\cite{Nieves:2012tt,Peng:2020hql,Sakai:2019qph}, and some phenomenological models~\cite{He:2019rva,Yalikun:2021bfm}. The second is to develop models, such as triangle diagrams~\cite{Faessler:2007gv,Dong:2008gb,Dong:2013iqa,Shen:2016tzq,Xiao:2019mvs,Lin:2019qiv}. The former can account for the masses and widths of exotic states, but the latter can only deal with the widths. An initial particle decaying into two final particles at the quark level can be described by three-point correlation functions in the QCD sum rules, which can extract the initial particle's couplings to the two final particles and then obtain the corresponding decay widths~\cite{Dias:2013xfa,Azizi:2021utt,Chen:2022sbf,Wang:2023sii}.       

A typical example is the two-body decays of the pentaquark states. From  HQSS, the  $S$-wave  $\bar{D}^{(\ast)}\Sigma_{c}^{(\ast)} \to J/\psi(\eta_{c}) N$ interactions are only related to the spin of the light quark $1/2$, denoted by one coupling:  $g_2=\langle \bar{D}^{(\ast)}\Sigma_{c}^{(\ast)} | 1_{H}\otimes 1/2_{L} \rangle= \langle \bar{D}^{(\ast)}\Sigma_{c}^{(\ast)} | 0_{H}\otimes 1/2_{L} \rangle $.  Similarly, one can express the $\bar{D}^{(*)}\Lambda_c \to J/\psi(\eta_{c}) N$ interactions  by a second parameter: $g_1= \langle \bar{D}^{(\ast)}\Lambda_{c} | 1_{H}\otimes 1/2_{L} \rangle =\langle \bar{D}^{(\ast)}\Lambda_{c}| 0_{H}\otimes 1/2_{L} \rangle$. As for the $\bar{D}^{(\ast)}\Sigma_{c}^{(\ast)} \to \bar{D}^{(*)}\Lambda_c$ interactions, they depend only on one coupling constant in the heavy quark limit. Therefore,  we parameterize the $\bar{D}^{(\ast)}\Sigma_{c}^{(\ast)} \to \bar{D}^{(*)}\Lambda_c$ potential by  one  coupling:   $C_b^{\prime}= \langle \bar{D}^{(\ast)}\Sigma_{c}^{(\ast)}  | \bar{D}^{(\ast)}\Lambda_{c} \rangle$.    The potentials of $J/\psi N \to J/\psi N  $, $J/\psi N  \to \eta_{c}N$ and $\eta_{c} N  \to \eta_{c}N$ are suppressed due to the OZI rule, which is also supported by lattice QCD simulations~\cite{Skerbis:2018lew}. 
Combining the two parameters $C_a$ and $C_b$ characterizing the  $\bar{D}^{(\ast)}\Sigma_{c}^{(\ast)} \to \bar{D}^{(\ast)}\Sigma_{c}^{(\ast)}$ interactions,  the $\bar{D}^{(*)}\Sigma_{c}^{(*)}$,  $\bar{D}^{*}\Lambda_c$, $\bar{D}\Lambda_c$,  $J/\psi p$, and $\eta_c p$ coupled-channel interactions are parameterised by the following  parameters $C_a$, $C_b$,  $C_b^{\prime}$, $g_{1}$, and $g_2$, which are  determined by reproducing the masses and widths of $P_c(4312)$, $P_c(4440)$, and $P_c(4457)$ in two scenarios A and B~\cite{Pan:2023hrk}.

 \begin{table}[ttt]
\centering
\caption{Pole positions (in units of MeV) of six hidden-charm pentaquark molecules and the  couplings to their constituents  in Scenario A and Scenario B.  }
\label{pcdecaywidths}
\scalebox{0.95}{
\begin{tabular}{c c c c c c c}
  \hline\hline
     Scenario   &    \multicolumn{6}{c}{A}  \\ \hline
     Name & $P_{c1}$ & $P_{c2}$ & $P_{c3}$ & $P_{c4}$ & $P_{c5}$ & $P_{c6}$ \\
     Molecule  &  $\bar{D}\Sigma_c $  & $\bar{D}\Sigma^{*}_c$ & $\bar{D}^{\ast}\Sigma_c$ & $\bar{D}^{\ast}\Sigma_c$  & $\bar{D}^{\ast}\Sigma_c^{\ast}$  & $\bar{D}^{\ast}\Sigma_c^{\ast}$   \\
     $J^P$ &  $\frac{1}{2}^-$  &  $\frac{3}{2}^-$ & $\frac{1}{2}^-$ & $\frac{3}{2}^-$  & $\frac{1}{2}^-$ & $\frac{3}{2}^-$  \\
     Pole (MeV) &  4310.6+3.5$i$  &4372.8 +2.7$i$  & 4440.6+8.6$i$ & 4458.4+0.7$i$ & 4500.0+9.9$i$ &  4513.2+7.7$i$  \\
     $g_{P_{c}\Sigma_{c}^{*}\bar{D}^{*}}$ & -  & - & - & - & 2.686 & 2.194 \\
     $g_{P_{c}\Sigma_{c}\bar{D}^{*}}$ &  - & - & 2.554 & 1.082 & 0.141 & 0.218 \\
     $g_{P_{c}\Sigma_{c}^{*}\bar{D}}$ &  - & 2.133 & - & 0.179 & - & 0.237 \\
     $g_{P_{c}\Sigma_{c}\bar{D}}$ &  2.089 & - & 0.254 & - & 0.139 & - \\
     $g_{P_{c}\Lambda_{c}\bar{D}^{*}}$ & 0.234 & 0.074 & 0.177 & 0.050 & 0.110 & 0.241 \\
     $g_{P_{c}\Lambda_{c}\bar{D}}$ &  0.014 & - & 0.158 & - & 0.207 & - \\
     $g_{P_{c}J/\psi N}$ &  0.251 & 0.454 & 0.584 & 0.103 & 0.434 & 0.532 \\
     $g_{P_{c}\eta_{c} N}$ &  0.420 & - & 0.261 & - & 0.527 & - \\\hline
    Scenario   &   \multicolumn{6}{c}{B}   \\ \hline
    Name & $P_{c1}$ & $P_{c2}$ & $P_{c3}$ & $P_{c4}$ & $P_{c5}$ & $P_{c6}$ \\
     Molecule  &  $\bar{D}\Sigma_c$  & $\bar{D}\Sigma^{*}_c$ & $\bar{D}^{\ast}\Sigma_c$ & $\bar{D}^{\ast}\Sigma_c$  & $\bar{D}^{\ast}\Sigma_c^{\ast}$  & $\bar{D}^{\ast}\Sigma_c^{\ast}$   \\
     $J^P$ &  $\frac{1}{2}^-$  &  $\frac{3}{2}^-$ & $\frac{1}{2}^-$ & $\frac{3}{2}^-$  & $\frac{1}{2}^-$ & $\frac{3}{2}^-$  \\
     Pole (MeV)  &  4309.9+4$i$  & 4365.8+6.2$i$ &  4458.4+4.5$i$ & 4441.4+1.1$i$ & 4521.6+7.5$i$ &  4522.5+3.7$i$  \\
     $g_{P_{c}\Sigma_{c}^{*}\bar{D}^{*}}$ & -  & - & - & - & 1.841 & 1.621 \\
     $g_{P_{c}\Sigma_{c}\bar{D}^{*}}$ &  - & - & 1.679 & 2.462 & 0.107 & 0.143 \\
     $g_{P_{c}\Sigma_{c}^{*}\bar{D}}$ &  - & 2.451 & - & 0.099 & - & 0.171 \\
     $g_{P_{c}\Sigma_{c}\bar{D}}$ &  2.072 & - & 0.161 & - & 0.131 & - \\
     $g_{P_{c}\Lambda_{c}\bar{D}^{*}}$ & 0.392 & 0.090 & 0.247 & 0.159 & 0.232 & 0.223 \\
     $g_{P_{c}\Lambda_{c}\bar{D}}$ & 0.020 & - & 0.191 & - & 0.281 & - \\
     $g_{P_{c}J/\psi N}$ &  0.263 & 0.704 & 0.277 & 0.168 & 0.314 & 0.312 \\
     $g_{P_{c}\eta_{c} N}$ & 0.413 & - & 0.164 & - & 0.328 & - \\   \hline\hline
\end{tabular}}
\end{table}

In Table~\ref{pcdecaywidths}, we present the pole positions of the hidden-charm pentaquark molecules and the couplings to their constituents.  From the obtained pole positions of $P_c(4312)$, $P_c(4440)$, and $P_{c}(4457)$, it is evident that scenario A, yielding results consistent with the experimental data, is preferred over Scenario B, which is quite different from the single-channel study~\cite{Liu:2019tjn}. Our study shows that the coupled-channel effects can help distinguish the two possible scenarios.  In a similar approach but without considering the $\bar{D}^{(*)}\Lambda_c$ channels,  Scenario A  is still slightly preferred over Scenario B~\cite{Xie:2022hhv}.  We note in passing that the chiral unitary model~\cite{Xiao:2019aya} also prefers scenario A.  We further note that the coefficients in the contact-range potentials are derived assuming the HQSS, while the  HQSS  breaking is not considered.  In Ref.~\cite{Yamaguchi:2019seo}, it was shown that the tensor term of the one-pion exchange potentials plays a crucial role in describing the widths of the pentaquark molecules,  while the $D$-wave potentials are neglected in this work.        Therefore, we can not conclude which scenario is more favorable. 
In Refs.~\cite{Burns:2021jlu,Burns:2022uiv}, Burns et al. proposed another case, named Scenario C, which corresponds to a particular case of  Scenario B, where the $J^P=1/2^-$  $\bar{D}^*\Sigma_c \to \bar{D}^*\Sigma_c $ potential is not strong enough to form a bound state. Therefore, $P_{c}(4457)$ is interpreted as a kinetic effect rather than a genuine state. From their values of $C_a$ and $C_b$~\cite{Burns:2022uiv}, the ratio  $C_b/C_a$ is determined to be around 0.5, which 
implies the emergence of a sizeable spin-spin interaction.   For the     $\bar{D}^{(*)}D^{(*)}$ and  $\bar{D}^{(*)}\Sigma_c^{(*)}$ systems, we obtain the ratio of $C_b/C_a$ in the range of $0.1-0.3$ using the light meson saturation mechanism~\cite{Liu:2019zvb,Liu:2020tqy}.    It is no surprise that such a large spin-spin interaction breaks the completeness of the multiplet picture of hidden-charm pentaquark molecules~\cite{Liu:2019tjn,Xiao:2019aya,Du:2019pij,Yamaguchi:2019seo,PavonValderrama:2019nbk,Lin:2019qiv,He:2019rva,Yalikun:2021bfm,Dong:2021juy,Zhang:2023czx}.

 From the couplings obtained in Table~\ref{pcdecaywidths}, the partial decay widths of pentaquark molecules, as well as the corresponding branching fractions, are predicted in Scenario A and Scenario B as shown in Table~\ref{pcpartialdecaywithspc}. One can see that the branching fractions of the decays $P_c \to \bar{D}^{(*)}\Lambda_c$ in Scenario A are larger than those  in Scenario B. In comparison, the branching fractions of the decays $P_c \to J/\psi p$ and  $P_c \to \eta_c p$ in Scenario A  are smaller than those in Scenario B. One can see that the two-body partial decays are uncertain due to the uncertainties of the  $\bar{D}^{(\ast)}\Sigma_{c}^{(\ast)} \to \bar{D}^{(*)}\Lambda_c$ interactions and  $\bar{D}^{(\ast)}\Sigma_{c}^{(\ast)} \to J/\psi(\eta_{c}) N$ interactions in the contact-range EFT approach. Such partial decay widths can not discriminate which scenario is better due to the slight differences in their values between Scenario A and Scenario B. As explained below,  
 the corresponding correlation functions~\cite{Liu:2023wfo}  can help determine the spins of the 
 $P_c(4440)$ and $P_c(4457)$.

 \begin{table}[ttt]
\centering
\caption{Two-body partial decay widths (in units of MeV) of hidden-charm pentaquark molecules as well as their branching fractions in Scenario A and Scenario B .  }
\label{pcpartialdecaywithspc}
\scalebox{0.90}{  
\begin{tabular}{cc c c c c c c}
  \hline \hline
     Scenario   &    \multicolumn{6}{c}{A}  \\ \hline
     Molecule  & $P_{c1}$ & $P_{c2}$ & $P_{c3}$ & $P_{c4}$ & $P_{c5}$ & $P_{c6}$ \\
     $\Gamma_{2}(\Sigma_c \bar{D}^{\ast})$ & -  & - & - & - & \makecell{0.50 \\(2.52 $\%$)} & \makecell{1.38 \\(8.87 $\%$)} \\
     $\Gamma_{3}(\Sigma^{*}_c \bar{D})$ &  - & - & - & \makecell{1.14\\(73.23 $\%$)} & - & \makecell{2.62 \\(16.87 $\%$)} \\
     $\Gamma_{4}(\Sigma_c \bar{D})$ &  - & - & \makecell{2.87 \\(16.59 $\%$)} & - & \makecell{1.04 \\(5.29 $\%$)} & - \\
     $\Gamma_{5}(\Lambda_c \bar{D}^{\ast})$ &  \makecell{0.83\\ (11.71 $\%$)} & \makecell{0.19\\ (3.48 $\%$)} & \makecell{1.44 \\(8.33 $\%$)} & \makecell{0.12 \\(7.82 $\%$)} & \makecell{0.65\\ (3.32 $\%$)} & \makecell{3.24\\ (20.81 $\%$)} \\
     $\Gamma_{6}(\Lambda_c \bar{D})$ &  \makecell{0.01\\ (0.14 $\%$)} & - & \makecell{1.60\\ (9.22 $\%$)} & - & \makecell{2.98\\ (15.14 $\%$)} & - \\
     $\Gamma_{7}(J/\psi N)$ &  \makecell{1.43\\ (20.22 $\%$)} & \makecell{5.16\\ (96.52 $\%$)} & \makecell{9.29\\ (53.64 $\%$)} & \makecell{0.29\\ (18.95 $\%$)} & \makecell{5.46\\ (27.73 $\%$)} & \makecell{8.31\\ (53.45 $\%$)} \\
     $\Gamma_{8}(\eta_{c} N)$ &  \makecell{4.81\\ (67.93 $\%$)} & - & \makecell{2.12 \\(12.23 $\%$)} & - & \makecell{9.06 \\(45.98 $\%$)} & - \\\hline
    Scenario   &   \multicolumn{6}{c}{B}   \\ \hline
     Molecule  &  $P_{c1}$ & $P_{c2}$ & $P_{c3}$ & $P_{c4}$ & $P_{c5}$ & $P_{c6}$ \\
     $\Gamma_{2}(\Sigma_c \bar{D}^{\ast})$ &  - & - & - & - & \makecell{0.36\\ (2.17 $\%$)} & \makecell{0.64\\ (8.30 $\%$)} \\
     $\Gamma_{3}(\Sigma^{*}_c \bar{D})$ &  - & - & - & \makecell{0.31 \\(13.63 $\%$)} & - & \makecell{1.41 \\(18.19 $\%$)} \\
     $\Gamma_{4}(\Sigma_c \bar{D})$ &  - & - & \makecell{1.23 \\(12.88 $\%$)} & - & \makecell{0.98\\ (5.92 $\%$)} & - \\
     $\Gamma_{5}(\Lambda_c \bar{D}^{\ast})$ &  \makecell{2.27 \\(26.71 $\%$)} & \makecell{0.26\\ (2.10 $\%$)} & \makecell{2.97 \\(30.92 $\%$)} & \makecell{1.17\\ (52.05 $\%$)} & \makecell{3.05 (18.48 $\%$)} & \makecell{2.82 \\(36.37 $\%$)} \\
     $\Gamma_{6}(\Lambda_c \bar{D})$ &  \makecell{0.02\\ (0.23 $\%$)} & - & \makecell{2.40 \\(25.03 $\%$)} & - & \makecell{5.65 \\(34.18 $\%$)} & - \\
     $\Gamma_{7}(J/\psi N)$ &  \makecell{1.57 \\(18.45 $\%$)} & \makecell{12.28 \\(97.90 $\%$)} & \makecell{2.13 \\(22.26 $\%$)} & \makecell{0.77 \\(34.33 $\%$)} & \makecell{2.92\\ (17.67 $\%$)} & \makecell{2.88\\ (37.14 $\%$)} \\
     $\Gamma_{8}(\eta_{c} N)$ & \makecell{ 4.65 \\(54.61 $\%$)} & - & \makecell{0.85 \\(8.86 $\%$)} & - & \makecell{3.57 \\(21.58 $\%$)} & - \\
  \hline  \hline
\end{tabular}}
\end{table}

 \begin{figure}[ttt]
\begin{center}
\includegraphics[width=5.0in]{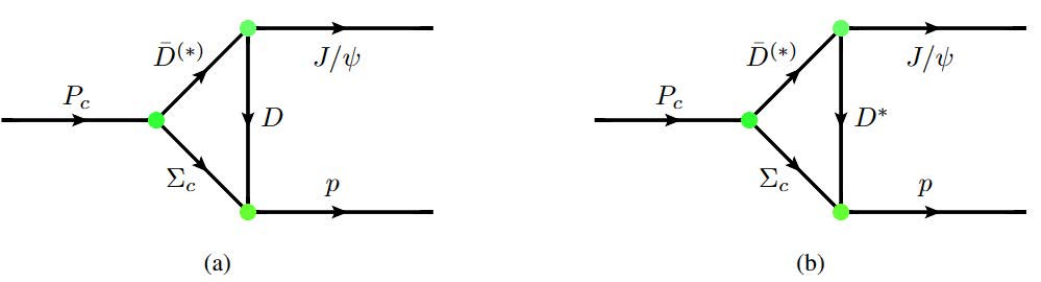}
\caption{  Triangle diagrams for the decays of  $P_c \to J/\psi p$.  }
\label{tripc}
\end{center}
\end{figure}

The triangle mechanism has been applied to study the decays of pentaquark molecules. In Ref.~\cite{Xiao:2019mvs}, assuming the three pentaquark states as $\bar{D}^{(*)}\Sigma_c$ molecules, we constructed the triangle diagrams connecting the $\bar{D}^{(*)}\Sigma_c$ channels  to the  $J/\psi p$ channels by exchanging the $D^{(*)}$ mesons as shown in Fig.~\ref{tripc}.       The partial widths of the decays $P_c \to J/\psi p$ and their ratios dependence on the cutoff  $\Lambda$  are shown, where the cutoff is from the form factor $(\Lambda^2-m^2)/(\Lambda^2-q^2)$.      Similarly, Lin et al. systematically investigated the decays of the $\bar{D}^{(\ast)}\Sigma_{c}^{(\ast)}$ molecules~\cite{Lin:2019qiv}, which considered more partial decay modes of the $\bar{D}^{(\ast)}\Sigma_{c}^{(\ast)}$  molecules. The latter work pointed out that the decays into the $\bar{D}^*\Lambda_{c}$ channel are the most important.   We note that the meson exchange theory has been tested for light meson exchanges but remains to be verified for heavy meson exchanges, especially when both heavy and light mesons can be exchanged.  
 The meson exchange theory dictates that charmed mesons are responsible for short-range interactions of 0.1 fm. In Ref.~\cite{Yamaguchi:2019djj},  the strength of the short-range potential provided by the quark potential model is much stronger than that provided by the heavy meson exchanges.  The origin of such short-range interactions remains unclear.

 \begin{figure}[!h]
\begin{center}
\includegraphics[width=5.1in]{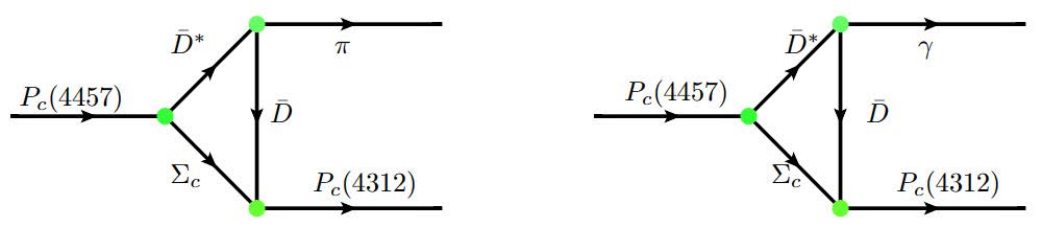}
\caption{ Triangle diagram of pionic and radiative  decays of $P_{c}(4457)$ to $P_{c}(4312)$ with the spin of $P_{c}(4457)$ being either 1/2 or 3/2.  }
\label{decayraditivepc}
\end{center}
\end{figure}

Pionic or radiative decays can relate members of the multiplet of hadronic molecules.   
We further investigated the radiative decays of the pentaquark molecules with the triangle diagrams. According to the LHCb measurements~\cite{Aaij:2019vzc}, the mass splitting between $P_{c}(4440)$ and $P_{c}(4312)$ is 128 MeV, which is less than the pion mass. Therefore, the decay of $P_{c}(4440) \to P_{c}(4312) \pi$ is forbidden  due to phase space.  The  mass splitting between $P_{c}(4457)$ and $P_{c}(4312)$ is 145 MeV, accordingly the   $P_{c}(4457)\to P_{c}(4312)\pi$ decay is allowed. Moreover, the radiative decays of  $P_{c}(4457)\to P_{c}(4312)\gamma$ and $P_{c}(4440)\to P_{c}(4312)\gamma$ are both allowed as shown in Fig.~\ref{decayraditivepc}. The widths of the pionic decays of $P_{c}(4457)$ to $P_c(4312)$ are estimated to be the order of hundreds of keV, while those of radiative decays are about $1$ keV. The ratio of the former to the latter agrees with the ratio of the pionic decay of $D^*\to D\pi$ to that of the radiative decay,  indicating that such a ratio can help verify the molecular nature of pentaquark states.   In the triangle diagram mechanism,  all the couplings are well constrained by experimental data, which largely reduces the uncertainties of the results. Therefore, such results are reliable. In addition to the above decays, the $\bar{D}^*\Sigma_c$ molecules can decay into $\bar{D}^*\Sigma_c$ and  $\bar{D}\Sigma_c$ molecules together with a $\pi$ and a photon~\cite{Guo:2023fih}. For the radiative  decays of $P_c(4457) \to P_c(4312)$, our results are consistent with the quark model results~\cite{Li:2021ryu}.

 \begin{figure}[!h]
\begin{center}
\includegraphics[width=5.1in]{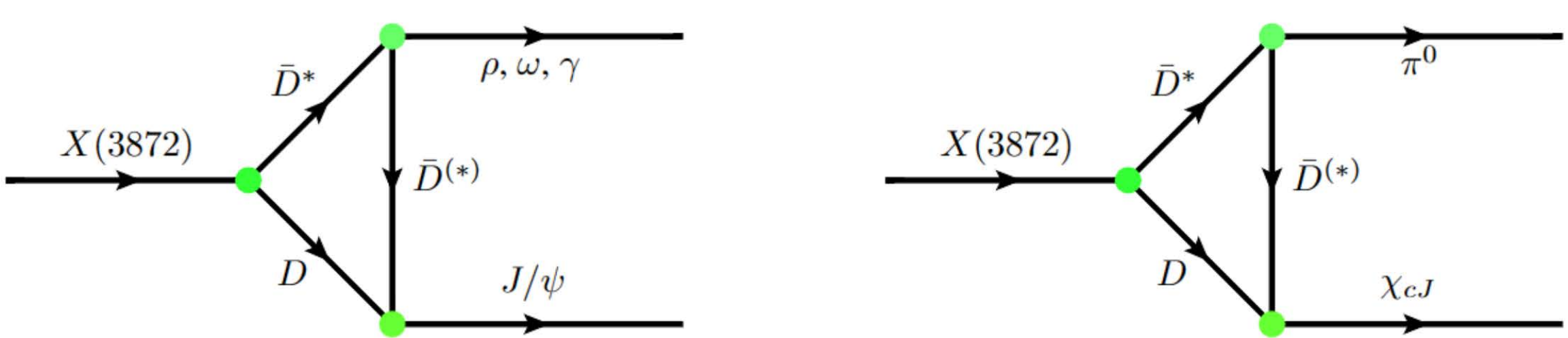}
\caption{ Triangle diagram of pionic and radiative  decays of $X(3872)$ to charmonium states.    }
\label{decayx3872}
\end{center}
\end{figure}

 With the triangle mechanism,  the decays of $X(3872)$ as a $\bar{D}^*D$ molecule have been extensively studied. Dong et al. investigated the radiative decays of $X(3872)$ to  $J/\psi$ and $\psi(2S)$ ~\cite{Dong:2009uf}  and pionic decays of $X(3872)$ to $\chi_{cJ}$($J=0,1,2$)~\cite{Dong:2009yp} as shown in Fig.~\ref{decayx3872}, where based on the triangle diagrams some more complex processes are explored.    Swanson  firstly claimed that the radiative decays of $X(3872)$ into $J/\psi$ and $\psi(2S)$ are quite sensitive to  $X(3872)$ as a pure $c\bar{c}$ state or a $\bar{D}^*D$ molecule~\cite{Swanson:2004pp}, which motivated a lot of theoretical studies~\cite{Dong:2008gb,Dong:2009uf,Wang:2010ej,Cardoso:2014xda,Badalian:2015dha,Cincioglu:2019gzd,Yu:2023nxk,Grinstein:2024rcu,Chen:2024xlw} and experimental studies~\cite{BaBar:2008flx,Belle:2011wdj,LHCb:2014jvf,LHCb:2024tpv}. The recent measurement of the LHCb Collaboration~\cite{LHCb:2024tpv} may disfavor the molecular interpretation for $X(3872)$. However,  Guo et al. argued that the size of such ratios is weakly sensitive to the long-range structure of $X(3872)$~\cite{Guo:2014taa}. The LHCb results, therefore, can not exclude the molecular interpretation for $X(3872)$.      Ref.~\cite{Wu:2021udi} constructed the triangle mechanism to study the decays of $X(3872)\to J/\psi \pi\pi$ and $X(3872)\to J/\psi \pi\pi\pi$ and  $\pi^0\chi_{cJ}$, suggesting that the isospin breaking effect can be attributed to the isospin breaking and the loop functions of the $\bar{D}^*D$ neutral and charged components, where the secondary decays of $\rho \to \pi \pi $ and $\omega \to \pi \pi \pi $ are considered in the triangle diagrams. Very recently, Wang et al. estimated the widths of  $X(3872)$ decaying into a pair of light mesons, showing that these branching ratios are around several percent of its total width~\cite{Wang:2022qxe}.  For the $X_2(4013)$  as the HQSS partner of $X(3872)$, the strong~\cite{Albaladejo:2015dsa} and radiative~\cite{Shi:2023mer}  decays have been investigated via the triangle diagrams, where the former decays into $\bar{D}D$ and $\bar{D}^*D$, and the latter radiatively decays into charmonium states.  In Ref.~\cite{Baru:2016iwj}, the $X_2$ width is estimated to be dozens of MeV once taking into account the nonperturbative pions~\cite{Baru:2016iwj}, while only several MeV for perturbative pions~\cite{Albaladejo:2015dsa}.    Because the inelastic potentials $\bar{D}^{(*)}D^{(*)} \to J/\psi \rho(\omega)$ in the heavy meson exchange would bring  large uncertainties to the couplings in the vertices,   it is difficult to construct a coupled-channel framework to study the two-body decays of $X(3872)$ and $X_2(4013)$.\footnote{ Very recently, Zhang et al. adopted the branching fractions of the decays $X(3872)\to J/\psi \rho$  and $X(3872)\to J/\psi \omega$  to determine the potentials $\bar{D}^{(*)}D^{(*)} \to J/\psi \rho(\omega)$  in the EFT approach, which largely reduces the uncertainty in the inelastic potentials~\cite{Zhang:2024fxy}.  }  
 In principle, the strong decays of  $J^{PC}=0^{++}$ $\bar{D}D$,  $J^{PC}=1^{+-}$ $\bar{D}^*D^*$,  and $J^{PC}=0^{++}$ $\bar{D}^*D^*$  bound states as the HQSS partners of the $J^{PC}=1^{++}$ $\bar{D}^*D$ state can proceed via the triangle diagrams as well, which have been investigated with the vector-vector interaction in the local hidden gauge framework~\cite{Molina:2009ct}. The radiative and pionic decays between the multiplet members of $\bar{D}^{(*)}D^{(*)}$ hadronic  molecules are naturally expected~\cite{Branz:2010rj}.

 \begin{figure}[ttt]
\begin{center}
\includegraphics[width=5.1in]{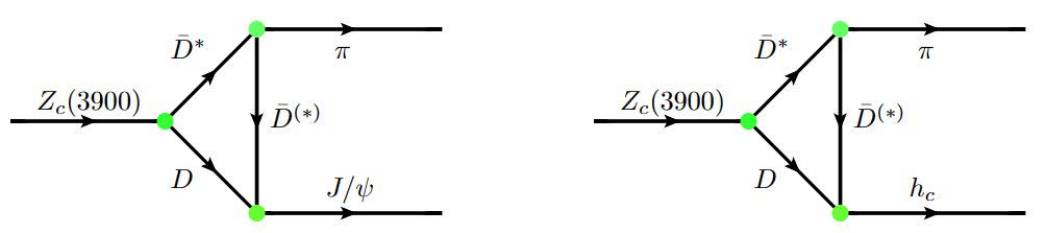}
\caption{ Triangle diagram for the pionic  decay of $Z_c(3900)$ to the charmonium states.    }
\label{decayx3900}
\end{center}
\end{figure}

 \begin{figure}[!h]
\begin{center}
\includegraphics[width=5.1in]{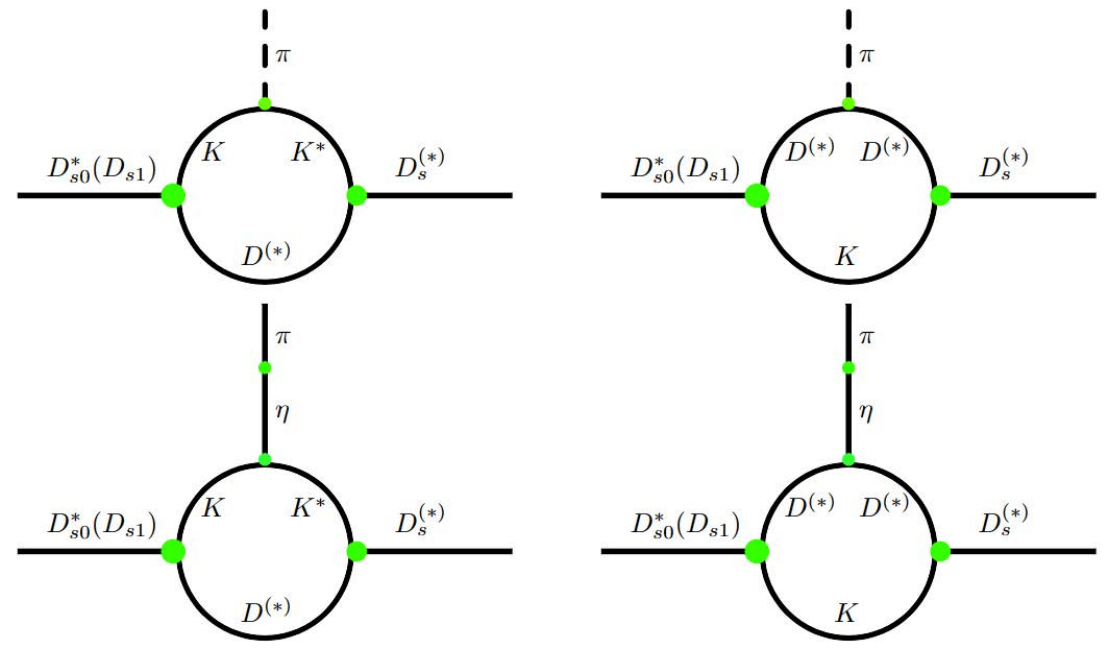}
\caption{ Decays of $D_{s0}^* \to D_s \pi$ and  $D_{s1} \to D_s^* \pi$~\cite{Faessler:2007gv,Faessler:2007us}.   }
\label{decayx2460}
\end{center}
\end{figure}

Next, we discuss the decays of the isovector doublet of $\bar{D}^*D$ and $\bar{D}^*D^*$ molecules. 
Employing the triangle diagram shown in Fig.~\ref{decayx3900},  Refs.~\cite{Li:2013xia,Xiao:2018kfx} estimated the decay width of $Z_c(3900)$  into a charmonium state plus a $\pi$, where the $Z_c(3900)$ couplings to the $\bar{D}^*D$ is estimated by saturating the width of $Z_c(3900)$.  It should be noted that identifying $Z_c(3900)$ as an  $\bar{D}^*D$ resonant state, the $\bar{D}^*D$ channel  contributes to the real and imaginary parts of the $Z_c(3900)$ pole, i.e., mass and width. Therefore, only saturating the width of $Z_c(3900)$ to determine   the $Z_c(3900)$ coupling to $\bar{D}^*D$ is not complete.     Similar decay modes of $Z_c(4020)$,  regarded as a $\bar{D}^*D^*$ molecule and the HQSS partner of $Z_c(3900)$,  were investigated through the triangle diagram~\cite{Li:2013xia,Qi:2023kwc}. In the molecular picture, the radiative  decays of  $Z_c(4020)$ and $Z_c(3900)$  were investigated using the triangle diagram mechanism~\cite{Chen:2015igx,Wang:2022aiu}.    Using the same approach, the decays of the  SU(3)-flavor partner of $Z_c(3900)$ have been studied~\cite{Wu:2021ezz}.   In Ref.~\cite{Xue:2017xpu}, Xue et al. took the triangle diagram mechanism to calculate the widtd of $Y(4260)$ as a $\bar{D}_1D$ bound state  decaying into $\bar{D}^*D^*$, yielding a width of around $11$ MeV.

The dominant decays of $D_{s0}^* \to D_s \pi$ and  $D_{s1} \to D_s^* \pi$ break the isospin symmetry. In Refs.~\cite{Faessler:2007gv,Faessler:2007us}, these decays proceed via the diagrams shown in Fig.~\ref{decayx2460}, named the ``Direct" diagrams and the   ``$\pi^0 -\eta$ mixing  " diagrams,  and their results indicate that the former decay modes play the dominant role. The widths of the  partial decays $D_{s0}^* \to D_s \pi$ and  $D_{s1} \to D_s^* \pi$    in this picture were estimated to be $80$~keV and $50\sim 80$~keV, respectively, consistent with the results of the EFT approach~\cite{Cleven:2014oka}, The EFT approach  including  the isospin breaking contributions up to next-to-leading order,  their  widths were shifted to be $132$ and $111$~keV~\cite{Fu:2021wde}.  In Ref.~\cite{Cleven:2014oka},  Lutz et al. estimated  the widths of the partial decays $D_{s0}^* \to D_s \pi$ and  $D_{s1} \to D_s^* \pi$ to be $76$~keV and $55$~keV, respectively, and both shifted to  $140$~keV once the chiral correction is considered.    In Ref.~\cite{Xiao:2016hoa}, the radiative and pionic decays of the  $D_{s1}$ to $D_{s0}^*$ were investigated via the triangle diagrams.

\subsection{Production mechanisms}

Studies of the production mechanisms of exotic states can help reveal their internal structure and provide useful guidance to experimental searches for their partners. Up to now, exotic states have often been discovered in inclusive and exclusive processes. The well-known exotic states, such as $D_{s0}^*(2317)$ and $T_{cc}$, are discovered in the inclusive process. The $X(3872)$ and $P_c$ states are discovered in the exclusive process.  The intermediate processes of inclusive productions of hadronic molecules are unclear, but they have been studied in statistical models or  Monte Carlo methods, such as PYTHIA, simulating these complex processes. Although the inclusive production of hadronic molecules suffers from relatively large uncertainties~\cite{Guo:2017jvc}, the experimental measurements as well as the developing theoretical methods can help reduce the  uncertainties~\cite{ExHIC:2010gcb,Guo:2014ppa,Esposito:2015fsa,ExHIC:2017smd,Braun-Munzinger:2018hat,Wu:2020zbx,Esposito:2020ywk,Zhang:2020dwn,Chen:2021akx,Yun:2022evm,Wu:2022wgn}.  For the exclusive processes, the production of hadrons in heavy hadron decays generally proceeds via weak interactions, such as $b$-flavored hadrons via the decay $b \to c\bar{c} s$ at the quark level as shown in Table~\ref{resultsprediction45},   which can be well described by the effective Hamiltonian approach. In addition, $J/\psi$  decays via the strong interaction, which can be well described by the non-relativistic QCD(NRQCD) and potential NRQCD. The uncertainties of exclusive productions of heavy hadrons can be reduced by abundant experimental data in $B(D)$ and $J/\psi$ decays. In this review, we mainly focus on the exclusive processes.

\begin{table*}[htp]
\centering
\caption{ Productions of exotic states in $b$-flavored hadron decays~\cite{HFLAV:2022esi}.  \label{resultsprediction45}}
\begin{tabular}{c c c c  c c c c c c c  }
\hline  \hline
    Decay modes   & ~~~~   Branching fractions  
         \\ \hline
$B^{+}\to D_{s0}^*(2317)^+ \bar{D}^{0} $    &~~~~$0.83^{+0.25}_{-0.15}\times 10^{-3}$    
         \\
$B^{+}\to D_{s0}^*(2317)^+ \bar{D}^{*0} $    &~~~~$(0.90\pm 0.68)\times 10^{-3}$    
         \\
$B^{+}\to D_{s1}(2460)^+ \bar{D}^{0} $    &~~~~$(3.28\pm 0.78)\times 10^{-3}$    
         \\
$B^{+}\to D_{s1}(2460)^+ \bar{D}^{*0} $    &~~~~$(10.5\pm 2.4)\times 10^{-3}$    
         \\
$B^{+}\to X(3872) K^+$    &~~~~$2.06^{+0.61}_{-0.49}\times 10^{-4}$    
         \\
$B^{+}\to X(3872) K^{*+}$    &~~~~$1.2^{+2.9}_{-2.7}\times 10^{-4}$         \\
$B^{+}\to X(3872) K^{0}\pi^+$    &~~~~$3.0^{+1.6}_{-1.1}\times 10^{-4}$    
         \\
$B_s^{0}\to X(3872) \phi$    &~~~~$(1.1\pm 0.4)\times 10^{-4}$    
         \\
$B^{+}\to X(3915) K^+$    &~~~~$(0.4\pm 1.6)\times 10^{-4}$    
         \\
$B^{+}\to (Y(3940)\to J/\psi \omega) K^+$    &~~~~$3.0^{+0.86}_{-0.67}\times 10^{-5}$    
         \\
$B^{+}\to (Y(4260)\to J/\psi \pi^+ \pi^-) K^+$    &~~~~$(2.00\pm 0.73)\times 10^{-5}$    
         \\
$B^{+}\to (\chi_{c1}(4140)\to J/\psi \phi) K^+$    &~~~~$0.66^{+0.37}_{-0.27}\times 10^{-5}$   \\ 
$B^{+}\to (\chi_{c1}(4274)\to J/\psi \phi) K^+$    &~~~~$0.31^{+0.24}_{-0.16}\times 10^{-5}$     \\ 
$\Lambda_b^0  \to  P_{c}(4380)^+ \pi^-$    &~~~~$0.162^{+0.095}_{-0.070}\times 10^{-4}$  
         \\
$\Lambda_b^0  \to  P_{c}(4380)^+ K^-$    &~~~~$3.2^{+7.4}_{-1.8}\times 10^{-4}$  
         \\
$\Lambda_b^0  \to  P_{c}(4457)^+ \pi^-$    &~~~~$0.051^{+0.032}_{-0.025}\times 10^{-4}$     \\
$\Lambda_b^0  \to  P_{c}(4457)^+ K^-$    &~~~~$1.5^{+2.6}_{-0.9}\times 10^{-4}$     \\ 
$\Lambda_b^0  \to  Z_c(4200)^{-} p$    &~~~~$0.24^{+0.14}_{-0.15}\times 10^{-4}$     \\ 
\hline  \hline
\end{tabular}
\end{table*}

From the perspective of the interactions, the production process of an exotic state in $b$-flavored hadrons can be factorized into short-distance and long-distance interactions, corresponding to weak and strong interactions. The weak interaction is encoded in the non-leptonic decays that can be described by the naive factorization approach, which generally includes the $T$  and $C$ diagrams of the topological diagrammatic approach~\cite{Cheng:2010ry,Li:2012cfa,Qin:2013tje}. The strong interaction is encoded in the final-state interactions described by the meson exchange theory. With these interactions determined,  the triangle diagram mechanism can account for the production of exotic states in heavy hadron decays.  Such a mechanism has been widely applied to study the branching fractions and CP violations of bottom hadrons~\cite{Du:1998ss,Cheng:2004ru,Lu:2005mx} and charmed baryon decays~\cite{Li:1996cj,Dai:1999cs,Ablikim:2002ep,Li:2002pj,Han:2021azw,Hsiao:2019ait,Cao:2023csx}.    This mechanism predicted large branching fractions for the weak decays of doubly charmed baryons, considering the final-state interactions, which provide valuable guidance for the experimental discovery of the $\Xi_{cc}$~\cite{Yu:2017zst}.  Being applied to the production of multiplets of hadronic molecules,  it will verify the molecular nature of these states and guide experimental searches for its partner's production in a similar process.

\subsubsection{Production rates of $DK$ and $D^*K$ molecules}
At first, we start with the productions of $D_{s0}^*(2317)$ and $D_{s1}(2460)$ in $B$ decays. Within the molecular picture, the productions of $D_{s0}^*(2317)$ and $D_{s1}(2460)$ in $b$-flavored hadron decays have been   discussed~\cite{Faessler:2007cu,Liu:2022dmm,Liu:2023cwk}. Assuming the production mechanism of $D_{s0}^*(2317)$ and $D_{s1}(2460)$ in $B$ decays are similar to those of ground-state mesons $D_{s}^{(*)}$ in $B$ decays,  the decays of $B \to \bar{D}^{(*)}D_{s0}^*(D_{s1})$ and  $B \to \bar{D}^{(*)}D_{s}^{(*)}$ proceed via the $W-$boson external emission at the quark level, which can be described by the naive factorization approach~\cite{Bauer:1986bm,Beneke:2000ry,Beneke:2001ev,Beneke:2003zv,Ali:1998eb,Cheng:2003sm}.    The decay constants of   $D_{s0}^*(2317)$ and $D_{s1}(2460)$  are estimated to be  $f_{D_{s0}^*}=58.74$ MeV  and  $f_{D_{s1}}=133.76$ MeV in the molecular picture~\cite{Liu:2023cwk}\footnote{  The effective Lagrangian approach was adopted to calculate the decay constants $f_{D_{s0}^*}$   and  $f_{D_{s1}}$ with small uncertainties, where the relevant couplings are completely determined by the experimental data, and the renormalization energy scales in the two-loop diagrams are determined by fitting their masses. }, which  were firstly calculated in the molecular picture in Ref.~\cite{Faessler:2007cu}.   Lattice QCD calculated the decay constants $f_{D_{s0}^*}=114(2)(0)(+5)(10)$ MeV  and  $f_{D_{s1}}=194(3)(4)(+5)(10)$ MeV with almost purely  molecular components~\cite{Bali:2017pdv}, larger than those of the phenomenological model~\cite{Faessler:2007cu,Liu:2023cwk}.   In particular, once the decay constants of $D_{s0}^*(2317)$ and $D_{s1}(2460)$ are obtained~\cite{Liu:2023cwk}, their production rates in $B_s$, $\Lambda_b$, and $\Xi_b$ decays can be straightforwardly calculated using the naive factorization approach as shown in Table~\ref{resultsbtodds123}. One can see that the production rates of $D_{s0}^*(2317)$ and $D_{s1}(2460)$ in $b$-flavored baryon decays are larger than those in $b$-flavored meson decays, which imply that these exotic states are likely to be detected in heavy baryons in the future. The obtained branching fractions of the decays $B \to \bar{D}^{(*)}D_{s0}^*(D_{s1})$ are smaller than the experimental data, which indicate that $D_{s0}^*(2317)$ and $D_{s1}(2460)$ can contain components other than the $DK$ and $D^*K$ components, such as the bare $c\bar{s}$ components.

\begin{table}[ttt]
\centering
\caption{Branching fractions of  ($10^{-3}$) of $D_{s0}^*(2317)$ and $D_{s1}(2460)$ in $b$-flavored hadron decays. \label{resultsbtodds123}
}
\begin{tabular}{c c | c  c c c c c c c}
  \hline \hline
   Decay modes    & $\mathrm{Branching ~ fractions}$   & ~~~~ Decay modes   &   $\mathrm{Branching~ fractions}$  
         \\ \hline 
     $B^{+} \to \bar{D}^{0}D_{s0}^{*}(2317)$    & ~~~~ $0.48$    & ~~        $B_{s}^{0} \to D_{s} D_{s0}^{\ast}(2317)$    & ~~~~ $0.47$  
        \\
                $B^{+} \to \bar{D}^{\ast0}D_{s0}^*(2317) $ & ~~~~ $0.39$    & ~~      $B_s^0 \to D_s^*D_{s0}^*(2317) $ & ~~~~ $0.27$     
         \\
     $B^{+} \to \bar{D}^{0}D_{s1}(2460) $ & ~~~~ $1.39$ & ~~      $B_s^0 \to {D}_sD_{s1}(2460) $ & ~~~~ $1.18$  
         \\  
  $B^{+} \to \bar{D}^{\ast0}D_{s1}(2460) $ & ~~~~ $4.36$ & ~~   $B_s^0 \to D_s^{\ast}D_{s1}(2460) $ & ~~~~ $4.11$     
         \\   \hline
 ~~$\Lambda_{b} \to \Lambda_{c}D_{s0}^*(2317)$    & ~~~~ $0.70$      &    ~~$\Xi_{b} \to \Xi_{c}D_{s0}^*(2317)$    & ~~~~ $0.58$    \\
      $\Lambda_{b} \to \Lambda_{c}D_{s1}(2460)$   & ~~~~ $4.34$  &         ~~$\Xi_{b} \to \Xi_{c}D_{s1}(2460)$   & ~~~~ $4.29$ \\
  \hline \hline
\end{tabular}
\end{table}

  \begin{table}[!h]
 \centering
 \caption{Decay constants of $D_{s0}^*(2317)$ and $D_{s1}(2460)$ as the excited states (in units of MeV). \label{Tab:FormFactor1decay} }
 \begin{tabular}{cccccccc}
 \hline\hline
 Decay Constants ~~~  & $f_{D_{s0}^*}$&   $f_{D_{s1}}$    \\
 \hline
 QCD sum rule~\cite{Wang:2015mxa}~~~  &$333\pm 20$~~~  & $345\pm 17$~~~  \\
 Quark model~\cite{Veseli:1996yg}~~~  &$110$~~~    & $233$~~~  \\
Salpeter method~\cite{Wang:2007av}~~~   &$112$~~~   & $219$~~~ \\ 
covariant light-front quark model~\cite{Verma:2011yw}~~~   &$74.4^{+10.4}_{-10.6}$~~~   & $159^{+36}_{-32}$~~~  \\ 
Ours~~~   &$115.71$~~~   & $265.74$~~~  \\ 
\hline\hline
 \end{tabular}
 \end{table}

According to the recent study~\cite{Yang:2021tvc}, the $DK$ and $D^*K$ molecular components account for  $70\%$ and $50\%$ of the total wave functions of $D_{s0}^*(2317)$ and $D_{s1}(2460)$. As a result, we have embodied other components, such as the $c\bar{s}$ configuration, into the total wave functions. With the experimental branching fractions of the decays of $B^{+}\to \bar{D}^0 D_{s0}^{*}(2317)^+$ and  $B^{+}\to \bar{D}^0 D_{s1}^{*}(2460)^+$, we can obtain the decay constants $f_{D_{s0}^*}=75.83$~MeV and $f_{D_{s1}}=199.75$~MeV. With the values of $f_{D_{s0}^*}^{M}=58.74$~MeV and $f_{D_{s1}}^{M}=133.76$~MeV in the molecular picture and their molecular components in the total wave function, i.e., $70\%$ and  $50\%$,  one can obtain the decay constants $f_{D_{s0}^*}^{E}=115.71$~MeV and $f_{D_{s1}}^{E}=265.74$~MeV, which correspond to the  $D_{s0}^*$ and $D_{s1}$ as pure excited $c\bar{s}$ states.   It should be noted that the obtained decay constants depend on the ratio of the molecular component to the bare component in their wave functions.  The estimated values for the decay constants of  $D_{s0}^*(2317)$ and $D_{s1}(2460)$  are consistent with other approaches shown in Table~\ref{Tab:FormFactor1decay}, which indicates that the $D_{s0}^*(2317)$ and $D_{s1}(2460)$ are composed of a $D^{(*)}K$ component and a $c\bar{s}$ component from the perspective of the branching fractions of the weak decays.

 \begin{figure}[ttt]
\begin{center}
\includegraphics[width=6.1in]{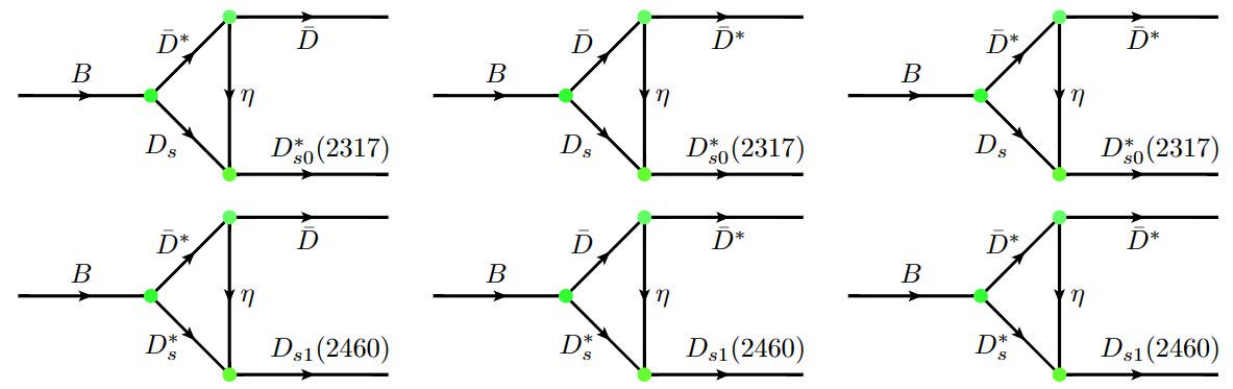}
\caption{ Triangle diagrams accounting for the decays of $B \rightarrow \bar{D}^{(*)} D_{s0}^*(2317)$ and $B \rightarrow \bar{D}^{(*)} D_{s1}(2460)$.    }
\label{Figds2317}
\end{center}
\end{figure}

In addition to the naive factorization approach, one can adopt the triangle diagram mechanism to investigate the above decays, where the $B$ meson firstly weakly decays into $D_{s}^{(*)}$ and $\bar{D}^{(*)}$, then the latter scatter into $\bar{D}^{(*)}$ and $\eta$, and finally the $D_{s}^{(*)}\eta$ interactions dynamically generate the $D_{s0}^*(2317)$ and $D_{s1}(2460)$ as shown in Fig.~\ref{Figds2317}. We note that other triangle diagrams with weak decays $B \to J/\psi(\eta_c) K$ were also considered in the study~\cite{Liu:2022dmm}, which actually leads to double counting since the decays  $B\to \bar{D}^{(*)}D_{s}^{(*)}$  contribute to the decays  $B \to J/\psi(\eta_c) K$.       
These Feynman diagrams can be calculated using the effective Lagrangian approach. In Ref.~\cite{Liu:2022dmm}, the effective Wilson coefficients are determined by reproducing the branching fractions of the weak decays, which are a bit smaller than the value of $a_{1}=1.07$ obtained at the charmed quark scale~\cite{Aaij:2011sn,Han:2021azw}.   In Table~\ref{results2317decay}, we present the branching fractions of the weak decays $B\to \bar{D}^{(\ast)}D_{s0}^*(2317)$ and $B\to \bar{D}^{(\ast)}D_{s1}(2460)$ obtained using the triangle diagram mechanism, in comparison with those obtained in the naive factorization approach~\cite{Liu:2023cwk}.   It should be noted that the couplings of the vertices in the triangle diagrams are determined by the experimental data or $SU(3)$ symmetry.   One can see that the results are of the same order of magnitude, but the difference can be as large as a factor of two. 

\begin{table}[!h]
\centering
\caption{Branching fractions ($10^{-3}$) of $B\to \bar{D}^{(\ast)}D_{s0}^*(2317)$ and $B\to \bar{D}^{(\ast)}D_{s1}(2460)$. \label{results2317decay}
}
\label{results1}
\begin{tabular}{c c c c c c c c c}
  \hline \hline
    Decay modes    &~~~~ Triangle Mechanism~\cite{Liu:2022dmm}  &~~~~ Naive Factorization~ \cite{Liu:2023cwk} 
         \\ \hline 
        $B^{+} \to \bar{D}^{0}D_{s0}^{*}(2317)^+ $   &~~~~ $0.55$   &~~~~ $0.48$  
                 \\     $B^{+} \to \bar{D}^{\ast0}D_{s0}^{*}(2317)^+ $  &~~~~ $0.59$    &~~~~$0.39$  
             \\\hline
                $B^{+} \to \bar{D}^{0}D_{s1}(2460)^+ $   &~~~~ $0.35$   &~~~~ $1.39$  
         \\                $B^{+} \to \bar{D}^{\ast0}D_{s1}(2460)^{+} $   &~~~~ $2.12$   &~~~~ $4.36$ 
         \\  
  \hline \hline
\end{tabular}
\end{table}

\subsubsection{Production rates of $\bar{D}^*D$ and $\bar{D}^*D^*$ molecules}

In this section, we focus on the production mechanism of the $\bar{D}^{*}D^{(\ast)}$ molecules.   
In Ref.~\cite{Guo:2013zbw}, Guo et al. proposed to produce $X(3872)$ as a  $\bar{D}D^{\ast}$ molecule via the radiative decay of the excited vector charmonium states. Later, using a similar approach, they investigated the production rates of $\bar{D}^{\ast}D^{\ast}$ molecules (the HQSS partners of  $\bar{D}^{\ast}D$ molecules)~\cite{Guo:2014ura}. In addition,  Dong et al. employed the effective Lagrangian approach to estimate the decay widths of $Y(4260) \to X(3872) \gamma$ in the molecular picture~\cite{Dong:2014zka}. Braaten et al. analysed the mass distribution of $X(3872)\gamma$ in the process of $e^{+}e^{-}\to X(3872)\gamma$  and found a narrow peak around 4.23 GeV due to the triangle singularity~\cite{Braaten:2019gfj,Braaten:2019gwc}, where $X(3872)$ is treated as a weakly bound state of $\bar{D}^{\ast0}D^{0}$  and the narrow peak structure around 4.23 GeV possibly corresponds to the $Y(4220)$ observed by the BESIII Collaboration~\cite{BESIII:2016bnd}. 
 The BESIII Collaboration  observed the radiative transition process $Y(4260)\to X(3872) \gamma$ in the process $e^{+}e^{-}\to X(3872)\gamma$~\cite{BESIII:2013fnz}.
 Such studies show that the isoscalar $\bar{D}D^{(\ast)}$  molecules can be produced in $e^{+}e^{-}$ collisions.

The $X(3872)$ can be produced in  $B$ decays.   In Refs.~\cite{Braaten:2004fk,Braaten:2004ai}, Braaten et al. proposed a mechanism to estimate the branching fractions of $B\to X(3872) K$, where the $B$ meson first weakly decays into $\bar{D}^* D K$,  and then  $X(3872)$ is dynamically generated via the $\bar{D}^*D$ rescattering.  Recently, they calculated the production rate of $X(3872)$ in the decay of $B\to X(3872)\pi K$ via the triangle diagram, where the $B$ meson first weakly decays into $\bar{D}^*D^*K$, then $\bar{D}^*$ or $D^*$ meson decays into $\bar{D}\pi$ or $D\pi$, and finally $X(3872)$ is  dynamically generated in the $\bar{D}^*D$ channel~\cite{Braaten:2019yua}.  Moreover, they obtained a narrow structure near the mass threshold of $\bar{D}^*D^*$ by analyzing the $X(3872)\pi$ mass distribution of the decay of $B\to X(3872)\pi K$ ~\cite{Braaten:2019yua},  which was assigned as a triangle singularity in other studies~\cite{Sakai:2020ucu,Yan:2022eiy}. Nevertheless, identifying $X(3872)$ as a pure $c\bar{c}$ charmonium,  Meng et al. calculated the decay of $B\to X(3872)K$ in the QCD factorization approach. They claimed that the branching ratio of $Br(B^+\to X(3872)K^{+})$ is equal to that of $Br(B^0\to X(3872)K^{0})$~\cite{Meng:2005er}. However, the Belle Collaboration estimated the  ratio to be~\cite{Belle:2011vlx}:
\begin{eqnarray}
\frac{Br(B^0\to X(3872)K^{0})}{Br(B^+\to X(3872)K^{+})}=0.50\pm0.14\pm0.04,
\end{eqnarray}
which shows large isospin breaking and can be understood in the molecular picture~\cite{Wang:2022xga,Liang:2023jxh}.

In Ref.~\cite{Chen:2013upa}, Chen et al. adopted the compositeness theorem proposed by Weinberg to study the near-mass-threshold lineshape of the $\bar{D}^0 D^{*0}$ mass distribution of $B \to \bar{D}^0 D^{*0} K  $ using the experimental data, where the short-range and long-range production mechanism of $X(3872)$  are accounted for by the short-distance vertex of the $B$ decay and the  $\bar{D}^0 D^{*0}$ rescattering into the $X(3872)$ via the weak decay  $B \to \bar{D}^0 D^{*0} K  $.  The non-vanishing value of  $Z=0.19$ indicated that $X(3872)$ contains a sizable molecular component. In Ref.~\cite{Wang:2015rcz},  Wang et al. investigated the short-distance  $c\bar{c}$ component of  $X(3872)$  in the semileptonic and nonleptonic  $B_c$ decays.

The $X(3872)$ production in heavy ion collisions has also been studied.    In Ref.~\cite{Bignamini:2009sk}, Bignamini et al. estimated the cross-sections of $X(3872)$ as a $\bar{D}^{*0}D^0$ molecule in $p\bar{p}$ collisions at Tevatron, which did not support the molecular interpretation for $X(3872)$. The inconsistency between experiment and theory in Ref.~\cite{Bignamini:2009sk}  disappears after considering the final-state interaction ~\cite{Artoisenet:2009wk,Albaladejo:2017blx}. In Refs.~\cite{Artoisenet:2009wk,Albaladejo:2017blx}, the estimated production of $X(3872)$ at the Large Hadron Collider   is also consistent with  the molecular interpretation for $X(3872)$.  In Refs.~\cite{ExHIC:2010gcb,ExHIC:2011say}, the authors suggested measuring the productions of exotic states such as $X(3872)$ in relativistic heavy ion collisions since the yields of exotic states estimated in the coalescence model are tied to their internal structure. 
In Ref.~\cite{Zhang:2020dwn},  Zhang et al.  assuming $X(3872)$ as either a molecule or a tetraquark state, computed the yield of $X(3872)$ in Pb-Pb collisions and claimed that the obtained physical observables with two order-of-magnitude difference in two different pictures can help reveal the internal structure of $X(3872)$, which are pretty different from the results of  Ref.~\cite{Chen:2021akx}.    The measurement of   the dependence on the productions of $X(3872)$ and $\psi(2S)$  on hadron multiplicities in $pp$ collisions by the  LHCb Collaboration~\cite{LHCb:2020sey} preferred to assign the $X(3872)$ as a compact tetraquark state and challenged the molecular interpretation~\cite{Esposito:2020ywk}. In Ref.~\cite{Braaten:2020iqw}, Braaten et al. argued that different from the proposal for the sensitive dependence of the $X(3872)$ breakup reaction rate on the  geometric cross-section,   the dependence on the productions of $X(3872)$ and $\psi(2S)$  on hadron multiplicities in $pp$ collisions can be described assuming $X(3872)$ as a hadronic molecule. Later, the productions  of $X(3872)$ and $\psi(2S)$ are measured in   $p P_b$~\cite{LHCb:2024bpb} and $P_b P_b$ collisions~\cite{CMS:2021znk}. The increasing ratio of the cross-sections $\sigma^{X(3872)}/\sigma^{\psi(2S)}$  from $pp$ to $p P_b$ to $P_b P_b$ collisions indicates that the $X(3872)$ exhibits different dynamics in the nuclear medium than the conventional charmonium $\psi(2S)$~\cite{Guo:2023dwf}.    In addition,  Kamiya et al. studied the correlation functions of $\bar{D}^*D$ in heavy ion collisions to investigate the nature of $X(3872)$~\cite{Kamiya:2022thy}.

 \begin{figure}[ttt]
\centering
\includegraphics[width=1.0\columnwidth]{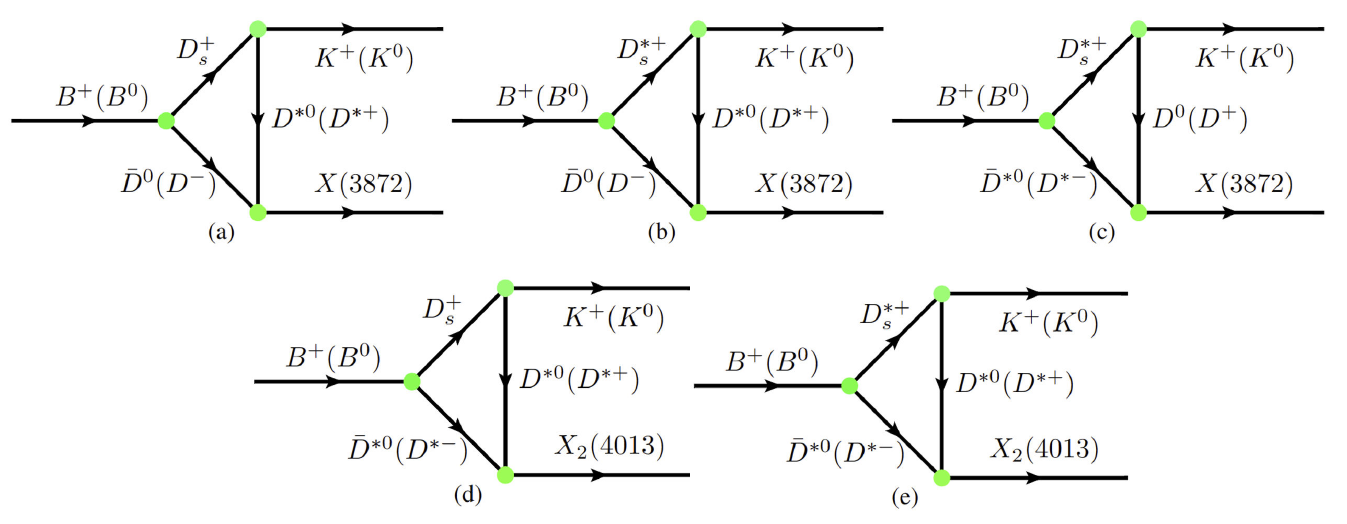}
\caption{\label{3872trinew}
Triangle diagrams accounting for (a-c) $B^{+}({B}^0) \to D_s^{(*)+}\bar{D}^{(*)0}(D^{(\ast)+})\to X(3872)K^{+}(K^0)$ and (d-e) $B^{+}({B}^{0}) \to D_s^{(*)+}\bar{D}^{(*)0}(D^{(\ast)+})\to X_2(4013)K^+(K^0)$.}
\end{figure}

In comparison with the $X(3872)$, 
the $Z_{c}(3900)$ does not have many production modes. The D0 Collaboration observed the signal of $Z_{c}(3900)$ in the $J/\psi\pi^{\pm}$ mass distribution 
in the semi-inclusive decays of $b$-flavored hadrons~\cite{D0:2018wyb} except that it is produced in electron-positron collisions.  Following the discovery in the semi-inclusive decays of $b$-flavored hadrons,  Wu et al. investigated the production of  $Z_{c}(3900)$ in the $B_c$ decay, which can be accessed at large colliders~\cite{Wu:2019vbk}. Based on the discovery of $Z_{c}(3900)$ via the $Y(4260)$, 
 many works describe the lineshape of $J/\psi \pi$ and $\pi\pi$ mass distributions and reproduced the $Z_{c}(3900)$  peak with the assumption that $Y(4260)$ is strongly coupled to a pair of charmed mesons~\cite{Chen:2013coa,Liu:2013vfa}, where $Z_{c}(3900)$ is regarded as a  kinetic effect rather than a genuine state.   Wang et al. found that the existence of a genuine $Z_c$ state is necessary to explain the BESIII experimental data because the effect of the triangle singularity seems prominent~\cite{Wang:2013cya,Cleven:2013mka}, which was demonstrated in Ref.~\cite{Guo:2014iya}.     
However, the production of $Z_{c}(3900)$ in $B$ decays remains unobserved. The Belle Collaboration observed no signal of $Z_{c}(3900)$ in the decay $B \to J/\psi  \pi K$~\cite{Belle:2014nuw} or in the decay  $B\to \bar{D}^*DK$~\cite{Belle:2008fma}.

Recently, we systematically studied the production rates of $\bar{D}^*D$ and $\bar{D}^*D^*$ molecules in $B$ decays in a unified framework. The production mechanism of $\bar{D}^*D$ and $\bar{D}^*D^*$ molecules via the triangle diagrams are shown in Fig.~\ref{3872trinew}. First,  the $B$ meson weakly decays into a pair of charmed mesons $D_{s}^{(\ast)}$ plus $\bar{D}^{(\ast)}$, which proceed via the $W-$emission mechanism at the quark level.  The strength of the $W-$ emission mechanism  is usually the largest~\cite{Chau:1982da,Chau:1987tk,Molina:2019udw}.  Then, the charmed-strange mesons $D_{s}^{(\ast)}$ scatter into the charmed mesons $D^{(\ast)}$ and a kaon.  Finally, the $\bar{D}D^*$ and $\bar{D}^*D^*$ molecules are dynamically generated via the final-state interactions. The isoscalar and isovector $\bar{D}D^*$ molecules refer to $X(3872)$ and $Z_c(3900)$, and isoscalar and isovector $\bar{D}^*D^*$ molecules to $X_2(4013)$ and $Z_c(4020)$.  The effective Lagrangian approach is adopted to calculate the production rates as shown in Table~\ref{resultsdwert}.

\begin{table}[ttt]
\centering
\caption{Branching fractions ($10^{-4}$) of $B^{+(0)}\to X(3872)/X_2(4013) K^{+(0)}$ and $B^{+(0)}\to Z_{c}(3900)/Z_{c}(4020) K^{+(0)}$ and ratios $\mathcal{B}(B^0\to)$/$\mathcal{B}(B^+\to)$. \label{resultsdwert}
}
\begin{tabular}{c| c c| c c }
\hline   \hline
    Decay modes    &~~~~ Our predictions  &~~~~ Exp.~\cite{ParticleDataGroup:2022pth}   &~~~ Ratio ~~~~   &    ~~~~ Exp. data~\cite{ParticleDataGroup:2022pth}
         \\ \hline 
        $B^{+} \to X(3872)K^+ $   &~~~~ $1.49\pm 0.62$   &~~~~ $2.1\pm 0.7$   & ~~~\multirow{2}{2.2cm}{ $0.62\pm0.13$}   &~~~~~\multirow{2}{1.8cm}{ $0.52\pm 0.26$}
                 \\   $B^{0} \to X(3872)K^0 $   &~~~~ $0.93\pm0.39$  &~~~~ $1.1 \pm 0.4$  
            & &
         \\ \hline  $B^{+} \to X_2(4013)K^+ $  &~~~~ $0.23\pm0.08$    &~~~~$-$   & ~~~\multirow{2}{2.2cm}{ $0.75\pm0.16$}  &~~~~~\multirow{2}{1.0cm}{ $-$}
         \\     $B^{0} \to X_2(4013)K^0 $  &~~~~ $0.17\pm0.06$   &~~~~ $-$  
         &   &
             \\\hline
        $B^{+} \to Z_{c}(3900)K^+ $   &~~~~ $0.21\pm 0.11$   &~~~~ $<4.7\times 10^{-5}$   & ~~~\multirow{2}{2.2cm}{ $0.63 \pm0.29$} &~~~~~\multirow{2}{1.0cm}{ $-$}
                 \\   $B^{0} \to Z_{c}(3900)K^0 $   &~~~~ $0.13\pm0.07$  &~~~~ $-$  & & 
      \\ \hline
   $B^{+} \to Z_{c}(4020)K^+ $  &~~~~  $0.0095\pm 0.0033$  &~~~~$<1.6\times 10^{-5} $   & ~~~\multirow{2}{2.2cm}{ $1.05\pm0.14$} &~~~~~\multirow{2}{1.0cm}{ $-$}
         \\     $B^{0} \to Z_{c}(4020) K^0 $  &~~~~ $0.0100\pm 0.0034$    &~~~~ $-$   & &   
\\ \hline \hline
\end{tabular}
\end{table}

Our results show that the branching ratios of the decays  $B^+ \to X(3872)K^+$ and  $B^0 \to X(3872)K^0$ are  $(0.63\sim 2.39)\times 10^{-4}$ and $(0.42 \sim 1.56)\times 10^{-4}$, consistent with the experimental data~\cite{ParticleDataGroup:2022pth}.  The branching ratios of  $B^+ \to Z_c(3900)K^+$ and  $B^0 \to Z_c(3900)K^0$ are of the order of $ 10^{-5}$,  smaller than the experimental upper limits~\cite{Belle:2015yoa}.   Moreover, we predicted the branching ratios of the decays $B\to X_2(4013) K$ and $B\to Z_{c}(4020) K$ 
to be the order of $10^{-5}$ and $10^{-6}$. We obtained a hierarchy for the production rates of all  $\bar{D}^*D$ and $\bar{D}^*D^*$ molecules in $B$ decays, which can help clarify the nature of $XZ$ states in these energy regions.

\begin{table}[!h]
    \centering
    \caption{Ratios of the couplings in particle basis to the couplings in isospin basis.}
    \begin{tabular}{c|cccc}
    \hline\hline
     Molecules    &~~~~${D}^{\ast+}D^-$  &~~~~${D}^{+}D^{\ast-}$ &~~~~${D}^{\ast0}\bar{D}^{0}$   &~~~~${D}^{0} \bar{D}^{\ast0}$ 
         \\ \hline
        $X(3872) $   &~~~~ $1/2$   &~~$-1/2$ &~~~~ $1/2$  &~~$-1/2$ \\
        $Z_c(3900)$   &~~~~ $1/2$    &~~~~ $1/2$  &~~$-1/2$  &~~$-1/2$
         \\ \hline  Molecules     &~~~~${D}^{\ast+}D^{\ast-}$ &~~~~${D}^{\ast0}\bar{D}^{\ast0}$ & &
         \\ \hline
        $X_2(4013) $   &~~~~ $1/\sqrt{2}$  &~~~~ $1/\sqrt{2}$  & &  \\
        $Z_c(4020)$  &~~~~ $1/\sqrt{2}$  &~~ $-1/\sqrt{2}$   &  & 
\\
    \hline\hline
    \end{tabular}
   \label{couplingsr3872}
\end{table}

\begin{figure}[!h]
\centering
\includegraphics[width=0.9\columnwidth]{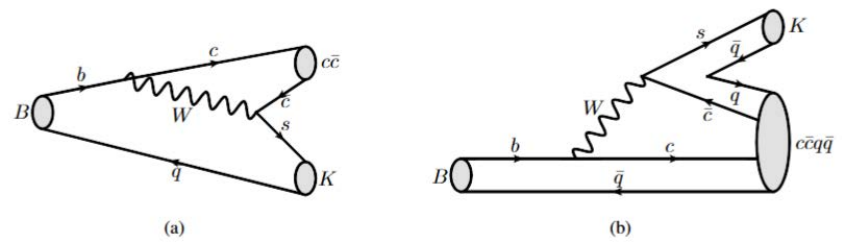}
\caption{\label{38723900trinew}  Short-distance component of $X(3872)$ allows its production directly via a $ c\bar{c}$ configuration in (a), while  $X(3872)$ and $Z_c(3900)$ can both be produced
via the long-distance $\bar{D}^*D$ component. 
 }
\end{figure}

We emphasize that the ratios of branching fractions are more certain than the absolute branching fractions in our model. We obtained  the ratio   $\mathcal{B}[B^+ \to X(3872) K^+]/\mathcal{B}[B^0 \to X(3872) K^0]=0.66 $, consistent with the experimental data~\cite{ParticleDataGroup:2022pth}. Our results show that the large isospin-breaking effect is attributed to the isospin breaking of the $\bar{D}^*D$ neutral and charged components. 
 Considering HQSS, we predict the ratio $\mathcal{B}[B^+ \to X_2(4013) K^+]/\mathcal{B}[B^0 \to X_2(4013) K^0]=0.35 $, which shows larger isospin breaking effects.  In addition,  our results show that the branching fractions of $\mathcal{B}[B\to Z_{c}(3900)K] $ are smaller than those of $\mathcal{B}[B\to X(3872)K]$  by one order of magnitude, which provides a natural explanation why $Z_c(3900)$ has not been observed in $B$ decays.  We note that only the amplitude of  Fig.~\ref{3872trinew}~(a)  and that  of Fig.~\ref{3872trinew}~(c) 
contribute to the decays  of the $B$ meson into  the $\bar{D}^*D$ molecules, while the contribution of Fig.~\ref{3872trinew}~(b) is accidentally small. The signs of the amplitude of  Fig.~\ref{3872trinew}~(a)  and that of Fig.~\ref{3872trinew}~(c)  depend on the relative sign between the charged and neutral components in the wave functions of the  $\bar{D}^*D$ molecules.  From Table~\ref{couplingsr3872}, one can see that the sign is opposite for the isoscalar molecules but the same for the isovector molecules. As the two amplitudes for the isoscalar molecules add constructively, but those for the isovector molecules add destructively, the $Z_{c}(3900)$ production rates in $B$  decays are lower than those of $X(3872)$ in $B$  decays. We note that Ref.~\cite{Zhao:2020qpd} proposed a different mechanism to explain why the $Z_c(3900)$ production rates in $B$ decays are lower than those of $X(3872)$.  As shown in Fig.~\ref{38723900trinew},  the $Z_{c}(3900)$ containing the $\bar{D}^*D$ component can not be produced directly in the short-distance process but can be produced in the long-distance process, while $X(3872)$ containing the $c\bar{c}$ and $\bar{D}^*D$ components can be produced via both the short-distance and long-distance processes. Therefore,  the $X(3872)$ can be observed, but $Z_c(3900)$ can not in high-energy productions.

We only considered the contribution of the $\bar{D}^*D^{(\ast)}$ channel to the $XZ$ states. However, other channels, such as $\bar{D}_sD_s$, $\bar{D}^*D^*$, and $\bar{D}_s^*D_s^*$, may play a role in forming the $X(3872)$~\cite{Gamermann:2009fv,Ji:2022vdj}. Moreover, the $X(3872)$ may contain a  $c\bar{c}$ component~\cite{Yamaguchi:2019vea}.   With only the $\bar{D}^*D$ contribution, we can well describe the production rates of $X(3872)$ in $B$ decays, which further supports that $X(3872)$ contains a sizable $\bar{D}^*D$ molecular component.   With the experimental data so far,  our predictions for the production rates of the other $\bar{D}^*D^{(\ast)}$ molecules in $B$ decays are  useful to verify the molecular nature of $Z_{c}(3900)$, $X_{2}(4013)$, and $Z_c(4020)$.

\begin{table}[!h]
\centering
\caption{Branching fractions of $B\to X_{s\bar{s}/q\bar{q}}/X(3960) K^{+}$ where $X_{s\bar{s}/q\bar{q}}$ is a bound state of  $D_s^+ D_s^-$ or $\bar{D}D$. \label{resultsddss}
}
\label{results}
\begin{tabular}{c c c c c c c c}
  \hline \hline
    Decay modes    &~~~~ Our results    &~~~~ Exp~\cite{ParticleDataGroup:2020ssz} 
         \\ \hline 
        $B^{+} \to X_{s\bar{s}} K^{+} $   &~~~~ $({ 2.1\sim 17.0})\times 10^{-4}$   &~~~~ $\mathrm{Br}(B^{+}\to \chi_{c0}(3915)K^{+})<2.8\times 10^{-4}$ 
         \\     $B^{+} \to (X_{q\bar{q}} \to \eta_{c}\eta)  K^{+} $    &~~~~ $({ 0.9\sim6.7})\times 10^{-4}$  &~~~~ $\mathrm{Br}(B^{+} \to \eta_{c}\eta K^{+})<2.2\times 10^{-4}$  
                  \\   \hline 
                  $B^{+} \to X(3960) K^{+} $    &~~~~ $<{ 2.4}\times 10^{-4}$  &~~~~ - 
\\
  \hline \hline
\end{tabular}
\end{table}

Similarly, another HQSS partner, i.e.,  the $\bar{D}D$ molecule,  can also be produced in $B$ decays, which proceed via the triangle diagram mechanism shown in Fig.~2 of Ref.~\cite{Xie:2022lyw}.  Since the existence of the  $\bar{D}D$ molecule still needs experimental verification, we show the branching fractions of the  $\bar{D}D$ molecule in $B$ decays as a function of the $\bar{D}D$ molecule mass. Our results show that its branching fraction is $10^{-4}$, likely to be observed in current experiments. More precise data for $B$ decays are   useful to verify the molecular nature of the $\bar{D}D$ molecule. Using the same framework, we calculated the branching fractions of the $\bar{D}_sD_s$ molecule as a bound or resonant state in $B$ decays. The results are shown in Table~\ref{resultsddss}.

 \begin{figure}[!h]
\begin{center}
\includegraphics[width=6.1in]{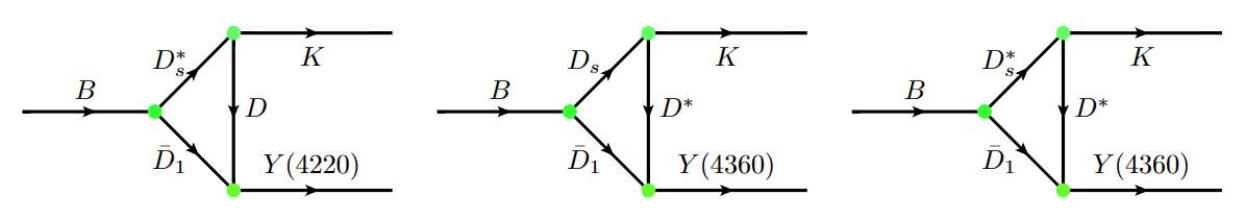}
\caption{ Triangle diagrams accounting for $B \rightarrow K Y(4220)$ and $B \rightarrow K Y(4360)$.   }
\label{production4260}
\end{center}
\end{figure}

Next, we will comment on the productions of $Y$ states in $B$ decays. As discussed above, the $Y(4220)$ and $Y(4360)$ can be viewed as a doublet of $\bar{D}D_1$ hadronic molecules. Similarly, one can study the productions of $\bar{D}D_1$ molecules in $B$ decays via the triangle diagram mechanism as shown in Fig.~\ref{production4260}. With the effective Lagrangian approach, one can obtain the branching fractions of the decays $B \to Y(4220) K$  and $B \to Y(4360) K$, which we plan as future work. 

From our above analysis, we conclude that the charmonium-like states (also known as the $XYZ$ states) can be produced in $B$ decays via the triangle diagrams because the $XYZ$ states have strong couplings to pairs of charmed mesons.

\subsubsection{Production rates of the pentaquark states}

\begin{figure}[!h]
\begin{center}
\begin{overpic}[scale=.62]{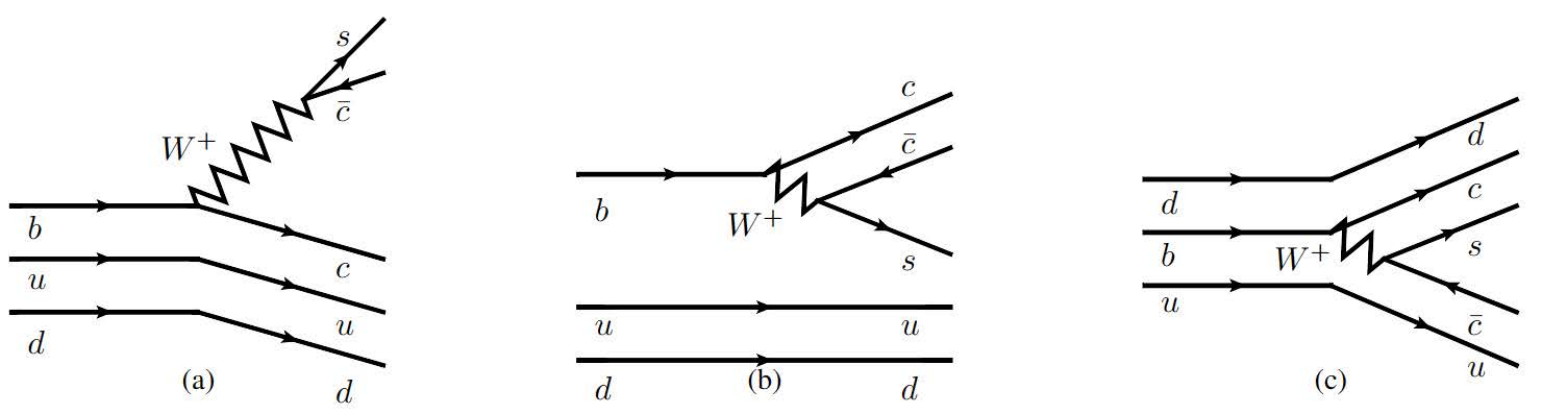}
\end{overpic}
\caption{Three possible topologies for the weak decay $b\to c \bar{c}s$.  }
\label{weakdecayquark}
\end{center}
\end{figure}

\begin{figure}[!h]
\begin{center}
\begin{overpic}[scale=.8]{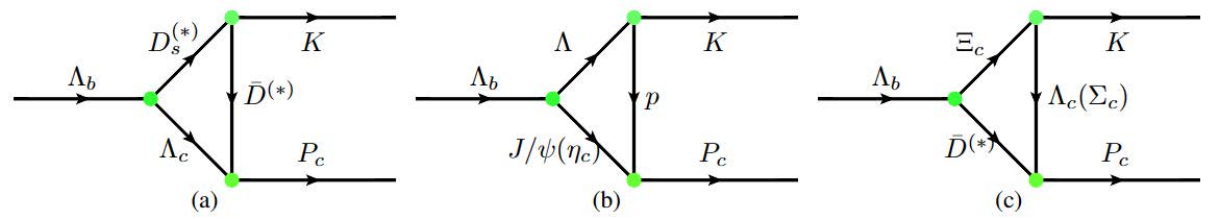}
\end{overpic}
\caption{The triangle diagram generating the pentaquark molecules based on the weak vertices of Fig.~\ref{weakdecayquark}. }
\label{triaburns}
\end{center}
\end{figure}

The pentaquark states were discovered in the $\Lambda_{b}$ decay in $pp$ collisions. To understand their production, it is necessary to understand the weak decays of $\Lambda_b$.  At the quark level, there are three possible topologies for the Cabibbo-favored weak decay ($b\to c \bar{c}s$) as shown in Fig.~\ref{weakdecayquark}, which correspond to the decays of $\Lambda_b \to \bar{D}_s\Lambda_c$,  $\Lambda_b \to J/\psi\Lambda$, and $\Lambda_b \to \bar{D}\Xi_c$ at the hadron level. Such weak decays proceed via the external W-emission mechanism shown in Fig.~\ref{weakdecayquark}~(a) and the internal W-conversion mechanism shown in Fig.~\ref{weakdecayquark}~(b) and (c).  The former is color-favored, while the latter is suppressed, indicating that topology (a) is responsible for producing the pentaquark states.   It should be noted that in terms of the power
counting rules of the soft-collinear effective theory, the ratio of the topological diagrams (b) to (c) has the following power counting relation $\frac{|C|}{|T|}\sim \mathcal{O}(\frac{\Lambda^{h}_{QCD}}{m_{Q}})$~\cite{Leibovich:2003tw,Mantry:2003uz,Jia:2024pyb}.  Since the factor $\mathcal{O}(\frac{\Lambda^{h}_{QCD}}{m_{Q}})$ is estimated to be $\mathcal{O}(1)$ in the charm sector, the order of the branching fractions of the weak decays  $\Lambda_c \to \Sigma^0 K^+$ and $\Lambda_c \to \Lambda^0 K^+$ is similar, which indicates that the color suppressed effect is small.  However, the factor is changed into  $\mathcal{O}(\frac{\Lambda^{h}_{QCD}}{m_{Q}})\sim 0.25$ in the bottom sector, which indicates that topology (b) is color suppressed. Therefore, the non-factorization contributions of the $C$-type topology is larger in the charm sector than in the bottom sector, indicating that the non-perturbative effect is quite prominent.     In particular, the topology of the decays $\Lambda_b \to \Sigma_c \bar{D}_s^{(*)} $ cannot be drawn at the quark level, unlike the decays $\Lambda_c \to \Sigma K^{(*)} $, implying that the branching fractions of the decays  $\Lambda_b \to \Sigma_c \bar{D}_s^{(*)} $ should be small. In Ref.~[192], the productions of pentaquark states are from the three-body weak decays $\Lambda_b \to \Sigma_c \bar{D}^{(*)} K$.   The topology diagrams of the decays $\Lambda_b \to \Sigma_c \bar{D}^{(*)} K$ can be drawn at the quark level, which is color suppressed in contrast to the weak decays $\Lambda_b \to \Lambda_c \bar{D}^{(*)} K$ as shown in Fig.~1 of Ref.~\cite{LHCb:2024fel}, i.e., $\mathcal{B}(\Lambda_b \to \Sigma_c \bar{D}^{(*)} K)/\mathcal{B}(\Lambda_b \to \Lambda_c \bar{D}^{(*)} K) \sim (\frac{|C|}{|T|})\sim 0.25$, consistent with recent measurements of the LHCb Collaboration~\cite{LHCb:2024fel}.

The branching fractions for topology (a) and (b) are  $\mathcal{B}(\Lambda_b \to \bar{D}_s\Lambda_c)=1.1\pm1.0\%$ and  $\mathcal{B}(\Lambda_b \to J/\psi\Lambda)=(3.72\pm1.07)\times10^{-4}$~\cite{Burns:2022uiv}. The branching fraction for topology (c) is naively estimated to be   $\mathcal{B}(\Lambda_b \to \bar{D}\Xi_c)=2.5\times10^{-4}$~\cite{Burns:2019iih}.  One can see that topologies (b) and (c) are smaller than topology (a) by a factor of $50$.    Based on these weak decay vertices, the pentaquark molecules can be generated via the triangle diagrams shown in Fig.~\ref{triaburns}.    The topologies (b) and  (c) can mix via topology (a), which can lead to double counting if they are considered simultaneously.  Since topology (b) and topology (c) in Fig.~(23) are non-factorized diagrams, the final-state interaction approach is one effective approach to deal with such non-factorized diagrams. Therefore, for the decays $\Lambda_b \to J/\psi(\eta_c)\Lambda$, in addition to the short-range contribution from the effective Wilson coefficient $a_2$,  the final-state interaction also contributes to the decays $\Lambda_b \to J/\psi(\eta_c)\Lambda$, proceeding via $\Lambda_b \to \bar{D}_s^{(*)} \Lambda_c  \to  J/\psi(\eta_c)\Lambda $ via the $D_s^{(*)}$ mesons exchange. In other words, the weak decays  $\Lambda_b \to \bar{D}_s^{(*)} \Lambda_c $  contribute to the long-range contributions of the  decays $\Lambda_b \to J/\psi(\eta_c)\Lambda$.     
Similarly, the decays  $\Lambda_b \to D_s^{(*)}\Lambda_c$ contribute to the long-range contribution of the decays of $\Lambda_b \to \bar{D}^{(*)}\Xi_c$.       Therefore,  topology (a) is usually adopted to produce the pentaquark molecules.

\begin{figure}[!h]
\begin{center}
\begin{overpic}[scale=.70]{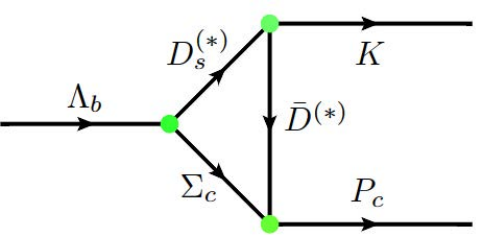}
\end{overpic}
\caption{Triangle diagram of $\Lambda_b\to D_s^{(\ast)}\Sigma_{c}\to P_{c}K$. }
\label{triawu345}
\end{center}
\end{figure}

    In Ref.~\cite{Wu:2019rog}, Wu et al. constructed the triangle mechanism where the $\Lambda_{b}$ baryon weakly decays into $D_{s}^{(\ast)}\Sigma_{c}$, then $D_{s}^{(\ast)}$ scatter into $\bar{D}^{(\ast)}K$, and finally the $\bar{D}^{(\ast)}\Sigma_{c}$ molecules are dynamically  generated  as shown in Fig.~\ref{triawu345}. We note that the  $\Lambda_b$ decay into $\Sigma_c^{(*)}$ is highly suppressed due to the fact the light quark pair transition between a symmetric and antisymmetric spin-flavor configuration is  forbidden~\cite{Falk:1992ws,Gutsche:2018utw},     which indicates that the production of pentaquark molecules is difficult (if not impossible) via the weak decays of $\Lambda_b \to {D}_s^{(*)}\Sigma_c$ \footnote{In Ref.~\cite{Wu:2019rog}, the $\Lambda_b \to \Sigma_c$ transition is assumed to be proportional to the  $\Lambda_b \to \Lambda_c$ transition, characterized by an unknown parameter $R$.   By reproducing the experimental production rates of the pentaquark molecules,    $R$ is found to be about 0.1. }.  
   For the vertices of the molecules, the  $\bar{D}^{(*)}\Sigma_c^{(*)}$, $\bar{D}^{(*)}\Lambda_c$, $J/\psi p$, and  $\eta_c p$ channels   contribute to the pentaquark molecules. However, the relative importance of each component remains uncertain, especially that of $\bar{D}^{(*)}\Lambda_c$.  According to  Ref.~\cite{Xiao:2019aya}, the $\bar{D}^{(*)}\Lambda_c$, $\eta_c p$,  and  $J/\psi p$ channels are less important. However,  Refs.~\cite{Yamaguchi:2019seo,He:2019rva} claim that the $\bar{D}^{(*)}\Lambda_c$ contributions are sizable. 
If  the weak vertices involve  $\mathcal{B}(\Lambda_b\to D_{s}^{(\ast)}\Lambda_{c})$, then the vertices generating the molecules  are small. On the other hand, if one takes the dominant vertices to generate the molecules, one needs to introduce the unknown weak vertex  $\mathcal{B}(\Lambda_b\to D_{s}^{(\ast)}\Sigma_{c})$. Based on the obtained couplings of the pentaquark molecules to their components in Scenario A and Scenario B, we employ the effective Lagrangian approach to calculate the branching fractions of the decays $\Lambda_b \to P_c K$ as shown in Table~\ref{branchingratiospv}. Since the pentaquark molecules are generated by the $S$-wave $\bar{D}^{(*)}\Lambda_c$  interactions, only the pentaquark molecules with $J^P=\frac{1}{2}^-$ and $J^P=\frac{3}{2}^-$ are produced in our model, which indicates that the production rate of the pentaquark molecule with $J^P=\frac{5}{2}^-$ in the $\Lambda_b$ decay is low.

\begin{table}[ttt]
\centering
\caption{Branching fractions ($10^{-6}$) of  $\Lambda_b$ decay into a $K$ meson and a hidden-charm pentaquark  molecule in Scenario A and Scenario B.  }
\begin{tabular}{c c c c c c c}
  \hline \hline 
     Scenario   &    \multicolumn{6}{c}{A}  \\ \hline
     Molecule  &${P_{c1}}$ & ${P_{c2}}$ & ${P_{c3}}$ & ${P_{c4}}$ & ${P_{c5}}$ & ${P_{c6}}$  \\
     $\mathcal{B}(\Lambda_b \to {P_{c}}K)$ &  35.18  & 1.49 &  15.30 & 0.48 & 6.37 &  9.01 \\\hline
    Scenario   &   \multicolumn{6}{c}{B}   \\ \hline
     Molecule  &${P_{c1}}$ & ${P_{c2}}$ & ${P_{c3}}$ & ${P_{c4}}$ & ${P_{c5}}$ & ${P_{c6}}$  \\
     $\mathcal{B}(\Lambda_b \to {P_{c}} K)$ & 98.88   & 2.27 & 27.23 & 5.21 & 21.69  & 7.43  \\
  \hline\hline 
\end{tabular}
\label{branchingratiospv}
\end{table}

The  products of  the branching fractions  $ \mathcal{B}(\Lambda_b^0\to   P_{c}^{+}  K^{-}) $ and the branching fractions $\mathcal{B}(P_{c}^{+} \to J/\psi p)$ have been measured:  
\begin{eqnarray}
\label{pcbranching}
\mathcal{B}(\Lambda_b^0\to { P_{c}(4312)^{+}} K^{-}) \cdot \mathcal{B}({ P_{c}(4312)^{+}}\to J/\psi p)&= 0.96^{+1.13}_{-0.39}\times 10^{-6}\,,  \nonumber\\ 
\mathcal{B}(\Lambda_b^0\to { P_{c}(4440)^{+}} K^{-}) \cdot \mathcal{B}({P_{c}(4440)^{+}}\to J/\psi p)&= 3.55^{+1.43}_{-1.24}\times 10^{-6}\,,   \\ \nonumber
\mathcal{B}(\Lambda_b^0\to { P_{c}(4457)^{+}} K^{-}) \cdot \mathcal{B}({P_{c}(4457)^{+}}\to J/\psi p)&= 1.70^{+0.77}_{-0.71}\times 10^{-6}\,.
\end{eqnarray}
To compare with the experiments, we have estimated the products  $ \mathcal{B}(\Lambda_b^0\to  P_{c}^+  K^{-}) \cdot \mathcal{B}(P_{c}^+\to J/\psi p)  $ in Scenario A and Scenario B  by first estimating the branching fractions of the decays $P_{c} \to J/\psi p$.  Since the main uncertainties of our results are from the couplings of the pentaquark molecules to the $\bar{D}^{(*)}\Lambda_c$
 and $J/\psi p$, we have adopted the couplings determined by the ChUA to calculate these values~\cite{Xiao:2020frg},  which actually correspond to our results of scenario A as shown in Table~\ref{branchingratiospv1sd1235}. Our results show that  the production rate of  $P_c(4457)$ in the $\Lambda_b$ decay in Scenario A, the production rate of $P_c(4312)$ in Scenario B, and the production rate of $P_{c}(4440)$ in the ChUA , deviate much from the experimental data, which indicates that the production rates of $P_c(4312)$, $P_c(4440)$, and $P_c(4457)$ in the $\Lambda_b$ decays can not be simultaneously  described in   the triangle diagram mechanism.   In addition to the uncertainties in the hidden-charm pentaquark molecules couplings to $\bar{D}^{(*)}\Lambda_c$ and $J/\psi p$, resulting in our results inconsistent with the experimental data, our production models need to be further improved. The LHCb Collaboration measured the ratio  of the  branching fractions    $\mathcal{B}(\Lambda_b \to \Sigma_c \bar{D}^{(*)} K)/\mathcal{B}(\Lambda_b \to \Lambda_c \bar{D}^{(*)} K)=0.282\pm0.016\pm0.016\pm0.005$, indicating that the three-body weak decays $\Lambda_b \to \Sigma_c \bar{D}^{(*)} K$ may also contribute to the productions of hidden-charm pentaquark molecules~\cite{LHCb:2024fel}. However, the relation between these two production mechanisms is not obvious. We plan to clarify this issue in the future.

\begin{table}[ttt]
\centering
\caption{Branching fractions ($10^{-6}$) of the decays   $\Lambda_b \to (P_{c} \to J/\psi p) K$  in Scenario A and Scenario B.  }
\begin{tabular}{c c c c c c c}
  \hline \hline 
     Scenario   &    \multicolumn{6}{c}{A}  \\ \hline
   Molecule  &$P_{c1}$ & $P_{c2}$ & $P_{c3}$ & $P_{c4}$ & $P_{c5}$ & $P_{c6}$  \\
     Ours &  7.11 & 1.44&  8.21& 0.09& 1.77& 4.82\\
     ChUA~\cite{Xiao:2020frg} &  1.82  & 8.62 & 0.13 & 0.83 & 0.04 & 2.36 \\
     Exp & 0.96  & - & 3.55 & 1.70 & - & - \\\hline
    Scenario   &   \multicolumn{6}{c}{B}   \\ \hline
      Molecule  &$P_{c1}$ & $P_{c2}$ & $P_{c3}$ & $P_{c4}$ & $P_{c5}$ & $P_{c6}$  \\
     Ours & 18.24 & 2.22& 6.06& 1.79& 3.83& 2.76\\
     ChUA~\cite{Xiao:2020frg} &  -  & - & - &  - & - & - \\
     Exp &  0.96  & - &  1.70 & 3.55 & - & - \\ 
  \hline \hline 
\end{tabular}
\label{branchingratiospv1sd1235}
\end{table}

 These  hidden-charm pentaquark  molecules are observed in the $J\psi p$ invariant mass distribution, and one can also expect to see them in the  $\bar{D}^* \Lambda_c$ invariant mass distribution.  
As a byproduct, with the branching fractions of the decays $P_c \to \bar{D}^{(*)}\Lambda_c$ we predicted the products of   $ \mathcal{B}(\Lambda_b\to   P_{c} K^{}) \cdot \mathcal{B}(P_{c} \to \bar{D}^{(*)}\Lambda_c )  $ in Scenario A and Scenario B as shown in Table~\ref{branchingratiospv1sd11}.   We can see that the branching fractions  of  the pentaquark molecules  in the decays   $\Lambda_b \to ( P_{c} \to J/\psi p) K$ and   $\Lambda_b \to ( P_{c} \to \bar{D}^* \Lambda_c) K$ are similar except for $ P_{c2}$. The branching fraction $\mathcal{B}[\Lambda_b \to ( P_{c2}\to \bar{D}^* \Lambda_c) K]$ is smaller than the branching fraction   $\mathcal{B}[\Lambda_b \to ( P_{c2} \to J/\psi p) K]$ by two orders of magnitude. 
We encourage experimental searches for these pentaquark states in the $\bar{D}^*\Lambda_c$ invariant mass distributions of the $\Lambda_b$ decays.

\begin{table}[!h]
\centering
\caption{Branching fractions($10^{-6}$) of the decays  $\Lambda_b \to ( P_{c} \to \bar{D}^*\Lambda_c) K$  in Scenario A and Scenario B.  }
\begin{tabular}{c c c c c c c}
  \hline \hline 
     Scenario   &    \multicolumn{6}{c}{A}  \\ \hline
      Molecule  &$P_{c1}$ & $P_{c2}$ & $P_{c3}$ & $P_{c4}$ & $P_{c5}$ & $P_{c6}$  \\
                    Ours &  4.12  & 0.05 & 1.27  & 0.00 & 0.21 & 1.87 \\    \hline
    Scenario   &   \multicolumn{6}{c}{B}   \\ \hline
       Molecule  &$P_{c1}$ & $P_{c2}$ & $P_{c3}$ & $P_{c4}$ & $P_{c5}$ & $P_{c6}$  \\
         Ours & 26.41   & 0.05 & 8.42  & 2.71 & 4.01 & 2.70 \\
  \hline \hline 
\end{tabular}
\label{branchingratiospv1sd11}
\end{table}

 The hidden-charm pentaquark states have only
 been observed in the exclusive $b$ decays in proton-proton collisions.  
 The production of pentaquark states in other processes has been proposed. In Refs.~\cite{Wang:2015jsa,HillerBlin:2016odx,Karliner:2015voa,Wang:2019krd,Wu:2019adv}, the authors claimed that the hidden-charm pentaquark states can be produced in the $J/\psi$ photoproduction off proton. This process could distinguish whether these pentaquark states are genuine or anomalous triangle singularities.   Moreover, it is suggested that the hidden-charm pentaquark states can be produced in $e^+e^-$ collisions~\cite{Li:2017ghe} and antiproton-deuteron collisions~\cite{Voloshin:2019wxx}. Based on Monte Carlo simulations, the inclusive production rates of these pentaquark states are estimated in proton-proton collisions~\cite{Chen:2021ifb,Ling:2021sld} and electron-proton collisions~\cite{Shi:2022ipx}, which are helpful for future experimental searches for the pentaquark states.

\section{Three-body hadronic molecules}
\label{three-body}

Nowadays, using nucleons as degrees of freedom and with the two-body $NN$ interaction determined by the nucleon-nucleon scattering data and the residual small $NNN$ interaction, ab initio calculations can reproduce most of the ground states and low-lying excited states of light nuclei~\cite{Wiringa:1991kp,Carlson:1997qn,Pudliner:1997ck}. Adding hyperons to the nuclear system, one ends up with hypernuclei. Following the same approach, the properties of hypernuclei can also be understood very well~\cite{Bando:1990yi,Hashimoto:2006aw,Hiyama:2009zz}. Such a picture can be extended to other three-body 
hadronic systems based on the molecular candidates discovered experimentally,  such as  $X(3872)$, $D_{s0}^*(2317)$, the $P_c$ pentaquark states, and $T_{cc}(3875)^+$ mentioned above.  The idea is that if one exotic hadron $C$ is a bound state of two conventional hadrons $A$ and $B$, then with the same interaction between $A$ and $B$ which leads to the formation of $C$, one can study the three-body $ABB$ and $AAB$ systems and check whether they can form bound/resonant states. If the $ABB$ or $AAB$ system binds, then an experimental (or lattice QCD) confirmation on the existence of this state can potentially verify the molecular picture of the exotic state $C$, i.e., it is dominantly a hadronic molecule of $A$ and $B$\footnote{ The subsystem potentials, resonant two-body potentials, and three-body force would contribute to generating the three-body hadronic molecules dynamically.   Compared to the studies in nuclear physics, no empirical information on the three-body interactions is available for the three identical baryons we have studied. Thus,  the impact of the three-body force on the three-body hadronic molecules is uncertain. As for the resonant two-body potentials, we can choose a three-body system in which the resonant two-body potentials are much suppressed and therefore, their effects,  to some extent, are under control.       }. 
Conversely, studies on multi-hadron states can also illuminate the relevant hadron-hadron interactions.  It should be noted that the binding of a three-body system is not only determined by the interactions but also by other factors such as their masses and quantum numbers. A small reduced mass can increase the kinematic energy barrier and make the systems difficult to bind. Quantum numbers such as spin and isospin in multi-body systems can also significantly affect the bindings of multi-hadron systems. In the case that $AA$ ($BB$) is not bound, the binding of $AB$ is not a sufficient but necessary condition for the binding of $AAB$ ($ABB$)~\cite{Wu:2022ftm}. 

There have been many studies of three-body systems containing nucleons and hyperons. For instance, Refs.~\cite{Garcilazo:1997mf,Garcilazo:1999hh,Mota:1999qp,Valcarce:2001in,Mota:2001ee} studied the possible existence of nonstrange tribaryons including $NNN$~\cite{Garcilazo:1999hh,Mota:2001ee}, $NN\Delta$~\cite{Garcilazo:1999hh,Mota:2001ee}, $N\Delta\Delta$~\cite{Mota:1999qp,Mota:2001ee} and $\Delta\Delta\Delta$~\cite{Garcilazo:1997mf,Valcarce:2001in,Mota:2001ee} systems.
The strange tribaryons have also been studied, including single-strangeness $\Lambda NN$ and $\Sigma NN$~\cite{Garcilazo:2007ss,Fernandez-Carames:2006uyg, Garcilazo:2022pgt}, double-strangeness $\Lambda \Lambda N$ and $\Xi NN$~\cite{Garcilazo:2015noa,Garcilazo:2016gkj,Garcilazo:2016ylj, Garcilazo:2020ofz,Garcilazo:2016ams}, and multi-strangeness $N\Xi\Xi$~\cite{Garcilazo:2016ams}, $\Omega NN$ and $\Omega\Omega N$~\cite{Garcilazo:2019igo,Zhang:2021vsf}.
On the experimental side, a strange tribaryon $S^0(3115)$ was reported in the $^4$He (stopped $K^-$, $p$) reaction, which mainly decays into $\Sigma NN$~\cite{Suzuki:2004ep}. In Ref.~\cite{Maezawa:2004va}, this strange tribaryon is explained as a  nona-quark  state. Other searches for strange tribaryons have also been performed~\cite{Sato:2006bw,Sato:2007sb,KEK-PSE549:2009ejj, Yim:2010zza}. 

A particularly relevant line of research is the studies on the clusters of different numbers of antikaons and nucleons~\cite{Dote:2008zz, Dote:2008xa, Barnea:2012qa, Ikeda:2010tk, Revai:2014twa, Wycech:2008zzb, Shevchenko:2007zz, Kanada-Enyo:2008wsu, Hyodo:2022xhp, Kezerashvili:2021ren}. The $\Lambda(1405)$ can be considered as a quasibound  $\Bar{K}N$ state~\cite{Dalitz:1959dn, Dalitz:1960du}, from which one can deduce that the $\bar{K}N$ interaction is strongly attractive. In such a case, it is natural to expect the existence of a $\bar{K}NN$ bound state. The $\Bar{K}\Bar{K}N$ system was found to bind with a binding energy of $10\sim33$ MeV~\cite{Kanada-Enyo:2008wsu}, and the $K\bar{K}N$ systems were also found bound in Ref.~\cite{Xie:2010ig} with the fixed-center approximation and in Ref.~\cite{Jido:2008kp} with a variational approach. In  Ref.~\cite{Hyodo:2022xhp}, multi-hadron states composed of an antikaon and different numbers of nucleons, i.e., $\Bar{K}NN$,  $\Bar{K}NNN$, $\Bar{K}NNNN$,  and $\Bar{K}NNNNN$, were found to bind with binding energies of $25\sim28$ MeV, $45\sim50$ MeV, $68\sim76$ MeV, and $70\sim81$ MeV.

\begin{figure}[!h]
\begin{center}
\begin{overpic}[scale=1]{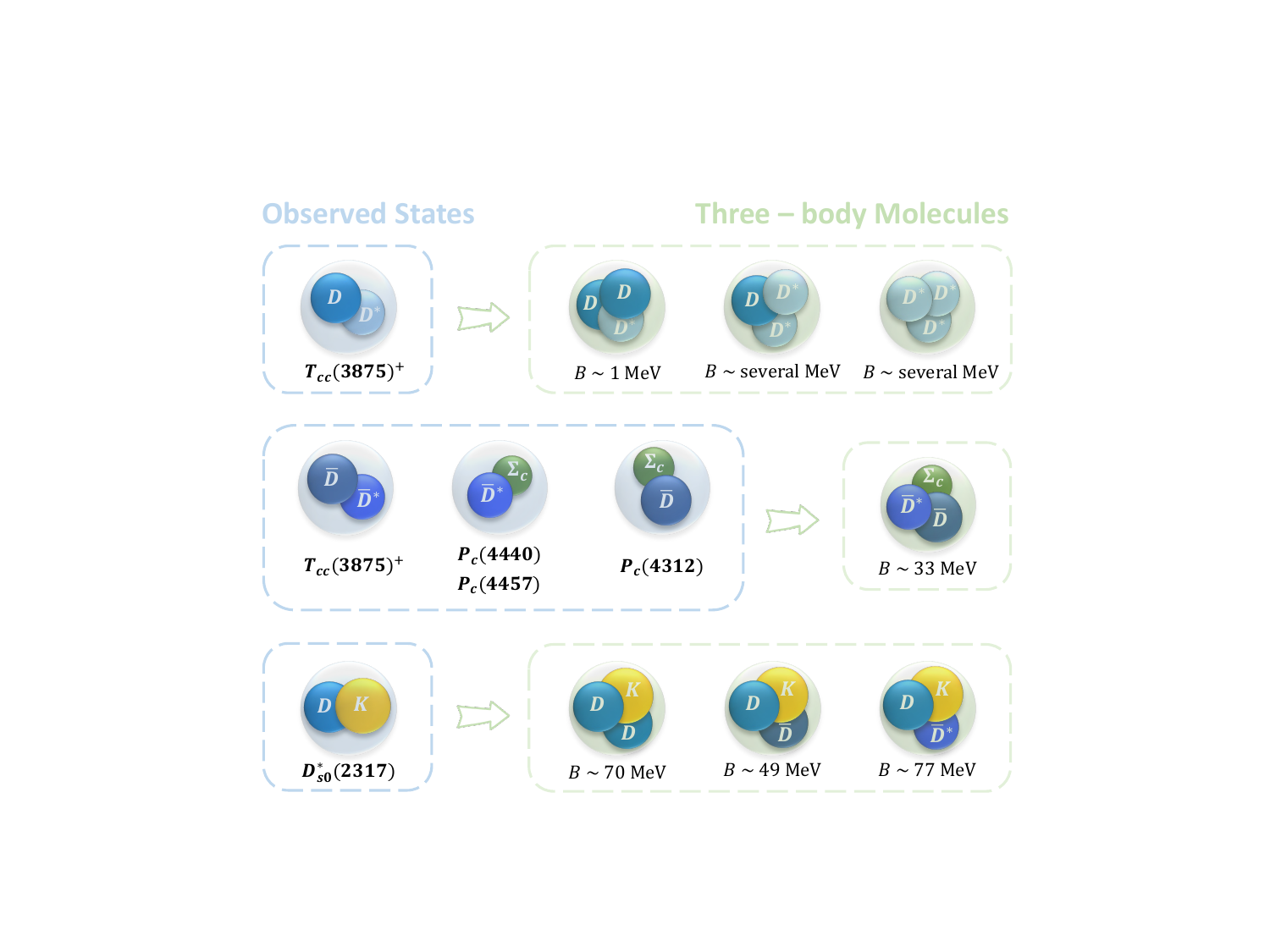}
\end{overpic}
\caption{Three-body hadronic molecules based on the assumption that the observed states are two-body hadronic molecules.}
\label{2-3 body}
\end{center}
\end{figure}

In this review, we focus on possible three-body systems that help reveal the nature of the many two-body exotic hadrons covered in this review, especially in the heavy flavor sector.  In the charm sector, there are many studies about multi-hadron systems. Most of them are based on the molecular nature assumptions of discovered exotic states, including $X(3872)$ as a $D\bar{D}^*$ molecule, $D_{s0}^*(2317)$ as a $DK$ molecule, the $P_c$ pentaquark states as $D\bar{\Sigma}_{c}^{(*)}$ molecules, and $T_{cc}(3875)^+$ as a $DD^*$ molecule. With these assumptions and the two-body hadron-hadron interactions, multi-hadron systems such as $D^{(*)}D^{(*)}(\bar{D}^{(*)})K$, $\bar{D}\bar{D}^{(*)}\Sigma_c$, and
$D^{(*)}D^{(*)}D^{(*)}$ have been studied and found to form bound/resonant states. Their masses, decays, and productions have been studied, which provide valuable information for future experimental searches. Extending the studies from the charm sector to the bottom sector is straightforward, employing heavy quark flavor symmetry. Since multi-hadron bound/resonant states in the bottom sector are more difficult to observe experimentally, we mainly focus on the charm sector(See Fig.~\ref{2-3 body}).

\subsection{$DD(\bar{D})K$ molecules}


As discussed in Sec.~\ref{multiplet molecules}, the $D_{s0}^*(2317)$ and $D_{s1}(2460)$ can be explained as $DK$ and $D^*K$ molecules.  Adding a $D$ meson to the $DK$ system, one can study the $DDK$ system and search for likely bound states by solving the Schr\"odinger equation via the Gaussian Expansion Method (GEM).  For the details of the GEM, one can refer to Appendix.E and Ref.~\cite{Hiyama:2003cu}.   The pairwise potentials are obtained via OBE model or EFT associated with a momentum cutoff or a typical radius to demonstrate that the potentials only apply within a certain scale. The Jacobi coordinates for the $DDK$ system are shown in Fig.~\ref{3-jacobi}, where the symbols $N_{1}$, $N_{2}$, and $N_{3}$ denote $D$, $D$, and $K$ mesons, respectively. In Ref.~\cite{Wu:2019vsy}, the $DD$ potential is described by the OBE model, where the cutoff $\Lambda$ in the form factor $\frac{\Lambda^2 - m^2}{\Lambda^2 - q^2}$ is set at 1 GeV by reproducing the $X(3872)$ pole~\cite{Liu:2019stu}. 
The $DK$ potential contains a leading-order attractive potential $C_L'e^{-(r/R_c)^2}$ and a next-to-leading order repulsive correction $C_S'e^{-(r/R_S)^2}$. In the particular calculation, we set $R_S=$ 0.5 fm, $C_S'=$ 0, 300, 1000, 3000 MeV for the repulsive correction. As for the leading-order attractive potential, we set $R_c=$ 1,2,3 fm, and $C_L'$ is determined by reproducing the $D_{s0}^*(2317)$ as a $DK$ bound state with a binding energy of 45 MeV. For all parameter sets, a $DDK$ bound state is found below the three-body mass threshold by $67 \sim 72$ MeV, also below the $DD_{s0}^*(2317)$ mass threshold. 
Therefore, if the $D_{s0}^*(2317)$ is assumed as a $DK$ bound state, a $DDK$ bound state is expected. In other words, the existence of a $DDK$ molecule can help verify the molecular nature of $D_{s0}^*(2317)$.
The $DDK$ system is found to bind with a binding energy of about 90 MeV in the three-body coupled-channel $DDK$, $DD_s\pi$, and $DD_s\eta$ system by solving the Faddeev equations in Ref.~\cite{MartinezTorres:2018zbl} and about 70 MeV in finite volume~\cite{Pang:2020pkl}, which are consistent with the results of Ref.~\cite{Wu:2019vsy}. In addition, Ref.~\cite{Debastiani:2017vhv} points out that the $KK$ repulsive interaction, which is of the same magnitude as the $DK$ attraction, prevents the $DKK$ system from binding.

Since the $DDK$ bound state carries double charm, it isn't easy to be produced experimentally.  If we replace one of the charmed $D$ mesons with an anti-charmed $\bar{D}$ meson, we have a hidden-charm system of $D\bar{D}K$, which is more likely to be produced experimentally. Therefore, we investigate the  three-body systems of $D\bar{D}K$ and $D\bar{D}^*K$ in the same approach~\cite{Wu:2020job}. One should note that according to chiral perturbation theory, the $\bar{D}K$ interaction in the isospin zero channel is only half that of the $DK$ interaction. Caution should be taken when one transforms the Weinberg-Tomozawa chiral potential from momentum space to coordinate space. In particular, no $\bar{D}K$ bound state with isospin zero~\cite{Pan:2023zkl} exists,  while it can form a virtual state~\cite{Albaladejo:2016lbb,Du:2017zvv,Cheung:2020mql}.  
As shown in Table~\ref{Results:BE}, two bound states $D\bar{D}K$ and $D\bar{D}^*K$ are obtained with the binding energies of 49 MeV and 77 MeV, which are also below the $D_{s0}^*(2317)\bar{D}$ and $D_{s1}(2460)\bar{D}$ mass thresholds, in agreement with Refs.~\cite{Ma:2017ery,Ren:2018pcd,Wei:2022jgc}.  The minimum quark contents of such molecules are the same as those of hidden-charm tetraquark states with strangeness $Z_{cs}$, while the parity of $Z_{cs}$ is different from that of the $D\bar{D}^{(\ast)}K$ molecules. Therefore, the $D\bar{D}^{(\ast)}K$ molecules  can be regarded as excited states of $Z_{cs}$. We should note that the existence of the $D\bar{D}K$ molecule is heavily tied to the $DK$ interaction, which, in other words, indicates that the existence of the  $D\bar{D}K$ molecule would verify or refute the molecular nature of $D_{s0}^*(2317)$. 
Therefore, we strongly recommend experimental searches for the $D\bar{D}K$ molecule, particularly at the Belle II experiment~\cite{Wu:2022wgn}.

\begin{table}[h]
    \setlength{\tabcolsep}{7pt}
    \caption{Masses and binding energies (in units of MeV) of the $D\bar{D}K$ and $D\bar{D}^*K$ bound states, in comparison with the results of other works.
    }
    \centering
    \begin{tabular}{c|c c c}
    \hline
    \hline
         & Ref.~\cite{Wu:2020job} & Ref.~\cite{Ma:2017ery}& Ref.~\cite{Ren:2018pcd}\\
    \hline
       $\frac{1}{2}(0^-)$ $D\bar{D}K$&$4181.2^{+2.4}_{-1.4}$($B_3\simeq48.9^{+1.4}_{-2.4}$)  & - &-\\  
       $\frac{1}{2}(1^-)$ $D\bar{D}^*K$&$4294.1^{+6.6}_{-3.1}$($B_3\simeq77.3^{+3.1}_{-6.6}$)&$4317.92^{+6.13}_{-6.55}$($B_3\simeq53.52^{+6.55}_{-6.13}$) & $4307\pm2$($B_3\simeq64\pm2$)\\
    \hline\hline
    \end{tabular}
    \label{Results:BE}
\end{table}

\begin{figure}[!h]
\begin{center}
\begin{overpic}[scale=1]{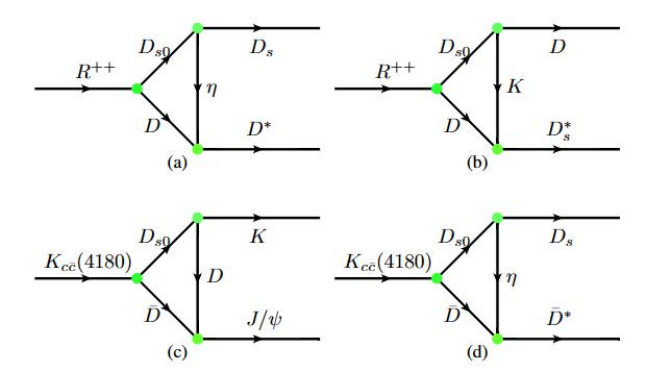}
\end{overpic}
\caption{Triangle diagram representing the decays of the $R^{++}$ state to $D_sD^*$(a) and $DD_s^*$(b) and the $K_{c\Bar{c}}(4180)$ state to $J/\psi K$(c) and $D_s\bar{D}^*$(d).   }
\label{DDKdecay}
\end{center}
\end{figure}

Identifying the   $D_{s0}^*(2317)$  as a $DK$ bound state, we predicted several relevant three-body bound states, e.g., $DDK$,  $D\bar{D}K$
and $D\bar{D}^*K$. We further investigated the strong decay of these three hadronic molecules to guide experimental searches for them.
Their decays are similar to the two-body molecules. That is to say; the three-body molecules can be viewed as quasi-two-body molecules. The proportion of each configuration can be described by their wave function. Since the $DD$ interaction is quite weak, the $DDK$ molecule can be regarded as a $DD_{s0}^*(2317)$ molecule.  The partial decay width of the $DD_{s0}^*(2317)$ molecule to three-body final states is small due to the small widths of its constituents. Nevertheless, the $DD_{s0}^*(2317)$ can scatter into $DD_{s}^*$ or $D^*D_s$, and the orbital angular momentum between the final states is $P$-wave. The contact-range approach cannot solve the $DD_{s0}^*(2317)-D^*D_s$ coupled-channel problem due to several unknown parameters. We resort to the triangle diagram depicting the  $DD_{s0}^*(2317)$ molecule transition to the $DD_{s}^*$ or $D^*D_s$ via exchanges of a $K$ meson or an $\eta$ meson as shown in Fig.~\ref{DDKdecay}. Using the effective Lagrangian, we estimated the widths of the $DDK$ molecule to be the order of several MeV~\cite{Huang:2019qmw}.  
As for the $D\bar{D}K$ molecule,  the $D_{s0}^*(2317)\bar{D}$ configuration accounts for $86.8\%$ of its wave function, and thus its three-body decay behavior is similar to that of $D_{s0}^*(2317)D$. For the two-body decay,  the $D\bar{D}K$ molecule can decay into $J/\psi K$ and $D_{s}\bar{D}^*$ via  exchanges of $D$ and $\eta$ mesons as shown in Fig.~\ref{DDKdecay}. Using the effective Lagrangian approach, we estimated the partial decay widths of $D\bar{D}K\to J/\psi K$ and  $D\bar{D}K\to \bar{D}_sD^*$ to be about 0.5 MeV and 0.2 MeV.

\begin{figure}[!h]
\begin{center}
\begin{overpic}[scale=1]{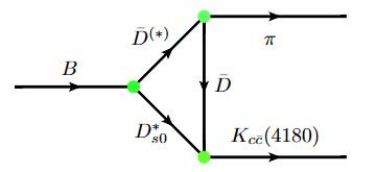}
\end{overpic}
\caption{The triangle diagram of $B \to \bar{D}^{(\ast)}D_{s0}^*\to K_{c\Bar{c}}(4180)\pi$.   }
\label{triawu}
\end{center}
\end{figure}

It is easy to realize that the $DDK$ molecule contains two charmed quarks, leading to low production yields. 
This is consistent with the null result of the Belle Collaboration~\cite{Belle:2020xca}. Similar to $T_{cc}$, the $DDK$ molecule will likely be detected in the inclusive process. 
In the exclusive process, the $DDK$ molecule can be produced in the semi-inclusive decays of $b$-flavored hadrons, such as the $B_c$ meson. Very recently, Li et al. estimated  the production rates of $DD^*$ and  $D_sD^*$ molecules in the $B_c$ decays~\cite{Li:2023hpk}.   Due to the low production rate of the $B_c$ meson, searches for the $DDK$ molecule via the $B_c$ decay is difficult. The hidden-charm $D\bar{D}K$ molecule can be observed in the exclusive process, such as the $B$ meson decay.   The $B$ meson first decays into $\bar{D}^{(*)}D_{s0}^*(2317)$, then $\bar{D}^{(*)}$ meson scatters into $\bar{D}\pi$, and finally the $K_{c\Bar{c}}(4180)$ is dynamically generated by the $\bar{D}D_{s0}^*(2317)$ interaction as shown in Fig.~\ref{triawu}. This should be explicitly studied in the future.

\subsection{$\bar{D}\bar{D}^{\ast}\Sigma_{c}$ molecules}


\begin{table}[!h]
    \setlength{\tabcolsep}{7pt}
    \centering
    \caption{Binding energies (in units of MeV), expectation values of the Hamiltonian (potential and kinetic energies) (in units of MeV) and root-mean-square radii (in units of fm) of the $\bar{D}\bar{D}^{*}\Sigma_{c}$ system obtained in the three cases detailed in the main text. \label{results1} }
    \begin{tabular}{ccccccccc}
    \hline\hline
     $I(J^{P})$& $B$ & $T$ & $V_{\bar{D}^{*}\Sigma_{c}}$ & $V_{\bar{D}\Sigma_{c}}$ & $V_{\bar{D}\bar{D}^{*}}$ & $r_{\bar{D}^{*}\Sigma_{c}}$ & $r_{\bar{D}\Sigma_{c}}$ & $r_{\bar{D}\bar{D}^{*}}$   \\\hline
     \multicolumn{9}{c}{Case I ~~~~~~$\Lambda_{P}=\Lambda_{T}=0.998$ GeV}\\\hline
     $1(\frac{1}{2}^{+})$ & 10.86 & 65.41 & -19.64 & -21.69 & -34.94 & 1.42 & 1.41 & 1.36 \\
     $1(\frac{3}{2}^{+})$ & 7.06 & 52.18 & -19.66 & -10.46 & -29.12 & 1.62 & 1.81 & 1.64 \\\hline
     \multicolumn{9}{c}{Case II ~~~~~~$\Lambda_{T}=0.998$ GeV\quad$\Lambda_{P}=1.16$  GeV}\\\hline
     $1(\frac{1}{2}^{+})$ & 37.24 & 116.16 & -41.53 & -72.44 & -39.43 & 1.00 & 0.88 & 1.03 \\
     $1(\frac{3}{2}^{+})$ & 29.63  & 92.50 & -81.32 & -21.67 & -19.15 & 0.91 & 1.36 & 1.40 \\\hline
     \multicolumn{9}{c}{Case III ~~~~~~$\Lambda_{P}=\Lambda_{T}=1.16$ GeV}\\\hline
     $1(\frac{1}{2}^{+})$ & 63.07 & 169.01 & -52.14 & -66.03 & -113.91 & 0.83 & 0.82 & 0.75 \\
     $1(\frac{3}{2}^{+})$ & 46.94 & 141.01 & -61.84 & -25.27 & -100.84 & 0.91 & 1.02 & 0.86\\
    \hline\hline
    \end{tabular}\\
\end{table}

Since the three pentaquark states $P_{c}(4312)$, $P_{c}(4440)$, and  $P_{c}(4457)$
are widely viewed as $\bar{D}^{(\ast)}\Sigma_{c}$ bound states,  it is interesting to investigate the three-body hadron system $\bar{D}\bar{D}^{\ast}\Sigma_{c}$, where the $\bar{D}\bar{D}^{\ast}$ interaction is determined by reproducing the mass of $T_{cc}(3875)^+$ as a $DD^*$ bound state.   The Jacobi coordinates for the $\bar{D}\bar{D}^{\ast}\Sigma_{c}$ system are shown in Fig.~\ref{3-jacobi}, where the symbols $N_{1}$, $N_{2}$, and $N_{3}$ denote $\bar{D}$, $\bar{D}^{\ast}$, and $\Sigma_{c}$, respectively.  The  $\bar{D}\bar{D}^{\ast}$ and $\bar{D}^{(\ast)}\Sigma_{c}$ interactions are described in the OBE model by exchanging light mesons $\pi$, $\rho$, $\sigma$, and $\omega$. To estimate the effect of the finite size of hadrons in the OBE model, a monopole form factor $\frac{\Lambda^2 -m^2}{\Lambda^2 -q^2}$ for the relevant vertices of charmed particles and exchanged light mesons was employed, which introduces the cutoff $\Lambda$. Since the OBE potentials for the $\bar{D}\bar{D}^{\ast}$ and $\bar{D}^{(\ast)}\Sigma_{c}$ two-body subsystems should be different, we choose two different cutoffs, $\Lambda_{P}$ for the $\bar{D}^{(\ast)}\Sigma_{c}$ system and $\Lambda_{T}$ for the $\bar{D}\bar{D}^{\ast}$ system.
The values of the cutoffs $\Lambda_{P}$ and $\Lambda_{T}$ are determined by reproducing the masses of ${P_{c}}^{\prime}$s and $T_{cc}(3875)^+$, yielding $\Lambda_{P} = 1.16$ GeV and $\Lambda_{T}=0.998$ GeV. 
To accommodate the so-induced uncertainties, we take three sets of cutoff values to search for three-body bound states:
Case I: $\Lambda_{T}$=$\Lambda_{P}$=0.998 GeV; Case II: $\Lambda_{T}$=$0.998$ GeV,~$\Lambda_{P}$=1.16 GeV; and Case III: $\Lambda_{T}$=$\Lambda_{P}$=1.16 GeV. One can see that we obtained two three-body bound states of $\bar{D}\bar{D}^{\ast}\Sigma_{c}$ with $I(J^{P})=1(1/2^{+})$ and $I(J^{P})=1(3/2^{+})$,  and of binding energies $37.24$ MeV and $29.63$ MeV below the $\bar{D}\bar{D}^{\ast}\Sigma_{c}$ mass threshold as shown in Table~\ref{results1}~\cite{Pan:2022xxz}.

\begin{figure}[ttt]
\begin{center}
\includegraphics[width=4.5in]{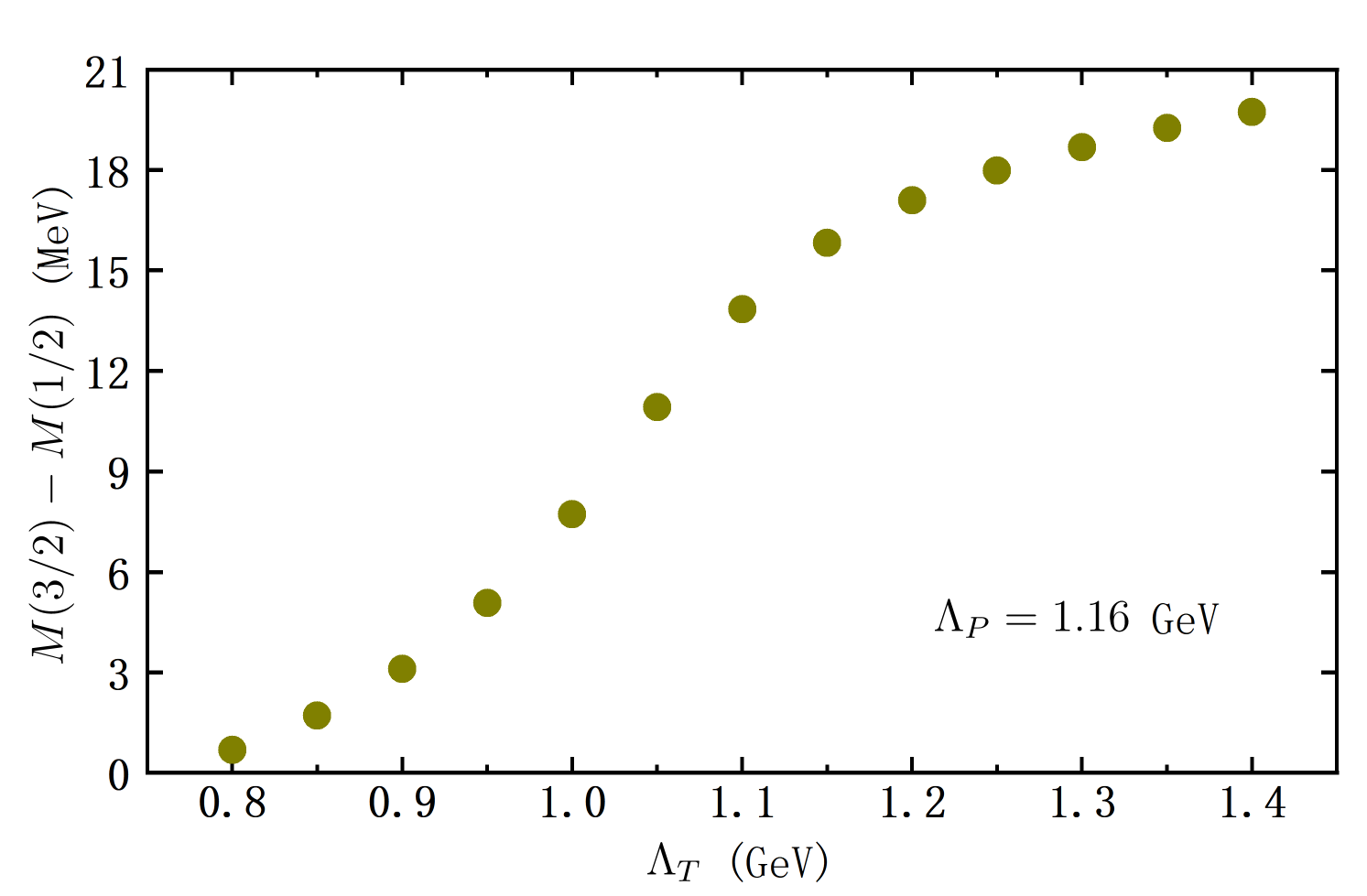}
\caption{ Mass splitting of the three-body $\bar{D}\bar{D}^{\ast}\Sigma_{c}$ doublet  as a function of the cutoff of the $\bar{D}\bar{D}^{\ast}$ potential for fixed $\Lambda_P=1.16$ GeV.   }
\label{mass splitting}
\end{center}
\end{figure}

The mass splitting of the three-body $\bar{D}\bar{D}^{\ast}\Sigma_{c}$ doublet in case III larger than that of case II implies that the strength of the $\bar{D}\bar{D}^{\ast}$  potential affects the mass splitting. The mass splitting as a function of the cutoff of the $\bar{D}\bar{D}^{\ast}$  potential is presented in Fig.~\ref{mass splitting}.  It is obvious that the mass splitting increases with the strength of the $\bar{D}\bar{D}^{\ast}$  potential. Interestingly, the positive mass splitting means that the mass of the spin 3/2 $\bar{D}\bar{D}^{\ast}\Sigma_{c}$  bound state is larger than that of its spin 1/2 counterpart. While in this case the mass of the $J^{P}=\frac{1}{2}^{-}$ $\bar{D}^{\ast}\Sigma_{c}$ system   is larger than that of the $J^{P}=\frac{3}{2}^{-}$ $\bar{D}^{\ast}\Sigma_{c}$ system~\cite{Liu:2019zvb}. That is, the mass splitting of the three-body $\bar{D}\bar{D}^{\ast}\Sigma_{c}$ doublet  is oppositely correlated to the mass splitting of the two-body
$\bar{D}^{\ast}\Sigma_{c}$ bound states, which offers a non-trivial way to check the molecular nature of the involved states. 

\begin{table}[!h]
    \setlength{\tabcolsep}{7pt}
    \centering
   \caption{ Weights of each Jacobian channel in the $\bar{D}\bar{D}^* \Sigma_{c}$ system for Case II\label{weights} }
    \begin{tabular}{cccc}
    \hline\hline
     $I(J^{P})\quad$& $\bar{D}^*\Sigma_{c}-\bar{D}\quad$ & $\bar{D}\Sigma_{c}-\bar{D}^*\quad$ & $\bar{D}\bar{D}^*-\Sigma_{c}$   \\\hline 
     $1(\frac{1}{2}^{+})$ & 19.18 $\%$ & 52.42 $\%$ & 28.40 $\%$ \\
     $1(\frac{3}{2}^{+})$ & 78.39 $\%$  & 8.09 $\%$ &  13.52 $\%$ \\\hline\hline
    \end{tabular}
\end{table}  

 Such $\bar{D}\bar{D}^{\ast}\Sigma_{c}$ bound states  can be regarded as suppositions of three  quasi two-body bound states, $P_{c2}(P_{c3})\bar{D}$, $P_{c1}\bar{D}^{\ast}$, and $\bar{T}_{cc}\Sigma_{c}$.  The weight of each configuration equals the weight of  the corresponding Jacobian channel.  They are tabulated in Table~\ref{weights} for Case II.
We treated $P_{c2/c3}$, $P_{c1}$, and $\Sigma_{c}$ as unstable particles compared with the $\bar{D}$, $\bar{D}^{\ast}$, and $\bar{T}_{cc}$ particles~\cite{ParticleDataGroup:2020ssz,LHCb:2021auc} so that these unstable particles can further decay into  two other particles. The decay modes of the three-body bound states
$\bar{D}\bar{D}^{\ast}\Sigma_{c}$ into $J/\psi p \bar{D}$, $J/\psi p \bar{D}^{\ast}$, and $\Lambda_{c}\pi \bar{T}_{cc}$,  
are  shown  in Fig.~\ref{decay}.  
The $\bar{D}\bar{D}^{\ast}\Sigma_{c}$ states will not couple to a pair of traditional hadrons since its minimum number of valence quarks is seven. As a result, they can only  decay into at least three traditional hadrons, which means the decay mechanism shown in Fig.~\ref{decay} should play a dominant role.

\begin{figure*}[ttt]
\begin{center}
\subfigure[$ ~$]{
\begin{minipage}[t]{0.31\linewidth}
\centering
\includegraphics[width=2.0in]{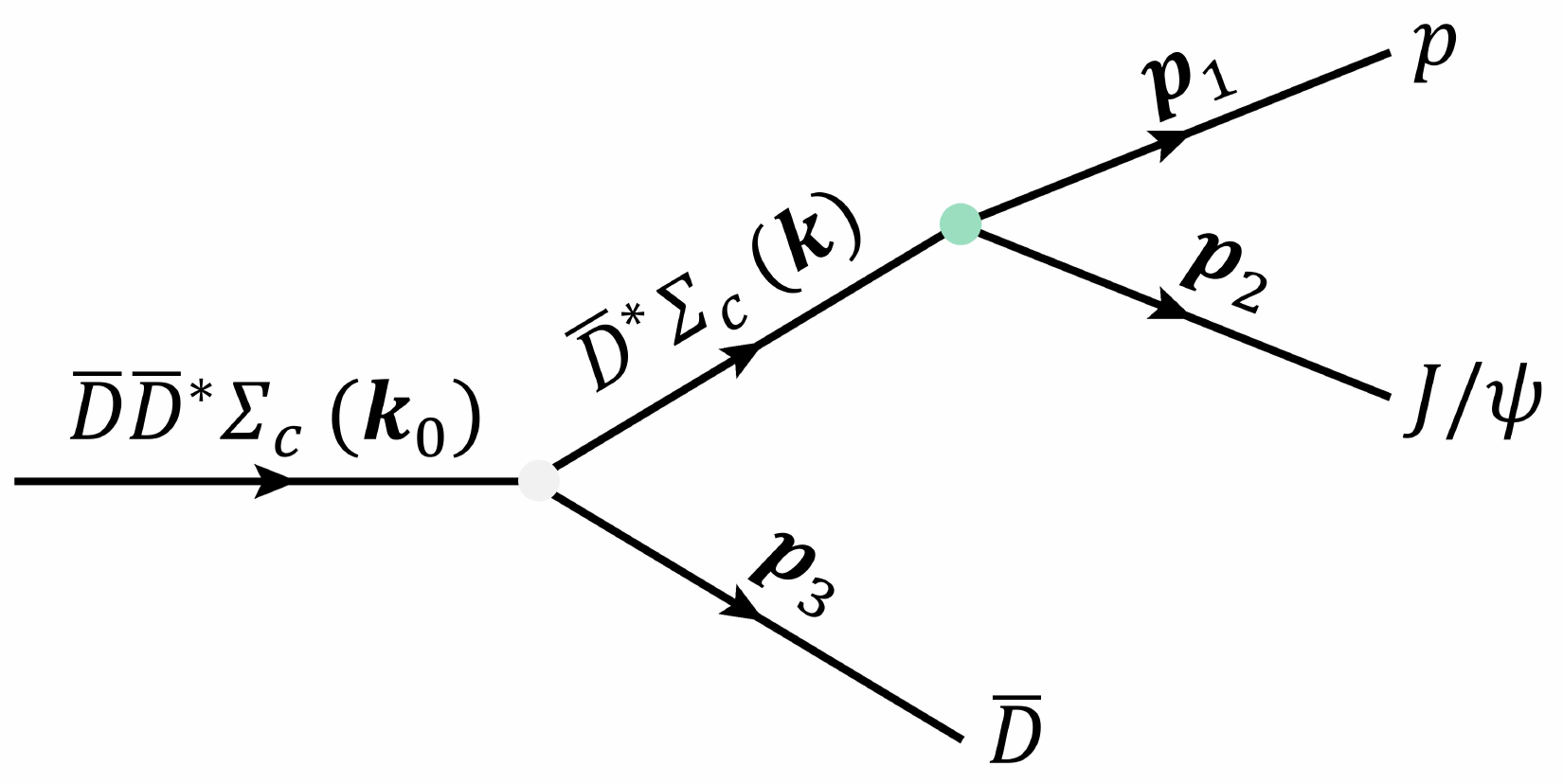}
\end{minipage}
}
\subfigure[$ ~$]{
\begin{minipage}[t]{0.31\linewidth}
\centering
\includegraphics[width=2.0in]{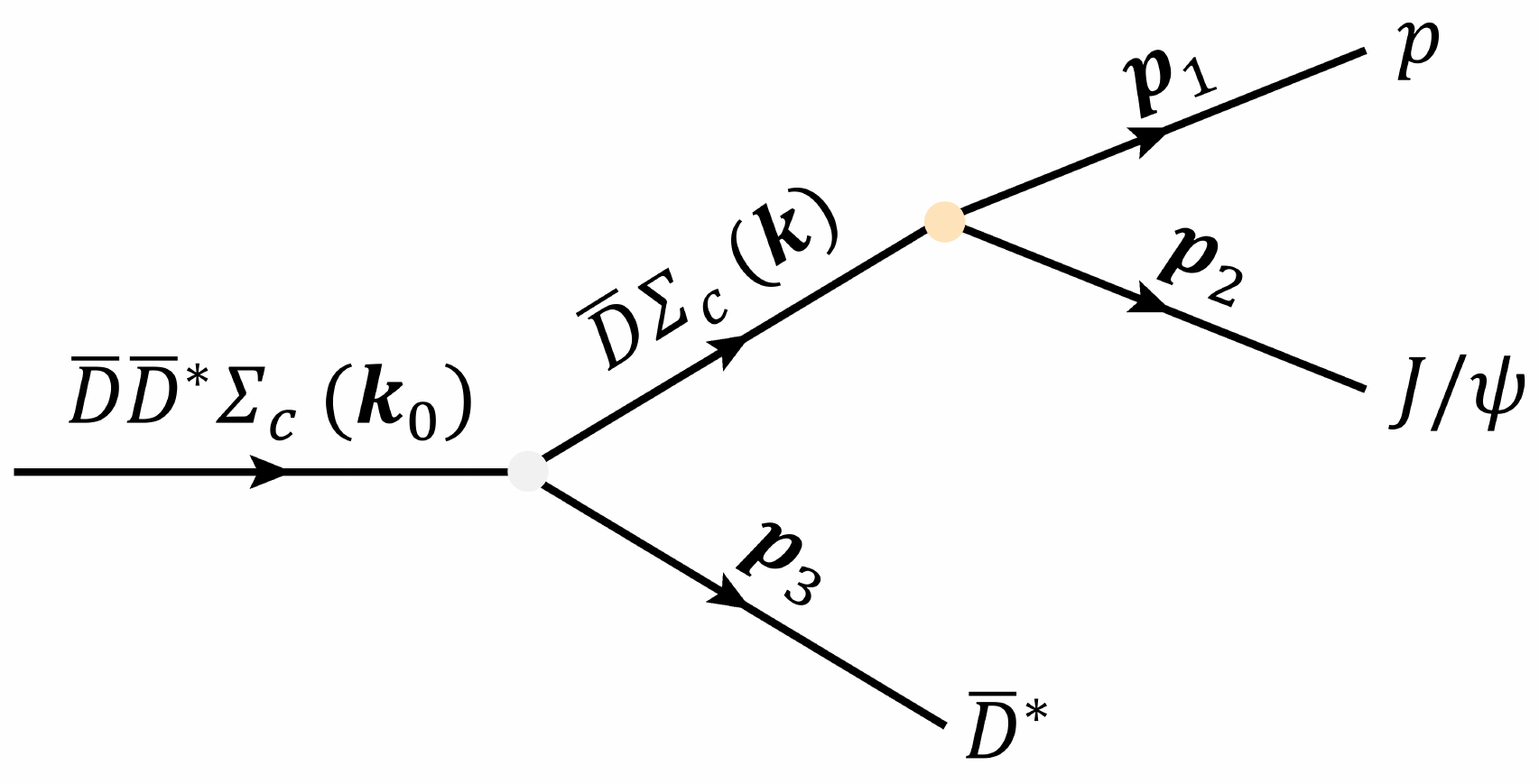}
\end{minipage}
}
\subfigure[$ ~$]{
\begin{minipage}[t]{0.31\linewidth}
\centering
\includegraphics[width=2.0in]{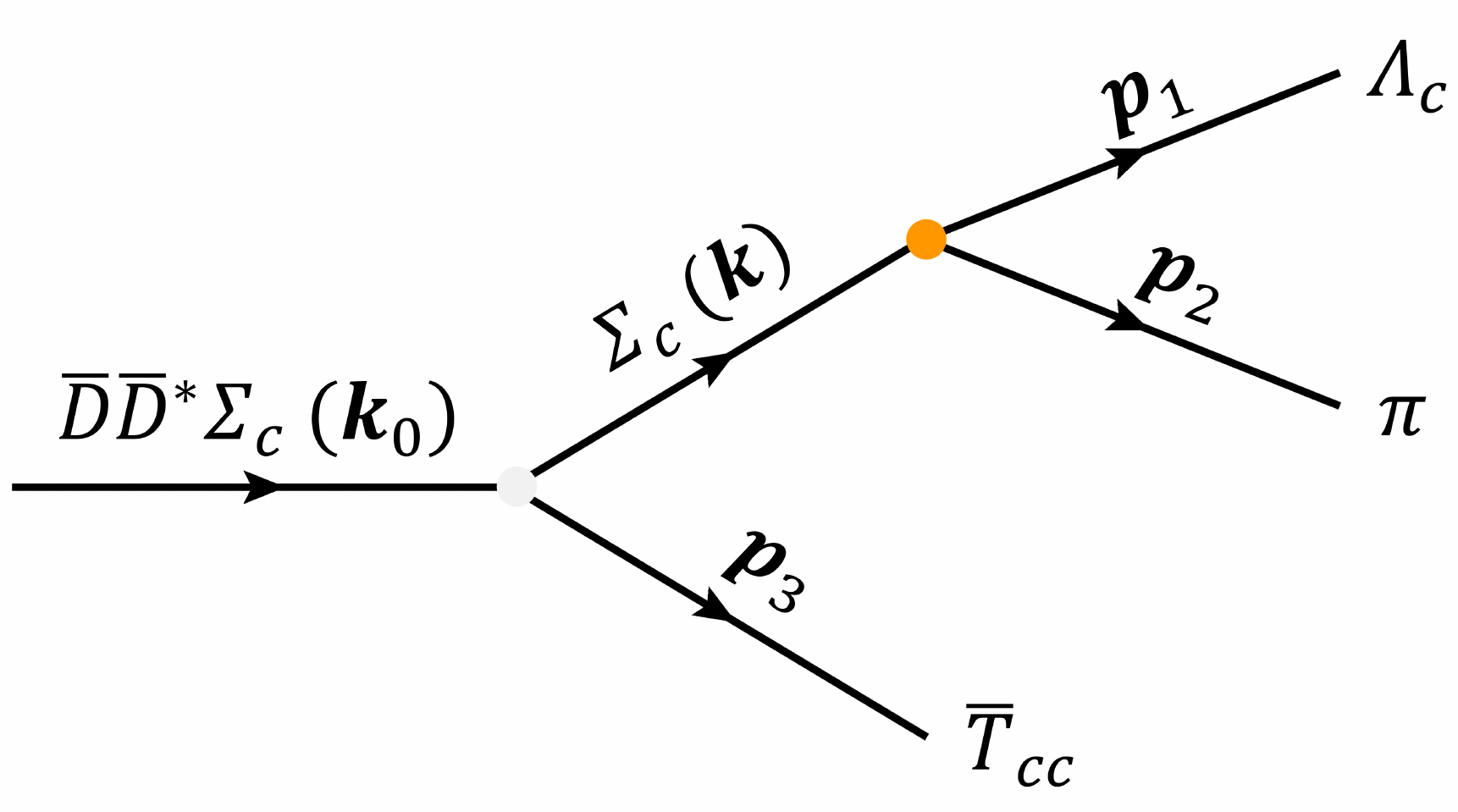}
\end{minipage}
}
\caption{ Tree level diagrams for the three-body   $\bar{D}\bar{D}^{\ast}\Sigma_{c}$ bound states decaying into $J/\psi p \bar{D}$(a), $J/\psi p \bar{D}^{\ast}$(b), and $\pi \Lambda_{c} \bar{T}_{cc}$(c).     \label{decay} }
\label{decay}
\end{center}
\end{figure*}


\begin{table}[!h]
\setlength{\tabcolsep}{5pt}
\centering
\caption{Partial decay widths of the $\bar{D}\bar{D}^{\ast}\Sigma_{c}$ molecules.   }
\label{results}
\scalebox{0.935}{
\begin{tabular}{ccccccc}
\hline\hline
\multirow{2}{*}{Modes}&\multicolumn{2}{c}{Fig.~\ref{decay}(a)} &\multicolumn{2}{c}{Fig.~\ref{decay}(b)} & \multicolumn{2}{c}{Fig.~\ref{decay}(c)}  \\
&$Hq_{1} \to J/\psi p\bar{D}  $& $Hq_{2}  \to J/\psi p  \bar{D}$ &$Hq_{1} \to J/\psi p \bar{D}^{\ast} $  &  $Hq_{2}  \to  J/\psi p \bar{D}^{\ast}$&$Hq_{1}  \to \bar{T}_{cc}\Lambda_{c}\pi$     &  $Hq_{2} \to \bar{T}_{cc}\Lambda_{c}\pi  $\\\hline
Value& 2.4 MeV  &  0.7 MeV& 2.0 MeV & 0.3 MeV& 0.05 keV& 0.5 keV    \\ 
  \hline\hline
\end{tabular}
}
\end{table}

The partial decay widths of the two three-body bound states, $Hq_{1}$ and $Hq_{2}$ are listed in Table~\ref{results},  where $Hq_{1}$ and $Hq_{2}$ denote $I(J^{P})=1(1/2^{+})$ and $I(J^{P})=1(3/2^{+})$ $\bar{D}\bar{D}^{\ast}\Sigma_{c}$ bound states, respectively. One can find that the partial widths of the bound state with $J=1/2$ decaying  into $J/\psi p \bar{D}$
and $J/\psi p \bar{D}^*$  are much larger than those of the  $J=3/2$ state, which indicates that the $J/\psi p\bar{D}$ and $J/\psi p\bar{D}^*$ decay modes can help us to discriminate the spins of the $Hq_{1}$ and $Hq_{2}$ molecules.  However, the partial widths of both  $Hq_{1}$ and $Hq_{2}$ decaying into $\bar{T}_{cc}\Lambda_{c}\pi$ are rather small  due to the small phase spaces. 
Note that there exist some uncertainties about the partial decay widths given in Table~\ref{results} because the partial widths of the three pentaquark states decaying into $J/\psi p$ are not precisely known~\cite{Xiao:2020frg}. However, we suggest to search for them in the $J/\psi p \bar{D}^{\ast}$ or $J/\psi p\bar{D}$ mass distributions.

\subsection{$D^{(*)}D^{(*)}D^{(*)}$ molecules}

The $T_{cc}(3875)^+$ discovered by the LHCb Collaboration can be understood as a $DD^*$ bound state~\cite{Liu:2019stu,Meng:2021jnw,Ling:2021bir,Dong:2021bvy,Feijoo:2021ppq,Du:2021zzh,Chen:2021vhg,Agaev:2021vur,Ren:2021dsi,Yan:2021wdl,Albaladejo:2021vln,Fleming:2021wmk}. Adding a charmed meson $D$ into the $DD^*$ system, it is meaningful to investigate the three-body system $DDD^*$, which carries the highest charm number. For such a three-body system, the Jacobi coordinates are shown in Fig.~\ref{3-jacobi}, where the symbols $N_{1}$, $N_{2}$, and $N_{3}$ denote $D$, $D$, and $D^*$, respectively. Using the GEM, we search for the likely $DDD^*$ bound state by solving the Schr\"{o}dinger equation. We take the OBE model to obtain the $DD$ and $DD^*$ interactions,  which will introduce a form factor $\frac{\Lambda^2 - m^2}{\Lambda^2 - q^2}$ with a cutoff $\Lambda$. The value of the cutoff is determined by reproducing the $T_{cc}(3875)^+$ as a $ D^*D$  bound state. Assuming the same cutoff in  $DD$ and $DD^*$ systems\footnote{ The form factors are generally embedded in the vertices of heavy meson couplings to light mesons, reflecting the composite nature of the heavy mesons and light mesons. The cutoff is determined by the finite size of the heavy mesons~\cite{Yasui:2009bz,Yamaguchi:2011xb}. In the heavy quark limit, the sizes of the $D$ and $D^*$ mesons are assumed to be the same thus the $DD$ and $DD^*$ systems share the same cutoff. }, we obtain the $DD$ and $DD^*$ interactions.   In addition, we consider the Coulomb interactions between the charmed mesons and then search for the  $DDD^*$ bound state.   
In Table~\ref{tab:resultsDDD*}, we present the binding energy of the $DDD^*$ system.  We obtain a $DDD^*$  bound state with a binding energy of around 1 MeV, denoted as $H_{ccc}$~\cite{Wu:2021kbu}. The quark contents of the $H_{ccc}$ state are the same as that of $\bar{D}T_{\bar{c}\bar{c}}$, which is a kind of hadronic molecule composed of a charmed meson and a doubly charmed tetraquark state. We find that the mass of  $\bar{D}T_{\bar{c}\bar{c}}$ is larger than that of  $H_{ccc}$. 
It is worth noting that the existence of the $\bar{D}T_{\bar{c}\bar{c}}$ molecule is dependent on the molecular nature of the pentaquark states~\cite{Pan:2022whr}, while the existence of $H_{ccc}$ is dependent on the molecular nature of $T_{cc}$~\cite{Wu:2021kbu}.  Such a state could, in principle, be observed with the upcoming LHC data and will unambiguously determine the nature of the $T_{cc}^+$ state.

\begin{table}[h!]
    \setlength{\tabcolsep}{7pt}
    \centering
    \caption{ Binding energies, RMS radii and Hamiltonian expectation values of the doubly charged $I(J^P)=\frac{1}{2}(1^-)$ $DDD^*$ state with $S$-wave OBE  and Coulomb interactions.  The uncertainties originate from finite heavy quark mass, 
    as explained in the main text.}\label{tab:resultsDDD*}
    \begin{tabular}{ccccccc}
    \hline\hline
          $\Lambda$(MeV) & $B$(MeV) & $r_{DD^{\ast}}$ & $r_{DD}$ & $\left \langle T \right \rangle$ & $\left \langle V_{DD^{\ast}} \right \rangle$ & $\left \langle V_{DD} \right \rangle$ \\ \hline
          976 & $0.16^{+0.01}_{-0.01}$ & $8.83^{+0.01}_{-0.23}$ & $10.74^{+0.09}_{-0.08}$ & $7.65^{+1.32}_{-0.52}$ & $-7.81^{+0.47}_{-1.27}$ & $-0.00^{+0.02}_{-0.07}$ \\
          998 & $1.09^{+0.17}_{-0.13}$& $4.50^{+0.83}_{-0.65}$ & $5.86^{+1.19}_{-0.92}$ & $23.65^{+2.85}_{-2.03}$ & $-24.14^{+2.45}_{-2.58}$ & $-0.60^{+0.31}_{-0.43}$ \\
          1013 & $2.22^{+0.27}_{-0.23}$ & $3.15^{+0.41}_{-0.33}$ & $4.04^{+0.59}_{-0.47}$ & $33.34^{+2.93}_{-2.76}$ & $-34.40^{+2.52}_{-2.62}$ & $-1.16^{+0.46}_{-0.56}$ \\ \hline\hline
    \end{tabular}\\
\end{table}

Considering HQSS, we also investigated the $D^{\ast}D^{\ast}D$ and $D^{\ast}D^{\ast}D^{\ast}$ systems~\cite{Luo:2021ggs}. Following the same strategy as Ref.~\cite{Wu:2021kbu}, two-body interactions are derived in the OBE model, and the cutoff $\Lambda$ is determined by reproducing the binding energy of $T_{cc}(3875)^+$. In addition, the $S$-$D$ mixing and coupled-channel effects are also considered. With the GEM, we find that the $I(J^P)=\frac{1}{2}(0^-,1^-,2^-)~D^{\ast}D^{\ast}D $ and $I(J^P)=\frac{1}{2}(0^-,1^-,2^-,3^-)~D^{\ast}D^{\ast}D^{\ast}$ systems can form loosely bound states with binding energies of a few MeV, which could be viewed as good hadronic molecular candidates. While, the  $I(J^P)=\frac{3}{2}(0^-,1^-,2^-)~D^{\ast}D^{\ast}D $ and $I(J^P)=\frac{3}{2}(1^-,2^-,3^-)~D^{\ast}D^{\ast}D^{\ast}$ systems are more difficult to bind. We suggest searching for these molecular states in the following decay modes: a) a double-charm molecular state and a charmed meson, b) three charmed mesons, and c) three charmed mesons with several pions and photons.

In Ref.~\cite{Bayar:2022bnc}, the authors investigated the $D^*D^*D^*$ system considering that two $D^*$'s generate a $I(J^P)=0(1^+)$ bound state~\cite{Dai:2021vgf}. By solving the Faddeev equation with the fixed center approximation, they found that the $D^*D^*D^*$ system is bound in the $I(J^P)=\frac{1}{2}(0^-,1^-,2^-)$ channels with binding energies of 53.1 MeV, 33.8 MeV, and 34.6 MeV. The absence of the $J^P=3^-$ and $I=\frac{3}{2}$ states is a consequence of the fact that the $D^*D^*$ binds only in $I(J^P)=0(1^+)$ in their model. In addition, the $I(J^P)=\frac{1}{2}(0^-)$ state has a width of $99.7$ MeV because of the $D^*D^* \to D^*D$ decays considered. It is interesting to see that in Ref.~\cite{Luo:2021ggs}, the authors mention that there is no bound-state solution with $\Lambda = 1$ GeV and $J^{P}=3^{-}$, while bound states are formed for the other $J^P$ configurations. Besides, assuming that $T_{cc}^{*}$, an isoscalar  heavy partner of the $T_{cc}(3875)^+$, exists below the $D^*D^*$ threshold, the  Efimov  effect may emerge in the $D^*D^*D^*$ system for reasonable molecular proportion and binding energy of the $T_{cc}^{*}$~\cite{Ortega:2024ecy}.

\begin{table}[!h]
\renewcommand{\arraystretch}{1}
\scalebox{0.83}{
\begin{threeparttable}
\centering
\caption{Summary for heavy-flavor three-body states. Energies are in units of MeV. \label{3body-research} }
\begin{tabular}{c cc c c}
  \hline \hline
    Components  &  $I(J^P)$ & Results (Method)   &  Decay modes    \\ \hline 
    $DNN$ & $\frac{1}{2}(0^-)$ & BS $\sim 3500-15i$ (FCA, V)~\cite{Bayar:2012dd} & $\Lambda_c \pi^- p, \Lambda_c p$~\cite{Bayar:2012dd} \\
    \makecell{$NDK,ND\Bar{K},$\\$ND\Bar{D}$} & $\frac{1}{2}(\frac{1}{2}^+)$ & BS $\sim 3050, 3150, 4400$ (FCA)~\cite{Xiao:2011rc} &  $\dagger$ \\
    $DD^*N$ & $\frac{1}{2}(\frac{1}{2}^+, \frac{3}{2}^+)$ & BS $\sim 4773.2, 4790.7 $ (GEM)~\cite{Luo:2022cun} & \makecell{$ T_{cc}p,DDp+\pi(\gamma),\Xi_{cc}+\pi(\gamma),$\\ charmed baryon + charmed meson}~\cite{Luo:2022cun} \\
    $DD^*N$ & $\frac{3}{2}(-)$ & difficult to form bound states  (GEM)~\cite{Luo:2022cun} & $\dagger$ \\
    $DK\Bar{K}$ & $\frac{1}{2}(0^{-})$  &\makecell{ $D$-like state $\sim 2845.5$ (FCA)~\cite{Debastiani:2017vhv},\\ $D$-like state $\sim 2900$ (QCDSR, $\chi$F)~\cite{MartinezTorres:2012jr} }&  $\pi\pi D$~\cite{Debastiani:2017vhv} \\
    $DKK$ & $\frac{1}{2}(0^{-})$ &no bound state (FCA)~\cite{Debastiani:2017vhv} &  $\dagger$ \\
    $\Bar{D}\Bar{K}\Sigma_c $ & $1(\frac{1}{2}^+)$ & BS $\sim 4738.6$ (GEM)~\cite{Wu:2021gyn} &  $D \Xi' , D_{s}\Sigma_{c}$~\cite{Wu:2021gyn} \\
    $D^{(*)} ~\mathrm{multi}~\rho$ & ... & several $D^{(*)}_{J}$ states (FCA)~\cite{Xiao:2012dw,Xiao:2016kak} &  $\dagger$ \\
    $\rho D\Bar{D}$ & $0(?),1(?)$ &BS $\sim 4241-10i,[4320-13i,4256-14i] $ (FCA)~\cite{Durkaya:2015wra} &  $\dagger$ \\
    $DDK$ &$\frac{1}{2}(0^{-})$& \makecell{ BS $\sim 4162$ (GEM)~\cite{Wu:2019vsy}, $4140$ ($\chi$F)~\cite{MartinezTorres:2018zbl},\\ $4160$ (FV)~\cite{Pang:2020pkl}$\qquad\qquad\quad$}  &  $DD_{s}^{*}, D^* D_s$~\cite{Huang:2019qmw} \\
    $D\Bar{D}K$ &$\frac{1}{2}(0^{-})$& BS $\sim 4181.2$ (GEM)~\cite{Wu:2020job}, $4191$ (FCA)~\cite{Wei:2022jgc} & $D_s \Bar{D}^* , J/\psi K$~\cite{Wu:2020job} \\
     $DD^{*}K$ &  $\frac{1}{2}(1^{-})$ & BS $\sim 4317.9$ (BO)~\cite{Ma:2017ery}, $4364.4$ (GEM)~\cite{Tan:2024omp} & $\dagger$  \\
    $D\Bar{D}^{*}K$ & $\frac{1}{2}(1^{-})$ & \makecell{BS $\sim 4294.1$ (GEM)~\cite{Wu:2020job}, $4317.9$ (BO)~\cite{Ma:2017ery}, \\$4307$ (FCA)~\cite{Ren:2018pcd}$~~\qquad\qquad\qquad$ }&  $D^{(*)}_{s}\Bar{D}^{(*)}, J/\psi K^*$~\cite{Ren:2019umd,Ma:2017ery} \\
    $D^{*}D^{*}\Bar{K}^{*}$ &  $\frac{1}{2}(0^{-},1^{-},2^{-})$ & \makecell{$4845-40i,\qquad\qquad\qquad\qquad$\\ BS $\sim[4850-46i,4754-50i],$ (FCA)~\cite{Ikeno:2022jbb}\\ $[4840-43i,4755-50i]\quad\qquad$ } & \makecell{$D^{*}D^{*}\Bar{K}^{*},$\\ $ D^{*}D^{(*)}\Bar{K}^{*},$\\$ [D^{*}D^{*}\Bar{K}^{*},D^{*}D^{(*)}\Bar{K}^{*}] $}~\cite{Ikeno:2022jbb} \\
     $\Bar{D}\Bar{D}^{*}\Sigma_{c}$ & $1(\frac{1}{2}^+, \frac{3}{2}^+)$ & BS $\sim 6292.3, 6301.5$ (GEM)~\cite{Pan:2022xxz} &   $J/\psi p \Bar{D}^{(*)}, \Bar{T}_{cc}\Lambda_{c}\pi$~\cite{Pan:2022xxz} \\
     $J/\psi K\Bar{K}$ & $0(1^-)$ & $Y(4260 )\sim 4150-45i$ ($\chi$F)~\cite{MartinezTorres:2009xb} & $\dagger$ \\
       $DDD^{*}$ & $\frac{1}{2}(1^{-})$ & BS $\sim 5742.2$ (GEM)~\cite{Wu:2021kbu} & $DDD\pi (\gamma)$\cite{Wu:2021kbu} \\
       $D\bar{D}D^{*}$ & $\frac{1}{2}(1^{-})$ & BS $\sim 5734.5$ (GEM)~\cite{Tan:2024omp} & $\dagger$ \\
       $DD^{*}D^{*}$ & $\frac{1}{2}(0^-,1^-,2^-)$ & several loosely bound states  (GEM)~\cite{Luo:2021ggs} &  charmed mesons + ...~\cite{Luo:2021ggs} \\
       $D^{*}D^{*}D^{*}$ & \makecell{$\frac{1}{2}(0^-,1^-,2^-,3^-)$\\$\frac{1}{2}(0^-,1^-,2^-)$} & \makecell{several loosely bound states  (GEM)\cite{Luo:2021ggs}\\ BS $\sim 5790.9-49.8i,5990.2,5989.4$ (FCA)~\cite{Bayar:2022bnc} }& charmed mesons + ...~\cite{Luo:2021ggs} \\
       $D^{*}D^{*}D^{(*)}$ & $\frac{3}{2}(-)$ & difficult to form bound states  (GEM)~\cite{Luo:2021ggs} &  $\dagger$ \\
       $D^*D^*\bar{D}$&$\frac{1}{2}(2^-)$& BS $\sim 5879$ (F)~\cite{Valderrama:2018sap} & $\dagger$\\
       $D^*D^*\bar{D}^*$&$\frac{1}{2}(3^-)$& BS $\sim 6019$ (F)~\cite{Valderrama:2018sap} & $\dagger$\\
       $\Omega_{ccc}\Omega_{ccc}\Omega_{ccc}$ & $?(\frac{3}{2}^+)$ &  no bound state (GEM)~\cite{Wu:2023eyd} &  $\dagger$ \\
        $\Xi_{cc}\Xi_{cc}\Bar{K}$ & $\frac{1}{2}(0^-)$ & BS $\sim 7641.8$ (GEM)~\cite{Wu:2021ljz} &   $\dagger$ \\
\hline \hline
\end{tabular}
\begin{tablenotes}[flushleft]
       \item {\footnotesize\textbf{Meaning of the abbreviations:}  BS - bound state, GEM - Gaussian expansion method, F - Faddeev equation, BO - Born-Oppenheimer approximation, FCA - Fixed center approximation, V - Variational approach, QCDSR - QCD sum rule, $\chi$F - chiral Faddeev, FV - finite volume}
    \end{tablenotes}
\end{threeparttable}
}
\end{table}

\begin{table}[!h]
\renewcommand{\arraystretch}{1}
\scalebox{1}{
\begin{threeparttable}
\centering
\caption{Summary for heavy-flavor three-body states. Energies are in units of MeV. \label{3body-research-b} }
\begin{tabular}{c cc c}
  \hline \hline
    Components  &  $I(J^P)$ & Results (Method)   &  Decay modes    \\ \hline 
    $BD\Bar{D},BDD$  & $\frac{1}{2}(1^{-})$ & BS $\sim 8955$, probable BS $\sim 8960$  (FCA)~\cite{Dias:2017miz} &  $\dagger$ \\
    \makecell{$DB^{*}\Bar{B}^{*},DB\Bar{B},$\\$D^{*}B^{*}\Bar{B}^{*},D^{*}B\Bar{B}$}  & $\frac{1}{2}(?)$ & BS $\sim 12384,12294,12520,12430$ (FCA)~\cite{Dias:2018iuy}  &  $\dagger$ \\
    $BBB^{*}$ & ... & probable BS (BO)~\cite{Ma:2018vhp} &   $\dagger$ \\
    $BB^{*}B^{*}-B^{*}B^{*}B^{*}$ & $\frac{1}{2}(2^-)$ & BS $\sim 15658$ (F)~\cite{Garcilazo:2018rwu} &  $\dagger$ \\
    $\rho B^{*}\Bar{B}^{*}$ & $1(3^-)$& a state $\sim 10987-40i$(FCA)~\cite{Bayar:2015zba} & $\dagger$ \\
    $BB^{*}\Bar{K},B\bar{B}^{*}\Bar{K}$ & $\frac{1}{2}(1^-)$& BS $\sim 11014$ (BO)~\cite{Ma:2017ery} & $\dagger$ \\
    $BB\Bar{K}$ & $\frac{1}{2}(0^-)$ & BS $\sim 10945$ (GEM)~\cite{Wu:2021ljz} & $\dagger$ \\
    $B\Bar{B}\Bar{K}$ & $\frac{1}{2}(0^-)$& BS $\sim 10659$ (FCA)~\cite{Ren:2018qhr} & $\dagger$ \\
    $B^{*}\Bar{B}^{*}\Bar{K}$ & $\frac{1}{2}(2^-)$& BS $\sim 10914$ (FCA)~\cite{Ren:2018qhr} & $\dagger$ \\
    $B\Bar{B}\Bar{K}^{*}$ & $\frac{1}{2}(1^-)$& BS $\sim 11002, 11264$ (FCA)~\cite{Ren:2018qhr} & $\dagger$ \\
    $B^{*}\Bar{B}^{*}\Bar{K}^{*}$ & $\frac{1}{2}(3^-)$& BS $\sim 11078, 11339$ (FCA)~\cite{Ren:2018qhr} & $\dagger$ \\
    $\Bar{B}^{*}\Bar{B}^{*}\Bar{K}^{*}$ & $\frac{1}{2}(0^-,1^-,2^-)$& BS $\sim 11413-88i, 11405-130i, 11491-126i $ (FCA)~\cite{Bayar:2023itf} & $\dagger$ \\
      $B^{*}B^{*}\Bar{K}$ & $\frac{1}{2}(0^+,1^+,2^+)$ &  BS $\sim 11040$ (F)~\cite{Valderrama:2018knt} &$\dagger$\\
      $\Xi_{bb}\Xi_{bb}\Bar{K}$ & $\frac{1}{2}(?)$ &  BS $\sim 20608, 20667$ (F)~\cite{Valderrama:2018knt} &$\dagger$\\
    $\Omega_{bbb}\Omega_{bbb}\Omega_{bbb}$ & $?(\frac{3}{2}^+)$ &   no bound state (GEM)~\cite{Wu:2023eyd} &  $\dagger$\\

\hline \hline
\end{tabular}
\begin{tablenotes}[flushleft]
       \item {\footnotesize\textbf{Meaning of the abbreviations:}  BS - bound state, GEM - Gaussian expansion method, F - Faddeev equation, BO - Born-Oppenheimer approximation, FCA - Fixed center approximation}
    \end{tablenotes}
\end{threeparttable}
}
\end{table}

\subsection{Other heavy-flavor three-body molecules}
Here, we briefly summarize studies of other heavy-flavor three-body bound states. With the same method mentioned above, i.e., GEM, many three-body systems were also studied based on the two-body interactions determined by experimentally observed states. In Ref.~\cite{Wu:2021gyn}, the $\bar{D}\bar{K}$ and $\bar{D}\Sigma_c$ interactions were determined by the  $\bar{D}_{s0}^*(2317)$ and $P_{c}(4312)$ states, and the $\Sigma_c \bar{K}$ interaction was related to the $N \bar{K}$ one via chiral symmetry.  The three-body $\Sigma_c \bar{D}\bar{K}$ system was found to form a bound state, named $P_{cs}^* (4739)$, with quantum numbers $I(J^P ) = 1(1/2^+)$ and a binding energy of about 77.8 MeV. They also found that the $P_{cs}^* (4739)$ state can decay into $\Xi' \bar{D}$ and $\bar{D}^*_s \Sigma_c$ with partial decay widths of a few tens of MeV. In Ref.~\cite{Luo:2022cun}, the $DD^*$, $DN$, and $D^*N$ interactions were described in the OBE model associated with the observed $T_{cc}^+$, $\Sigma_{c}(2800)$, and $\Lambda_{c}(2940)$ states. Two $DD^* N$ bound states with $I(J^P) = \frac{1}{2}(\frac{1}{2}^+)$ and $\frac{1}{2}(\frac{3}{2}^+)$ were found with only $S$-wave pairwise interactions. The conclusion remains unchanged, considering the $S$-$D$ mixing and coupled-channel effects. On the contrary, the $I(J^P) = \frac{3}{2}(\frac{1}{2}^+)$ and $\frac{3}{2}(\frac{3}{2}^+)$ $DD^* N$ systems are difficult to bind. Utilizing heavy quark symmetry, Ref.~\cite{Wu:2021ljz} studied the $\Xi_{cc}\Xi_{cc}\Bar{K}$ and $BB\Bar{K}$ with $I(J^P)=\frac{1}{2}(0^-)$. The $B(\Xi_{cc})\Bar{K}$ two-body interactions were derived in chiral perturbation theory, and the interactions between two identical particles were described in the OBE model. The $\Xi_{cc}\Xi_{cc}\Bar{K}$ and $BB\Bar{K}$ bound states with binding energies of about 92 MeV and 109 MeV were found. 

In the fixed center approximation (FCA) to the Faddeev equation, a three-body system is treated as a third particle scattering with a two-body cluster, which can be associated with an experimentally observed state. Due to its simplicity, it has been employed to study many three-body systems containing heavy quarks. Considering the $\Lambda_c(2595)$ resonance as a $DN$ cluster, the scattering among 
two nucleons and a $D$ meson generates a $I(J^P)=\frac{1}{2}(0^-)$ $DNN$ bound state with a mass of about 3500 MeV and a width of 20 $\sim$ 40 MeV~\cite{Bayar:2012dd}. A variational method (V) was employed to study the $DNN$ system ending with similar results~\cite{Bayar:2012dd}. In Ref.~\cite{Xiao:2011rc}, considering the scattering of a nucleon on the $D\Bar{D}$ cluster, which corresponds to the hidden-charm resonance $X(3700)$, three relatively narrow bound or quasibound $I(J^P)=\frac{1}{2}(\frac{1}{2}^+)$ $NDK$, $ND\Bar{K}$, $ND\Bar{D}$ states with energies of 3050 MeV, 3150 MeV, and 4400 MeV were found. Considering the $D_1(2420)$ as a $\rho D$ cluster, and following the same strategy, the $\rho D\Bar{D}$ and $D$-multi$\rho$ systems were  studied~\cite{Durkaya:2015wra, Xiao:2016kak}. In Ref.~\cite{Durkaya:2015wra}, one $I=0$ $\rho D\Bar{D}$ bound state with a mass of around 4241 MeV and a width of about 20 MeV was found. While in isospin $I=1$, two bound states were found, one with a mass around 4320 MeV and a width about 25 MeV in the $\rho - X(3700)$ scattering, and the other with a mass around 4256 MeV and a width about 28 MeV in the $\Bar{D} - D_1(2420)$ scattering.   In Ref.~\cite{Xiao:2016kak}, assuming the $f_{2}(1270)$ as a $\rho\rho$ cluster, several $D_{J}$-like states were predicted in the $D$-multi$\rho$ few-body systems. Several $D^{*}_{J}$-like states were also predicted  in the $D^*$-multi$\rho$ few-body systems assuming $D^*_{2}(2460)$ as a  $D^* \rho$ cluster~\cite{Xiao:2012dw}. Assuming $T_{cc}$ and $X(2900)$ states are $D^* D^*$ and $D^* \Bar{K}^*$ bound states, several $D^* D^* \Bar{K}^*$ bound states were obtained with quantum numbers $I(J^P)=\frac{1}{2}(0^-,1^-,2^-)$, one state for $J^P=0^-$, two states for $J^P=1^-,2^-$, with binding energies of $56\sim 151$ MeV and widths of $80\sim 100$ MeV~\cite{Ikeno:2022jbb}. The $BD$ system analogous to the bound $DK$ system was found in Ref.~\cite{Sakai:2017avl}. Therefore, Ref.~\cite{Dias:2017miz} studied the $BD\Bar{D}$ and  $BDD$ systems considering the multiple rescattering of the $D(\Bar{D})$ meson with the $BD$ cluster. They obtained a $BD\Bar{D}$ bound state with a mass of about $8928\sim 8985$ MeV. For $BDD$, some clues of a bound state in the energy region $8935 \sim 8985$ MeV were found, and the results are sensitive to the theoretical uncertainties. Based on the bound $B\Bar{B}$ and $B^{*}\Bar{B}^{*}$ systems in isospin $I = 0$ ~\cite{Ozpineci:2013zas}, Ref.~\cite{Dias:2018iuy} studied the $D^{(*)}B^{(*)}\Bar{B}^{(*)}$ systems considering that the clusterized $B\Bar{B}(B^{*}\Bar{B}^{*})$ systems interact with a third particle $D^{(*)}$. As a result, they found four three-body bound states $DB^{*}\Bar{B}^{*}$, $D^{*}B^{*}\Bar{B}^{*}$, $DB\Bar{B}$, and $D^{*}B^{*}\Bar{B}^{*}$ with binding energies of about $20 \sim 30$ MeV. In addition, they found resonant bumps above the $D^{(*)}[B^{(*)}\Bar{B}^{(*)}]$ thresholds with widths of about 10 MeV. Again, the results were sensitive to the theoretical uncertainties. 

In addition, other methods for heavy-flavor three-body systems exist. In Ref.~\cite{MartinezTorres:2012jr},  a $D$-like meson with a mass of 2890 MeV and a width of about 55 MeV was predicted treating $DK\Bar{K}$ as $Df_{0}(980)$ in two methods, i.e., QCD sum rule and chiral Faddeev equation.  Similar results were  obtained in Ref.~\cite{Debastiani:2017vhv} by the FCA, and evidence for a $I(J^P)=\frac{1}{2}(0^-)~DK\Bar{K}$ state with a mass of about $2833\sim 2858$ MeV, mainly made of $Df_{0}(980)$, was found. In Ref.~\cite{MartinezTorres:2009xb}, they studied the $J/\psi \pi\pi$ and $J/\psi K\Bar{K}$ coupled channels solving the chiral Faddeev equations to investigate the existence of the $J^{PC}=1^{--}$, $Y(4260)$ resonance. They obtained a peak of around 4150 MeV with a width of about 90 MeV. All heavy-flavor three-body states studied are collected in Table~\ref{3body-research}  and Table~\ref{3body-research-b}.

Lattice QCD simulations have also been widely employed in studies of three-body systems, but most of them focused on the light-flavor sector~\cite{Draper:2023boj, Blanton:2021mih, Romero-Lopez:2021zdo, Romero-Lopez:2022usb, Horz:2022glt,  Mai:2018djl, Mai:2021nul, Blanton:2021llb, Alexandru:2020xqf, Mai:2021lwb, Bulava:2022ovd, Draper:2023xvu, Hansen:2019nir}. In the heavy-flavor sector, very few studies have been performed~\cite{Pang:2020pkl}. We hope to see more lattice QCD studies of heavy-flavor three-body systems.

We must note that the studies of three-body hadronic systems can not only help reveal the nature of the relevant sub-two-body interactions such that one can verify the nature of two-body hadronic molecules but also are essential by themselves because the three-body molecules are the beginning of new kinds of periodic tables composed of (unstable) mesons and hadrons, beyond what is well known in nuclear and hypernuclear physics.  This perspective has recently been emphasized in, e.g., Ref.~\cite{Wu:2022ftm}. In particular, it has been shown that four-body systems of kaons and $D$ mesons are likely to exist~\cite{Wu:2019vsy,Pan:2023zkl}.

\section{Femtoscopic correlation functions for exotic hadrons}\label{sec:CF}
Traditionally, hadron-hadron interactions were studied in scattering experiments. Using such techniques, we have learned a lot about the nuclear force. However, such experiments are difficult for unstable particles because of their short lifetime and the lack of suitable targets. In the last few years, femtoscopy, which measures two particle momentum correlation functions in high-energy proton-proton (pp), proton-nucleus (pA), and nucleus-nucleus (AA) collisions, has made remarkable progress in probing the strong interactions between various pairs of hadrons~\cite{Fabbietti:2020bfg}. Historically, the femtoscopic technique can be traced back to the 1950s, when Hanbury Brown and Twiss used photon interferometry to measure the apparent angular diameter of stars~\cite{HanburyBrown:1956bqd}. Subsequently, this method was applied in heavy-ion collisions and used to help explore properties like the size of the emitting source or the time dependence of the emission process~\cite{Goldhaber:1960sf, Gyulassy:1979yi, Zajc:1984vb, Fung:1978eq, Wiedemann:1999qn}. In the earlier studies, the efforts focused on the analysis of $\pi$ or $K$ pairs, where the quantum statistics and Coulomb force determine the behavior of the correlations. 

Thanks to the small size of the particle emission source, the abundant rare hadrons produced in relativistic heavy-ion collisions, and the excellent capabilities of detectors to identify particles and measure their momenta, the measurements of momentum correlation functions have become possible in revealing the precise dynamics of the strong interactions between pairs of hadrons. Recently, the strong interactions among $\pi^\pm K^\pm$~\cite{ALICE:2020mkb}, $\pi^\pm K_S^0$~\cite{ALICE:2023eyl}, $K_S^0K_S^0$~\cite{ALICE:2021ovd}, $K_S^0K^\pm$~\cite{ALICE:2017jto, ALICE:2018nnl, ALICE:2021ovd}, $K^{\pm}p$~\cite{ALICE:2019gcn, ALICE:2022yyh}, $\phi p$~\cite{ALICE:2021cpv}, $K^{\pm}\Lambda$~\cite{ALICE:2020wvi, ALICE:2023wjz}, $pp$~\cite{ALICE:2018ysd}, $p\Lambda^-$~\cite{ALICE:2018ysd, ALICE:2021njx}, $p\Sigma^0$~\cite{ALICE:2019buq, ALICE:2021njx}, $p\Xi^-$~\cite{ALICE:2019hdt}, $p\Omega^-$~\cite{ALICE:2020mfd}, $\Lambda\Lambda$~\cite{ALICE:2018ysd, ALICE:2019eol}, $\Lambda\Xi^-$~\cite{ALICE:2022uso}, and baryon-antibaryon~\cite{ALICE:2019igo, ALICE:2021cyj} pairs have been determined by the ALICE Collaboration at the LHC. Meanwhile, the strong interactions among $p\Xi^-$~\cite{Isshiki:2021bqh}, $p\Omega^-$~\cite{STAR:2018uho}, $\Lambda\Lambda$~\cite{STAR:2014dcy, Isshiki:2021bqh}, $\Xi^-\Xi^-$~\cite{Isshiki:2021bqh} and antiproton-antiproton~\cite{STAR:2015kha} pairs have been measured by the STAR Collaboration at RHIC. More recently, the femtoscopic technique has been used to understand the genuine three-body interaction for $ppK^\pm$~\cite{ALICE:2023gxp}, $ppp$~\cite{ALICE:2022boj}, and $pp\Lambda$~\cite{ALICE:2022boj}. In addition, it is worthwhile noting that the recent measurement of the $pD^-$, $\pi^\pm D^\pm$, and $K^\pm D^{(*)\pm}$ correlation functions demonstrated the potential to access the charm sector in experiments~\cite{ALICE:2022enj, ALICE:2024bhk}.

On the other hand, the femtoscopy studies have also triggered a large number of related theoretical studies~\cite{Morita:2014kza, Morita:2016auo, Ohnishi:2016elb, Morita:2019rph, Kamiya:2019uiw, Haidenbauer:2018jvl, Ogata:2021mbo, Kamiya:2021hdb, Haidenbauer:2021zvr, Liu:2022nec, Molina:2023jov, Molina:2023oeu, Sarti:2023wlg} in the light quark (u, d, s) sector. The authors in Ref.~\cite{Molina:2023jov} evaluated the $K\Sigma-K\Lambda-\eta N$ interactions in the chiral unitary approach. They demonstrated that the corresponding correlation functions could illuminate the relation between the $N^*(1535)$ state and these coupled channels, emphasizing the need to analyze the data in terms of coupled channels to avoid misleading results. In Ref.~\cite{Molina:2023oeu}, we conducted a model-independent analysis of the $\bar{K}^0K^+$ correlation functions obtained from pp collisions at 13 TeV. The data imply the existence of the $a_0$ resonance, namely, $a_0(980)$. Detailed discussions of the correlation functions involving heavy charmed and bottom quarks are given in the following subsections.

\subsection{Experimental and theoretical basics of Femtoscopy}
The critical observable in Femtoscopy is the momentum correlation function. It is defined as the ratio of the Lorentz-invariant two-particle spectrum to the product of single-particle inclusive momentum spectra~\cite{Koonin:1977fh, Lednicky:1981su, Bauer:1992ffu, Heinz:1999rw, Lisa:2005dd, Wiedemann:1999qn, ExHIC:2017smd},
\begin{align}\label{Eq:def}
  C(\boldsymbol{p}_1,\boldsymbol{p}_2)=\frac{E_1E_2{\rm d}N_{12}/({\rm d}^3p_1{\rm d}^3p_2)}{(E_1{\rm d}N_1/{\rm d}^3p_1)\cdot(E_2{\rm d}N_2/{\rm d}^3p_2)}=\frac{P(\boldsymbol{p}_1,\boldsymbol{p}_2)}{P(\boldsymbol{p}_1)\cdot P(\boldsymbol{p}_2)},
\end{align}
where $\boldsymbol{p}_i$ ($i=1,2$) is the momentum of each particle, and $E_i$ ($i=1,2$) is the energy of particle $i$. The above definition can also be regarded as the ratio of the probability of simultaneously measuring two particles with momenta $\boldsymbol{p}_1$ and $\boldsymbol{p}_2$ to the product of the single-particle probabilities.

Experimentally, the correlation function can be obtained rather straightforwardly  using the so-called mixed-event technique~\cite{Fabbietti:2020bfg}, which is computed as,
\begin{align}\label{Eq:exp}
  C(k)=\mathcal{N}\frac{N_{\rm same}(k)}{N_{\rm mixed}(k)},
\end{align}
where the relative momentum of two particles $k=|\boldsymbol{p}_1-\boldsymbol{p}_2|/2$, $N_{\rm same}(k)$ and $N_{\rm mixed}(k)$ represent the same and different event $k$ distributions, respectively. The normalization constant $\mathcal{N}$, which denotes the corrections of experimental effects\footnote{ The  $\mathcal{N}$ term is the correction to the experimental effects: the normalization of the $k^*$ distribution of pairs from mixed events,  the effects produced by finite experimental resolution,  and the influence of residual correlations. The correction dependence on the momentum is estimated experimentally.  As shown in Ref.~\cite{ALICE:2020mfd},  the finite experimental momentum resolution modifies the measured correlation functions by at most $8\%$ at low momentum $k$.   }, is usually evaluated in the high-momentum region (such as $k\in[500, 800]$ MeV/c),    where the impact of final-state interactions is negligible, and thus the correlation function approaches unity.   Since the number of pairs available from mixed events is much higher than the number of pairs produced in the same collision used in ${N_{\rm same}(k)}$, the mixed-event distribution, ${N_{\rm mixed}(k)}$, has to be scaled down. It is commonly expected that the correlated pairs are no longer correlated to each other in the high energy region $500<k<800$~MeV, which dictates that the correlation function approaches $C(k)=1$ and further determines the size of the term   $\mathcal{N}$. 

Theoretically, the two-particle momentum correlation function can be computed by the Koonin–Pratt (KP) formula~\cite{Koonin:1977fh, Pratt:1990zq},
\begin{subequations}
  \begin{align}
    C(\boldsymbol{p}_1,\boldsymbol{p}_2)
    &=\frac{\int{\rm d}^4x_1{\rm d}^4x_2~S_1(x_1,\boldsymbol{p}_1)S_2(x_2,\boldsymbol{p}_2)~|\Psi^{(-)}(\boldsymbol{r},\boldsymbol{k})|^2}{\int{\rm d}^4x_1{\rm d}^4x_2~S_1(x_1,\boldsymbol{p}_1)S_2(x_2,\boldsymbol{p}_2)}\label{Eq:KP1}\\
    &\simeq\int{\rm d}\boldsymbol{r}~S_{12}(r)~|\Psi^{(-)}(\boldsymbol{r},\boldsymbol{k})|^2,\label{Eq:KP2}
  \end{align}
\end{subequations}
where $S_i(x_i,\boldsymbol{p}_i)~(i=1,2)$ is the single-particle source function of particle $i$. $\Psi^{(-)}$ denotes the relative wave function with the relative coordinate $\boldsymbol{r}$ and momentum $\boldsymbol{k}$, in which the effects of final-state interactions are embedded. Eq.~\eqref{Eq:KP2} can be derived by integrating the c.m. coordinates from Eq.~\eqref{Eq:KP1} neglecting the time difference of the particle emission and the momentum dependence of the source function.    The resonance source model, introduced by the ALICE Collaboration, has seen widespread applications in femtoscopic analyses as a data-driven approach~\cite{ALICE:2021cpv,ALICE:2019buq,ALICE:2020mfd,ALICE:2021njx,ALICE:2021cyj,ALICE:2022enj}. This model is rooted in the assumption that all hadrons in small collision systems originate from a shared emission source. In this framework, the source comprises a Gaussian core responsible for emitting all primordial particles, augmented by an exponential tail stemming strongly decaying resonances~\cite{ALICE:2020ibs,ALICE:2023sjd}. Notably, the core size, denoted as $R$, exhibits a distinct scaling with the transverse mass ($m_{\rm T}$)~\cite{ALICE:2020ibs,ALICE:2023sjd}, a phenomenon often linked to collective effects like radial flow, as previously documented in heavy-ion collisions. This scaling has been accurately replicated in the single-particle emission model CECA by incorporating spatial-momentum correlations~\cite{Mihaylov:2023pyl}, utilizing both thermal and transport models to estimate resonance yields and delineate decay kinematics~\cite{Pierog:2013ria,Andronic:2017pug}. Considering the negligible impact of strong decays on heavy flavor hadrons, as evidenced by the $\bar{D}N$ femtoscopic analysis~\cite{ALICE:2022enj}, it is reasonable to adopt a Gaussian source function $S_{12}(r)=\exp\left[-r^2/(4R^2)\right]/(2\sqrt{\pi}R)^3$ in the first theoretical studies. In future experiments, the source size $R$ can be accurately determined by measuring the $m_{\rm T}$ of the particle pair of interest and anchoring it to the proton-proton correlation data.   Due to the dominant role of $S$-wave interactions in the low-momentum region, only the $S$-wave final-state interactions are usually assumed to modify the relative wave function. For a non-identical two-particle system experiencing only strong interactions, the relative wave function in the two-body outgoing state can be written as,
\begin{align}\label{Eq:WF}
  \Psi^{(-)}_S(\boldsymbol{r},\boldsymbol{k})=e^{i\boldsymbol{k}\cdot\boldsymbol{r}}-j_0(kr)+\psi_0(r,k),
\end{align}
where the spherical Bessel function $j_0$ represents the $l = 0$ component of the non-interacting wave function, and $\psi_0$ denotes the $l = 0$ scattering wave function affected by the strong interaction. Substituting the relative wave function \eqref{Eq:WF} into the KP formula, the correlation function then becomes 
\begin{align}\label{Eq:CF}
  C(k)\simeq1+\int_0^\infty4\pi r^2{\rm d}r~S_{12}(r)~\left[|\psi_0(r,k)|^2-|j_0(kr)|^2\right].
\end{align}

Before using exact scattering wave functions to evaluate correlation functions, it is useful to introduce an analytical model developed by Lednick${\rm \acute{y}}$ and Lyuboshits (LL)~\cite{Lednicky:1981su}. It has been widely used to extract strong interactions from experimental data~\cite{Fabbietti:2020bfg}. This model obtains correlation functions using the asymptotic wave function and the effective range correction. Then, correlation functions are given in terms of the scattering amplitude $f(k)=1/(-1/a_0+r_{\rm eff}k^2/2-ik)$ with the scattering length $a_0$ and the effective range $r_{\rm eff}$~\footnote{Here a negative (positive) scattering length corresponds to a weakly attractive potential (repulsive potential or attractive potential capable of generating a bound state).},
\begin{align}
  C_{LL}(k)=1+\frac{|f(k)|^2}{2R^2}F_3\left(\frac{r_{\rm eff}}{R}\right)+\frac{2{\rm Re}f(k)}{\sqrt{\pi}R}F_1(2kR)-\frac{{\rm Im}f(k)}{R}F_2(2kR),
\end{align}
where $F_1(2kR)=\int_0^{2kR}{\rm d}x\exp[x^2-(2kR)^2]/(2kR)$, $F_2(2kR)=(1-\exp[-(2kR)^2])/(2kR)$, and $F_3(r_{\rm eff}/R)=1-r_{\rm eff}/(2\sqrt{\pi}R)$.
The LL  model usually does not consider the Coulomb interaction and coupled-channel effects.

In general, the exact scattering wave function can be obtained by solving the Schr\"odinger equation in coordinate space~\cite{Mihaylov:2018rva, Kamiya:2019uiw} or the LS equation in momentum space~\cite{Haidenbauer:2018jvl, Liu:2023uly}. For our purpose, it is convenient to first obtain the reaction amplitude $T$ by solving the LS equation and then derive the scattering wave function using the relation $|\psi\rangle=|\varphi\rangle+G_0T|\varphi\rangle$, where $G_0$ and $|\varphi\rangle$ represent the free propagator and the free wave function, respectively. More specifically, we use the following coupled-channel scattering equation to obtain the reaction amplitude,
\begin{align}\label{Eq:Kadyshevsky}
  T_{\nu'\nu}(k',k)=V_{\nu'\nu}\cdot f_{\Lambda_F}(k',k)+\sum_{\nu^{\prime\prime}}\int_0^\infty\frac{{\rm d}k^{\prime\prime}k^{\prime\prime2}}{8\pi^2}\frac{V_{\nu'\nu^{\prime\prime}}\cdot f_{\Lambda_F}(k',k^{\prime\prime})\cdot T_{\nu^{\prime\prime}\nu}(k^{\prime\prime},k)}{E_{1,\nu^{\prime\prime}}E_{2,\nu^{\prime\prime}}(\sqrt{s}-E_{1,\nu^{\prime\prime}}-E_{2,\nu^{\prime\prime}}+i\epsilon)},
\end{align}
where $\sqrt{s}=E_{1,\nu}(k)+E_{2,\nu}(k)$, and $E_{1(2),\nu}(k)=\sqrt{k^2+M_{1(2),\nu}^2}$. As shown in Eq.~\eqref{Eq:Kadyshevsky}, to avoid ultraviolet divergence in numerical evaluations, we multiply the potential $V_{\nu'\nu^{(\prime\prime)}}$ with a Gaussian regulator $f_{\Lambda_F}(k,k')=\exp[-(k/\Lambda_F)^{2}-(k'/\Lambda_F)^{2}]$ to suppress high-momentum contributions~\cite{Liu:2019tjn, Liu:2023uly}, where $\Lambda_F$ is a cutoff parameter to be determined. We can then compute the scattering wave function with the half-off-shell $T$-matrix in the following way,
\begin{align}\label{Eq:Fourier_Bessel}
  \psi_{\nu'\nu}(r,k)=\delta_{\nu'\nu}j_0(kr)+\int_0^\infty\frac{{\rm d}k'k^{\prime2}}{8\pi^2}\frac{T_{\nu'\nu}(k',k)\cdot j_0(k'r)}{E_{1,\nu^{\prime}}E_{2,\nu^{\prime}}(\sqrt{s}-E_{1,\nu^{\prime}}-E_{2,\nu^{\prime}}+i\epsilon)},
\end{align}
where $j_0$ is the spherical Bessel function of angular momentum $l=0$. The single-channel scattering wave function is called $\psi_0$ ($\nu'=\nu$). Using the above-obtained wave function, we can consider coupled-channel effects by replacing the modulus squared in Eq.~\eqref{Eq:CF} with
\begin{align}\label{Eq:Coupling}
  |\psi_0(r,k)|^2\rightarrow\sum_{\nu'}\omega_{\nu'}|\psi_{\nu'\nu}(r,k)|^2,
\end{align}
where the sum runs over all possible coupled channels, and $\omega_{\nu'}$ is the weight for the individual components of the multi-channel wave function. In addition, the contribution from the Coulomb interaction has to be considered for systems of two charged particles, which is expected to play a significant role in the low-momentum region~\cite{Haidenbauer:2020kwo, Liu:2022nec}. Using the Vincent–Phatak method~\cite{Vincent:1974zz, Holzenkamp:1989tq}, one can treat the Coulomb force in momentum-space.

It is worthwhile to note that in Ref.~\cite{Vidana:2023olz}, a formalism was developed that allows one to factorize the scattering amplitudes outside the integrals in the formulae (see Eq.~\eqref{Eq:Kadyshevsky} and Eq.~\eqref{Eq:Fourier_Bessel}). The integrals explicitly involve the range of the strong interaction. Following this approach, the $T$-matrix can be converted into an algebraic BS equation based on the on-shell approximation,
\begin{align}\label{Eq:osLS}
  T=\frac{V}{1-VG},
\end{align}
for the single-channel case, where $G$ is the loop function of the intermediate particles,
\begin{align}\label{Eq:osLF}
  G(\sqrt{s})=\int_0^{q_{\rm max}}\frac{{\rm d}^3k'}{(2\pi)^3}\frac{E_1(k')+E_2(k')}{2E_1(k')E_2(k')}\frac{1}{s-[E_1(k')+E_2(k')]^2+i\epsilon}.
\end{align}
Eq.~\eqref{Eq:osLS} in the cutoff regularization can be justified using dispersion relations~\cite{Oller:2000fj}, but can be equally obtained using a separable potential
$V(k',k)=V\cdot\theta(q_{\rm max}-k')\cdot\theta(q_{\rm max}-k)$. Then, the scattering wave function can be expressed with the $T$-matrix by,
\begin{align}\label{Eq:osWF}
  \Psi^{(-)}_S(\boldsymbol{r},\boldsymbol{k})=e^{i\boldsymbol{k}\cdot\boldsymbol{r}}+T(k,k)\cdot\theta(q_{\rm max}-k)\cdot \widetilde{G}(r,\sqrt{s}),
\end{align}
where $\widetilde{G}(r,\sqrt{s})$ function is given by,
\begin{align}\label{Eq:osLFnew}
  \widetilde{G}(r,\sqrt{s})=\int_0^{q_{\rm max}}\frac{{\rm d}^3k'}{(2\pi)^3}\frac{E_1(k')+E_2(k')}{2E_1(k')E_2(k')}\frac{j_0(k'r)}{s-[E_1(k')+E_2(k')]^2+i\epsilon}.
\end{align}
In practice, Eq.~\eqref{Eq:osLS} can be easily generalized to coupled channels as $T=[1-VG]^{-1}V$, where $V$ is the transition potential $V_{ij}$ between the channels $i$ and $j$, and $G$ is the diagonal loop function $G\equiv {\rm diag}[G_i(\sqrt{s})]$, with $G_i(\sqrt{s})$ the loop function of each particular channel $i$. Finally, the coupled-channel correlation function can be written explicitly as,
\begin{align}\label{Eq:osCF}
  C(k)=&1+\theta(q_{\rm max}-k)\int_0^\infty4\pi r^2{\rm d}r~S_{12}(r)~\times\nonumber\\
  &\left[\left|j_0(kr)+T_{ii}\cdot\widetilde{G}_{(i)}(r,\sqrt{s})\right|^2+\sum_{j\neq i}\omega_j\left|T_{ji}\cdot\widetilde{G}_{(j)}(r,\sqrt{s})\right|^2-|j_0(kr)|^2\right].
\end{align}
The factor $\theta(q_{\rm max}-k)$ is inoperative for relative momentum $k$ of interest in correlation functions, with values of $k$ smaller than $q_{\rm max}$.

\subsection{Some general features of correlation functions}
In this subsection, we review some general features of correlation functions concerning studies of hadronic molecules. A fundamental physical picture for femtoscopy is that relativistic heavy-ion collisions generate particle sources from which hadron-hadron pairs emerge with relative momentum $k$ and can undergo final-state interactions before being detected. Consequently, $k$ is reduced or increased via an attractive or repulsive interaction. In other words, the magnitude of the correlation function in the low-momentum region will be above unity for an attractive interaction, whereas between zero and unity for a repulsive interaction.

For transparency and without loss of generality, we work with the square-well model and study four potentials: repulsive, weakly attractive, moderately attractive, or strongly attractive. The scattering wave function is obtained analytically by solving the stationary Schr\"odinger equation $-\frac{\hbar^2}{2\mu}\nabla^2\psi+V_0\theta(d-r)\psi=E\psi$, where the reduced mass $\mu$ is chosen as $470$ MeV, the range parameter $d$ is set at  $2.5$ fm, and the depth parameter $V_0$ is set at $25$, $-10$, $-25$, and $-75$ MeV for a repulsive potential, a weakly attractive potential not strong enough to generate a bound state, a moderately attractive potential capable of generating a shallow bound state, and a strongly attractive potential yielding a deep bound state, respectively~\footnote{Here we refer to a bound state as a shallow bound state if its binding energy can be described by the effective-range expansion up to $q^2$, and otherwise as a deep bound state.}.

\begin{figure}[htpt]
\begin{center}
\includegraphics[width=6.5in]{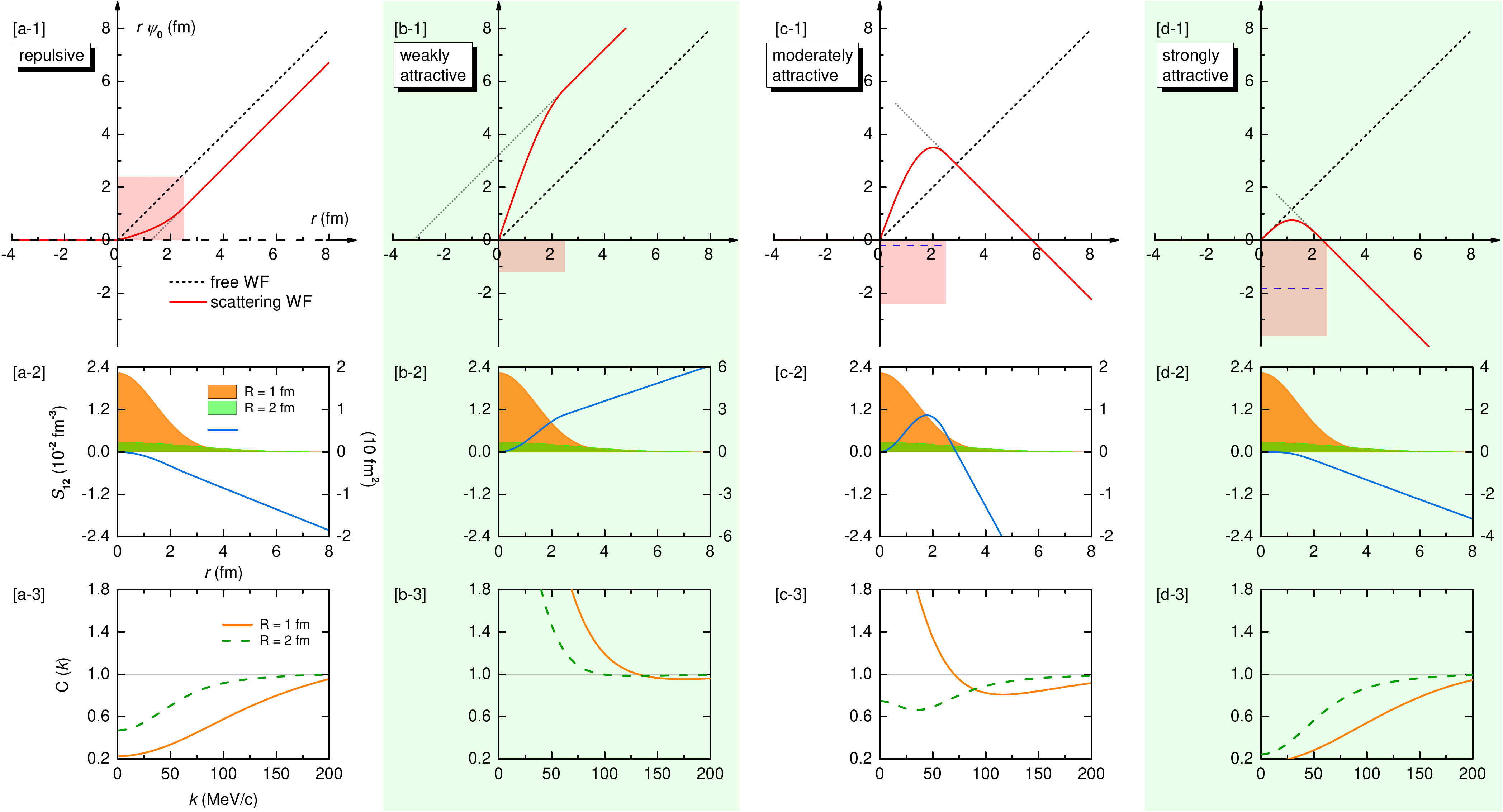}
\caption{Scattering wave functions, source functions, and correlation functions for the four different square-well potentials, namely, (a) a repulsive potential, (b) a weakly attractive potential, (c) a moderately attractive potential, and (d) a strongly attractive potential~\cite{Liu:2023uly}.}
\label{Fig:CF_properties}
\end{center}
\end{figure}

In Fig.~\ref{Fig:CF_properties}, panels (a1-d1) show the products of the relative distance $r$ and the  $S$-wave function $\psi_0$ for the relative momentum $k\simeq3$ MeV/c. According to Eq.~\eqref{Eq:CF}, the correlation function depends on two factors, namely, the difference between the free and scattering wave functions squared, i.e., $\Delta\equiv r^2(|\psi_{0}|^2-|j_{0}|^2)$, and the source function $S_{12}$, shown as the blue solid line and colored regions in panels (a2-d2). The comparison between the free and scattering $r\cdot\psi_0$ can be captured by the sign of $\Delta$, which is directly related to the properties of the correlation functions in the low-momentum region. As the source size $R$ increases, the magnitude of the Gaussian source function decreases rapidly, and its tail becomes longer, reducing the corresponding correlation function. The final correlation functions are displayed in panels (a3-d3). One can conclude that (a) for a repulsive potential, the correlation functions are between zero and unity for different $R$; (b) for a weakly attractive potential, they are above unity for different $R$; (c) for a moderately attractive potential, the low-momentum correlation function is above unity for small $R$ while below unity for large $R$; and (d) for a strongly attractive potential, they are between zero and unity for different $R$~\cite{Liu:2023uly}. The above observations based on the square-well model are consistent with the aforementioned physical picture and the analysis performed in the Lednicky-Lyuboshitz model~\cite{ExHIC:2017smd}, but more intuitive.

The above conclusions can also be verified using Eq.~(\ref{Eq:osCF}) in the single-channel case with the potential of the form $V(k)=a+bk^2$, where $a$ and $b$ are (real) low-energy constants (LECs) to be determined. The advantage of this framework is that it can be easily extended to study correlation functions in the presence of virtual or resonant states. Here, for the resonant state case, the real part of the pole position with respect to the threshold $\Delta M=M-m_1-m_2$ is set at 2.5, 5, and 10 MeV, respectively; the imaginary part of the pole position (half-width) $\Gamma/2$ is set at 1 and 10 MeV, respectively. For the virtual state case, $\Delta M$ is set at $-2.5$, $-5$, and $-10$ MeV, respectively. The corresponding LECs and correlation functions can be fixed and calculated with these pole settings.

\begin{figure*}[htbp]
  \centering
  \includegraphics[width=0.98\textwidth]{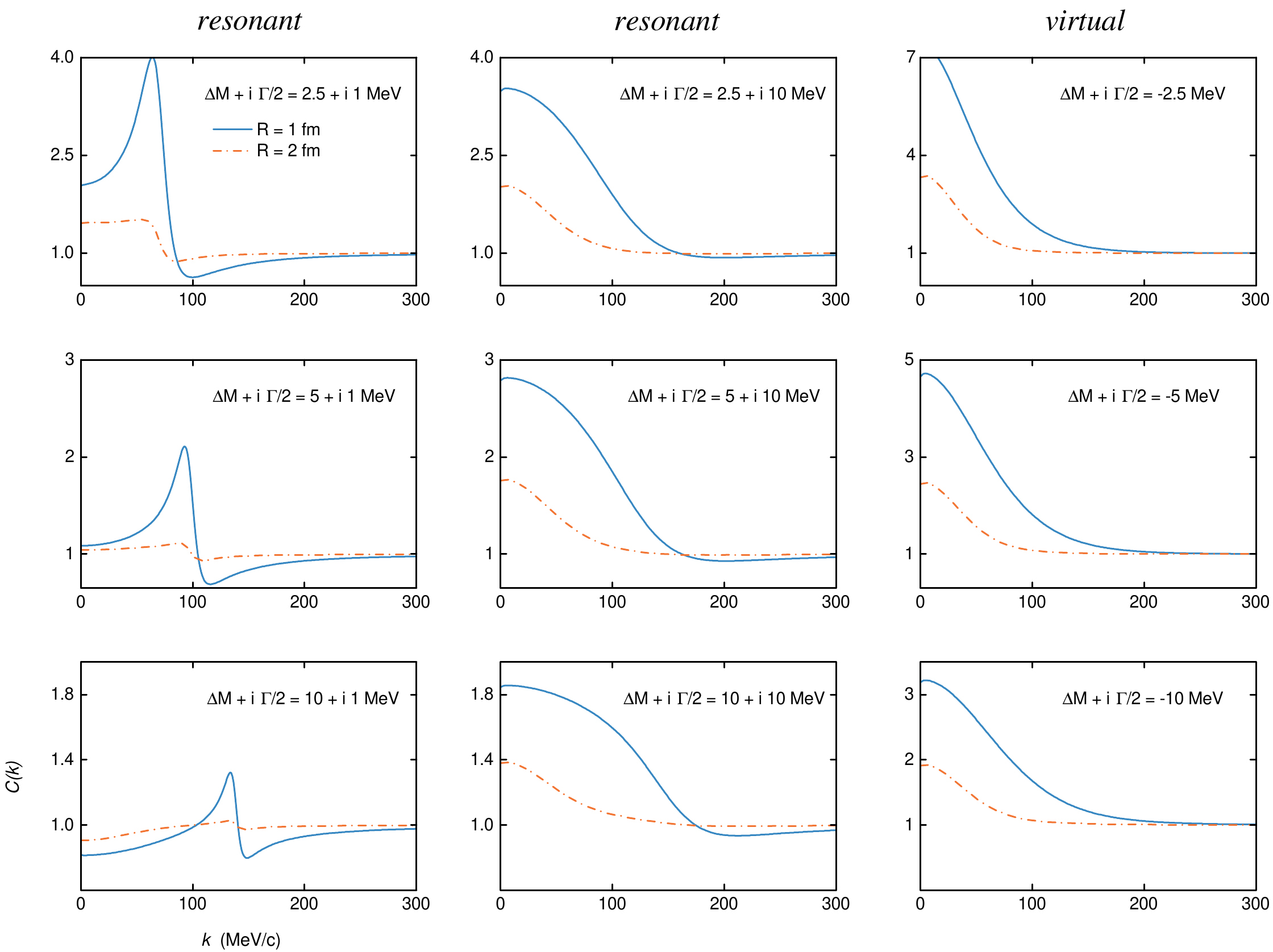}
  \caption{Correlation functions corresponding to potentials capable of generating a resonant (left and middle panels) or virtual  (right panels) state with the pole positions $\Delta M + i\Gamma/2$ specified in the respective plots. The source sizes are set at $R = 1$ and $2$ fm, the masses $m_1$ and $m_2$ are set at 1.8 and 2 GeV, and the sharp cutoff $q_{\rm max}$ is fixed to $1$ GeV~\cite{Liu:2024nac}.}\label{Fig:CF_properties2}
\end{figure*}

As shown in Fig.~\ref{Fig:CF_properties2}, for a resonant state characterized by a narrow width (see left panels), the correlation function exhibits a pronounced enhancement (peak) at momentum slightly below $k_R$, where $k_R$ is the c.m. momentum corresponding to the resonance mass. Conversely, a suppression (valley) is observed slightly above $k_R$. However, for a resonant state characterized by a larger width (see middle panels), the peak transforms into an overall enhancement in the low-momentum region, and the valley structure in the high-momentum region becomes less pronounced. In addition, the enhancement and suppression weaken as $k_R$ moves towards the high-momentum region or the source size increases.

In the context of virtual states (see right panels), the correlation functions consistently exceed unity for varying source sizes, indicating the presence of a purely attractive interaction. Meanwhile, as the virtual-state pole moves away from the threshold, the strength of the attractive interaction decreases, reducing the correlation function.

\subsection{Correlation functions for $D_{s0}^*(2317)$, $D_0^*(2300)$, $D_1(2420)$, and $D_1(2430)$}
\begin{figure}[htbp]
\centering
\begin{minipage}[t]{0.46\textwidth}
\includegraphics[width=0.98\textwidth]{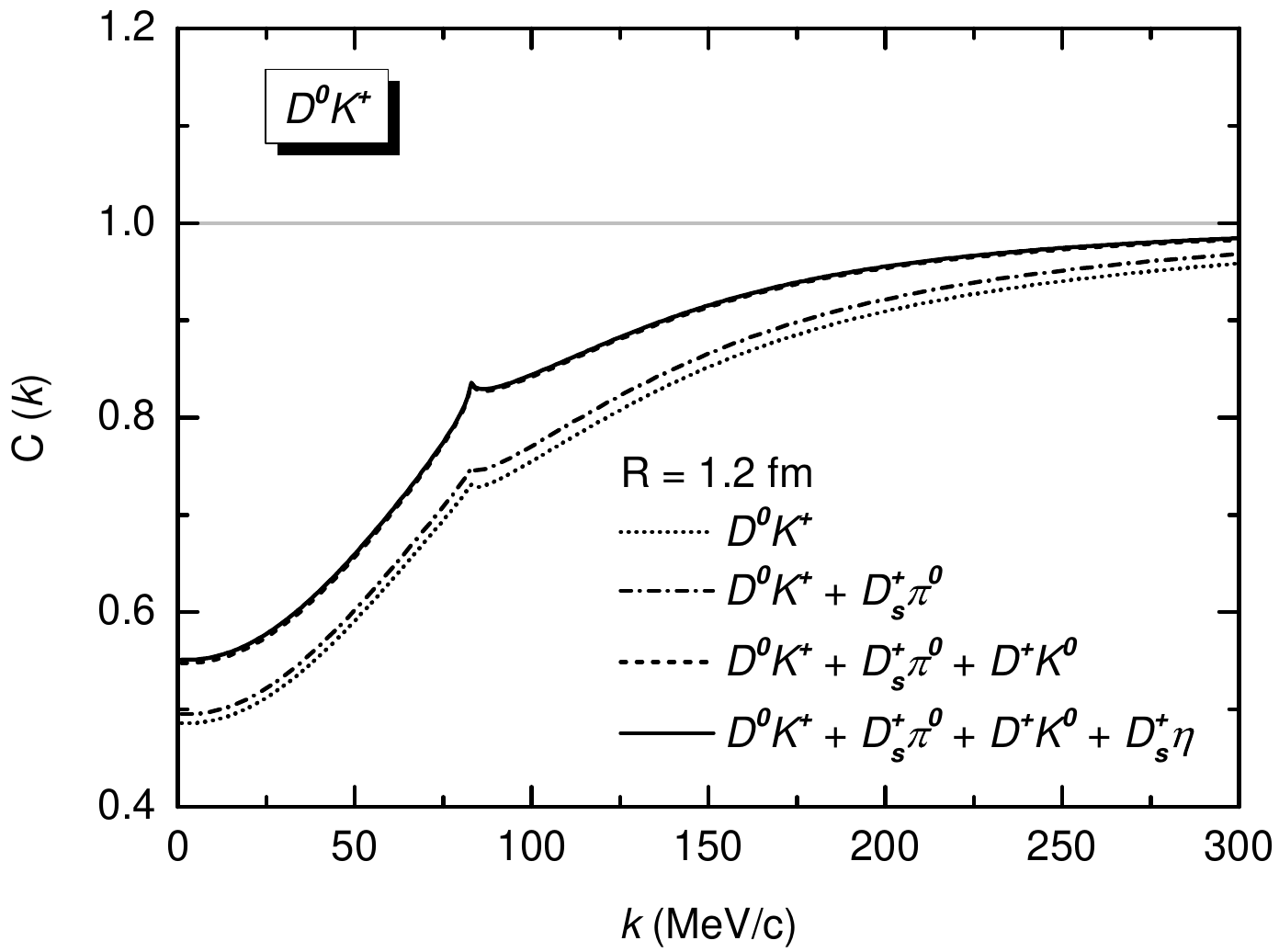}
\end{minipage}
\begin{minipage}[t]{0.46\textwidth}
\includegraphics[width=0.98\textwidth]{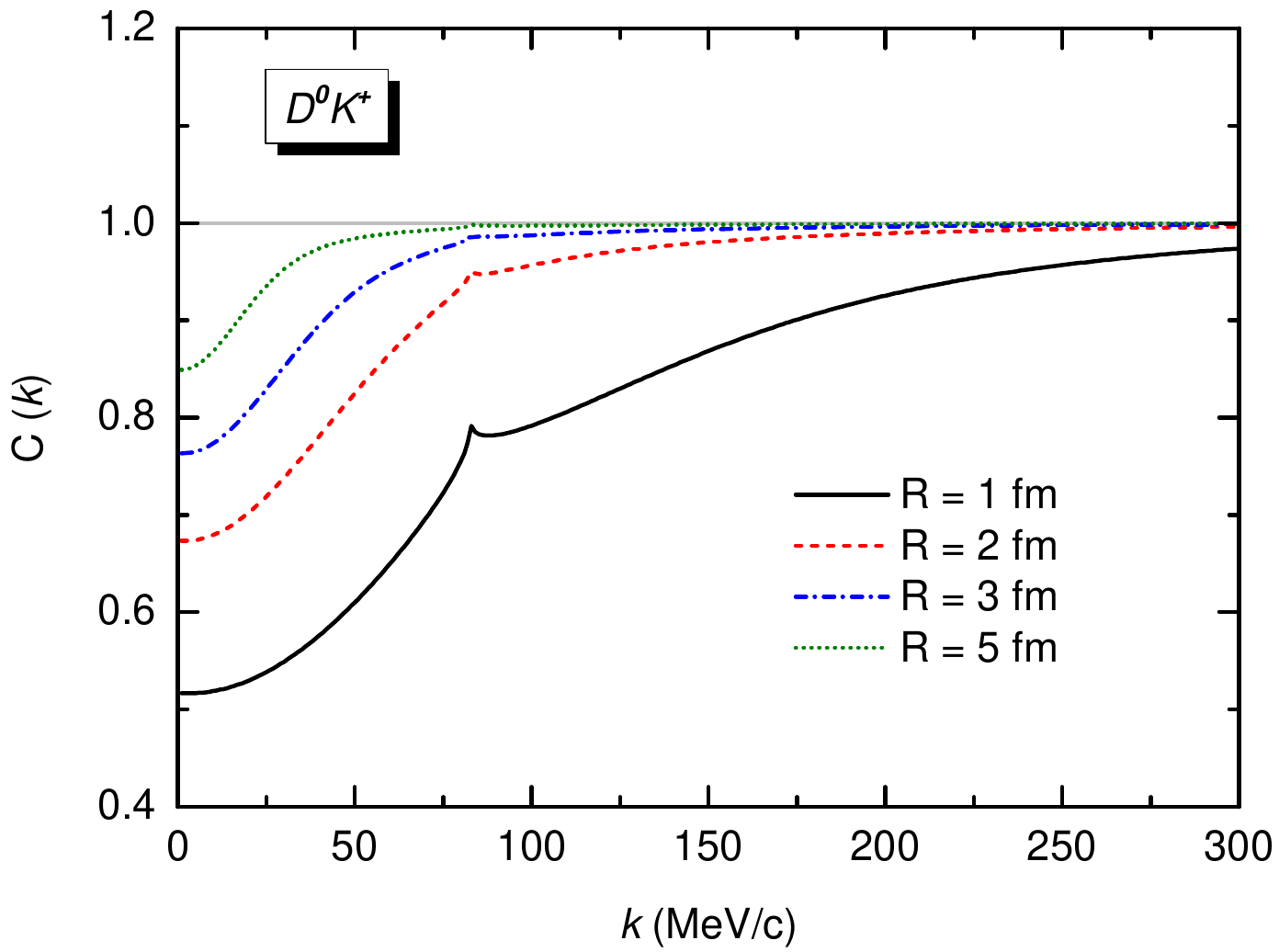}
\end{minipage}
\caption{(left panel) $D^0K^+$ correlation function as a function of the relative momentum $k$. (right panel) Source size dependence of the $D^0K^+$ correlation function~\cite{Liu:2023uly}.}
\label{Fig:CF_Ds0star}
\end{figure}

The $DK$ interaction in isospin zero is attractive to such an extent that a bound state can be generated~\cite{Gamermann:2006nm, Guo:2006fu, Altenbuchinger:2013vwa}, i.e., $D_{s0}^*(2317)$. It will be interesting to confirm the attractive nature of the $DK$ interaction directly. For such a purpose, we studied the $DK$ correlation function for the first time, which, if measured, can be used to verify or refute the hadronic molecular picture of $D_{s0}^*(2317)$. We first evaluated the $DK$ coupled-channel interaction in the leading order unitarized heavy-meson chiral perturbation theory and calculated the corresponding correlation function~\cite{Liu:2023uly}. As shown in Fig.~\ref{Fig:CF_Ds0star},  the inelastic coupled-channel contribution, which is mainly from the $D^0K^+-D^+K^0$ transition, can be sizable and lead to a cusp-like structure in the $D^0K^+$ correlation function around the $D^+K^0$ threshold. We found that the source size dependence of the $DK$ correlation function is very different from that of moderately strong attractive interactions, which can be utilized to verify the nature of $D_{s0}^*(2317)$ as a deeply bound $DK$ state.   It should be noted that if the $D_{s0}^*(2317)$ contains a smaller $DK$ molecular 
component, the  $DK$ scattering length would be smaller, and correspondingly the 
correlation function would change. The quantitative estimation for the variation of the $DK$ correlation function, considering other components of  $D_{s0}^*(2317)$, will be examined in the future.  The above results are also confirmed in Ref.~\cite{Albaladejo:2023pzq}, where the interactions between a heavy pseudoscalar boson and a Nambu-Goldstone boson are derived from the next-to-leading-order unitarized heavy-meson chiral perturbation theory.

Once experimental data are available, one can derive the corresponding strong interactions from those correlation functions, which is the so-called inverse problem in Femtoscopy studies~\cite{Ikeno:2023ojl, Albaladejo:2023wmv, Feijoo:2023sfe, Li:2024tof}. In Ref.~\cite{Ikeno:2023ojl}, the authors deal with the inverse problem of extracting information from the $D^0K^+$, $D^+K^0$ and $D_s^+\eta$ correlation functions. In the absence of such data, they used the synthetic data extracted from an interaction model based on the local hidden
gauge approach. Subsequently, they made no specific assumption on the potential $V_{ij}$ except for the isospin symmetry and $V_{33}(3\equiv D_s\eta)=0$, the latter is quite general since the vertex $\eta\eta V$ vanishes in most models. Therefore, the interaction matrix $V$ gets a bit simplified to,
\begin{align}
V=
\left( \begin{matrix}
V_{11}& V_{12}& V_{13} \\
& V_{11}& V_{13} \\
& & 0 \\
\end{matrix} \right).
\end{align}
where the indices $i=1, 2, 3$ represent the $D^0K^+$, $D^+K^0$ and $D_s^+\eta$ channels, respectively. 
To account for possible missing coupled channels and possible contribution of a non-molecular component, they introduced energy-dependent parts in the potential, such that,
\begin{subequations}
\begin{align}
V_{11}&=V_{11}'+\frac{\alpha}{M_V^2}(s-\bar{s}), \\
V_{12}&=V_{12}'+\frac{\beta}{M_V^2}(s-\bar{s}), \\
V_{13}&=V_{13}'+\frac{\gamma}{M_V^2}(s-\bar{s}),
\end{align}
\end{subequations}
where $\bar{s}$ is the energy squared of the $D^0K^+$ threshold. The $T$-matrix with the three coupled channels is evaluated through $T=[1-VG]^{-1}V$. Using Eq.~\eqref{Eq:osCF}, there are eight parameters to describe the synthetic correlation functions, namely, $V_{11}', V_{12}', V_{13}', \alpha, \beta, \gamma, q_{\rm max}$ and $R$. The question is whether one can determine these parameters from the three correlation functions. From the fits obtained with this set of free parameters, the authors of Ref.~\cite{Ikeno:2023ojl} found that the inverse problem can determine the existence of a bound state, its isospin nature, the compositeness, and the scattering length and effective range of all three channels. In other words, different magnitudes tied to the interaction of the three channels can be evaluated from the correlation functions with reasonable accuracy.

\begin{figure}[htbp]
\centering
\begin{minipage}[t]{0.46\textwidth}
\includegraphics[width=0.98\textwidth]{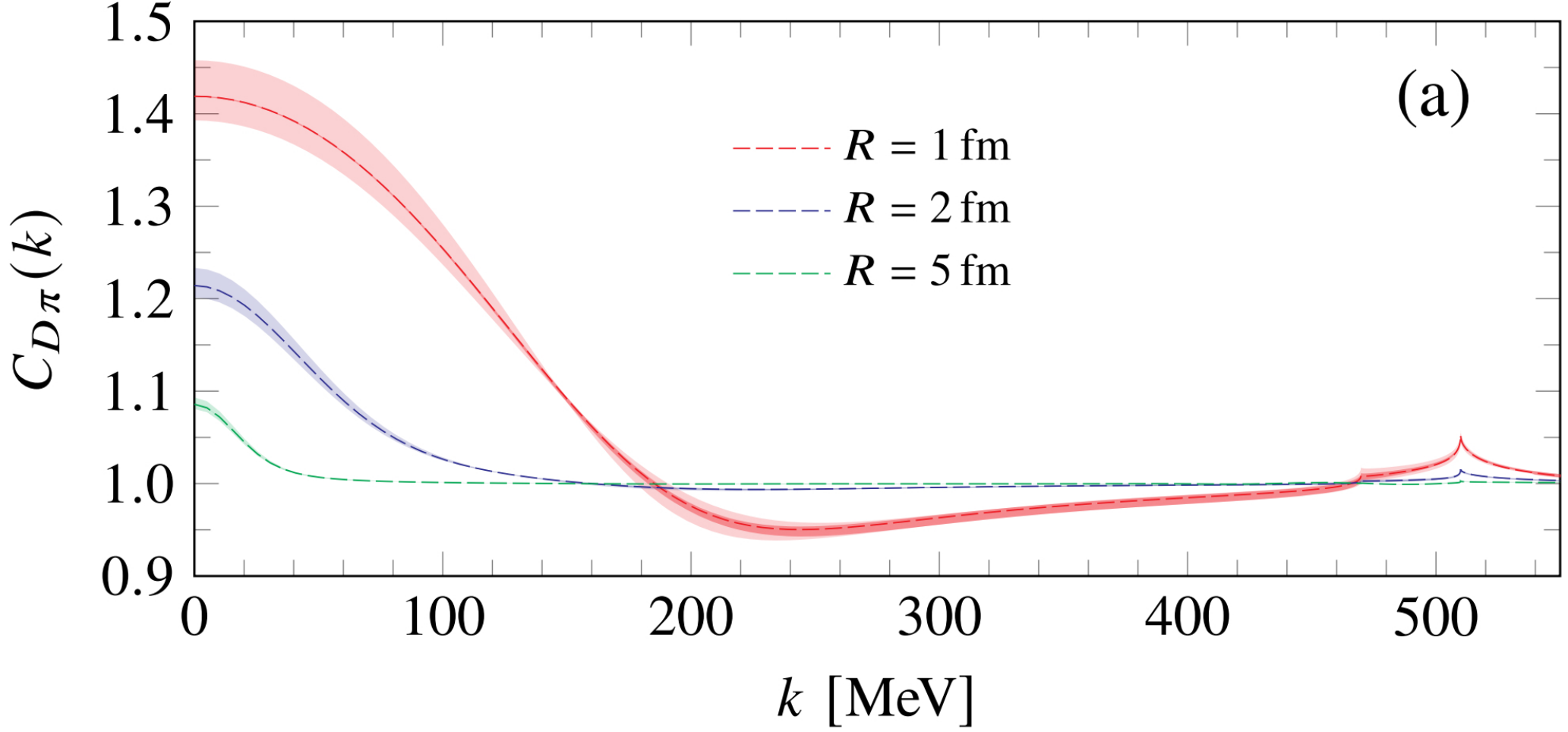}
\end{minipage}
\begin{minipage}[t]{0.46\textwidth}
\includegraphics[width=0.98\textwidth]{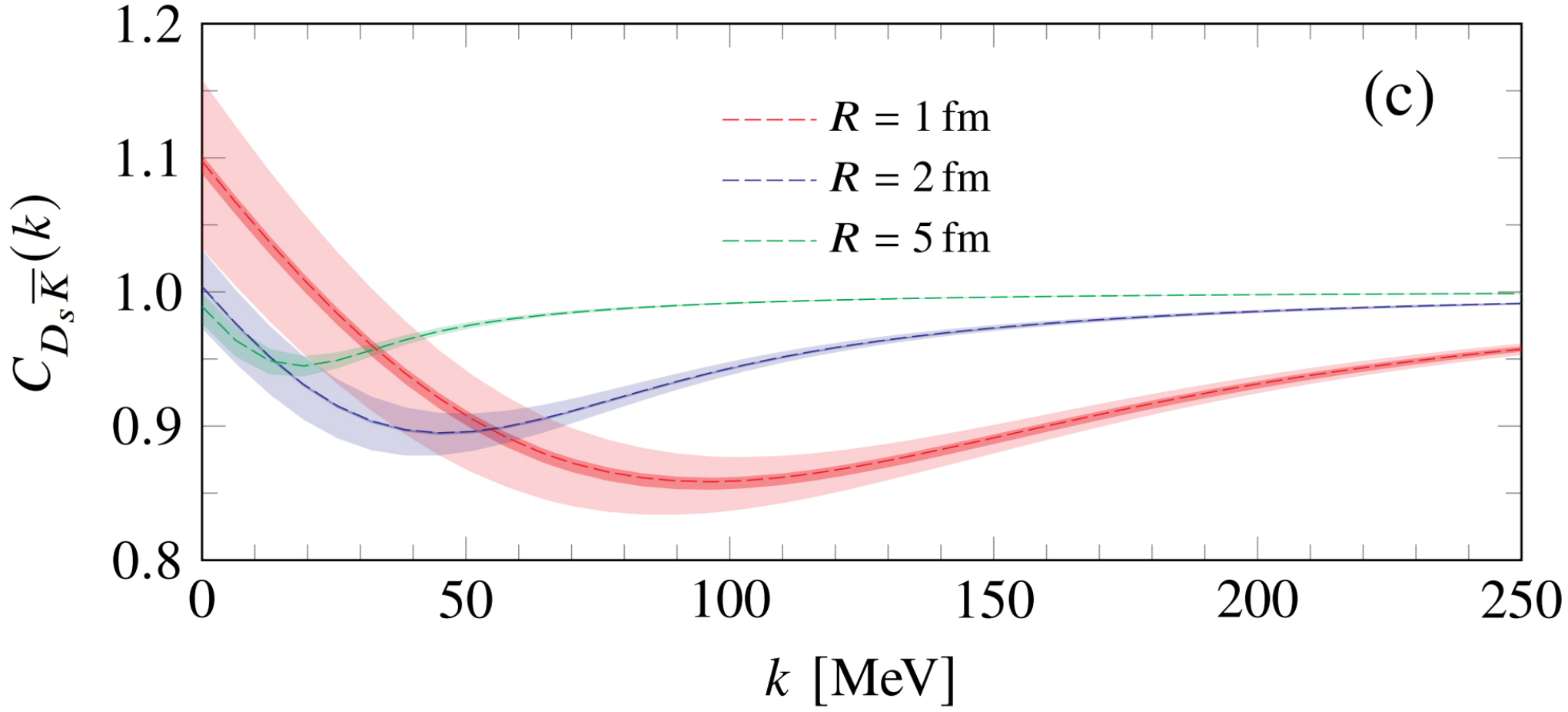}
\end{minipage}
\caption{Correlation functions for the $D\pi$ (left panel) and $D_s\bar{K}$ (right panel) channels with $I=I_z=1/2$ as a function of their c.m. momentum $k$ for different source sizes.  \\  Source: Figures  taken from Ref.~\cite{Albaladejo:2023pzq}.}
\label{Fig:CF_D0star}
\end{figure}

For the system of lightest pseudoscalar open-charm mesons and Goldstone bosons with $(S,I)=(0,1/2)$, Ref.~\cite{Albaladejo:2023pzq} predicted the corresponding correlation functions and argued that the effect of the two-pole structure around 2300 MeV can be seen in the $D\pi$, $D\eta$, and $D_s\bar{K}$ correlation functions. As shown in Fig.~\ref{Fig:CF_D0star}, two distinct minima can be observed in the correlation functions $C_{D\pi}$ and $C_{D_s\bar{K}}$, especially for a small collision system, which are produced by the lower and higher $D_0^*(2300)$ poles, respectively. 

\begin{figure}[htbp]
\centering
\begin{minipage}[t]{0.46\textwidth}
\includegraphics[width=0.98\textwidth]{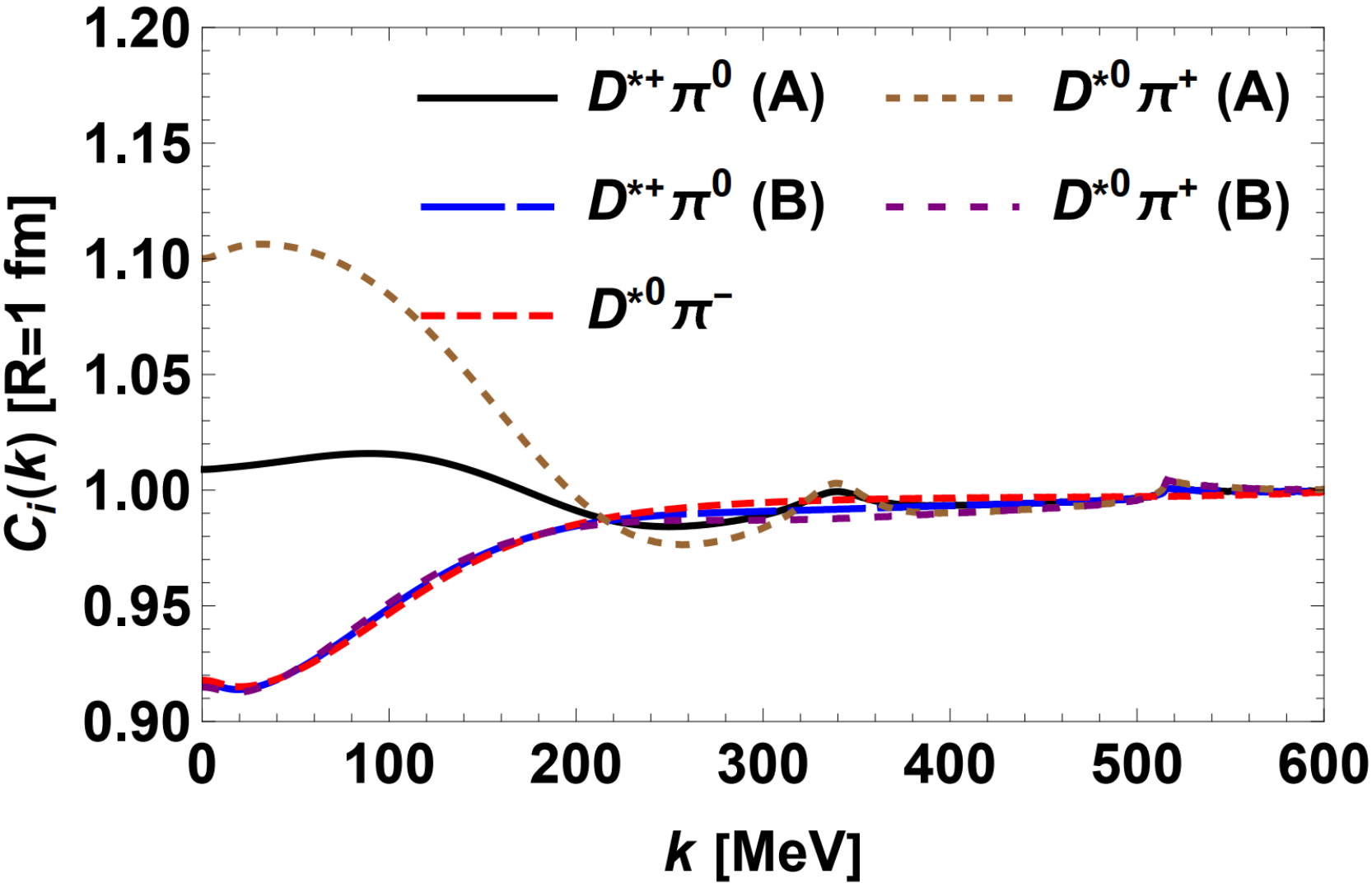}
\end{minipage}
\begin{minipage}[t]{0.46\textwidth}
\includegraphics[width=0.98\textwidth]{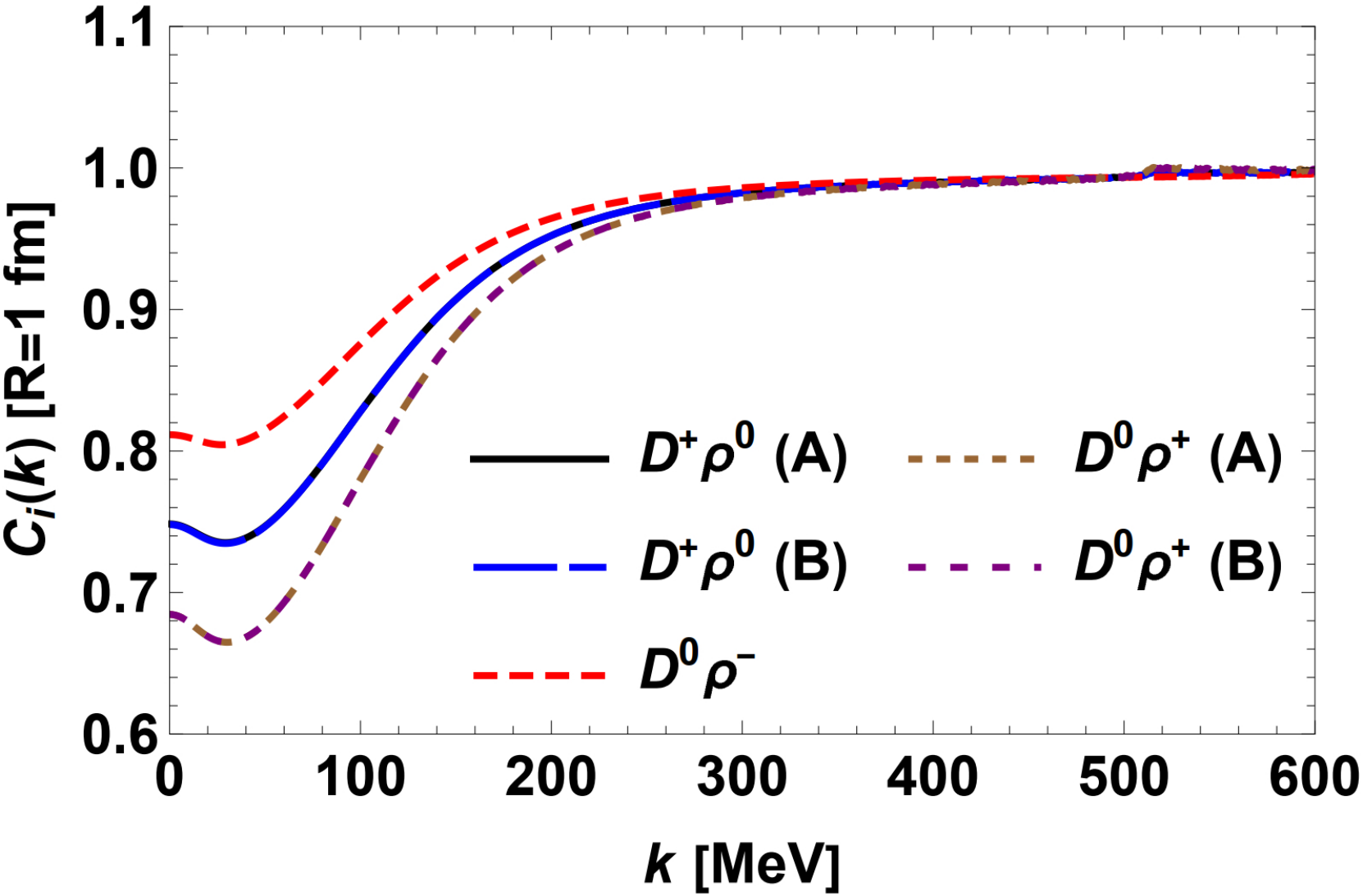}
\end{minipage}
\caption{Correlation functions for the physical $D^*\pi$ (left panel) and $D\rho$ (right panel) channels in models A and B, for $R = 1$ fm.  \\  Source:  Figures taken from Ref.~\cite{Khemchandani:2023xup}.}
\label{Fig:CF_D1}
\end{figure}

Two lightest open-charm axial mesons exist, i.e., $D_1(2430)$ and $D_1(2420)$,  whose masses are very similar, but the former has a larger width. In Ref.~\cite{Khemchandani:2023xup}, the authors developed two models, which can produce compatible properties for the two lightest $D_1$ states but result in different scattering lengths: one in agreement with the findings of lattice QCD~\cite{Lang:2022elg} (referred to as model A) and the other in agreement with the estimation obtained using the $D\pi$ results from the ALICE Collaboration~\cite{4715876, 4921802, 5363368} (referred to as model B).  It should be noted that similar to the two states associated with  $D_0^*(2300)$,   the broad $D_1(2430)$ can also correspond to two states ~\cite{Guo:2006rp,Du:2017zvv}, the lower state consistent with the result of Model A of Ref.~\cite{Khemchandani:2023xup}.   As shown in Fig.~\ref{Fig:CF_D1}, they presented the correlation functions for both cases and found that $C_{D^*\pi}$ and $C_{D\rho}$ can be used to test both models and might encode sufficiently identifiable signatures of the $D_1(2430)$ and $D_1(2420)$ states.

It is worthwhile to emphasize that except the $pD^-$ correlation measured by the ALICE Collaboration~\cite{ALICE:2022enj}, results have been reported by the ALICE Collaboration for $\pi^\pm D^{(*)\pm}$ and $K^\pm D^{(*)\pm}$~\cite{ALICE:2024bhk}.  Within the
current uncertainties, there is a tension between the $\pi^\pm D^{(*)\pm}$ scattering lengths extracted by the ALICE Collaboration and the results of   lattice QCD~\cite{Mohler:2012na},   unitarized ChPT~\cite{Guo:2009ct,Liu:2012zya,Yao:2015qia,Guo:2018tjx,Huang:2021fdt,Torres-Rincon:2023qll}, and  ChPT~\cite{Liu:2011mi,Guo:2018kno}.     This discrepancy indicates the necessity for more high-precision femtoscopic experiments and rigorous analyses. Soon, the precision of these measurements will improve with the data taken during the LHC Run 3.     In Ref.~\cite{Torres-Rincon:2023qll}, based on the next-to-leading order chiral potential, the authors calculated the correlation function of $D$ mesons and light mesons using an off-shell T-matrix approach to access the two-meson wave function. They predicted the correlation functions involving charged $D^+$, $D^{*+}$, $D_s^+$ and $D_s^{*+}$ with $\pi^\pm$ and $K^\pm$. Their results are similar to those of other theoretical models and comparable to the experimental data. However, the predicted $D^+\pi^-$ correlation function significantly differs from the preliminary ALICE results.  This puzzle must be solved by considering extra theoretical modifications or waiting for more definite experimental results.

\subsection{Correlation functions for $T_{cc}(3875)$ and $X(3872)$}
\begin{figure}[htbp]
\centering
\begin{minipage}[t]{0.46\textwidth}
\includegraphics[width=0.98\textwidth]{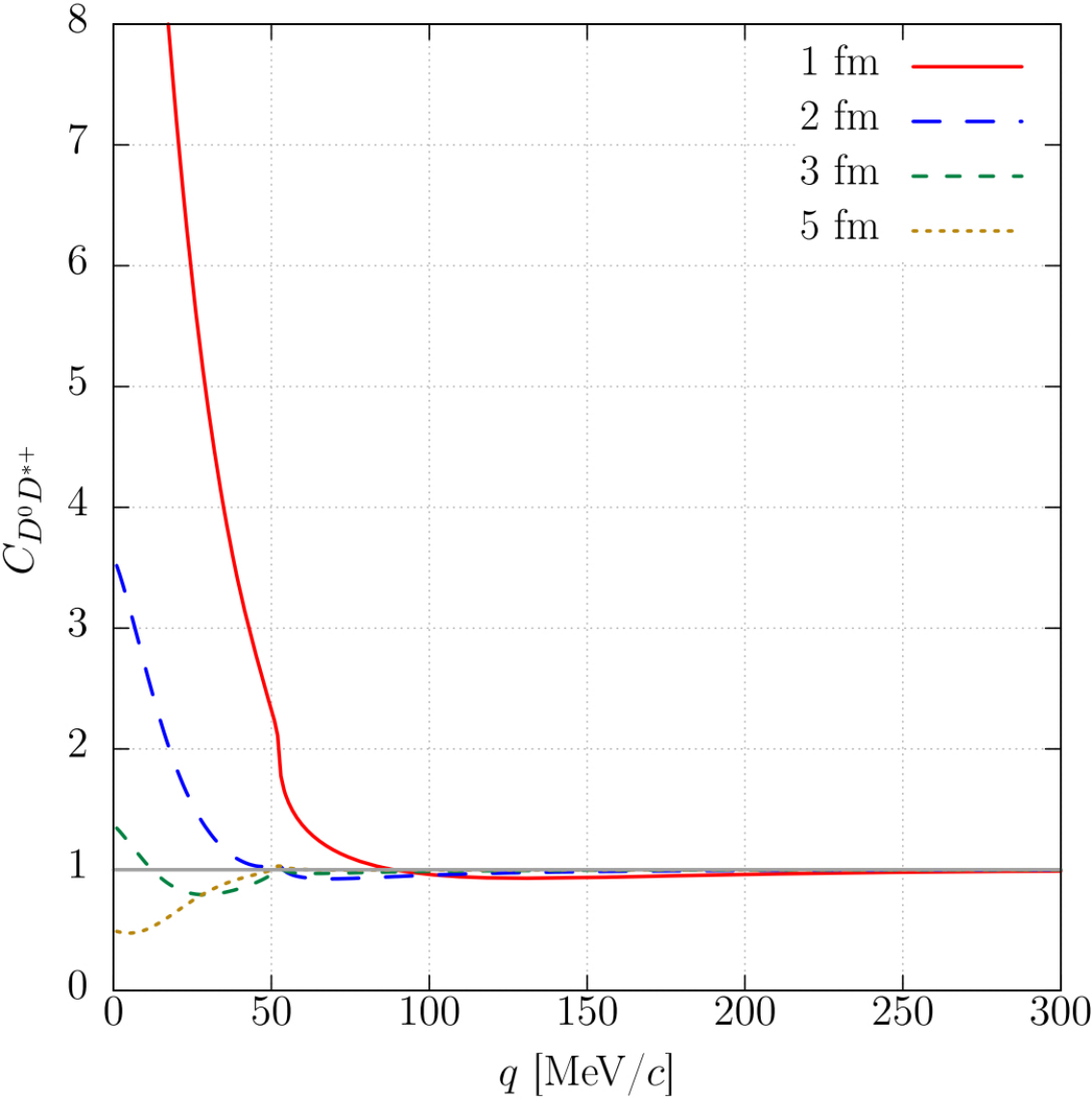}
\end{minipage}
\begin{minipage}[t]{0.46\textwidth}
\includegraphics[width=0.98\textwidth]{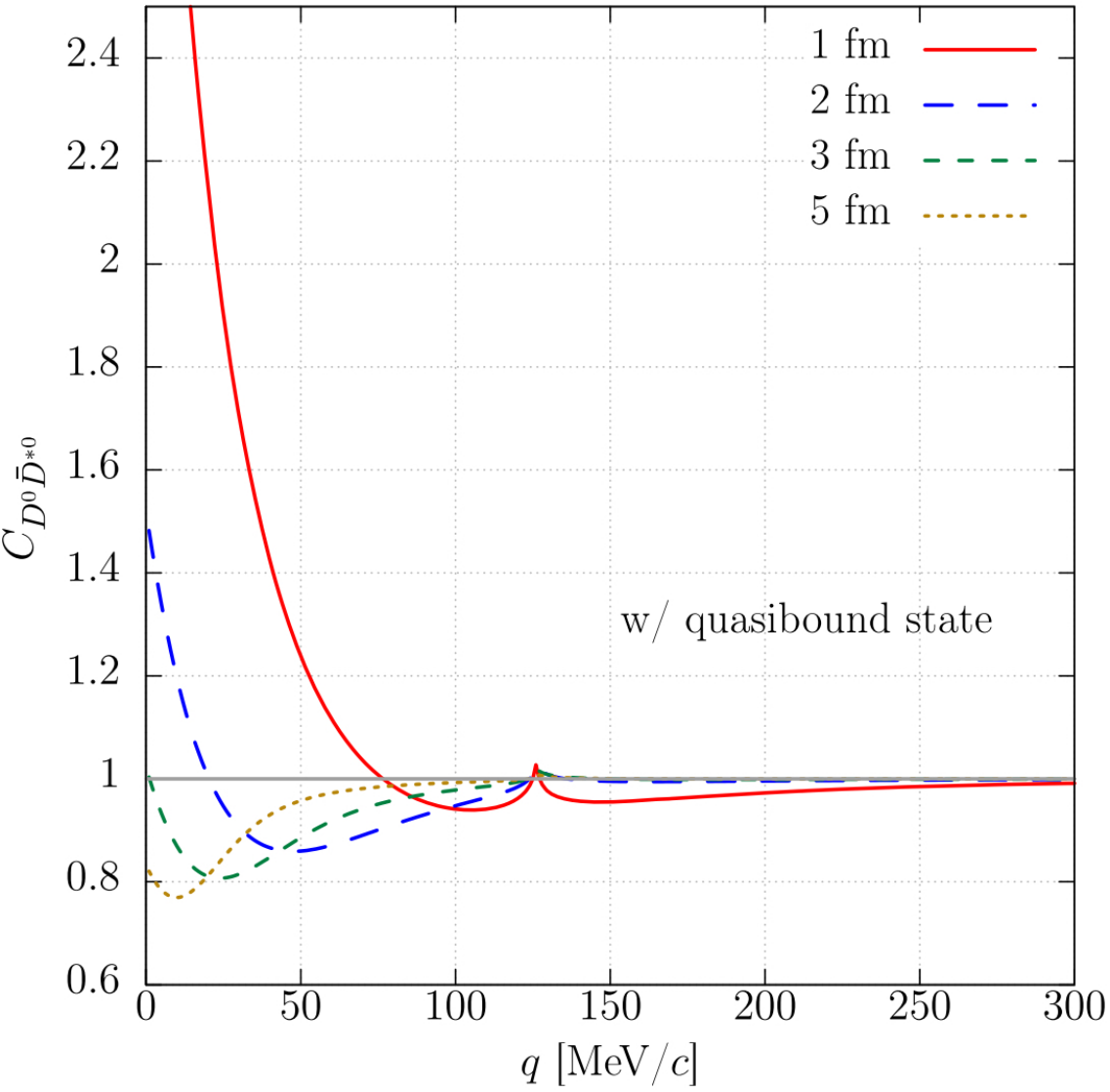}
\end{minipage}
\caption{Correlation functions of the $D^0D^{*+}$ (left panel) and $D^0\bar{D}^{*0}$ (right panel) pairs for source sizes $R=1$, $2$, $3$, and $5$ fm.  \\  Source:  Figures taken from Ref.~\cite{Kamiya:2022thy}.}
\label{Fig:CF_TX}
\end{figure}

In this subsection, we discuss the potential of the femtoscopy study in understanding the nature of the $T_{cc}(3875)$ and $X(3872)$ states. Ref.~\cite{Kamiya:2022thy} studied the correlation functions of the $D^0D^{*+}$ ($D^+D^{*0}$) and $D^0\bar{D}^{*0}$ ($D^+D^{*-}$) channels in connection with the $T_{cc}(3875)$ and $X(3872)$ states. They first constructed one-range Gaussian potentials for the $DD^*$ and $D\bar{D}^*$ channels, which reproduce the empirical information, and then predicted the $D^0D^{*+}$ ($D^+D^{*0}$) and $D^0\bar{D}^{*0}$ ($D^+D^{*-}$) correlation functions, as shown in Fig.~\ref{Fig:CF_TX}. The source size dependence is typical to the system with a shallow bound state for both correlation functions: enhancement in the small source case and suppression in the large source case.

A different method formulated in momentum space, rather than in coordinate space, was used in Ref.~\cite{Vidana:2023olz} to obtain the $D^0D^{*+}$ and $D^+D^{*0}$ correlation functions, which are in qualitative agreement with Ref.~\cite{Kamiya:2022thy}. In addition, the authors investigated the inverse problem of determining the $D^0D^{*+}$ and $D^+D^{*0}$ interactions from the pseudo correlation functions in Ref.~\cite{Albaladejo:2023wmv}. With a coupled-channel unitary scheme, which has the freedom
to accommodate missing channels relevant to the interaction and some genuine components of non-molecular nature, they demonstrated that from these synthetic data, one can extract the existence of the $T_{cc}(3875)$ bound state, the probabilities of each channel, and the scattering lengths and effective ranges for the coupled channels.

\subsection{Correlation functions for $Z_c(3900)$ and $Z_{cs}(3985)$}

As mentioned above, $Z_c(3900)$ and $Z_{cs}(3985)$ are often interpreted as the $D\bar{D}^*$ and $D\bar{D}_s^*$ resonant states from the experimental analyses based on the BW parametrization~\cite{BESIII:2013ris, Belle:2013yex, BESIII:2020qkh}. However, various theoretical re-analyses of the same data suggest they can be either resonances or virtual states~\cite{Du:2022jjv, Ji:2022uie, Yan:2023bwt, Chen:2023def}. Recently, we demonstrated that the femtoscopic correlation function can tell whether $Z_c(3900)$ and $Z_{cs}(3985)$ are resonances or virtual states~\cite{Liu:2024nac}. 

\begin{figure}[htpt]
\begin{center}
\includegraphics[width=0.98\textwidth]{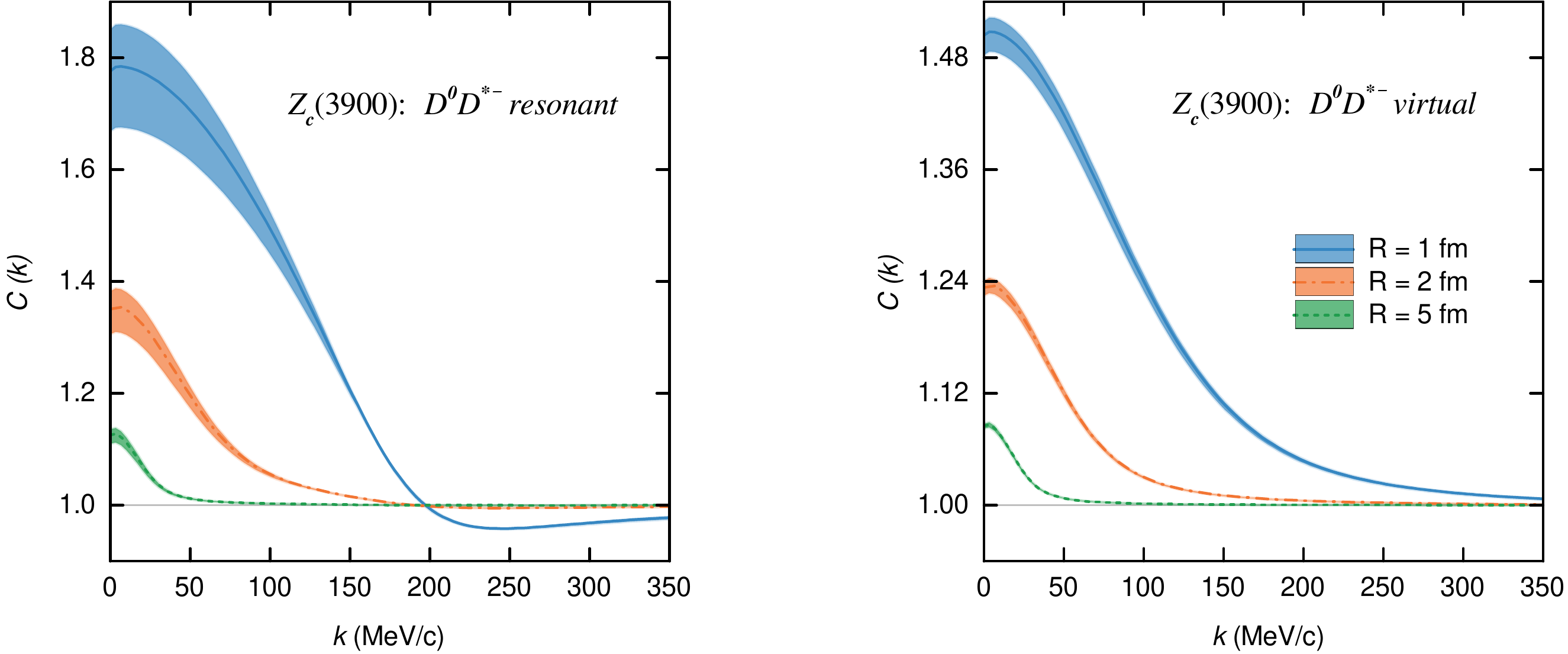}
\caption{$D^0D^{*-}$ correlation function as a function of the relative momentum $k$ for different source sizes $R = 1$, $2$, and $5$ fm. The results are obtained with the energy-dependent potential, which describes the $Z_c(3900)$ as a resonant state (left panel), and with the constant contact-range potential, which interprets the $Z_c(3900)$ as a virtual state (right panel), respectively. The bands reflect the variation of the sharp cutoff in the range of $q_{\rm max} = 0.8-1.2$ GeV, where the corresponding parameters $a$ and $b$ are redetermined by reproducing the pole position~\cite{Liu:2024nac}.}
\label{Fig:CF_Zc}
\end{center}
\end{figure}

\begin{figure}[htpt]
\begin{center}
\includegraphics[width=0.98\textwidth]{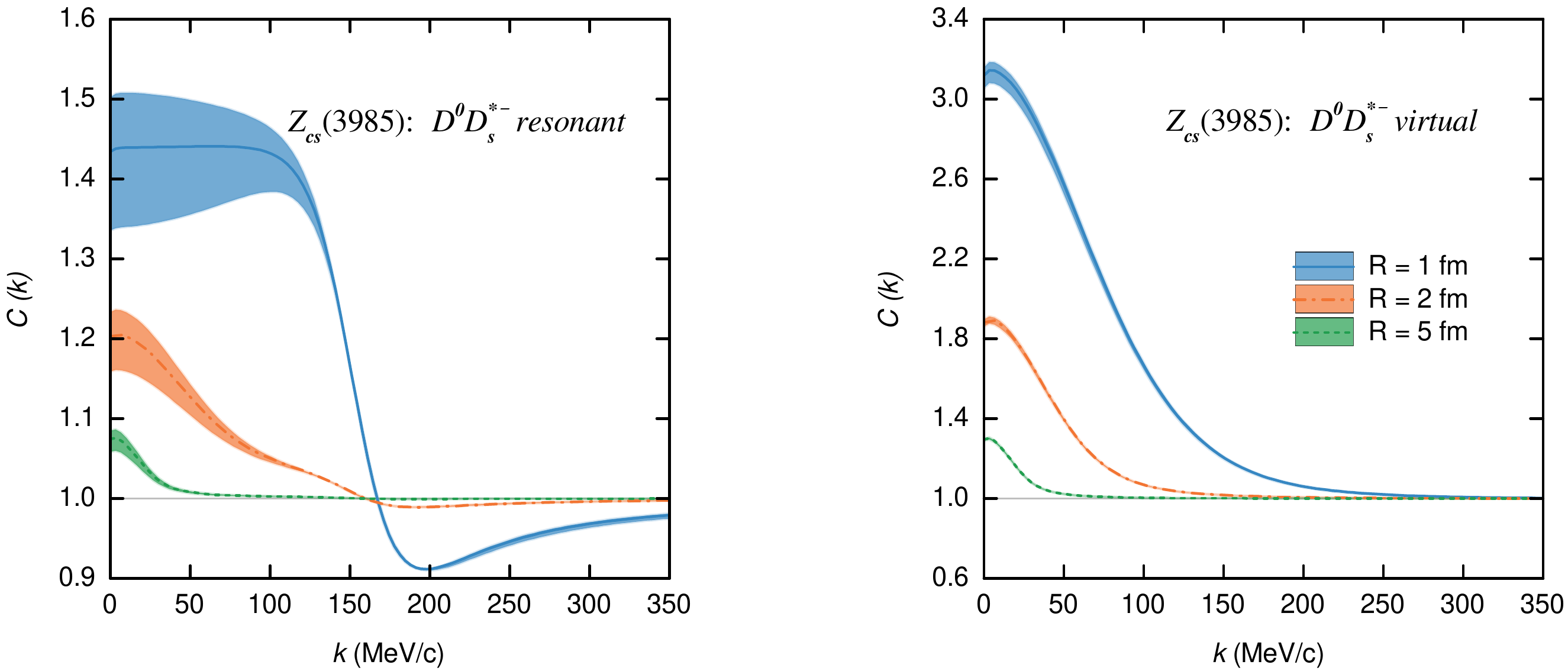}
\caption{$D^0D_s^{*-}$ correlation function as a function of the relative momentum $k$ for different source sizes $R = 1$, $2$, and $5$ fm. The results are obtained with the energy-dependent potential, which describes the $Z_{cs}(3985)$ as a resonant state (left panel), and with the constant contact-range potential, which interprets the $Z_{cs}(3985)$ as a virtual state (right panel), respectively. The bands reflect the variation of the sharp cutoff in the range of $q_{\rm max} = 0.8-1.2$ GeV, where the corresponding parameters $a$ and $b$ are redetermined by reproducing the pole position~\cite{Liu:2024nac}.}
\label{Fig:CF_Zcs}
\end{center}
\end{figure}

In Ref.~\cite{Liu:2024nac}, we first evaluated the $D^0D^{*-}$/$D^0D_s^{*-}$ interactions by precisely reproducing the pole positions of $Z_c(3900)$/$Z_{cs}(3985)$ in the resonant and virtual state scenarios, respectively. Subsequently, we computed the corresponding $D^0D^{*-}$/$D^0D_s^{*-}$ correlation functions. As shown in Fig.~\ref{Fig:CF_Zc}, the $D^0D^{*-}$ correlation function exhibits a pronounced enhancement in a wide range of the relative momentum $k$ in the virtual state scenario. Conversely, in the resonant state scenario, the $D^0D^{*-}$ correlation function remains enhanced in the low-momentum region but exhibits a suppression in the high-momentum region. This feature serves as a crucial criterion for distinguishing between the two scenarios. In addition, as the source size $R$ decreases, the difference in the correlation functions between these two scenarios becomes more obvious, reflecting the short-range nature of the strong interaction.

Similarly, as depicted in Fig.~\ref{Fig:CF_Zcs}, the $D^0D_s^{*-}$ correlation functions exhibit analogous trends to those observed for $D^0D^{*-}$ results. However, due to the narrower width of $Z_{cs}(3985)$, the suppression in the high-momentum region becomes more significant in the resonant state scenario. Moreover, the distinct behavior of the $D^0D_s^{*-}$ correlation function at zero momentum offers an additional means of distinguishing between the resonant and virtual state scenarios for the $Z_{cs}(3985)$ state.

\subsection{Correlation functions for $P_c(4312)$, $P_c(4440)$, and $P_c(4457)$}
As mentioned above, the spins of the pentaquark states $P_c(4440)$ and $P_c(4457)$ play a decisive role in unraveling their nature but remain undetermined experimentally. There have been a large number of theoretical studies trying to distinguish their spins from various perspectives~\cite{Liu:2019tjn, Xiao:2019mvs, Wu:2019rog, Wang:2019krd, Meng:2019ilv, Pan:2019skd, PavonValderrama:2019nbk, Lin:2019qiv, Zhu:2019iwm, Du:2019pij, Du:2021fmf, Li:2021ryu, Xie:2020niw, Ling:2021sld, Ling:2021lmq, Ozdem:2021ugy, Shi:2022ipx, Xie:2022hhv, Burns:2022uiv, Zhang:2023czx}. Recently, we proposed to discriminate the spins of $P_c(4440)$ and $P_c(4457)$ with femtoscopic correlation functions~\cite{Liu:2023wfo}.

\begin{figure}[htpt]
\begin{center}
\includegraphics[width=0.98\textwidth]{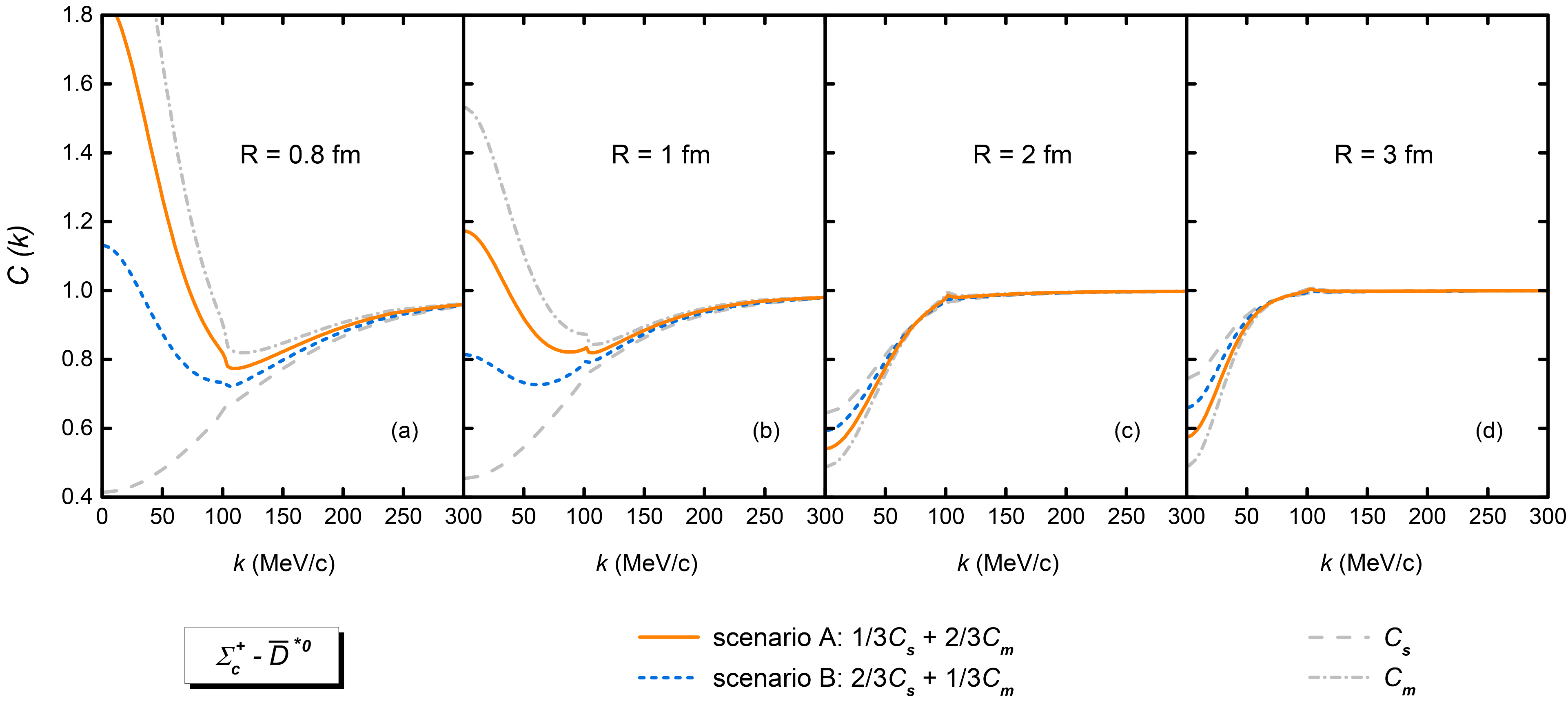}
\caption{Spin-averaged $\Sigma_c^+\bar{D}^{*0}$ correlation function as a function of the relative momentum $k$ for different source sizes $R=0.8, 1, 2$ and $3$ fm, respectively. The blue short-dashed lines represent the spin-averaged results in scenario A, where $P_c(4440)$ and $P_c(4457)$ have $J^P=(1/2)^-$ and $J^P=(3/2)^-$ respectively, while the orange solid lines represent the results in scenario B where $P_c(4440)$ and $P_c(4457)$ have $J^P=(3/2)^-$ and $J^P=(1/2)^-$~\cite{Liu:2023wfo}.}
\label{Fig:Pc4450}
\end{center}
\end{figure}

\begin{figure}[htpt]
\begin{center}
\includegraphics[width=0.42\textwidth]{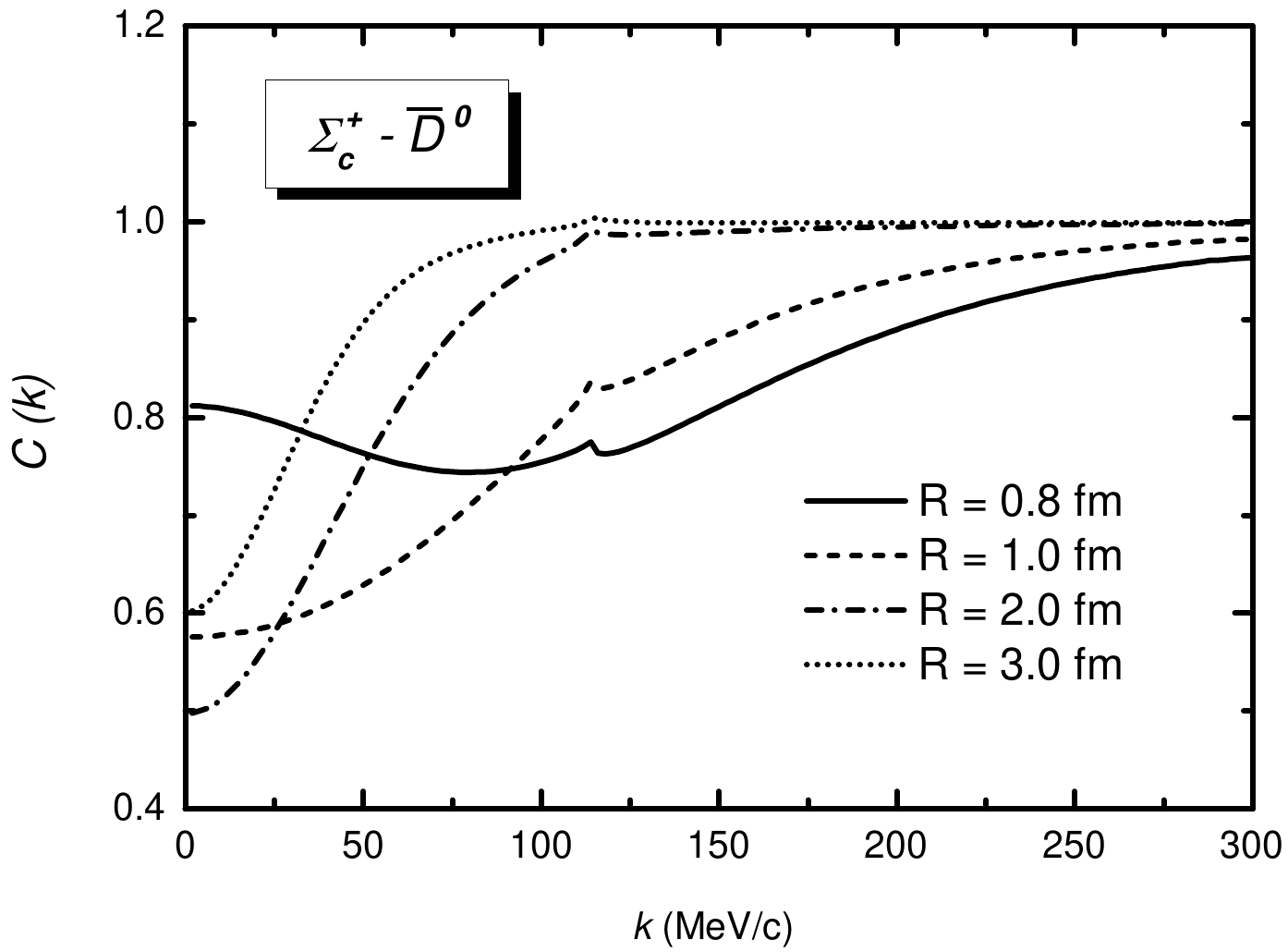}
\caption{Source size ($R$) dependence of the $\Sigma_c^+\bar{D}^0$ correlation function~\cite{Liu:2023wfo}.}
\label{Fig:Pc4312}
\end{center}
\end{figure}

In Ref.~\cite{Liu:2023wfo}, assuming $P_c(4440)$ and $P_c(4457)$ are $\Sigma_c\bar{D}^*$ bound states, we first evaluated their interactions without specifying their spin by reproducing the masses of $P_c(4440)$ (strongly attractive) and $P_c(4457)$ (moderately attractive) in the resonance saturation model. Then, we calculated the corresponding $\Sigma_c^+\bar{D}^{*0}$ correlation functions for the first time. In Fig.~\ref{Fig:Pc4450}, the $\Sigma_c\bar{D}^*$ correlation functions $C_s$ ($C_m$) calculated with the moderately (strongly) attractive interactions are shown as the dash-dotted (dashed) lines, which are in remarkable agreement with the aforementioned general features of correlation functions, namely, a moderately attractive interaction and a strongly attractive one may result in entirely different low-momentum behaviors. The spin-averaged $\Sigma_c\bar{D}^*$ correlation functions and their source size dependence are also shown in Fig.~\ref{Fig:Pc4450}. We found that the low-momentum behaviors of the spin-averaged results are pretty different in the two spin assignments, especially for a small collision system, which can be used to determine the spins of $P_c(4440)$ and $P_c(4457)$.   In addition to the previously mentioned data-driven approach, the authors suggest using the source size dependence of the $\Sigma_c^+\bar{D}^0$ correlation function to determine the source size of $\Sigma_c^+\bar{D}^{*0}$, where these two systems are assumed to be characterized by a common emitting source due to the rather similar masses of $\Sigma_c^+\bar{D}^0$ and $\Sigma_c^+\bar{D}^{*0}$ pairs. Once the $\Sigma_c^+\bar{D}^0$ interaction is determined reasonably from the $P_c(4312)$, the space-time dimension of the emitting source is the only unknown factor in the femtoscopic study.   Similarly, we evaluated the $\Sigma_c^+\bar{D}^0$ interaction by reproducing the $P_c(4312)$ mass and predicted the corresponding correlation function in Fig.~\ref{Fig:Pc4312}, which exhibits an apparent and nonmonotonic source size dependence. It is worth noticing that the strategy proposed can be applied to decipher the nature of other hadronic molecules and thus deepen our understanding of the non-perturbative strong interaction.

\subsection{Correlation functions for $T_{bb}$}
Analogous to the $T_{cc}(3875)$ in the charm sector, a $T_{bb}$ state in the bottom sector is expected to exist~\cite{Ohkoda:2012hv, Li:2012ss, Bicudo:2012qt, Bicudo:2015kna, Wang:2018atz, Yu:2019sxx, Zhao:2021cvg, Dai:2022ulk}. In Ref.~\cite{Dai:2022ulk}, the authors argued that the $T_{bb}$ state could be built up from the $B^0B^{*+}$ and $B^+B^{*0}$ channels and is predicted to be a molecular state of these components with isospin $I = 0$, a binding energy of 21 MeV and a width of 14 eV in the local hidden gauge approach. More recently, the $B^0B^{*+}$ and $B^+B^{*0}$ correlation functions had been predicted with the input from the above local hidden gauge approach~\cite{Feijoo:2023sfe}, where the interaction between the bottom mesons is generated by exchanging vector mesons. 

\begin{figure}[htbp]
\centering
\begin{minipage}[t]{0.46\textwidth}
\includegraphics[width=0.98\textwidth]{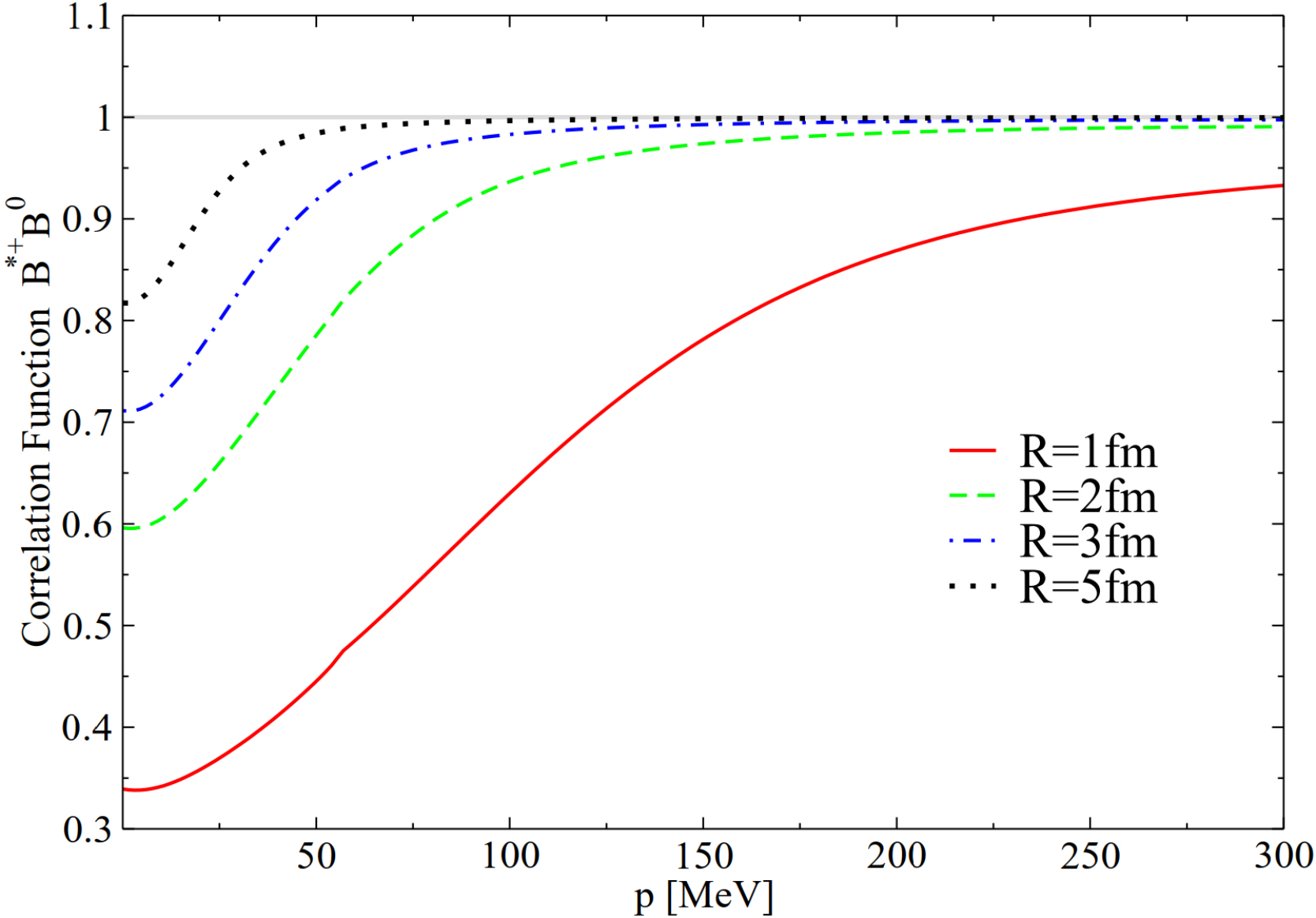}
\end{minipage}
\begin{minipage}[t]{0.46\textwidth}
\includegraphics[width=0.98\textwidth]{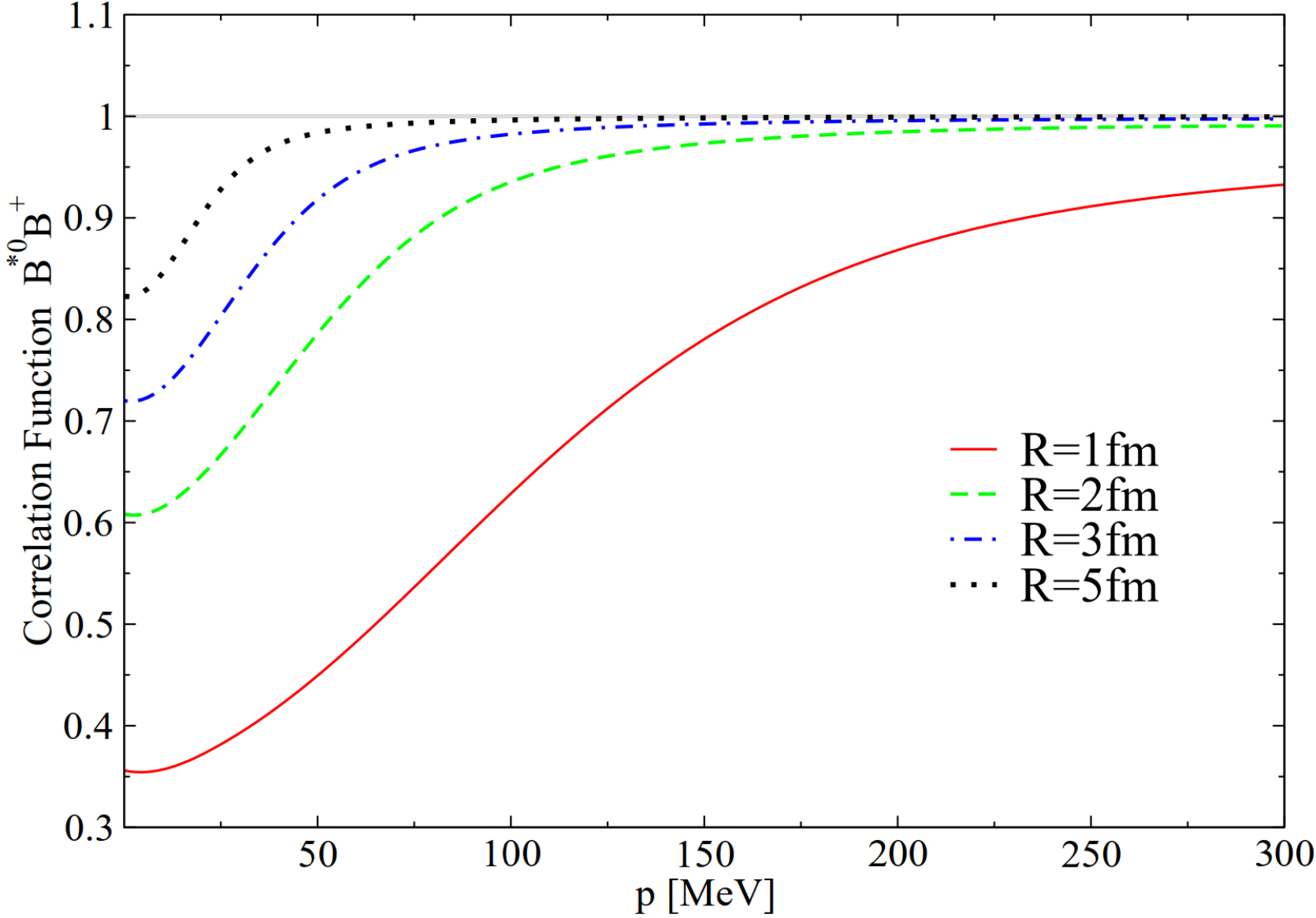}
\end{minipage}
\caption{$B^0B^{*+}$ and $B^+B^{*0}$ correlation functions for different source sizes.  \\ Source:  Figures   taken from Ref.~\cite{Feijoo:2023sfe}.}
\label{Fig:Tbb}
\end{figure}

Fig.~\ref{Fig:Tbb} showed the $B^0B^{*+}$ and $B^+B^{*0}$ correlation functions for different source sizes $R$ obtained in Ref.~\cite{Feijoo:2023sfe}. These correlation functions are all suppressed in a wide range of the relative momentum $p$ for both small and large source sizes, which reflects the existence of the $T_{bb}$ state. The magnitude of the correlation functions changes appreciably with $R$.  Ref.~\cite{Feijoo:2023sfe} also addressed the inverse problem of determining the low-energy observables related to this interaction from the knowledge of correlation functions. For this, they parametrized the potential in a very general form, which explicitly considers the freedom of non-molecular nature. Using the parameters obtained by fitting to the synthetic data, the authors obtained a bound state with a binding energy of 20.62 MeV, in excellent agreement with the known result of 21 MeV. These results show that it is promising for future experimental measurements of the $B^0B^{*+}$ and $B^+B^{*0}$ correlation functions to reveal the nature of the $T_{bb}$ state.

More recently, Ref.~\cite{Li:2024tof} studied the $B^0D^+$, $B^+D^0$ correlation functions. The $B^0D^+$ and $B^+D^0$ system was shown to develop a $BD$ bound state by about $40$ MeV, using inputs consistent with the $T_{cc}(3875)$ state~\cite{Vidana:2023olz, Albaladejo:2023wmv}. The inverse problem was also studied, determining scattering observables from the correlation functions, including the existence of the bound state and its molecular nature. All correlation functions involving heavy charmed and bottom quark studied are summarized in Table~\ref{Tab:CF_sum}.

\begin{table}[htpt]
\centering
\setlength{\tabcolsep}{7.6pt}
\caption{Studies of correlation functions of heavy-flavor systems.}\label{Tab:CF_sum}
\begin{tabular}{ccccc}
\hline
\hline
{\rm State} &{\rm Mass}  &{\rm Width}  &{\rm S-wave threshold}  &{\rm Correlation Function pair}  \\
 &${\rm MeV/c^2}$  &${\rm MeV/c^2}$  &${\rm MeV/c^2}$  &  \\
\hline
$D_{s0}^*(2317)$ &$2317.8\pm0.5$~\cite{ParticleDataGroup:2022pth}  &$<3.8$  &$DK(+45.2)$  &$D^0K^+/D^+K^0$~\cite{Liu:2023uly, Albaladejo:2023pzq, Ikeno:2023ojl}  \\
 &  &  &$D_s\eta(+198.4)$  &$D_s^+\eta$~\cite{Ikeno:2023ojl}  \\
$D_0^*(2300)$ &$2343\pm10$~\cite{ParticleDataGroup:2022pth}  &$229\pm16$  &$D\pi(-336.9)$  &$D\pi~(I=1/2)$~\cite{Albaladejo:2023pzq}  \\
 &  &  &$D\eta(+72.9)$  &$D\eta~(I=1/2)$~\cite{Albaladejo:2023pzq}  \\
 &  &  &$D_s\bar{K}(+120.3)$  &$D_s\bar{K}~(I=1/2)$~\cite{Albaladejo:2023pzq}  \\
&$2092.4$ (lower)~\cite{Torres-Rincon:2023qll}  &$259$  &$D\pi(-86.3)$  &$D\pi~(I=1/2)$~\cite{Torres-Rincon:2023qll}  \\
 &$2647.2$ (higher)~\cite{Torres-Rincon:2023qll}  &$529.6$  &$D_s\bar{K}(-183.9)$  &$D_s\bar{K}~(I=1/2)$~\cite{Torres-Rincon:2023qll}  \\
$D_1(2420)$ &$2422.1\pm0.6$~\cite{ParticleDataGroup:2022pth}  &$31.3\pm1.9$  &$D\rho(+221.2)$  &$D\rho(I=1/2)$~\cite{Khemchandani:2023xup}  \\
$D_1(2430)$ &$2412\pm9$~\cite{ParticleDataGroup:2022pth}  &$314\pm29$  &$D^*\pi(-264.8)$  &$D^*\pi(I=1/2)$~\cite{Khemchandani:2023xup}  \\
&$2233.6$ (lower)~\cite{Torres-Rincon:2023qll}  &$261.6$  &$D^*\pi(-86.4)$  &$D^*\pi(I=1/2)$~\cite{Torres-Rincon:2023qll}  \\
&$2719.2$ (higher)~\cite{Torres-Rincon:2023qll}  &$660.2$  &$D_s^*\bar{K}(-112.01)$  &$D_s^*\bar{K}(I=1/2)$~\cite{Torres-Rincon:2023qll}  \\
$T_{cc}(3875)$ &$3874.83\pm0.11$~\cite{ParticleDataGroup:2022pth}  &$0.41\pm0.17$  &$DD^*(+2.3)$  &$D^0D^{*+}/D^+D^{*0}$~\cite{Kamiya:2022thy, Vidana:2023olz, Albaladejo:2023wmv}  \\
$X(3872)$ &$3871.65\pm0.06$~\cite{ParticleDataGroup:2022pth}  &$1.19\pm0.21$  &$D\bar{D}^*(+5.5)$  &$D^0\bar{D}^{*0}/D^+D^{*-}$~\cite{Kamiya:2022thy}  \\
$Z_c(3900)$ &$3887.1\pm2.6$~\cite{ParticleDataGroup:2022pth}  &$28.4\pm2.6$  &$D\bar{D}^*(-12)$  &$D^0D^{*-}$~\cite{Liu:2024nac}  \\
$Z_{cs}(3985)$ &$3988\pm5$~\cite{ParticleDataGroup:2022pth}  &$13\pm4$  &$D\bar{D}_s^*(-10.96)$  &$D^0D_s^{*-}$~\cite{Liu:2024nac}  \\
$P_c(4312)$ &$4311.9_{-0.9}^{+7.0}$~\cite{ParticleDataGroup:2022pth}  &$10\pm5$  &$\Sigma_c\bar{D}(+5.6)$  &$\Sigma_c^+\bar{D}^0$~\cite{Liu:2023wfo}  \\
$P_c(4440)$ &$4440_{-5}^{+4}$~\cite{ParticleDataGroup:2022pth}  &$21_{-11}^{+10}$  &$\Sigma_c\bar{D}^*(+19.2)$  &$\Sigma_c^+\bar{D}^{*0}$~\cite{Liu:2023wfo}  \\
$P_c(4457)$ &$4457.3_{-1.8}^{+4.0}$~\cite{ParticleDataGroup:2022pth}  &$6.4_{-2.8}^{+6.0}$  &$\Sigma_c\bar{D}^*(+2.2)$  &$\Sigma_c^+\bar{D}^{*0}$~\cite{Liu:2023wfo}  \\
$T_{bb}~[{\rm predicted}]$ &$10583$~\cite{ParticleDataGroup:2022pth}  &$1.4\times10^{-5}$  &$BB^*(+21.2)$  &$B^0B^{*+}/B^+B^{*0}$~\cite{Feijoo:2023sfe}  \\
$T_{\bar{b}c}~[{\rm predicted}]$ &$7110.41$~\cite{ParticleDataGroup:2022pth}  &$-$  &$BD(+37.1)$  &$D^0B^+/D^+B^0$~\cite{Li:2024tof}  \\
\hline
\hline
\end{tabular}
\end{table}

\section{ Summary and Outlook}
\label{sum}

Since 2003, many states beyond the conventional $q\bar{q}$ mesons and $qqq$ baryons have been discovered. These discoveries enrich hadron spectra and provide an invaluable opportunity to understand the non-perturbative strong interaction. These so-called exotic states have been proposed to be hadronic molecules, compact multiquark states, kinetic effects, or their mixtures.  Among them, the hadronic molecular picture is one of the most prevailing interpretations since most exotic states are near the mass thresholds of pairs of conventional hadrons.  Their mass spectrum, decay widths, and production rates have been thoroughly investigated within the molecular picture,  most of which agree with existing data. 

In this work, we first reviewed the experimental and theoretical status on
several candidates for hadronic molecules containing heavy quarks. These include:  $D_{s0}^{*}(2317)$, $D_{s1}(2460)$,  $X(3872)$, $T_{cc}(3875)$, $Z_{c}(3900)$, $Z_{c}(4020)$, $Z_{cs}(3985)$, $P_{c}(4312)$, $P_{c}(4440)$, and $P_{c}(4457)$. These states are arguably the most studied among the exotic hadrons discovered. We provided a concise introduction to the experiments where they were observed and commented on the relevance of these experiments in deciphering their nature.  We also summarized related theoretical studies on their properties, particularly those in the molecular picture. Here, we want to point out the connection between the unquenched quark model and the hadronic molecular picture. 
The unquenched quark model treats these states as mixtures of conventional $q\bar{q}$/$qqq$ hadrons and hadronic molecules~\cite{Ni:2023lvx,Deng:2023mza}. These studies supported the molecular picture. They only differ in the terminology. In the unquenched quark model, one explicitly treats the $q\bar{q}$/$qqq$ and $MM$/$MB$ degrees of freedom. The (modified)  $^3 P_0$  model provides the couplings between the two components. On the other hand, in the hadronic molecular picture, only hadron-hadron interactions are explicitly considered. The existence of a $q\bar{q}$ or $qqq$ core and the effects of missing coupled channels are hidden in the relevant parameters, such as those in the regulator function. The relative importance of these components can be quantified using the Weinberg compositeness. For this, there exist a large number of works; see Refs.~\cite{Hyodo:2013nka,Aceti:2014ala,Sekihara:2014kya,Li:2021cue,Baru:2021ldu,Song:2022yvz} for example.

Now, the crucial question is to confirm or refute the molecular nature of these exotic states. This review discussed three novel approaches that can help achieve such an ambitious goal. 

First, one can derive from the molecular picture the underlying two-body hadron-hadron interactions. These interactions can be extended to other systems adopting heavy quark spin and flavor symmetry, SU(3)-flavor symmetry, and heavy antiquark diquark symmetry to predict multiplets of hadronic molecules. The discovery of these multiplets will provide a highly nontrivial verification of the hadronic molecular picture. Historically, this led to more confidence in the power of SU(3) flavor symmetry and the quark model's validity. Recent experimental discoveries of the three pentaquark states, $P_c(4312)$, $P_c(4440)$, $P_c(4457)$, the pentaquark and tetraquark states with strangeness, $P_{cs}(4459)$, $P_{cs}(4338)$, and $Z_{cs}(985)$, indeed hinted at the existence of multiplets of hadronic molecules. Much future work needs to be done regarding the predictions of multiplets. For instance, all the symmetries are broken at the level of a few tens of percent. Such symmetry-breaking effects need to be studied in more detail because they can lead to the nonexistence of the predicted hadronic molecules. In addition, the discovery channels of these predicted states need to be identified, and whether they can be discovered at current or future experimental facilities should be examined. One more subtle issue is that compact quark models also predicted the existence of multiplets, see, e.g., Refs.~\cite{Maiani:2021tri,Ferretti:2021zis}. Therefore, even if they are discovered, a careful analysis of future experimental discoveries may be needed. It can well be that only by combining all three approaches summarized in the present review (and others not covered here) can one finally decipher the nature of the many exotic hadrons described in this review.

Second, assuming these exotic states are two-body hadronic molecules, one can derive their potential by reproducing their masses.  With these potentials, one can study the existence of three-body hadronic molecules. The three-body hadronic molecules are made of three hadrons and thus have unique quantum numbers. They can form new kinds of matter in addition to nuclei and hypernuclei.  If discovered, they can help verify the molecular nature of these exotic states. Therefore, we strongly encourage experimental searches for such three-body hadronic molecules.  One notes that most previous studies focused on the heavy flavor sector. This is because, in the light flavor sector,  the production/annihilation of light quark-antiquark pairs is easier from the energy perspective and, therefore, can occur more frequently.
In addition, the kinematic effects of light hadrons are more significant, and therefore, relativistic effects need to be considered in many cases.  As a result, light-quark sectors are more difficult to deal with.
As discussed in this review, extensive studies on three-body hadronic molecules have been conducted, and significant progress has been made in the heavy flavor sector. Many three-body bound/resonant states, including their masses, intrinsic quantum numbers, and decay widths, have been predicted. 
These theoretical studies are valuable but still insufficient. More systematic and in-depth studies are needed. Most predictions of three-body hadronic molecules are based on assumptions about the existence of sub-two-body molecules, from which the underlying two-body hadron-hadron interactions are derived. We know that three-body forces are important in nuclear physics and essential for precisely describing light nuclei (at least in nonrelativistic frameworks). 
Previous studies have neglected three-body forces. They need to be studied, and their effects must be carefully evaluated in the future. The main difficulty lies in the lack of experimental constraints on three-body hadronic interactions. One also needs to figure out where to find the predicted three-body molecular states. 
Most previous studies mainly focused on their masses and decay widths. The ultimate way to verify all these theoretical studies is to discover them experimentally. One must study their production rates in various physical processes, inclusive or exclusive, in either hadron-hadron collisions or electron-positron colliders.

Third, we reviewed the application of the femtoscopic technique in providing direct experimental information on two-body hadron-hadron interactions. There were well-developed theoretical approaches to achieve such a purpose using the experimentally measured correlation functions.  Using these interactions, one can, in principle, check whether they are strong enough to generate two-body bound states or resonances. Thus, one can check the hadronic molecular picture for the many discovered exotic states. In the past few years, we have seen increasing experimental and theoretical interests and the amount of work performed.  Nevertheless, more work is still needed. For instance, although a spherical Gaussian source is often adopted in previous studies, 
there are still discussions on other choices, such as the Cauchy~\cite{Oliver2017} or the 
 L\'evy~\cite{Csanad:2024hva} distributions. Second, most previous studies rely on prior theoretical information to extract the underlying hadron-hadron interaction. It may be helpful to develop fully model-independent approaches to derive these interactions. The first experimental analyses have appeared, which require further theoretical investigation. Third, recent experimental and theoretical studies of three-hadron interaction using the femtoscopic technique have been reported~\cite{ALICE:2022boj,ALICE:2023gxp,ALICE:2023bny}. This will provide an unprecedented opportunity for us to derive three-body interactions without complicated medium effects. This will also provide essential information for the studies of three-body hadronic molecules. Nevertheless, more theoretical studies are needed to exploit the experimental techniques fully.

Finally, even the most favorable hadronic molecular candidates can contain a small $q\bar{q}/qqq$ or compact multiquark component.  How the existence of such components (even as small as 1 percent) affects the above studies assuming these states as pure hadronic molecules remains to be investigated. However, we do not anticipate they will qualitatively affect the conclusions drawn. 

To summarize, the experimental discoveries of the so-called exotic hadrons and the subsequent intensive theoretical studies have significantly enriched our understanding of the non-perturbative strong interaction.  Nevertheless, a unified and complete picture of these hadrons is still missing, though the molecular picture seems to prevail. The three novel approaches reviewed in this work can help verify the molecular picture in a highly nontrivial way. We further point out future directions to advance the three approaches, our understanding of hadron spectra, and the underlying non-perturbative strong interaction. 
\section{Acknowledgments}

We thank Eulogio Oset for carefully reading the draft and the many valuable comments on the first version of this manuscript. We are very grateful for our collaborators: Atsushi Hosaka, Xiang Liu,  Manuel Pavon Valderrama,  Ju-Jun Xie, Yin Huang, Si-Qiang Luo,  Mario Sánchez Sánchez,  Eulogio Oset, Xi-Zhe Ling, Fang-Zheng Peng, En Wang, Qi Wu, and Jia-Ming Xie. This work is partly supported by the National Key R\&D Program of China under Grant No. 2023YFA1606703, and the National Natural Science foundation of China under Grants No. 12435007, No. 11735003 and No.11975041. Ming-Zhu Liu acknowledges support from the National Natural Science Foundation of
China under Grant No.12105007. Jun-Xu Lu acknowledges support from the National Natural Science Foundation of China under Grant No.12105006. 
Ya-Wen Pan acknowledges support from the China Scholarship Council scholarship and the Academic Excellence Foundation of BUAA for PhD Students. 
Zhi-Wei Liu acknowledges support from the National Natural Science Foundation of China under Grant No.12405133, No.12347180, China Postdoctoral Science Foundation under Grant No.2023M740189, and the Postdoctoral Fellowship Program of CPSF under Grant No.GZC20233381
Tian-Wei Wu acknowledges support from the National Natural Science Foundation of China under Grant No.12405108.

\appendix

\section{Contact-range potentials}
This section summarizes the contact-range potentials constrained by various symmetries. These serve as the basis for the predictions of multiplets of hadronic molecules. 

\subsection{$\bar{D}^{(\ast)}\Sigma_{c}^{(\ast)}$ heavy anti-meson and heavy baryon systems}
In the EFTs, the leading-order interactions between pairs of hadrons include two terms: one-pion exchange and contact range, while the one-pion exchange can be treated perturbatively~\cite{Lu:2017dvm}. The following briefly explains how to construct the contact-range potentials for relevant systems constrained by the HQSS. For a pair of charmed mesons and baryons, the $\bar{D}^{(\ast)}\Sigma_{c}^{(\ast)}$ interaction results in seven states~\cite{Liu:2018zzu,Liu:2019tjn,Du:2019pij}. According to the HQSS, the spin wave function of the $\bar{D}^{(\ast)}\Sigma_{c}^{(\ast)}$ pairs  can be written as follows, in terms of the spins of the heavy quarks $s_{1H}$ and $s_{2H}$ and those of the light quark(s) (often referred to as brown
muck~\cite{Isgur:1991xa,Flynn:1992fm}) $s_{1L}$ and $s_{2L}$, where 1 and 2 denote $\bar{D}^{(*)}$ and $\Sigma_c^{(*)}$, respectively, via the following spin coupling formula,
\begin{eqnarray}
\label{9j}
&&|s_{1l}, s_{1h}, j_{1}; s_{2l}, s_{2h},j_{2}; J\rangle =
  \\ \nonumber &&  
  \sqrt{(2j_{1}+1)(2j_{2}+1)(2s_{L}+1)(2s_{H}+1)}\left(\begin{matrix}
s_{1l} & s_{2l} & s_{L} \\
s_{1h} & s_{2h} & s_{H} \\
j_{1} & j_{2} & J%
\end{matrix}\right)|s_{1l},
s_{2l}, s_{L}; s_{1h}, s_{2h},s_{H}; J\rangle.
\end{eqnarray}
More explicitly, for the seven $\bar{D}^{(\ast)}\Sigma_{c}^{(\ast)}$ states, the decompositions read
\begin{eqnarray}
|\Sigma_{c}\bar{D}(1/2^{-})\rangle &=& \frac{1}{2}0_{H}\otimes
{1/2}_{L}+ \frac{1}{2\sqrt{3}}1_{H}\otimes {1/2}_{L}+
\sqrt{\frac{2}{3}}1_{H}\otimes {3/2}_{L},   \\ \nonumber
|\Sigma_{c}^{\ast}\bar{D}(3/2^{-})\rangle &=&
-\frac{1}{2}0_{H}\otimes {3/2}_{L}+ \frac{1}{\sqrt{3}}1_{H}\otimes
{1/2}_{L}+ \frac{\sqrt{\frac{5}{3}}}{2}1_{H}\otimes {3/2}_{L},  \\ \nonumber
|\Sigma_{c}\bar{D}^{\ast}(1/2^{-})\rangle &=&
\frac{1}{2\sqrt{3}}0_{H}\otimes {1/2}_{L}+ \frac{5}{6}1_{H}\otimes
{1/2}_{L}-\frac{\sqrt{2}}{3}1_{H}\otimes {3/2}_{L},  \\ \nonumber
|\Sigma_{c}\bar{D}^{\ast}(3/2^{-})\rangle &=&
\frac{1}{\sqrt{3}}0_{H}\otimes {3/2}_{L}- \frac{1}{3}1_{H}\otimes
{1/2}_{L}+ \frac{\sqrt{5}}{3}1_{H}\otimes {3/2}_{L},   \\ \nonumber
|\Sigma_{c}^{\ast}\bar{D}^{\ast}(1/2^{-})\rangle &=&
\sqrt{\frac{2}{3}}0_{H}\otimes {1/2}_{L}-
\frac{\sqrt{2}}{3}1_{H}\otimes {1/2}_{L}-\frac{1}{3}1_{H}\otimes
{3/2}_{L},                                 \\ \nonumber
|\Sigma_{c}^{\ast}\bar{D}^{\ast}(3/2^{-})\rangle &=&
\frac{\sqrt{\frac{5}{3}}}{2}0_{H}\otimes {3/2}_{L}+
\frac{\sqrt{5}}{3}1_{H}\otimes {1/2}_{L}- \frac{1}{6}1_{H}\otimes
{3/2}_{L},  \\ \nonumber
|\Sigma_{c}^{\ast}\bar{D}^{\ast}(5/2^{-})\rangle &=& 1_{H}\otimes
{3/2}_{L},
\end{eqnarray}
where $0_{H}$ and $1_{H}$ are the total spin of a pair charmed and anticharmed quarks, i.e., $1/2_{1H} \otimes 1/2_{2H}=0_H \oplus 1_H$, and the total spin of  light quark and diquark are  $1/2_{1l} \otimes 1_{2l}=1/2_L \oplus 3/2_L$.    
In the heavy quark limit, the  $S$-wave $\bar{D}^{(\ast)}\Sigma_{c}^{(\ast)}$ interactions are independent of the spin of the heavy quark. Therefore, the potentials can be parameterized by two coupling constants  describing interactions between light quarks of spin 1/2 and 3/2, respectively, i.e.,  $F_{1/2}=\langle 1/2_{L} | V| 1/2_{L} \rangle$  and $F_{3/2}=\langle 3/2_{L} | V| 3/2_{L} \rangle$:
\begin{eqnarray}
V_{\Sigma_{c}\bar{D}}(1/2^{-}) &=& \frac{1}{3}F_{1/2L}+
\frac{2}{3}F_{3/2L},   \\ \nonumber
V_{\Sigma_{c}^{\ast}\bar{D}}(3/2^{-}) &=&
\frac{1}{3}F_{1/2L}+
\frac{2}{3}F_{3/2L}, \\ \nonumber
V_{\Sigma_{c}\bar{D}^{\ast}}(1/2^{-}) &=&
\frac{7}{9}F_{1/2L}+
\frac{2}{9}F_{3/2L}, \\ \nonumber
V_{\Sigma_{c}\bar{D}^{\ast}}(3/2^{-}) &=&
\frac{1}{9}F_{1/2L}+
\frac{8}{9}F_{3/2L},  \\ \nonumber
V_{\Sigma_{c}^{\ast}\bar{D}^{\ast}}(1/2^{-}) &=&\frac{8}{9}F_{1/2L}+
\frac{1}{9}F_{3/2L},   \\ \nonumber
V_{\Sigma_{c}^{\ast}\bar{D}^{\ast}}(3/2^{-}) &=&\frac{5}{9}F_{1/2L}+
\frac{4}{9}F_{3/2L},  \\ \nonumber
V_{\Sigma_{c}^{\ast}\bar{D}^{\ast}}(5/2^{-}) &=& F_{3/2L},
\end{eqnarray}
which can be rewritten as a combination of $C_{a}$ and $C_{b}$, i.e., $F_{1/2} = C_a-2C_b$ and  $F_{3/2} = C_a+C_b$~\cite{Liu:2019tjn}. The potentials for the $\bar{D}^{(\ast)}\Sigma_{c}^{(\ast)}$ system are displayed in Table~\ref{tab:potential}. In the heavy quark limit, the $S$-wave $\bar{D}^{(\ast)}\Lambda_c\to \bar{D}^{(\ast)}\Lambda_c$ interactions are parameterised by one  coupling constant, i.e.,  $F_{1/2L}^{\prime}=\langle 1/2_{L}^{\prime} | V| 1/2_{L}^{\prime} \rangle$: 
\begin{eqnarray}
V_{\bar{D}\Lambda_{c}}(1/2^{-})= V_{\bar{D}^{*}\Lambda_{c}}(1/2^{-})=V_{\bar{D}^{*}\Lambda_{c}}(3/2^{-})=F_{1/2L}^{\prime},
\end{eqnarray}
where the parameter $F_{1/2L}^{\prime}$ can be rewritten as $C_a^{\prime}$.   

The  potentials of $J/\psi N \to J/\psi N  $, $J/\psi N  \to \eta_{c}N$ and $\eta_{c} N  \to \eta_{c}N$ are suppressed  due to the OZI rule, which is also supported by  lattice QCD simulations~\cite{Skerbis:2018lew}. In this work, we set $V_{J/\psi(\eta_{c}) N \to J/\psi(\eta_{c}) N }=0$.

In the following, the coupled-channel potentials are constructed using the same approach.    In the heavy quark limit, the $\bar{D}^{(\ast)}\Sigma_{c}^{(\ast)} \to J/\psi(\eta_{c}) N$ and  $\bar{D}^{(\ast)}\Sigma_{c}^{(\ast)} \to J/\psi \Delta$ potentials are allowed, while the  $\bar{D}^{(\ast)}\Sigma_{c}^{(\ast)} \to J/\psi \Delta$ potentials are suppressed due to  isospin symmetry breaking.
From  HQSS, the  $\bar{D}^{(\ast)}\Sigma_{c}^{(\ast)} \to J/\psi(\eta_{c}) N$ interactions  are only related to  the spin of the light quark $1/2$, denoted by one coupling:  $g_2=\langle \bar{D}^{(\ast)}\Sigma_{c}^{(\ast)} | 1_{H}\otimes 1/2_{L} \rangle= \langle \bar{D}^{(\ast)}\Sigma_{c}^{(\ast)} | 0_{H}\otimes 1/2_{L} \rangle $.  Similarly, we can express the $\bar{D}^{(*)}\Lambda_c \to J/\psi(\eta_{c}) N$ interactions  by another parameter: $g_1= \langle \bar{D}^{(\ast)}\Lambda_{c} | 1_{H}\otimes 1/2_{L} \rangle =\langle \bar{D}^{(\ast)}\Lambda_{c}| 0_{H}\otimes 1/2_{L} \rangle$. As for the $\bar{D}^{(\ast)}\Sigma_{c}^{(\ast)} \to \bar{D}^{(*)}\Lambda_c$ interactions, they are   dependent on  only one  coupling  constant in the heavy quark limit. Therefore,  we parameter the $\bar{D}^{(\ast)}\Sigma_{c}^{(\ast)} \to \bar{D}^{(*)}\Lambda_c$ potential by  one  coupling:   $C_b^{\prime}= \langle \bar{D}^{(\ast)}\Sigma_{c}^{(\ast)}  | \bar{D}^{(\ast)}\Lambda_{c} \rangle$.

In the heavy quark limit, the contact-range potentials of $\bar{D}^*\Sigma_{c}^*-\bar{D}^*\Sigma_{c}-\bar{D}\Sigma_{c}-\bar{D}^*\Lambda_{c}-\bar{D}\Lambda_{c}-J/\psi N-\eta_{c}N$ system with $J^{P}=1/2^-$ can be expressed as 
\begin{widetext}
\begin{equation}
    V^{J=1/2}   =   \\   \\ 
    \begin{pmatrix}
  C_a-\frac{5}{3}C_b  &  -\frac{\sqrt{2}}{3}C_b&-\sqrt{\frac{2}{3}}C_b&\sqrt{\frac{2}{3}}C_b^{\prime}&\sqrt{2}C_b^{\prime}&   -\frac{\sqrt{2}}{3}g_{2}  &\sqrt{\frac{2}{3}}g_{2} \\ -\frac{\sqrt{2}}{3}C_b &C_{a}-\frac{4}{3}C_b&\frac{2}{\sqrt{3}}C_b&-\frac{2}{\sqrt{3}}C_b^{\prime}&C_b^{\prime}&  \frac{5}{6}g_{2} & \frac{1}{2\sqrt{3}}g_{2}\\
  -\sqrt{\frac{2}{3}}C_b&    \frac{2}{\sqrt{3}}C_b&C_{a}&C_b^{\prime}& 0 &\frac{1}{2\sqrt{3}}g_{2}   &\frac{1}{2}g_{2} \\
\sqrt{\frac{2}{3}}C_b^{\prime}&  -\frac{2}{\sqrt{3}}C_b^{\prime}&C_b^{\prime}&  C_{a}^{\prime} & 0 & \frac{1}{2}g_{1}  & \frac{\sqrt{3}}{2}g_{1}\\
\sqrt{2}C_b^{\prime} &  C_b^{\prime}&0  &0 &C_a^{\prime}&\frac{\sqrt{3}}{2}g_{1}&-\frac{1}{2}g_{1} \\
-\frac{\sqrt{2}}{3}g_{2}  & \frac{5}{6}g_{2}&\frac{1}{2\sqrt{3}}g_{2}&\frac{1}{2}g_{1}&\frac{\sqrt{3}}{2}g_{1} &0&0  \\
\sqrt{\frac{2}{3}}g_{2} &\frac{1}{2\sqrt{3}}g_{2}&\frac{1}{2}g_{2}&\frac{\sqrt{3}}{2}g_{1}&-\frac{1}{2}g_{1}&0&0\end{pmatrix}
\label{contact11}
\end{equation}
\end{widetext}
and the contact potentials of $\bar{D}^*\Sigma_{c}^*-\bar{D}^*\Sigma_{c}-\bar{D}\Sigma_{c}^*-\bar{D}^*\Lambda_{c}- J/\psi N$ system with $J^{P}=3/2^-$
are written as  
\begin{equation}
    V^{J=3/2}=\begin{pmatrix}
        C_a-\frac{2}{3}C_{b}& -\frac{\sqrt{5}}{3}C_b & \sqrt{\frac{5}{3}}C_{b} & \sqrt{\frac{5}{3}}C_{b}^{\prime}&\frac{\sqrt{5}}{3} g_{2}\\
   -\frac{\sqrt{5}}{3}C_b & C_a+\frac{2}{3}C_{b}& \frac{1}{\sqrt{3}}C_{b} & \frac{1}{\sqrt{3}}C_{b}^{\prime}&-\frac{1}{3} g_{2}\\
  \sqrt{\frac{5}{3}}C_{b}&   \frac{1}{\sqrt{3}}C_{b}&C_{a}&-C_{b}^{\prime}&\frac{1}{\sqrt{3}} g_{2}\\
 \sqrt{\frac{5}{3}}C_{b}^{\prime} &   \frac{1}{\sqrt{3}}C_{b}^{\prime}&-C_{b}^{\prime} &  C_a^{\prime}&g_{1}\\
   \frac{\sqrt{5}}{3} g_{2} &  -\frac{1}{3}g_{2}&\frac{1}{\sqrt{3}} g_{2} & g_{1}&0\end{pmatrix}
   \label{contact22}
\end{equation}

In terms of HADS, one can extend the $\bar{D}^{(\ast)}\Sigma_{c}^{(\ast)}$ system to the  $\Xi_{cc}^{(\ast)}\Sigma_{c}^{(\ast)}$  and $\bar{D}^{(\ast)}T_{\bar{c}\bar{c}}^{(\ast)}$ systems, which indicate that the low energy constants of these systems are the same in the heavy quark limit~\cite{Pan:2019skd,Pan:2022whr}. 
With a similar approach, we can easily derive the contact-range potentials for the  $\Xi_{cc}^{(\ast)}\Sigma_{c}^{(\ast)}$ and  $\bar{D}^{(\ast)}T_{\bar{c}\bar{c}}^{(\ast)}$  systems, which are tabulated in Table~\ref{tab:potential}.

\begin{table}[ttt]
  \centering \caption{The lowest-order contact-range potentials
    for the heavy antimeson - heavy baryon, doubly heavy
    baryon - heavy baryon and compact doubly heavy tetraquark states-heavy antimeson systems,
    which depend on two unknown coupling constants $C_{a}$ and $C_{b}$.
\label{tab:potential}}
\begin{tabular}{ccc|ccc|ccccc}
  \hline\hline ~State~~~& ~~$J^{P}$~~~~~~~ &~~~~V ~~~~~ &  ~State~~~& ~~~~$J^{P}$~~~~~~~ &~~~~V ~~~~ &  ~~State~~~& ~~~~$J^{P}$~~~~~~~ &~~~~V ~~~~\\

  \hline \multirow{2}{0.8cm}{$\bar{D}\Sigma_{c}$} &
  \multirow{2}{0.8cm}{$1/2^{-}$} &\multirow{2}{0.8cm}{$C_{a}$}
  & \multirow{2}{0.8cm}{$\Xi_{cc}\Sigma_{c}$}
  & $0^{+}$  & $C_{a}+\frac{2}{3}C_{b}$  & $\bar{D}T_{\bar{c}\bar{c}}^{0}$ & $0^-$ & $C_a$      \\
  & & & & $1^{+}$    & $C_{a}-\frac{2}{9}C_{b}$     & $\bar{D}^*T_{\bar{c}\bar{c}}^{1}$ & $0^-$ &$C_{a}-C_{b}$        \\
  \hline
  \multirow{2}{0.8cm}{$\bar{D}\Sigma_{c}^{\ast}$}&
  \multirow{2}{0.8cm}{$3/2^{-}$} &\multirow{2}{0.8cm}{$C_{a}$}
  &\multirow{2}{0.8cm}{$\Xi_{cc}\Sigma_{c}^{\ast}$}
  & $1^{+}$  & $C_{a}+\frac{5}{9}C_{b}$     & $\bar{D}T_{\bar{c}\bar{c}}^{1}$ & $1^-$ &$C_{a}$          \\
  & & & & $2^{+}$    & $C_{a}-\frac{1}{3}C_{b}$   & $\bar{D}^*T_{\bar{c}\bar{c}}^{0}$ & $1^-$ &$C_{a}$             \\
  \hline
  \multirow{2}{0.8cm}{$\bar{D}^{\ast}\Sigma_{c}$}&$1/2^{-}$
  &$C_{a}-\frac{4}{3}C_{b}$ &
  \multirow{2}{0.8cm}{$\Xi_{cc}^{\ast}\Sigma_{c}$}  & $1^{+}$  & $C_{a}-\frac{10}{9}C_{b}$   & $\bar{D}^*T_{\bar{c}\bar{c}}^{1}$ & $1^-$ &$C_{a}-\frac{1}{2}C_b$   
\\
& $3/2^{-}$ &$C_{a}+\frac{2}{3}C_{b}$
 & & $2^{+}$  & $C_{a}+\frac{2}{3}C_{b}$    & $\bar{D}^*T_{\bar{c}\bar{c}}^{2}$ & $1^-$ &$C_{a}-\frac{3}{2}C_b$   
\\
\hline
\multirow{4}{0.8cm}{$\bar{D}^{\ast}\Sigma_{c}^{\ast}$} & $1/2^{-}$
&$C_{a}-\frac{5}{3}C_{b}$  &
\multirow{4}{0.8cm}{$\Xi_{cc}^{\ast}\Sigma_{c}^{\ast}$}  &$0^{+}$
& $C_{a}-\frac{5}{3}C_{b}$    & $\bar{D}T_{\bar{c}\bar{c}}^{2}$ & $2^-$ &$C_{a}$   
\\
& \multirow{2}{0.8cm}{$3/2^{-}$}
& \multirow{2}{1.5cm}{$C_{a}-\frac{2}{3}C_{b}$} &
& $1^{+}$ & $C_{a}-\frac{11}{9}C_{b}$    & $\bar{D}^*T_{\bar{c}\bar{c}}^{1}$ & $2^-$ &$C_{a}+\frac{1}{2}C_{b}$   
\\  & &  &
& $2^{+}$ & $C_{a}-\frac{1}{3}C_{b}$    & $\bar{D}^*T_{\bar{c}\bar{c}}^{2}$ & $2^-$ &$C_{a}-\frac{1}{2}C_{b}$  
\\
& {$5/2^{-}$}    &{$C_{a}+C_{b}$}    &   & $3^{+}$ & $C_{a}+C_{b}$     & $\bar{D}^*T_{\bar{c}\bar{c}}^{2}$ & $3^-$ &$C_{a}+C_{b}$     \\
\hline\hline
\end{tabular}
\end{table}

\begin{table}[!h]
\caption{SU(3)-flavor structure of the potentials for the heavy meson-baryon
  molecules, where the heavy meson belongs to a SU(3)-flavor triplet and
  the heavy baryon to a sextet.
}
\label{tab:su3}
\begin{tabular}{ccccc}
\hline \hline
  Molecule & $I$ & $S$ & $V$ & $V_{\rm eigen}$ \\
  \hline
  $\bar{P} \Sigma_Q$ & $\tfrac{1}{2}$ & \phantom{+}$0$ & $V^O$ & $-$\\
  $\bar{P} \Sigma_Q$ & $\tfrac{3}{2}$ & \phantom{+}$0$ & $V^D$ & $-$\\
  $\bar{P} \Xi_Q'$ & $0$ & $-1$& $V^O$ & $-$\\
  $\bar{P} \Xi_Q'-\bar{P}_s \Sigma_Q$ & $1$ & $-1$ &
  $\begin{pmatrix}
    \frac{1}{3} V^O + \frac{2}{3} V^D & -\frac{\sqrt{2}}{3}\,(V^O - V^D) \\
    -\frac{\sqrt{2}}{3}\,(V^O - V^D) &  \frac{2}{3} V^O + \frac{1}{3} V^D \\ 
  \end{pmatrix}$ & $\{ V^O, V^D \}$ \\
  $\bar{P} \Omega_Q-\bar{P}_s \Xi_Q'$ & $\tfrac{1}{2}$ & $-2$ &
  $\begin{pmatrix}
    \frac{1}{3} V^O + \frac{2}{3} V^D & -\frac{\sqrt{2}}{3}\,(V^O - V^D) \\
    -\frac{\sqrt{2}}{3}\,(V^O - V^D) &  \frac{2}{3} V^O + \frac{1}{3} V^D \\ 
  \end{pmatrix}$ & $\{ V^O, V^D \}$ \\
  $\bar{P}_s \Omega_Q$ & $0$ & $-3$ & $V^D$ & $-$ \\
  \hline \hline 
\end{tabular}
\end{table}

If we apply the SU(3)-flavor symmetry to the $\bar{D}^{(\ast)}\Sigma_{c}^{(\ast)}$ system, it will result in another mulitplet. For simplicity, we use the generic denotation $P^{(\ast)}$ for the heavy mesons and  $\Sigma_{Q}^{(\ast)}$ for the heavy baryons. In addition, we use the notation $P_{s}^{(\ast)}$ for the heavy mesons with $S=-1$ and $\Xi_{Q}^{\prime(\ast)}$ $(\Omega_{Q}^{(\ast)})$  heavy baryons with $S=-1(S=-2)$. In terms of the SU(3)-flavor symmetry, the ($\bar{P}^{(\ast)}$, $\bar{P}_{s}^{(\ast)}$) heavy anti-mesons and ($\Sigma_{Q}^{(\ast)}$, $\Xi_{Q}^{\prime(\ast)}$,  $\Omega_{Q}^{(\ast)}$) heavy baryons
belong to the $3$ and $6$ representation of the SU(3)-flavor group. Such mesons and baryons can be decomposed into $3\otimes6=8\oplus10$, i.e., the octet and decuplet representations, where the SU(3) Clebsch-Gordan coefficients can be found in Ref.~\cite{Kaeding:1995vq}. Their potentials can be decomposed into a linear combination of an octet and a decuplet contribution. 
\begin{eqnarray}
V=\lambda_{O}V_{O}+\lambda_{D}V_{D}
\end{eqnarray}
with $\lambda_{O}$ and $\lambda_{D}$ numerical coefficients and $V_{O}$ and $V_{D}$ the octet and decuplet components. Table~\ref{tab:su3} presents the heavy antimeson-baryon potentials decomposed into the octet and decuplet representations~\cite{Peng:2019wys}. 

In addition, the heavy baryons belong to the ``6'' representation, the heavy baryons belong to the ``3'' representation $(\Lambda_{Q},\Xi_{Q})$, and the heavy anti-meson would be decomposed into singlet and octet representations according to the SU(3)-flavor symmetry. According to the HQSS, the $\bar{P}^{(\ast)}B_{3}$ potentials are parameterised by one parameter~\cite{Liu:2020hcv} 
\begin{eqnarray}
V_{\bar{P}^{(\ast)}B_{3}}=C_{a}.
\end{eqnarray}

\subsection{$\bar{D}^{(\ast)}D^{(\ast)}$ heavy anti-meson and heavy meson systems }

The contact-range potentials of the
$\bar{D}^{(\ast)}D^{(\ast)}$ system can also be derived using the same approach. Under HQSS, the $\bar{D}^{(\ast)}D^{(\ast)}$ system results in six states and are parameterized by two constants $C_a$ and $C_b$~\cite{Guo:2013sya}.   With HADS, one can extend the $\bar{D}^{(\ast)}D^{(\ast)}$ system to the $\Xi_{cc}^{(\ast)}D^{(\ast)}$ and $\Xi_{cc}^{(\ast)}\bar{\Xi}_{cc}^{(\ast)}$ systems, where the contact-range potentials $C_{a}$ and $C_{b}$ in these three systems are the same~\cite{Liu:2020tqy}.  Table~\ref{tab:potentialhh} shows the potentials of these three systems.   

\begin{table}[!h]
  \centering \caption{Contact-range potentials
    for the heavy meson-heavy anti-meson, heavy meson-doubly heavy baryon, and doubly heavy
    baryon-doubly heavy anti-baryon systems depending on two unknown coupling constants, $C_{a}$ and $C_{b}$.
    These coupling constants can be determined from the sum of $C_{a}$ and $C_{b}$ by reproducing the mass of
    $X(3872)$ and the ratio of $C_{a}$ and $C_{b}$ by the light-meson saturation approach. \label{tab:potentialhh}}
\begin{tabular}{ccc|ccc|ccccc}
  \hline\hline State& $J^{PC}$ &V   & State~~~   &$J^{P}$~~~~~   &V & State~~~   &$J^{PC}$~~~   &V \\

  \hline\multirow{2}{0.8cm} {$D\bar{D}$} &
 \multirow{2}{0.8cm}{$0^{++}$} &\multirow{2}{0.8cm}{$C_{a}$}
  & \multirow{2}{0.8cm}{$D\Xi_{cc}$}
  & \multirow{2}{0.8cm}{$\frac{1}{2}^{-}$}  & \multirow{2}{0.8cm}{$C_{a}$}  &  $\bar{\Xi}_{cc}\Xi_{cc}$  & $0^{-+}$ & $C_{a}-\frac{1}{3}C_{b}$       \\
 &
 &
  &
  &  &   &  $\bar{\Xi}_{cc}\Xi_{cc}$  & $1^{--}$ & $C_{a}+\frac{1}{9}C_{b}$      \\  \hline
 \multirow{4}{2.0cm}{$D^{\ast}\bar{D}$/$D\bar{D}^{\ast}$}&
  &
  &
  &  &   & \multirow{4}{2.4cm}{$\bar{\Xi}_{cc}^{\ast}\Xi_{cc}/\bar{\Xi}_{cc}\Xi_{cc}^{\ast}$}  & $1^{-+}$ & $C_{a}+C_{b}$        \\
  &
 {$1^{++}$} &{$C_{a}+C_{b}$}
  &\multirow{2}{0.8cm}{$D^{\ast}\Xi_{cc}$}
  & $\frac{1}{2}^{-}$  & $C_{a}+\frac{2}{3}C_{b}$  &  & $1^{--}$ & $C_{a}+\frac{1}{9}C_{b}$          \\
  & {$1^{+-}$} &{$C_{a}-C_{b}$} & & $\frac{3}{2}^{-}$    & $C_{a}-\frac{1}{3}C_{b}$   &    & $2^{-+}$ & $C_{a}+C_{b}$          \\

   &
 &
  &
  &  &   &   & $2^{--}$ & $C_{a}-\frac{5}{3}C_{b}$      \\
  \hline

   &
 &
  &
 $D\Xi_{cc}^{\ast}$ & $\frac{3}{2}^{-}$ & $C_{a}$  & \multirow{4}{1.2cm}{$\bar{\Xi}_{cc}^{\ast}\Xi_{cc}^{\ast}$}   & $0^{-+}$ & $C_{a}-\frac{5}{3}C_{b}$      \\
\multirow{3}{0.8cm}{$D^{\ast}\bar{D}^{\ast}$} & $0^{++}$
&$C_{a}-2C_{b}$  &
\multirow{3}{0.8cm}{$D^{\ast}\Xi_{cc}^{\ast}$}  &$\frac{1}{2}^{-}$
& $C_{a}-\frac{5}{3}C_{b}$  &    &  $1^{--}$  &  $C_{a}-\frac{11}{9}C_{b}$
\\
& {$1^{+-}$}
& {$C_{a}-C_{b}$} &
& $\frac{3}{2}^{-}$ & $C_{a}-\frac{2}{3}C_{b}$  &    &  $2^{-+}$  &  $C_{a}-\frac{1}{3}C_{b}$
\\  & $2^{++}$ & {$C_{a}+C_{b}$} &
& $\frac{5}{2}^{-}$ & $C_{a}+C_{b}$  &    &  $3^{--}$  &  $C_{a}+C_{b}$
\\
\hline\hline
\end{tabular}
\end{table}

If  we apply the SU(3)-flavor symmetry to the $\bar{P}^{(\ast)}P^{(\ast)}$ system,   we have a singlet and an octet irreducible group representation 
\begin{eqnarray}
3 \otimes \bar{3} = 1 \oplus 8.
\end{eqnarray}
The $\bar{P}^{(\ast)}P^{(\ast)}$ potential in terms of SU(3)-flavor symmetry  are written as 
\begin{eqnarray}
V=\lambda_{s}V_{s}+\lambda_{o}V_{o},
\end{eqnarray}
with $V_{s}$ and $V_{o}$ the singlet and octet components . For the potentials of  the $D^{(\ast)}\bar{D}^{(\ast)}$ and $D_{s}^{(\ast)}\bar{D}_{s}^{(\ast)}$ in$I=0$, it would be decomposed into   
\begin{eqnarray}
V(D^{(\ast)}\bar{D}^{(\ast)}(I=0))&=& \frac{1}{3}V_{o}+\frac{2}{3}V_{s}, \\ \nonumber V(D_{s}^{(\ast)}\bar{D}_{s}^{(\ast)}(I=0))&=& \frac{2}{3}V_{o}+\frac{1}{3}V_{s}.
\end{eqnarray}

Within HQSS, the $S$-wave heavy mesons form the doublet states ($P,P^{\ast}$). As the orbital angular momentum changes to $L=1$, there will be two cases under HQSS.  First, the orbital angular momentum $L=1$ couples to the spin of the light quark, leading to total angular momenta $l=1/2$ or $l=3/2$, and then they couple to the spin of the heavy quark, finally resulting in two doublet states ($D_{0},D_{1}$) and  ($D_{1},D_{2}$). One should note that the widths of ($D_0,D_1$) are large, so they are not good candidates for hadronic molecules. However, the widths of ($D_1,D_2$) are rather narrow, which are good candidates for hadronic molecules. One example is the $Y(4260)$ which is proposed as a $\bar{D}D_{1}$ bound state~\cite{Ding:2008gr,Wang:2013cya,Li:2013yla}. Therefore, we also show the contact-range potentials of the $\bar{P}T$ system constrained by HQSS. In the heavy quark limit, their brown mucks have $s_{1l}=1/2$ and   $s_{2l}=3/2$, which indicates that the $\bar{P}T$ potentials  are only dependent on the two parameters $F_{1}$ and $F_{2}$, i.e.,  $F_{2}=\langle 2_{L} | V| 2_{L} \rangle$  and $F_{1}=\langle 1_{L} | V| 1_{L} \rangle$. 
  One should note that the $\bar{P}T$  system has direct and cross potentials, resulting in four parameters: $F_{1d}$, $F_{2d}$, $F_{1c}$, and $F_{2c}$.    The cross potentials are related to their charge parity. Following Ref.~\cite{Peng:2022nrj}, the C-parity of the $\bar{P}T$ system is $C=\eta(-1)^{S-S_P-S_T}$ with  $|\bar{P}T(\eta)\rangle= \frac{1}{\sqrt{2}}(|\bar{P}T\rangle+\eta|{P}\bar{T}\rangle)$, where $S$ is the total spin of the $\bar{P}T$  system,
$S_P$ and $S_{T}$ are the spins of meson P and T, and $\eta=\pm1$.   At first, we parameterized the potentials of  the $\bar{P}T$  system via the parameters above, in agreement with Ref.~\cite{Guo:2017jvc}.  Next we   rewrite the  parameters   $F_{1d}$, $F_{2d}$, $F_{1c}$, and $F_{2c}$  as a combination of  the direct potentials  $D_a$ and $D_b$ and the cross potentials $C_{a}$ and $C_{b}$.  Assuming  $\frac{5}{8}F_{2d}+\frac{3}{8}F_{1d}=D_{a}$ for the $D\bar{D}_1$ system and $F_{2d}=D_{a}+D_{b}$ for the $D^{\ast}D_{2}~(J=3)$ system,   we obtain $F_{1d}=D_a-\frac{5}{3}D_{b}$ and $F_{2d}=D_a+D_b $.  Similarly, assuming $\frac{5}{8}F_{2c}+\frac{1}{8}F_{1c}=C_{a}$ for the $D\bar{D}_1$ system and  $\frac{1}{8}F_{2c}-\frac{3}{8}F_{1c}=C_b$ for the $\bar{D}D_{2}^*$ system,  we  obtain    $F_{1c}=\frac{1}{2}C_a-\frac{5}{2}C_b$  and $F_{2c}=\frac{3}{2}C_a+\frac{1}{2}C_b$.     With these relationships, the $\bar{P}H$ potentials are 
\begin{eqnarray}
\label{DbarD1}
V_{\bar{D}D_1}(1^{-+}) &=& \frac{5}{8}F_{2d}+
\frac{3}{8}F_{1d}+\frac{5}{8}F_{2c}+
\frac{1}{8}F_{1c}=D_{a}+C_{a},   \\ \nonumber
V_{\bar{D}D_1}(1^{--}) &=& \frac{5}{8}F_{2d}+
\frac{3}{8}F_{1d}-\frac{5}{8}F_{2c}-
\frac{1}{8}F_{1c}=D_{a}-C_{a},   \\ \nonumber
V_{\bar{D}D_2^*}(2^{-+}) &=& \frac{5}{8}F_{2d}+
\frac{3}{8}F_{1d}+\frac{1}{8}F_{2c}-
\frac{3}{8}F_{1c}=D_{a}+C_{b},   \\ \nonumber
V_{\bar{D}D_2^*}(2^{--}) &=& \frac{5}{8}F_{2d}+
\frac{3}{8}F_{1d}-\frac{1}{8}F_{2c}+
\frac{3}{8}F_{1c}=D_{a}-C_{b},\\ \nonumber
V_{\bar{D}^{\ast}D_1 }(0^{-+}) &=& 
 F_{1d}-
 F_{1c}=D_{a}-\frac{5}{3}D_{b}-\frac{1}{2}C_{a}+\frac{5}{2}C_{b},\\ \nonumber
V_{\bar{D}^{\ast}D_1 }(0^{--}) &=& 
 F_{1d}+
 F_{1c}=D_{a}-\frac{5}{3}D_{b}+\frac{1}{2}C_{a}-\frac{5}{2}C_{b},\\ \nonumber
V_{\bar{D}^{\ast}D_1 }(1^{-+}) &=& 
 \frac{11}{16}F_{1d}+ \frac{5}{16}F_{2d}+ \frac{5}{16}F_{2c}-\frac{7}{16}
 F_{1c}=D_{a}-\frac{5}{6}D_{b}+\frac{1}{4}C_{a}+\frac{5}{4}C_{b},\\ \nonumber
V_{\bar{D}^{\ast}D_1 }(1^{--}) &=& 
 \frac{11}{16}F_{1d}+ \frac{5}{16}F_{2d}- \frac{5}{16}F_{2c}+\frac{7}{16}
 F_{1c}=D_{a}-\frac{5}{6}D_{b}-\frac{1}{4}C_{a}-\frac{5}{4}C_{b},\\ \nonumber
V_{\bar{D}^{\ast}D_1 }(2^{-+}) &=& 
 \frac{1}{16}F_{1d}+ \frac{15}{16}F_{2d}+ \frac{3}{16}F_{2c}-\frac{1}{16}
 F_{1c}=D_{a}+\frac{5}{6}D_{b}+\frac{1}{4}C_{a}+\frac{1}{4}C_{b},\\ \nonumber
V_{\bar{D}^{\ast}D_1 }(2^{--}) &=& 
 \frac{1}{16}F_{1d}+ \frac{15}{16}F_{2d}- \frac{3}{16}F_{2c}+\frac{1}{16}
 F_{1c}=D_{a}+\frac{5}{6}D_{b}-\frac{1}{4}C_{a}-\frac{1}{4}C_{b},
 \\ \nonumber
V_{\bar{D}^{\ast}D_2^* }(1^{-+}) &=& 
 \frac{15}{16}F_{1d}+ \frac{1}{16}F_{2d}+ \frac{1}{16}F_{2c}+\frac{5}{16}
 F_{1c}=D_{a}-\frac{3}{2}D_{b}+\frac{1}{4}C_{a}-\frac{3}{4}C_{b}, \\ \nonumber
V_{\bar{D}^{\ast}D_2^* }(1^{--}) &=& 
 \frac{15}{16}F_{1d}+ \frac{1}{16}F_{2d}- \frac{1}{16}F_{2c}-\frac{5}{16}
 F_{1c}=D_{a}-\frac{3}{2}D_{b}-\frac{1}{4}C_{a}+\frac{3}{4}C_{b}, \\ \nonumber
V_{\bar{D}^{\ast}D_2^* }(2^{-+}) &=& 
 \frac{9}{16}F_{1d}+ \frac{7}{16}F_{2d}- \frac{5}{16}F_{2c}-\frac{9}{16}
 F_{1c}=D_{a}-\frac{1}{2}D_{b}-\frac{3}{4}C_{a}+\frac{5}{4}C_{b},\\ \nonumber
V_{\bar{D}^{\ast}D_2^* }(2^{--}) &=& 
 \frac{9}{16}F_{1d}+ \frac{7}{16}F_{2d}+ \frac{5}{16}F_{2c}+\frac{9}{16}
 F_{1c}=D_{a}-\frac{1}{2}D_{b}+\frac{3}{4}C_{a}-\frac{5}{4}C_{b},
 \\ \nonumber
V_{\bar{D}^{\ast}D_2^* }(3^{-+}) &=&F_{2d}+ F_{2c}=D_{a}+D_{b}+\frac{3}{2}C_{a}+\frac{1}{2}C_{b}, \\ \nonumber
V_{\bar{D}^{\ast}D_2^* }(3^{--}) &=&F_{2d}- F_{2c}=D_{a}+D_{b}-\frac{3}{2}C_{a}-\frac{1}{2}C_{b},
\end{eqnarray}
which indicate that the  $\bar{P}H$ potentials are dependent on four parameters. We should note that the potentials of Ref.~\cite{Peng:2022nrj} differ from ours by a factor of $2/3$ or $3/2$ due to the different conventions.

\subsection{$\Sigma_{Q}\bar{\Sigma}_{Q}$ heavy baryon and heavy anti-baryon systems }
 
Since the spin of the light quark of the $\Sigma_{Q}$ baryons is 1, the spin of the light quarks of the $\Sigma_{Q}\bar{\Sigma}_{Q}$  can be 0, 1, or 2, which dictates that the $\Sigma_{Q}\bar{\Sigma}_{Q}$ potentials can be parameterized by three parameters denoted as $C_{0}$, $C_{1}$, and $C_{2}$. In the following, we derive the $\Sigma_{Q}\bar{\Sigma}_{Q}$  contact-range potentials in detail.  The C parity of the $\Sigma_{Q}\bar{\Sigma}_{Q}^{\ast}$ system is defined as $C=-\eta(-1)^{L+S}$ with the wave function $\frac{1}{\sqrt{2}}|\Sigma_{Q}\bar{\Sigma}_{Q}^{\ast}+\eta \bar{\Sigma}_{Q} {\Sigma}_{Q}^{\ast} \rangle$. The contact interactions of $\Sigma_{Q}^{(\ast)}\bar{\Sigma}_{Q}^{(\ast)}$  can be expressed in three terms, i.e., $\mathcal{L}=C_{0}+S_{1}\cdot S_{2}C_{1}+Q_{ij}\cdot Q_{ij}C_{2}$, where $C_{1}$ is dependent on the spin-spin term,
and $Q_{ij}\cdot Q_{ij}$ is related to the tensor-tensor interactions. The three parameters $C_{1}$, $C_{2}$, and $C_{3}$ can be reexpressed by three parameters $F_{0}$, $F_{1}$, and $F_{2}$. Assuming that $F_{2}=C_{0}+C_{1}+C_{2}$, $\frac{2}{3}F_{1}+\frac{1}{3}F_{0}=C_{0}-\frac{4}{3}C_{1}$, and $\frac{20}{27}F_{2}+\frac{2}{9}F_{1}+\frac{1}{27}F_{0}=C_{0}+\frac{4}{9}C_{1}$, we obtain $F_{0}=C_{0}-2C_{1}+10C_{2}$, $F_{1}=C_{0}-C_{1}-5C_{2}$, and $F_{0}=C_{0}+C_{1}+C_{2}$. As a result, one can parameterize the $\Sigma_{Q}^{(\ast)}\bar{\Sigma}_{Q}^{(\ast)}$  potentials by $C_{0}$, $C_{1}$, and $C_{2}$ as follows.
\begin{eqnarray}
V_{\Sigma_{Q}\bar{\Sigma}_{Q}}(J^{PC}=0^{-+})&=& \frac{2}{3}F_{1}+\frac{1}{3}F_{0}=C_{0}-\frac{4}{3}C_1,   \\ \nonumber
V_{\Sigma_{Q}\bar{\Sigma}_{Q}}(J^{PC}=1^{--})&=& \frac{20}{27}F_{2}+\frac{2}{9}F_{1}+\frac{1}{27}F_{0}=C_{0}+\frac{4}{9}C_1,     \\ \nonumber
V_{\bar{\Sigma}_{Q}^*{\Sigma}_{Q}}(J^{PC}=1^{--})  &=&\frac{5}{27}F_{2}+\frac{2}{9}F_{1}+\frac{16}{27}F_{0}=C_{0}-\frac{11}{9}C_1+5C_2,  \\ \nonumber
V_{\bar{\Sigma}_{Q}^*{\Sigma}_{Q}}(J^{PC}=1^{-+})  &=&F_{1}=C_{0}-C_1-5C_2,
  \\ \nonumber
V_{\bar{\Sigma}_{Q}^*{\Sigma}_{Q}}(J^{PC}=2^{-+})  &=& \frac{2}{3}F_{2}+\frac{1}{3}F_{1}=C_0+\frac{1}{3}C_1-C_2,  \\ \nonumber
V_{\bar{\Sigma}_{Q}^*{\Sigma}_{Q}}(J^{PC}=2^{--})  &=& F_{2}=C_0+C_1+C_2, \\ \nonumber
V_{\Sigma_{Q}^*\bar{\Sigma}_{Q}^*}(J^{PC}=0^{-+})&=& \frac{1}{3}F_{1}+\frac{2}{3}F_{0}=C_0-\frac{5}{3}C_1+5C_2,
\\ \nonumber
V_{\Sigma_{Q}^*\bar{\Sigma}_{Q}^*}(J^{PC}=1^{--}) &=&\frac{2}{27}F_{2}+\frac{10}{27}F_{0}+\frac{5}{9}F_{1}=C_0-\frac{11}{9}C_1+C_2,
\\ \nonumber
V_{\Sigma_{Q}^*\bar{\Sigma}_{Q}^*}(J^{PC}=2^{-+}) &=&\frac{1}{3}F_2+\frac{2}{3}F_{1}=C_0-\frac{1}{3}C_1-3C_2,
\\ \nonumber
V_{\Sigma_{Q}^*\bar{\Sigma}_{Q}^*}(J^{PC}=3^{--}) &=&F_{2}=C_0+C_1+C_2.
\end{eqnarray}

\section{Light-meson saturation approach}

The contact-range potentials of a given system are dependent on several couplings. 
The best approach to determine these couplings is to fit the experimental data.
However, for the $\bar{D}^{(\ast)}D^{(\ast)}$ system there exists only one molecular candidate $X(3872)$, which can only determine the sum of  $C_{a}$ and $C_{b}$.   We, therefore, resort to the light-meson saturation approach to estimate the ratio of $C_{a}$ and $C_{b}$ and determine the value of $C_{a}$ and $C_{b}$~\cite{Liu:2020tqy}.
The light meson saturation approach can only estimate the  relative strength and the sign of these couplings 
from the hypothesis that they are saturated by the exchange of light mesons, in particular, the vector mesons
$\rho$ and $\omega$,  and scalar meson $\sigma$~\cite{Peng:2020xrf,Peng:2021hkr}. The proportionality constant is unknown and depends on the details of the renormalization procedure.

\begin{table}[!h]
\caption{Couplings of the light mesons of the OBE model
  ($\pi$, $\sigma$, $\rho$, $\omega$) to the heavy-meson
  and heavy-baryon fields.
  For the magnetic-type coupling of the $\rho$ and $\omega$ vector mesons
  we have used the decomposition
  $f_{V} = \kappa_{V}\,g_{V}$, with $V=\rho,\omega$.
  $M$ refers to the mass scale (in units of MeV) involved
  in the magnetic-type couplings~\cite{Liu:2019zvb}.
}
\label{tab:couplingsDS}
\begin{tabular}{cc|cc|cc}
  \hline \hline
  Coupling  & Value for $P$/$P^*$ &   Coupling  & Value for $\Sigma_Q$/$\Sigma_Q^*$  &   Coupling  & Value for $D_1$/$D_2$ \\
  \hline
  $g_1$ & 0.60  & $g_2$ & 0.84 & $g_3$ & 0.90\\\
  $g_{\sigma 1}$ & 3.4 &  $g_{\sigma 2}$ & 6.8 &  $g_{\sigma 3}$ & 3.4  \\
  $g_{\rho 1}$ & 2.6  & $g_{\rho 2}$ & 5.8 & $g_{\rho 2}$ & 2.9\\
  $g_{\omega 1}$ & 2.6 & $g_{\omega 2}$ & 5.8 & $g_{\omega 2}$ & 2.9\\
  $\kappa_{\rho 1}$ & 2.3 &$\kappa_{\rho 2}$ & 1.7 &$\kappa_{\rho 2}$ & 4.6 \\
  $\kappa_{\omega 1}$ & 2.3  &   $\kappa_{\omega 2}$ & 1.7  &   $\kappa_{\omega 2}$ & 4.6 \\
  $M$ & 940  &  &  &  &  \\
  \hline \hline
\end{tabular}
\end{table}

The contact-range potentials of the $\bar{D}^{(\ast)}\Sigma_{c}^{(\ast)}$ system can be parameterized by two couplings $C_a$ and $C_b$, which are saturated by the sigma and vector meson exchanges: 
\begin{eqnarray}
C_{a}^{sat}(\Lambda\sim m_{\sigma},m_{V})&\propto& C_{a}^{S}+ C_{a}^{V},    \\ \nonumber
C_{b}^{sat}(\Lambda\sim m_{\sigma},m_{V})&\propto& C_{b}^{V},
\end{eqnarray}
where the cutoff $\Lambda\sim m_{\sigma},m_{V}$ implies that the saturation works at an EFT cutoff close to the masses of exchanged mesons, i.e., $0.6\sim 0.8$ GeV. The values of saturated couplings are expected to be proportional to the potentials of light-meson exchanges in the OBE model once we have removed the spurious Dirac-delta potential~\cite{Liu:2019zvb}.

Thus, $C_{a}$ and $C_{b}$ can be written as
\begin{eqnarray}
C_{a}^{sat(\sigma)}(\Lambda\sim m_{\sigma})&\propto& -\frac{g_{\sigma1}g_{\sigma2}}{m_{\sigma}^2},    \\ \nonumber
C_{a}^{sat (V)}(\Lambda\sim m_{V})&\propto& \frac{g_{v1}g_{v2}}{m_{v}^2}(1+\vec{\tau}_{1}\cdot\vec{T}_{2}),  \\ \nonumber
C_{b}^{sat (V)}(\Lambda\sim m_{V})&\propto&\frac{f_{v1}f_{v2}}{6 M^2}(1+\vec{\tau}_{1}\cdot\vec{T}_{2}),
\label{123}
\end{eqnarray}
where $g_{\sigma_{i}}$ denote the charmed meson and charmed baryon coupling to the sigma meson, $g_{vi}$ and $f_{vi}$ denote electric-type and magnetic-type couplings between charmed mesons as well as charmed baryons and light vector mesons, and $M$ is a mass scale to render $f_{vi}$ dimensionless, see, e.g., Table \ref{tab:couplingsDS}.
With the above preparations, we calculate the ratio of $C_{a}^{sat}$ and $C_{b}^{sat}$ as 
\begin{eqnarray}
\frac{C_{b}^{sat}}{C_{a}^{sat}}=0.124.
\end{eqnarray}
The couplings of $C_a$ and  $C_b$ can be fully determined by reproducing the masses of $P_{c}(4440)$ and $P_{c}(4457)$. In scenario A, the $P_{c}(4440)$ and $P_{c}(4457)$ are assumed as the $\bar{D}^*\Sigma_{c}$ bound states with $J^P=1/2^-$ and $J^P=3/2^-$, while in scenario B they are assumed to have  $J^P=3/2^-$ and $J^P=1/2^-$.  At the cutoff $\Lambda=0.75$ GeV,  we obtain $C_a=-30.8$~GeV$^{-2}$   and $C_b=5.1$~GeV$^{-2}$ for scenario A and   $C_a=-34.2$~GeV$^{-2}$   and $C_b=-5.1$~GeV$^{-2}$ for scenario B. The ratio of $C_a$ and  $C_b$ in scenario A and scenario B can be estimated as 
\begin{eqnarray}
(\frac{C_{b}}{C_{a}})^{A}&=&-0.166,     \\ \nonumber
(\frac{C_{b}}{C_{a}})^{B}&=&0.149.  
\end{eqnarray}
We can see that the ratio of $C_a$ and  $C_b$ in scenario B coincides with the value of the light meson saturation approach, which shows that the OBE model supports the results of scenario B of the EFT, in agreement with our numerical calculation~\cite{Liu:2019tjn,Liu:2019zvb}.  In other words, we can see that the light meson saturation approach works well, which inspires us to apply the light meson saturation approach to other systems.

 We estimate the couplings of $C_{a}$ and $C_{b}$ of the $\bar{D}^{(\ast)} {D}^{(\ast)}$ system  by the sigma and vector meson exchange saturation.
 $C_{a}$ and $C_{b}$ can be written as
\begin{eqnarray}
C_{a}^{sat(\sigma)}(\Lambda\sim m_{\sigma})&\propto& -\frac{g_{\sigma1}^2}{m_{\sigma}^2},    \\ \nonumber
C_{a}^{sat (V)}(\Lambda\sim m_{V})&\propto& -\frac{g_{v1}^2}{m_{v}^2}(1-\vec{\tau}_{1}\cdot\vec{\tau}_{2}),  \\ \nonumber
C_{b}^{sat (V)}(\Lambda\sim m_{V})&\propto& -\frac{f_{v1}^2}{6 M^2}(1-\vec{\tau}_{1}\cdot\vec{\tau}_{2}).
\label{123}
\end{eqnarray}
Then, we obtain the ratio of $C_a$ and  $C_b$  
\begin{eqnarray}
\frac{C_{b}^{sat}}{C_{a}^{sat}}=0.347.
\end{eqnarray}
Such a ratio is larger than that of the $\bar{D}^{(\ast)}\Sigma_{c}^{(\ast)}$ system, which indicates that the spin-spin term plays a more important role in the $\bar{D}^{(\ast)} {D}^{(\ast)}$ system.  As for the ${D}^{(\ast)} {D}^{(\ast)}$ system, we obtain  the ratio of $C_a$ and  $C_b$ via the G-parity transformation 
\begin{eqnarray}
\frac{C_{b}^{sat}}{C_{a}^{sat}}=0.246.
\end{eqnarray}

For the $\bar{D}^{(\ast)}D_{1,2}$ system, the contact-range potentials are parameterized by four parameters, $D_{a}$, $D_b$, $C_a$ and $C_b$, e.g., 
\begin{eqnarray}
V=D_a + \sigma_1\cdot S_2 D_b  +  \Sigma_1\cdot\Sigma_2 C_a +  Q_{1ij}\cdot Q_{2ji} C_b,
\end{eqnarray}
where $\sigma$ and $S$ demote the operator of spin $1/2$ and $1$. $\Sigma$ denotes the operator of spin 1 to 1/2 transition. The tensor operator $Q$ is defined as 
$Q_{ij}=(\sigma_i\Sigma_j+\Sigma_j\sigma_i)/2$.  
 Using the light meson saturation approach,  $D_{a}$, $D_{b}$, $C_{a}$, and $C_{b}$ can be saturated by~\cite{Peng:2021hkr} 
\begin{eqnarray}
D_{a}^{sat(\sigma)}(\Lambda\sim m_{\sigma})&\propto& -\frac{g_{\sigma1}g_{\sigma3}}{m_{\sigma}^2},    \\ \nonumber
D_{a}^{sat (V)}(\Lambda\sim m_{V})&\propto& \frac{g_{v1}g_{v3}}{m_{v}^2}(-1+\vec{\tau}_{1}\cdot\vec{\tau}_{2}),  \\ \nonumber
D_{b}^{sat (V)}(\Lambda\sim m_{V})&\propto& \frac{f_{v1}f_{v3}}{6 M^2}(-1+\vec{\tau}_{1}\cdot\vec{\tau}_{2}), \\ \nonumber
C_{a}^{sat (V)}(\Lambda\sim m_{V})&\propto& -\frac{f_{v}^{\prime2}}{4 M^2}\frac{\omega_{V}^2+\frac{1}{3}\mu_{V}^2}{\mu_{V}^2}(-1+\vec{\tau}_{1}\cdot\vec{\tau}_{2}), \\ \nonumber
C_{b}^{sat (V)}(\Lambda\sim m_{V})&\propto& -\frac{h_{v}^2}{80 M^4}\mu_{V}^2(-1+\vec{\tau}_{1}\cdot\vec{\tau}_{2}), 
\label{DD1}
\end{eqnarray}
 where $f_{v}^{\prime}$ and $h_{v}^{\prime}$ represent the electric dipole and magnetic quadrupole couplings of the vector meson. The couplings $\mu_{V}=\sqrt{m_V^2-\omega_V^2}$ and $\omega_V=m_{D_{1,2}}-m_{D^{(*)}}$.   The subscript of ``3'' denotes the couplings between $D_1/D_2$ and light mesons. The relevant couplings are $g_{\sigma3}=3.4$,
  $g_{v3}=2.9$, $f_{v3}=4.6\cdot2.9$, $f_{v}^{\prime}=3.1\cdot 2.9$, and $h_{v}=10.7\cdot 2.9$~\cite{Peng:2021hkr}.

\section{Effective Lagrangians}
In this section, we present the effective Lagrangians relevant to the studies covered in this review.
The Lagrangians describing the interactions between charmed mesons and $\sigma$, $\rho,\omega$ mesons read   
\begin{eqnarray}
  \mathcal{L}_{HH \pi} &=& \frac{g_1}{\sqrt{2} f_{\pi}}\, \langle
        H^{\dagger}
        \vec{\sigma} \cdot \nabla ( \vec{\tau} \cdot \vec{\pi})
        H \,  \rangle , \\
        \mathcal{L}_{HH \sigma} &=& {g}_{\sigma 1}\, \langle H^{\dagger} \sigma H \rangle
        \, , \\
        \mathcal{L}_{HH \rho} &=&
                {g}_{\rho 1}\, \langle H^{\dagger} \vec{\tau} \cdot \vec{\rho}_0 H  \rangle
                - \frac{f_{\rho 1}}{4 {M_1}} \langle  \epsilon_{ijk}
                H^{\dagger}\, \sigma_{k} \vec{\tau}  \cdot
                ( \partial_i \vec{\rho}_j - \partial_j \vec{\rho}_i )\, H
                \,  \rangle, \\
                \mathcal{L}_{HH \omega} &=&
                {g}_{\omega 1}\,\langle H^{\dagger} \, {\omega}_0 \, H \rangle
                - \frac{f_{\omega 1}}{4 {M_1}} \langle \epsilon_{ijk}
                H^{\dagger}\, \sigma_{ k} \,
                ( \partial_i {\omega}_j - \partial_j {\omega}_i )\, H  \rangle
                \, ,
\end{eqnarray}
where $g_{1}$, $g_{\sigma1}$, $g_{v1}$, and $g_{v1}$ represent the couplings between a charmed meson and the light mesons, and $H$ refers to the superfield, i.e.,  $H=D +D^*\cdot \sigma$, satisfying the HQSS.

The Lagrangians describing the interactions between charmed baryons and $\sigma$, $\rho,\omega$ mesons read  
\begin{eqnarray}
  \mathcal{L}_{SS \, \pi} &=& \frac{g_2}{\sqrt{2} f_{\pi}}\,
        ( \vec{S}^{\dagger} \times \vec{S} )
        \cdot \nabla ( \vec{T} \cdot \vec{\pi})
      \, , \\
        \mathcal{L}_{ST \, \pi} &=& \frac{g_3}{\sqrt{2} f_{\pi}}\,
         \vec{S}^{\dagger}
        \cdot \nabla  \vec{\pi} \, T
       , \\
        \mathcal{L}_{SS \, \sigma} &=& {g}_{\sigma 2}\, \vec{S}^{\dagger}
        \sigma \vec{S} \, , \\
        \mathcal{L}_{SS \, \rho} &=&
                {g}_{\rho 2}\,\vec{S}^{\dagger} \,
                \vec{T} \cdot \vec{\rho}_0 \, \vec{S}
                - \frac{f_{\rho 2}}{4 M_2} \epsilon_{ijk}
                ( \vec{S}^{\dag} \times \vec{S})_{k} \vec{T} \cdot
                ( \partial_i \vec{\rho}_j - \partial_j \vec{\rho}_i ) \, ,
                 \\
        \mathcal{L}_{SS \, \omega} &=&
                {g}_{\omega 2}\,\vec{S}^{\dagger} \,
                {\omega}_0 \, \vec{S}
                - \frac{f_{\omega 2}}{4 M_2} \epsilon_{ijk}
                 ( \vec{S}^{\dag} \times \vec{S})_{k}\,
                ( \partial_i {\omega}_j - \partial_j {\omega}_i ) \, ,                   \\
        \mathcal{L}_{TT \, \omega} &=&
                {g}_{\omega 3}\,{T}^{\dagger} \,
                {\omega}_0 \, {T} \, ,
                \\
        \mathcal{L}_{ST \, \rho} &=&
  -\frac{f_{\rho 3}}{4 M_3} \epsilon_{ijk}
                 \vec{S}^{\dag}_{k} 
                ( \partial_i \vec{\rho}_j - \partial_j \vec{\rho}_i ) T \, ,
\end{eqnarray} 
where the superfield $\vec{S}$ is $\vec{S}=\frac{1}{\sqrt{3}}\vec{\sigma}\vec{\Sigma}_{c} +\vec{\Sigma}_{c}^*$ in the heavy quark limit.  

As for the excited charmed mesons, the superfield for $D_{1}$ and $D_{2}$ in the heavy quark limit is $G^{i}=D_{2}^{ij}\sigma^{j}+\sqrt{\frac{2}{3}}D^{i}_{1}+ i\sqrt{\frac{1}{6}}\varepsilon_{ijk} D_{1}^{j} {\sigma}^{k}$. The corresponding  Lagrangians  are written as 
\begin{eqnarray}
  \mathcal{L}_{GG\pi} &=& \frac{g_3}{\sqrt{2} f_{\pi}}\, \langle
        G^{i\dagger}
        \vec{\sigma} \cdot \nabla ( \vec{\tau} \cdot \vec{\pi})
        G_{i} \,  \rangle , \\
   \mathcal{L}_{GH\pi} &=& \frac{g_3^{\prime}}{\sqrt{2}f_{\pi}}\, \langle
        G^{i\dagger}\cdot \nabla_{i}
        \vec{\sigma} \cdot \nabla ( \vec{\tau} \cdot \vec{\pi})
        \,  \rangle , \\       
        \mathcal{L}_{GG\sigma} &=& {g}_{\sigma 3}\, \langle G^{i\dagger} \sigma G_{i} \rangle
        \, , \\
        \mathcal{L}_{GG v} &=&
                {g}_{v 3}\, \langle G^{i\dagger} \vec{\tau} \cdot \vec{\rho}_0 G_{i}  \rangle
                - \frac{f_{v 3}}{4 {M_1}} \langle  \epsilon_{ijk}
                G^{\beta\dagger}\, \sigma_{k} \vec{\tau}  \cdot
                ( \partial_i \vec{\rho}_j - \partial_j \vec{\rho}_i )\, G_{\beta}
                \,  \rangle, \\
                \mathcal{L}_{GH v} &=&
                - \frac{f_{v 3}^{\prime}}{4 {M_1}} \langle 
                G^{i\dagger}\, \sigma_{ j} \,
                ( \partial_i {\omega}_j - \partial_j {\omega}_i )\, H  \rangle
                \, .
\end{eqnarray}
The coupling ${g}_{\sigma 3}$ is estimated to be ${g}_{\sigma 3}=\frac{1}{3}g_{\sigma NN}$, which equals the ground-state charmed meson couplings to the $\sigma$ meson.

\section{Lippmann-Schwinger equation}

Hadronic molecules are bound states of $q\bar{q}$ mesons and $qqq$ baryons. They manifest themselves as poles in the corresponding scattering amplitudes. As a result, it is necessary to study the scattering amplitudes $T$, defined by the $S$ matrix, i.e., $S=1+2i\rho T$. Since the $S$ matrix conserves causality and the probability of current, it satisfies analyticity and unitarity. 
From the optical theorem $\rm{Im}~T=T^{\dag}\rho T$ and the unitarity condition, one can obtain the imaginary part of the $T$ matrix  $\rm{Im}~T^{-1}=-\rho$ for $\sqrt{s}>m_{1}+m_{2}$. One can obtain the $T$ matrix by solving the Bethe-Salpeter (BS)  equation or LS equation with hadron-hadron potentials as inputs, i.e., $T=V+VGT$. As a result,  the phase space is equal to $\rho=\rm{Im}~G$.      The analytical continuation of the scattering amplitude  $T$ to the unphysical sheet has consequences only in the loop functions. In terms of the  Schwartz reflection theorem~\cite{Roca:2005nm},  an analytic function $f[z]$ in the complex plane 
obeys $f[z^*]^*=f[z]$. Applying this theorem to the loop function, one has $G(\sqrt{s}-i\varepsilon)=[G(\sqrt{s}+i\varepsilon)]^*$. With the relationship  $[G(\sqrt{s}+i\varepsilon)]^*=G(\sqrt{s}+i\varepsilon)-i2\mathrm{Im}~G(\sqrt{s}+i\varepsilon) $,  one can obtain $G(\sqrt{s}-i\varepsilon)=G(\sqrt{s}+i\varepsilon)-i2\mathrm{Im}~G(\sqrt{s}+i\varepsilon) $. Since the loop function at the end of the first Riemann sheet is equal to that at the beginning of the second Riemann sheet, i.e., $G_{II}(\sqrt{s}+i\varepsilon)=G_{I}(\sqrt{s}-i\varepsilon)$, the loop function in the second Riemann sheet is $G_{II}(\sqrt{s}+i\varepsilon)=G_{I}(\sqrt{s}+i\varepsilon)-i2\mathrm{Im}~G_{I}(\sqrt{s}+i\varepsilon)$. Therefore, the loop function in the second Riemann sheet is  $G_{II}(\sqrt{s})=G_{I}(\sqrt{s})-2i~\mathrm{Im}~G_{I}(\sqrt{s})$.  Moving from the first sheet to the second sheet, one needs to cross the unitarity cut and change the sign of the imaginary part of the loop function. The convention for the imaginary part is 
\begin{eqnarray}
\mbox{Sheet I}~~\mbox{Im} G>0;  ~~~~  \mbox{Sheet II}~~    \mbox{Im} G<0;
\end{eqnarray}

As for the BS equation, it can be written as  
\begin{eqnarray}
T=V+VGT,
\end{eqnarray}
where $V$ is the coupled-channel potential determined by the contact-range EFT approach described in the previous section, and $G$ is the two-body propagator.  
To avoid the ultraviolet divergence in evaluating the loop function $G$ and keep the unitarity of the $T$ matrix, we include a regulator of Gaussian form { $F(q^2,k)=e^{-2q^2/\Lambda^2}/e^{-2k^2/\Lambda^2}$} in the integral as
\begin{eqnarray}
G(\sqrt{s})= 2m_{1}\int \frac{d^{3}q}{(2\pi)^{3}}\frac{\omega_{1}+\omega_{2}}{2~\omega_{1}\omega_{2}} \frac{F(q^2,k)}{(\sqrt{s})^2-(\omega_1+\omega_2)^2+i \varepsilon}
\label{loopfunction},
\end{eqnarray}
where   $\sqrt{s}$ is  the total energy in the center-of-mass (c.m.) frame of $m_{1}$ and $m_{2}$,  $\omega_{i}=\sqrt{m_{i}^2+q^2}$ is the energy of the particle,  $\Lambda$ is the momentum cutoff,  and   the c.m. momentum $k$ is ,
\begin{eqnarray}
k=\frac{\sqrt{{s}-(m_{1}+m_{2})^2}\sqrt{s-(m_{1}-m_{2})^2}}{2{\sqrt{s}}}.
\end{eqnarray}
  We denote the loop function in the physical and unphysical sheets  as  
$G_{I}(\sqrt{s})$ and $G_{II}(\sqrt{s})$, which are related by $G_{II}(\sqrt{s})=G_{I}(\sqrt{s})-2i\mathrm{Im}~G_{I}(\sqrt{s})$~\cite{Oller:1997ti,Roca:2005nm}.   The imaginary part  of Eq.~(\ref{loopfunction})  can be derived via the Plemelj-Sokhotski formula, i.e., $\mathrm{Im}~G(\sqrt{s})=-\frac{2m_1 }{8\pi} \frac{k}{\sqrt{s}}$, and then Eq.~(\ref{loopfunction})  in the unphysical sheet is written as
\begin{eqnarray}
G_{II}(\sqrt{s})=G_{I}(\sqrt{s}){+~i\frac{2m_1 }{4\pi} \frac{k}{\sqrt{s}}}.
\end{eqnarray}

 The loop function can also be regularized in other methods, such as the momentum cut-off scheme and the dimensional regularization scheme~\cite{Oset:1997it,Jido:2003cb,Wu:2010jy,Hyodo:2011ur,Debastiani:2017ewu}.  The loop function parameterized in the dimensional regularization scheme  is 
 \begin{eqnarray}
G&=&\frac{2M_{l}}{16\pi^2 }\{a(\mu)+ln\frac{m^2}{\mu^2}+\frac{M^2-m^2+s}{2s}ln\frac{M^2}{\mu^2}\\ \nonumber &-&\frac{p_{cml}(\sqrt{s})}{\sqrt{s}}[ln(s-(M^2-m^2)+2\sqrt{s}p_{cml}(\sqrt{s}))   \\ \nonumber
&+&ln(s+(M^2-m^2)+2\sqrt{s}p_{cml}(\sqrt{s}))    \\ \nonumber
&-&ln(-s-(M^2-m^2)+2\sqrt{s}p_{cml}(\sqrt{s}))   \\ \nonumber
&-&ln(-s+(M^2-m^2)+2\sqrt{s}p_{cml}(\sqrt{s}))] \},
\end{eqnarray}
where $a(\mu)$ is the subtraction constant, $\mu$ is the regularisation scale, and  $p_{cm}$ is the c.m. momentum of a given system~\cite{Oller:1998zr}.

The loop function in the cut-off scheme is regularised as
\begin{eqnarray}
{(2\pi)^{3}}\frac{\omega_{1}+\omega_{2}}{2~\omega_{1}\omega_{2}} \frac{1}{(\sqrt{s})^2-(\omega_1+\omega_2)^2+i \varepsilon}
\label{loopfunctioncutoff},
\end{eqnarray}
where $q_{max}$ is a cutoff momentum~\cite{Oset:1997it}. Generally speaking, these two approaches are equivalent~\cite{Oller:1998hw}.

The loop function in Eq.(\ref{loopfunction}) is relativistic. Its non-relativistic form  is 
\begin{eqnarray}
G(\sqrt{s})=\int \frac{d^{3}q}{(2\pi)^{3}} \frac{F(q^2,k)}{{\sqrt{s}}-m_{1}-m_{2}-q^{2}/(2\mu_{12})+i \varepsilon}
\label{loopfunction},
\end{eqnarray}
where the c.m. momentum is  $k=\sqrt{2\mu_{12}(\sqrt{s}-m_{1}-m_{2})}$. As for the non-relativistic case, the amplitude is obtained by the LS equation.  In the unphysical sheet, the non-relativistic loop function changes to 
\begin{eqnarray}
G_{II}(\sqrt{s})=G_{I}(\sqrt{s})+~i\frac{\mu_{12} k}{\pi}F(k^2, k).
\end{eqnarray}

\begin{figure}[ttt]
\begin{center}
\begin{tabular}{cc}
\begin{minipage}[t]{0.5\linewidth}
\begin{center}
\begin{overpic}[scale=.6]{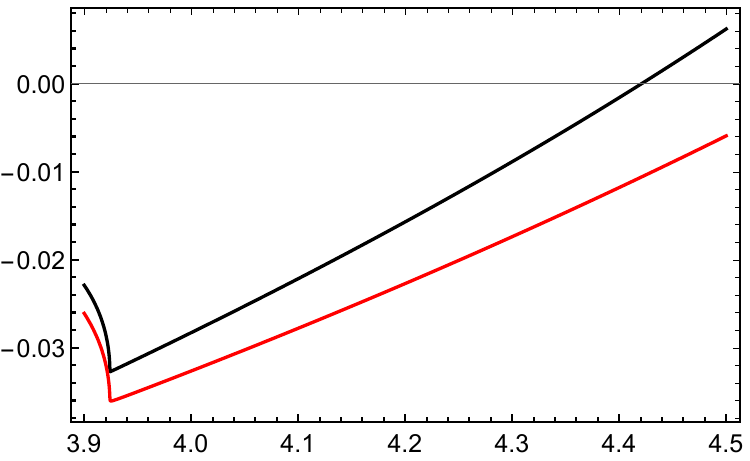}

\put(45,-5){$\sqrt{s}$~{\footnotesize (GeV)}}

\end{overpic}
\end{center}
\end{minipage}
&
\begin{minipage}[t]{0.5\linewidth}
\begin{center}
\begin{overpic}[scale=.6]{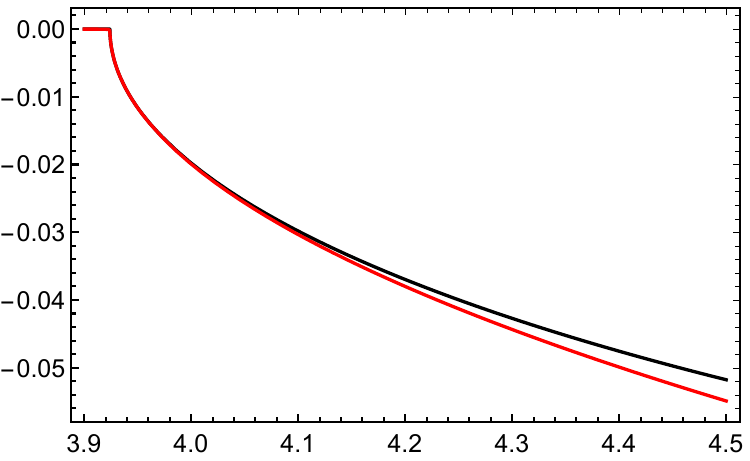}

\put(45,-5){$\sqrt{s}$~{\footnotesize (GeV)}}

\end{overpic}
\end{center}
\end{minipage}
\\  \\
\end{tabular}
\caption{ Real (left) and imaginary (right) parts of the $\eta_{c} N$ loop function as a function of the c.m energy $\sqrt{s}$. The black and red lines represent those of the relativistic and non-relativistic propagators.      }
\label{loopfunction}
\end{center}
\end{figure}

 The amplitudes $T$ in the relativistic and non-relativistic cases are related via $T_{r}=4m_{1}m_{2}T_{nor}$. 
 In Fig.~1, we show the real part (left panel) and imaginary part (right panel) of the $\eta_{c}N$ loop function in both the  relativistic (black) and non-relativistic (red) scenarios as a function of the c.m. energy. One can see that the difference between the real parts is more visible than that between the imaginary parts. One should be careful in dealing with coupled-channel problems, especially for hadrons containing light quarks.

Once the poles generated by the coupled-channel interactions are found, one determines the
 couplings between the molecular states and their constituents from the residues of the corresponding poles, 
\begin{eqnarray}
g_{i}g_{j}=\lim_{{\sqrt{s}}\to {\sqrt{s_0}}}\left({\sqrt{s}}-{\sqrt{s_0}}\right)T_{ij}(\sqrt{s}),
\end{eqnarray}
where $g_{i}$ denotes the coupling of channel $i$ to the dynamically generated state, and ${\sqrt{s_0}}$ is the pole position. 

Recent studies have shown that elementary bare states can couple to hadronic molecules, resulting in the mixing of states composed of a bare state and a hadronic molecule.   To  estimate the proportion of each component,  one often turns to the Weinberg compositeness theorem, which is parameterized by a parameter $Z$~\cite{Weinberg:1962hj}
\begin{eqnarray}
 Z= 1- \int d \alpha |\langle \alpha | d \rangle |^2,~~~ Z=\sum_{n}|\langle n | d \rangle |^2, 
\end{eqnarray}
where $| \alpha \rangle$ and  $| n \rangle$ represent the eigenstates of the continuum and discrete elementary particle state in a free Hamiltonian $H_0$, and $| d \rangle$ represents the physical state in the total Hamiltonian $H$ with the normalization of $\sum_n | n \rangle \langle n| + \int d \alpha | \alpha \rangle \langle \alpha | =1$ and $\langle d| d \rangle =1$.     
 $Z$ is the probability of finding an elementary component in the physical state corresponding to the field renormalization constant. Here, $Z=0$ implies that the physical state is a  pure hadronic molecule, and $0<Z<1$ indicates an elementary component inside the physical state.  With the relationship $| d\rangle=[H-H_0]^{-1}V |d \rangle $, one can obtain 
 \begin{eqnarray}
1-Z= Re\left[\int d \alpha \frac{|\langle \alpha |V| d \rangle |^2}{(E_{\alpha}+B)^2}\right], 
\end{eqnarray}
where $d\alpha= \frac{4\pi p^2 dp}{(2\pi)^3}=\frac{\mu^{3/2}}{\sqrt{2}\pi^2}E^{1/2}dE$. The coupling between $| d \rangle$ and $|\alpha\rangle$ can be  characterised by an effective coupling constant $g$,  
 \begin{eqnarray}
g^2= \frac{2\pi \sqrt{2\mu B}}{\mu^2}(1-Z), 
\end{eqnarray}
which encodes the structure information of the composite system.  For a scattering process in the low energy approximation, the scattering length $a$ and effective range $r_0$ are related
\begin{eqnarray}
 a= \frac{2(1-Z)}{2-Z}\frac{1}{\sqrt{2\mu B}}, ~~~~~~~  r_0=-\frac{Z}{1-Z}\frac{1}{\sqrt{2\mu B}}.      
\end{eqnarray}
On the other hand, 
the applicability of the Weinberg compositeness rule can be extended by taking into account unstable hadrons, the CDD (Castillejo-Dalitz-Dyson) pole, and higher order corrections~\cite{Hyodo:2011qc,Hyodo:2013iga,Sekihara:2014kya,Guo:2015daa,Kamiya:2015aea,Kamiya:2016oao,Sekihara:2016xnq,Kinugawa:2022fzn,Albaladejo:2022sux}.

In general,   hadron-hadron scattering amplitudes can be characterized by the  scattering length $a$ and effective range $r_0$ using the effective range expansion(ERE) 
\begin{eqnarray}
 f(k)=\frac{1}{k cot \delta - ik}\approx \frac{1}{-\frac{1}{a}+\frac{1}{2}r_0 k^2 - i k},   \label{ERE Formula} 
\end{eqnarray}
where the typical momentum is defined as $k=\sqrt{2\mu E}$. Such a hadron-hadron scattering amplitude can also be described by the above $T$ matrix, e.g.,  $T=(1-VG)^{-1}V$.  Based on the couplings $g_{i}$ obtained above, one can further arrive at  the probability of finding a state in a specific channel~\cite{Aceti:2012dd} 
\begin{eqnarray}
 P_{i}=-g_{i}^2 \frac{\partial G_{ii}(\sqrt{s})}{\partial \sqrt{s}},     
\end{eqnarray}
and then obtain  $Z=1-\sum_{i} P_{i}$, which can be used to estimate the proportion of hadronic molecular components. In Ref.~\cite{Song:2022yvz}, the interaction range is considered to extend the applicability of the Weinberg compositeness rule originally obtained in the limit of small energy and a zero range interaction. Another extension of Weinberg's formula is proposed in Ref.~\cite{Li:2021cue}.      
 
In case of no poles (either bound, resonant, or virtual) around the mass threshold of $m_1$ and $m_2$,  one can study the corresponding scattering length, which is obtained from the above  scattering amplitude $T(\sqrt{s})$  
\begin{eqnarray}
a_{\sqrt{s}=m_1+m_2}&=&\frac{\mu}{2\pi}T_{non}(\sqrt{s}), \\ \nonumber
a_{\sqrt{s}=m_1+m_2}&=&\frac{1}{8\pi \sqrt{s}}T_{r}(\sqrt{s}),
\end{eqnarray}
where $T_{non}$ and $T_{r}$ represent the scattering amplitudes with nonrelativistic and relativistic propagators, respectively. 

One should note that the above discussions only focused on separable potentials, e.g., contact-range potentials. Considering the potentials as a function of the momentum, one can not convert the LS equation to a simple algebraic equation, which needs to be solved numerically~\cite{PavonValderrama:2019nbk,Du:2019pij}.

\section{The Gaussian-Expansion Method}\label{GEMcal}

Many methods have been proposed to solve few-body systems, e.g., the GEM~\cite{Kamimura:1988zz, Hiyama:2003cu}, the stochastic variational method (SVM)~\cite{Varga:1995dm,Varga:1996jr,Usukura:1998zz}, the hyperspherical harmonic variational method (HH)~\cite{Viviani:1994pm}, the Green function Monte Carlo (GFMC)~\cite{Bakker:1995pg},  the diffusion Monte Carlo method (DMC)~\cite{Gordillo:2020sgc,Gordillo:2021bra,Ma:2022vqf,Ma:2023int}, the no-core shell model (NCSM)~\cite{Navratil:1998mr,Navratil:1999pw,Navratil:2000ww}, the resonating group method (RGM)~\cite{Entem:2000mq}, the Fadeev equation~\cite{Garcilazo:2017ifi}, the Fixed center approximation (FCA)~\cite{Deloff:1999gc,Roca:2010tf}, Complex scaling (CS)~\cite{Moiseyev:1998gjp,Myo:2014ypa}, the QCD sum rule (QSR)~\cite{Shifman:1978bx,Colangelo:2000dp}, the Born-Oppenheimer (BO) approximation~\cite{Born:1927rpw}, etc.
Kamimura proposed the GEM for the non-adiabatic three-body calculations of muonic molecules and muon-atomic collisions in 1988~\cite{Kamimura:1988zz}. Due to its many advantages such as fast convergence and high precision~\cite{Hiyama:2003cu, Hiyama:2018ivm}, the GEM has been widely applied to investigate few-body systems in nuclear physics~\cite{Hiyama:2003cu} and hadron physics~\cite{Hiyama:2005cf,Yoshida:2015tia,Pan:2023zkl,Wu:2022ftm,Pan:2022xxz,Luo:2021ggs,Wu:2021dwy,Wu:2020job,Wu:2019vsy}. Combining the GEM with the Complex scaling method, which Aguilar, Combes, and Balslev proposed in 1971 originally~\cite{Aguilar:1971ve, Balslev:1971vb}, one can treat bound states, resonant states, and continuum states on an equal footing. For comprehensive benchmark studies of these few-body methods, one can refer to Refs~\cite{Kamada:2001tv,Meng:2023jqk}. 

The Schr\"odinger equation of the three-body system reads
 \begin{equation}\label{schd}
   \hat{H}\Psi_{JM}^{total}=E\Psi_{JM}^{total} \, ,
 \end{equation}
with the corresponding Hamiltonian
\begin{equation}\label{hami}
  \hat{H}=\sum_{i=1}^{3}\frac{p_i^2}{2m_i}-T_{c.m.}+\sum_{1=i<j}^{3}V(r_{ij}) \, ,
\end{equation}
where $T_{c.m.}$ is the kinetic energy of the center of mass and $V(r_{ij})$ is the potential between particle $i$ and particle $j$. 
\begin{figure}[!h]
    \centering
    \includegraphics[width=15cm]{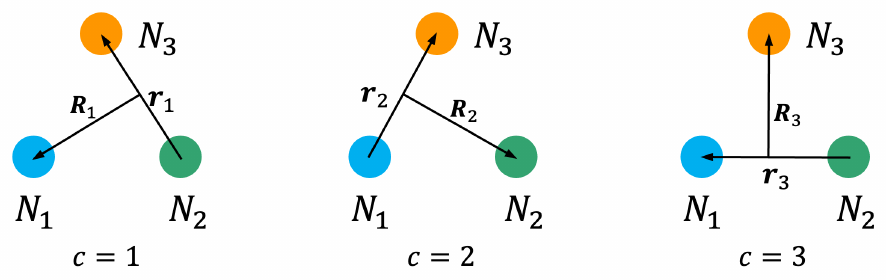}\\
    \caption{Three permutations of the Jacobi coordinates for a three-body system}\label{3-jacobi}
\end{figure}

The three Jacobi coordinates for a three-body system are shown in Fig.~\ref{3-jacobi}. The total wave function is a sum of the amplitudes of the three  rearrangements of the Jacobi coordinates, i.e., of the channels ($c = 1-3$) shown in Fig.~\ref{3-jacobi}
\begin{equation}\label{Sch}
    \Psi_{JM}^{total}=\sum_{c,\alpha}C_{c,\alpha}\, \Psi_{JM,\alpha}^{c}(\mathbf{r}_c,\mathbf{R}_c) \, ,
\end{equation}
where $\alpha=\{nl,NL,\Lambda,tT,s\Sigma\}$ and $C_{c,\alpha}$ are the expansion coefficients. Here $l$ and $L$ are the orbital angular momenta for the coordinates $r$ and $R$,  $s$ and $t$ are the spin and isospin of the two-body subsystem in each channel, $\Lambda$, $T$, and $\Sigma$ are the total orbital angular momentum, isospin, and spin. $n$ and $N$ are the numbers of Gaussian basis functions corresponding to coordinates $r$ and $R$, respectively.

The wave function of  each channel has the following form
  \begin{equation}\label{dd}
    \Psi_{JM,\alpha}^{c}(\mathbf{r}_c,\mathbf{R}_c) =
    H_{t,T}^c\otimes[\chi_{s, \Sigma}^c \otimes\Phi_{lL,\Lambda}^c]_{JM} \, ,
  \end{equation}
where $H_{t,T}^c$ is the isospin wave function, $\chi_{s, \Sigma}^c$ the spin wave function, and $\Phi_{lL,\Lambda}^c$ the orbital wave function. The total isospin wave function reads 
\begin{equation}\label{isospinwave}
    \begin{split}
          H_{t_1,T}^{c=1}& =[[\eta(N_2)\eta(N_3)]_{t_1}\eta(N_1)]_{T} \, , \\
          H_{t_2,T}^{c=2}& =[[\eta(N_1)\eta(N_3)]_{t_2}\eta(N_2)]_{T} \, , \\
          H_{t_3,T}^{c=3}& =[[\eta(N_2)\eta(N_1)]_{t_3}\eta(N_3)]_{T} \, ,
    \end{split}
\end{equation}
where $\eta$ is the isospin wave function of each particle. The spin wave function can be obtained in a way similar to  the isospin wave function.
The orbital wave function $\Phi_{lL,\Lambda}^c$ is given in terms of the Gaussian basis functions
\begin{equation}\label{nj}
    \Phi_{lL,\Lambda}^c(\mathbf{r}_c,\mathbf{R}_c)=[\phi_{n_cl_c}^{G}(\mathbf{r}_c)\psi_{N_cL_c}^{G}(\mathbf{R}_c)]_{\Lambda} \, ,
\end{equation}
with
\begin{equation}\label{nj}
\begin{split}
    &\phi_{nlm}^{G}(\mathbf{r}_c)=N_{nl}r_c^le^{-\nu_n r_c^2} Y_{lm}({\hat{r}}_c) \, ,\\
    &\psi_{NLM}^{G}(\mathbf{R}_c)=N_{NL}R_c^Le^{-\lambda_n R_c^2} Y_{LM}({\hat{R}}_c) \, .
    \end{split}
\end{equation}
Here $N_{nl}(N_{NL})$ are the normalization constants of the Gaussian basis, and the range parameters $\nu_n$ and $\lambda_n$ are given by
\begin{equation}\label{vn}
    \begin{split}
       \nu_n &=1/r_n^2,~~~~~~~ r_n=r_{min}a^{n-1}~~~~~~~~~~~ (n=1 - n_{max}) \, , \\
       \lambda_N &=1/R_N^2,~~~~ R_N=R_{min}A^{N-1}~~~~~~~ (N=1 - N_{max}) \, ,
    \end{split}
\end{equation}
in which $\{n_{max},r_{min},a$ or $r_{max}\}$ and  $\{N_{max},R_{min},A$ or $R_{max}\}$ are Gaussian basis parameters.

With the basis expansion, the Schr\"odinger equation of this system is transformed into a generalized matrix eigenvalue problem:
\begin{equation}\label{eigenvalue problem}
  [T_{\alpha \alpha'}^{ab}+V_{{\alpha \alpha'}}^{ab}-EN_{\alpha \alpha'}^{ab}]\, C_{b,\alpha'} = 0 \, .
\end{equation}
Here, $T_{\alpha \alpha'}^{ab}$ is the kinetic matrix element, $V_{\alpha \alpha'}^{ab}$ is the potential matrix element, and $N_{\alpha \alpha'}^{ab}$ is the normalization matrix element. $a$ and $b$ are Jacobi channels ranging from 1 to 3. The eigenenergy $E$ and coefficients $C_{b,\alpha'}$ are determined by the Rayleigh-Ritz variational principle.

\bibliography{reference}

\end{document}